\documentclass[natbib,smallcondensed]{svjour3}       
\usepackage{graphicx}
\usepackage[colorlinks=true,citecolor=blue,urlcolor=blue,breaklinks]{hyperref}
\usepackage{amsmath}
\usepackage{amssymb}
\usepackage{bm}
\usepackage{multirow}
\usepackage{booktabs}
\usepackage{xspace}
\usepackage{xcolor}        
\usepackage{url}           
\newcommand{\Horava}{Ho{\v{r}}ava\xspace}
\newcommand{\Lemaitre}{Lema\^\i{}tre\xspace}
\newcommand{\Painleve}{Painlev{\'e}\xspace}
\newcommand{\mathbold}[1]{\mbox{\boldmath $#1$}}
\DeclareMathOperator{\tr}{tr}
\DeclareMathOperator{\arctanh}{arctanh}
\DeclareMathOperator{\diag}{diag}
\newcommand{\g}{\kappa} 
\newcommand{\half}{\frac{1}{2}} 

\newcommand{\Vext}{V_\mathrm{ext}}
\newcommand{\csound}{c_\mathrm{s}}
\renewcommand{\d}{\mathrm{d}}

\newcommand{\im}{\mathrm{i}}
\newcommand{\A}{\mathbf{A}}
\newcommand{\F}{\mathbf{F}}
\newcommand{\f}{\mathbf{f}}
\newcommand{\n}{\mathbf{n}}
\newcommand{\K}{\mathbf{K}}
\renewcommand{\k}{\mathbf{k}} 
\newcommand{\V}{\mathbf{V}}
\newcommand{\vbf}{\mathbf{v}}
\newcommand{\x}{\mathbf{x}}
\newcommand{\bnabla}{\mathbold{\nabla}}
\renewcommand{\L}{\mathcal{L}}
\renewcommand{\O}{\mathcal{O}}

\definecolor{emerald}{rgb}{0.31, 0.78, 0.47}
\definecolor{purple}{rgb}{1,0,1}
\definecolor{lime}{HTML}{A6CE39} 

\newcommand{\blue}[1]{{\color{blue} #1}}

\newcommand{\cbs}[1]{{\color{emerald}{#1}}}

\newcommand{\orcidicon}{\includegraphics[width=0.32cm]{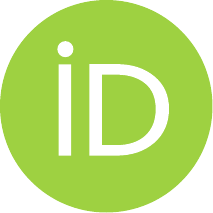}}
\newcommand\orcidCarlos{{\href{https://orcid.org/0000-0002-2134-377X}{\orcidicon}}}
\newcommand\orcidStefano{{\href{https://orcid.org/0000-0002-7632-7443}{\orcidicon}}}
\newcommand\orcidMatt{{\href{https://orcid.org/0000-0003-1088-6485}{\orcidicon}}}

\journalname{Living Reviews in Relativity}

\begin{document}

\title{Analogue gravity%
\thanks{Major revision, updated and expanded. \\
This article is a revised version of \url{https://doi.org/10.12942/lrr-2011-3}.}}

\author{
Carlos Barcel{\'o}{\,\orcidCarlos} \and
Stefano Liberati{\,\orcidStefano} \and
Matt~Visser{\,\orcidMatt}
}
\institute{Carlos Barcel{\'o} \at 
        Instituto de Astrof{\'\i}sica de Andaluc{\'\i}a (IAA-CSIC)\\
        Glorieta de la Astronom\'ia,\\ 
        18008 Granada, Spain\\
        \email{carlos@iaa.es}
\and
       Stefano Liberati \at
       SISSA -- International School for Advanced Studies\\
       Via Bonomea 265, 34136 Trieste, Italy,\\
       IFPU -- Institute for Fundamental Physics of the Universe\\ 
       Via Beirut 2, 34014 Trieste, Italy, and\\
       INFN -- Istituto Nazionale di Fisica Nucleare\\ 
       Sezione di Trieste, Trieste, Italy\\
       \email{liberati@sissa.it}
\and
       Matt Visser \at
       School of Mathematics and Statistics\\
       Victoria University of Wellington; PO Box 600\\
       Wellington 6140, New Zealand\\
       \email{matt.visser@sms.vuw.ac.nz}
       \vspace{-30pt}
}
\date{\vspace{-10pt}\blue{\bf Version of Friday 29 November 2024; \LaTeX-ed \today}} 

\maketitle

\begin{abstract}
Analogue gravity is a research programme that explores analogues of general relativistic gravitational fields within other physical systems, particularly but not exclusively in condensed matter systems, with the aim of gaining new insights into related problems. Analogue models of gravity boast a long and distinguished history, dating back to the early years of general relativity. This review article delves into the history, aims, results, and future prospects of various analogue models. We begin by presenting a particularly simple example of an analogue model, then traverse the rich history and complex array of models discussed in the literature. The last decade has witnessed significant and sustained advances in analogue gravity, resulting in hundreds of published articles, workshops, and books. The future of the analogue gravity programme looks promising, with rapid technological advances on the experimental front and the potential for analogue models to inspire innovative approaches to the problem of quantum gravity on the theoretical front. Most of all, these recent years have seen the rise of an unprecedented collaboration and interplay between different communities that we believe will set a new standard for interdisciplinary research in the years to come.
\end{abstract}
\keywords{Analogue gravity \and Analogue spacetimes \and Gravity}

\clearpage
\setcounter{tocdepth}{3}
\tableofcontents



\newpage

\listoffigures 

\newpage
\definecolor{emerald}{rgb}{0.31, 0.78, 0.47}
\newcommand{\ggm}[1]{{\color{emerald}{#1}}}

\section{Introduction}
\label{S:introduction}

\begin{quote}
And I cherish more than anything else the Analogies, my most trustworthy
masters. They know all the secrets of Nature, and they ought to be
least neglected in Geometry.
\null\hfill -- Johannes Kepler
\end{quote}

\begin{figure}[hptb]
    \centerline{\includegraphics[width=0.9\textwidth]{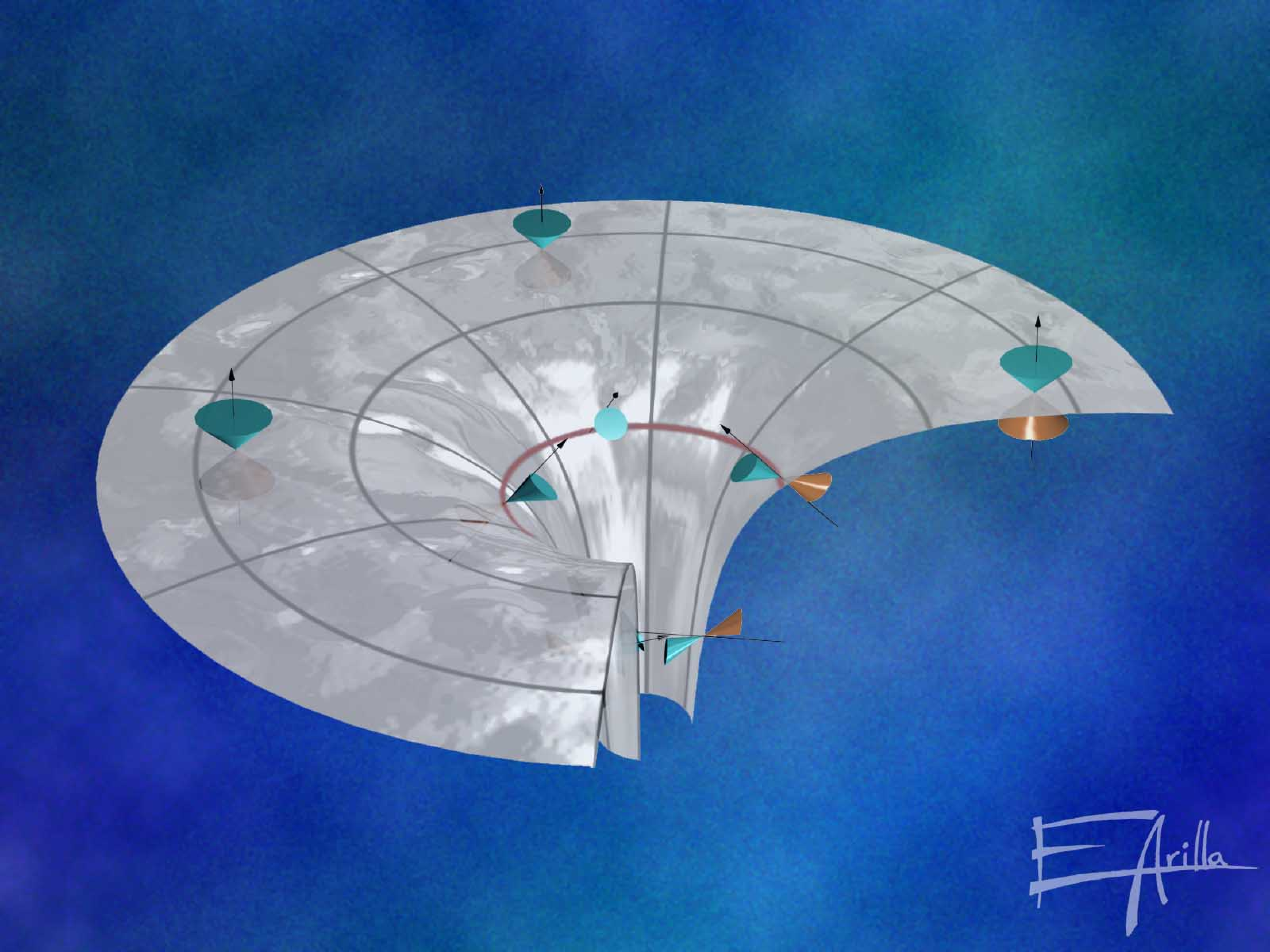}}
\caption[Cascading sound cones]{Artistic impression of cascading
    sound cones (in the geometrical acoustics limit) forming an
    acoustic black hole when supersonic flow tips the sound cones
    past the vertical.}
\label{F:cascade}
\end{figure}

\begin{figure}[hptb]
    \centerline{\includegraphics[width=0.9\textwidth]{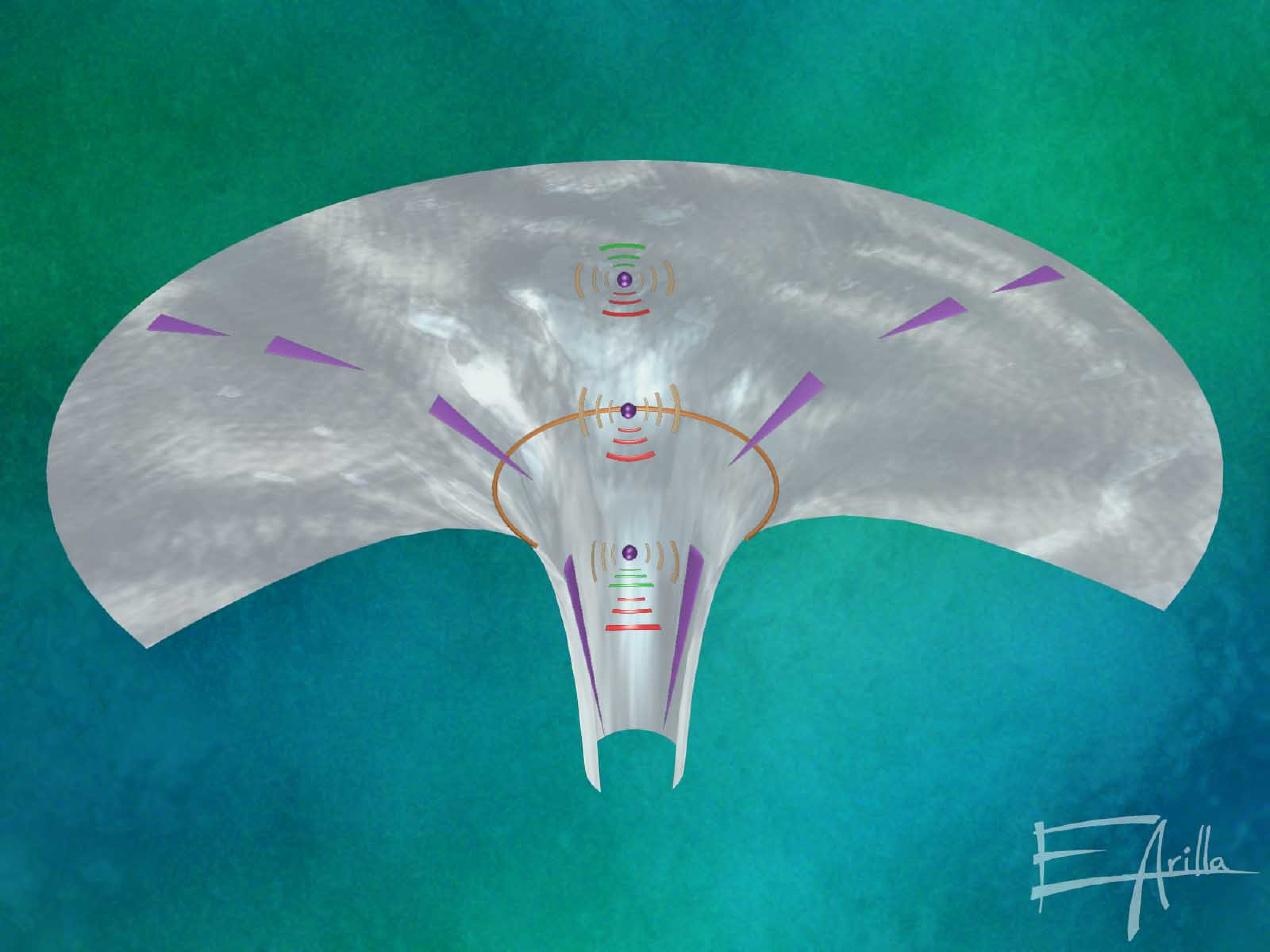}}
\caption[Trapped waves]{Artistic impression of trapped waves (in
    the physical acoustics limit) forming an acoustic black hole when
    supersonic flow forces the waves to move downstream.}
\label{F:ondas}
\end{figure}

\enlargethispage{25pt}
Analogies have played a very important role in physics and mathematics
-- they provide new ways of looking at problems that permit
cross-fertilization of ideas among different branches of science. A
carefully chosen analogy can be extremely useful in focusing
attention on a specific problem, and in suggesting unexpected routes
to a possible resolution. 
In this review article we shall focus on a specific analogy popularly 
known as ``analogue gravity'', (though one might make a case for the 
alternative phrase ``analogue spacetime''), which has proved to be an 
extremely rich research theme.
This research line started with the identification of several sharp 
analogies (typically but not always based on condensed matter physics) 
to probe aspects of the physics of curved spacetime -- and in particular 
to probe aspects of curved space quantum field theory, and to obtain 
lessons of potential relevance on the road towards a theory of quantum gravity.
Of course, analogy is not identity, and it is clear that the analogue 
models considered in the literature are not completely equivalent to 
general relativity -- we only need that the analogue gravity model 
(in order to be interesting from the gravitational perspective) should 
capture and reflect a sufficient number of relevant features of general relativity 
(or sometimes special relativity).\footnote{{Sigmund Freud once said ``Analogies, 
it is true, decide nothing, but they can make one feel more at home."}} 
Moreover, it is precisely the imperfection of those models what makes them 
really interesting --- in that they can provide a coherent framework 
for (counter-factual) alternatives to standard general relativity {and so allow us 
to test the robustness of some theoretical expectations w.r.t.~to new physics}.\footnote{Just don't take the analogy too far. We have (thankfully rarely) seen arguments to the effect that manipulating an analogue spacetime metric will lead to real physical changes in the gravitational field. No, this is simply not viable. }

The most well-known of these analogies is based on the realization that 
sound waves in a moving fluid provide an analogue system for light waves 
in a curved spacetime. A supersonic fluid flow can then generate a ``dumb hole'',
the acoustic analogue of a ``black hole'', and the analogy can be extended 
all the way to theoretically proving the necessity of phononic Hawking radiation 
from the acoustic horizon (at least in situations where a second quantization
of the sound wave field is meaningful).
This particular example provides (at least in principle) a concrete, 
and most paradigmatic, laboratory model for curved-space quantum field theory 
in a realm that was expected, even 25 years ago, to be technologically
accessible to experiment. (As we shall see below, this expectation has been, 
at present, more than borne out).

\enlargethispage{20pt}
After an initial period, essentially driven by theoretical analysis and speculation, experimentalists started to take on the challenge of testing these ideas around 2008. As of the writing of these lines, in 2024, there are a significant number of experimental groups which regularly perform experiments with analogue gravity motivation.
Numerous MSc and PhD thesis have already been defended within these groups. 
The initial motivation of analogue gravity has been enlarged and the main emphasis somewhat shifted (see e.g. the discussion in~\cite{pittphilsci20365}).

Currently, the interest in analogue gravity is not only to provide (indirect) experimental validation of gravitational effects difficult to observe in their native form,\footnote{A point that has sometimes been viewed as somewhat controversial, albeit see (for instance) the discussion presented in~\cite{Dardashti:2017kfw,Evans:2019uhg}.}
but analogue gravity has also been used:\\
---  i) to study and experimentally demonstrate the existence of {more general} phenomena {(such as, for instance, super-radiance and/or quasi-normal modes)} which like Hawking radiation appear to be more general than the process discovered in the gravitational realm; this can also be seen as analysis of the robustness of particular effects by showing how they survive the specifics of the different model systems; \\
---  ii) to open-ended {exploration of} different but specific model systems for their own sake, but also with the idea of generating cross-fertilization between their physics and that which might lie underneath gravitational systems in regimes beyond the general relativistic.     

From this perspective, the plethora of analogue models and systems, 
besides the already mentioned acoustic analogue,  are useful 
for these or many other reasons --- some of the analogue models are interesting for experimental reasons, others are useful for the way they provide new light on perplexing theoretical questions. The information flow is bi-directional, and sometimes insights developed within the context of general relativity can be used to understand aspects of the analogue system.

The list of analogue models is extensive, and in this review we will seek to do justice both to the key models, and to the key features of those models.
More importantly we will provide an introduction and an overview of the amazing body of developments performed in the 43 year period since the analogue gravity revolution started.

\subsection{Overview}

In the following sections we shall:
\begin{itemize}
\item Discuss the flowing fluid analogy in some detail.
\item Summarise the history and motivation for various analogue models.
\item Discuss the many physics issues various researchers have addressed.
\item Provide a representative catalogue of extant models.
\item Discuss the main physics results obtained to date, 
both on the theoretical and experimental sides.
\item Outline some of the many possible directions for future research.
\item Summarise the current state of affairs.
\end{itemize}
By that stage the interested reader will have had a quite thorough
introduction to the ideas, techniques, and hopes of the analogue gravity
programme.

\subsection{Motivations}
\enlargethispage{25pt}
The motivation for these investigations (both historical and current) is
rather mixed. In modern language, the reasons to investigate analogue
models are:

\begin{itemize}
\item Partly to develop an observational window on curved-space quantum fields.
\item Partly to use condensed matter to gain insight into classical general relativity.
\item Partly to use condensed matter to gain insight into curved-space quantum field theory.
\item Partly to use classical general relativity to gain insight into condensed matter physics.
\item Partly to gain insight into new and radically-different ways of dealing with ``quantum/emergent gravity''.
\item Partly to understand the nature and robustness of phenomena appearing in many systems including the gravitational realm.
\item Partly to perform open-ended explorations of new phenomenology in systems with analogue gravity behaviours.
\end{itemize}

\subsection{Going further}
\label{SS:going-further-1}

Apart from this present review article, and the references contained herein,
there are several key items that stand out as starting points for any deeper
investigation: 
\begin{itemize}

\item The book ``Artificial Black Holes'' edited by \citet{Novello:2002qg}.
\item The archival website for the ``Analogue models'' workshop:
\begin{itemize}
\item \url{http://www.sms.victoria.ac.nz/~visser/Analog/}
\end{itemize}
(This turn-of-the-century workshop made ``analogue spacetimes'' mainstream.)
\item The book ``The Universe in a Helium Droplet'' by \citet{Volovik:2003fe}. 
\item The \textit{Physics Reports} article ``Superfluid analogies of
      cosmological phenomena'', by \citet{Volovik:2000ua}. 
\item The review article by \citet{Balbinot:2006ua} 
      (focussing largely on back-reaction and short-distance effects). 
\item The \textit{Lecture Notes in Physics} volume on ``Quantum Analogies'' 
      edited by \citet{Unruh:LNP}. 
\item The book based on the 2011 Como summer school on analogue gravity~\citep{Faccio:2013kpa} 
      presenting an overview of the ideas and problems underlying research in analogue gravity together with a description of several viable experimental settings.
\item The book based on the Belem conference \citep{30-years} marking the 30-year anniversary of Unruh's 1981
      article \citep{Unruh:1981cg}.
\item The review by \citet{Kellay:2017} provides an overview of what can be done with soap bubbles 
      and soap films.
\item More recently the 2020 article on ``The next generation of analogue gravity experiments'' 
      by \citet{Jacquet:2020bar} focuses on experimental prospects.
\item A more philosophical survey can be found in the 2021 article 
     ``The latest frontier in analogue gravity: New roles for analogue experiments'' 
     by \citet{pittphilsci20365}.
\item The Nature Perspectives article \citep{Braunstein:2023jpo} gives an up to date (2023) overview.

\end{itemize}

\clearpage
\section{Acoustics: The Simplest Example of an Analogue Spacetime}
\label{S:simple}

Acoustics in a moving fluid is the simplest and cleanest example of an
analogue model \citep{Unruh:1981cg, Visser:1993ub, Visser:1998qn,
Visser:1997ux}. The basic physics is simple, the conceptual framework
is simple, and specific computations are often simple (whenever, that
is, they are not impossibly hard).\footnote{The need for a certain
degree of caution regarding the allegedly straightforward physics of
simple fluids might be inferred from the fact that the Clay Mathematics
Institute is currently offering a US\$~1,000,000 Millennium Prize for
significant progress on the question of existence and uniqueness of
solutions to the Navier--Stokes equation. 
See~\url{http://www.claymath.org/millennium/} for details.} 
 
\subsection{Background}
 
The basic physics is this: Low amplitude sound waves in fluids can usefully be viewed 
as linearized fluctuations on a suitable fluid background. 
A moving fluid will drag sound waves along with it, 
and if the speed of the fluid ever becomes supersonic, then in
the supersonic region sound waves will never be able to fight their way back
upstream \citep{Unruh:1981cg, Visser:1993ub, Visser:1998qn, Visser:1997ux}.
This implies the existence of a ``dumb hole'', a region from which
sound can not escape.\footnote{In correct English, the word ``dumb''
means ``mute'', as in ``unable to speak''. The word ``dumb'' does
\emph{not} mean ``stupid'', though even many native English speakers
tend to attribute to the term this derogatory meaning.} Of course this sounds very similar, at the level of a
non-mathematical \emph{verbal} analogy, to the notion of a ``black
hole'' in general relativity. The real question is whether this verbal
analogy can be turned into a precise \emph{mathematical} and
\emph{physical} statement -- it is only after we have a precise
mathematical and physical connection between (in this example) the
physics of acoustics in a fluid flow and at least some significant
features of general relativity that we can claim to have an ``analogue
model of (some aspects of) gravity''. 

\begin{figure}[htpb]
    \centerline{\includegraphics[width=0.6\textwidth]{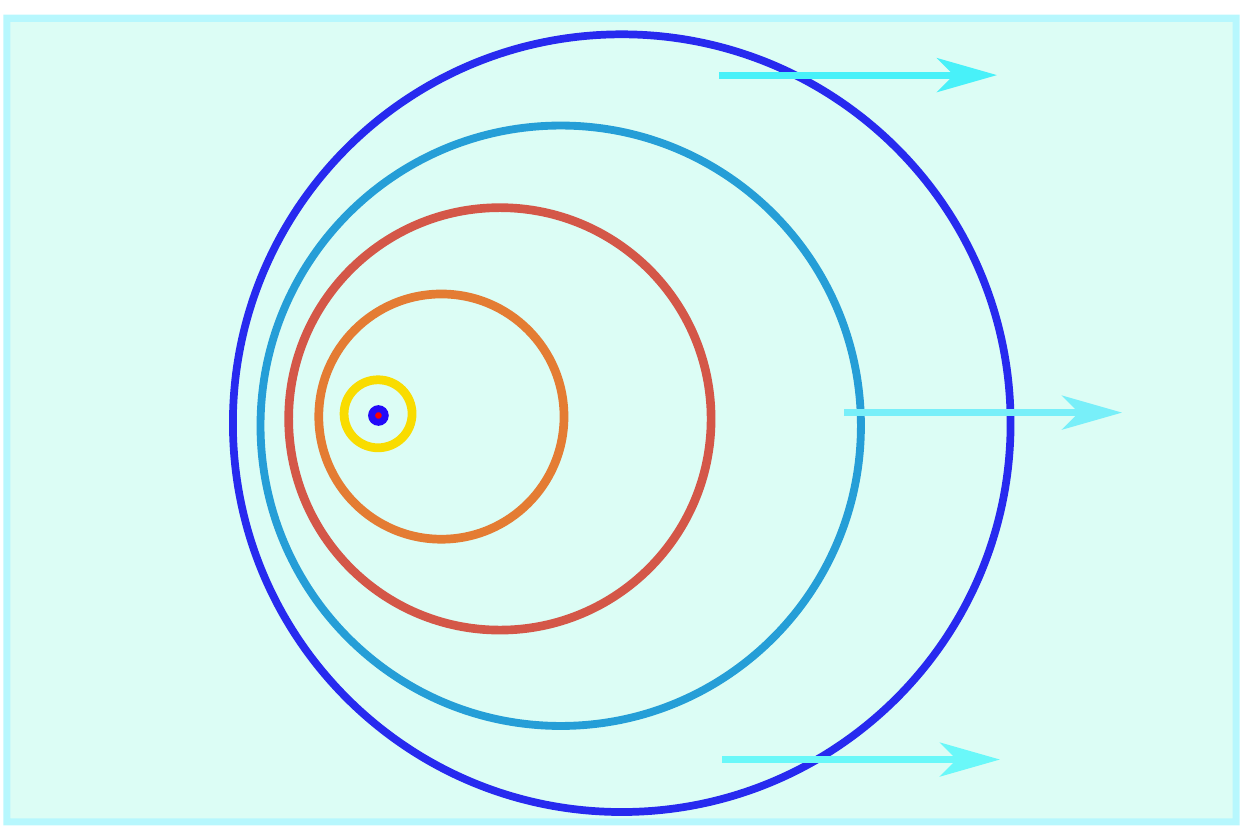}}
\caption[Moving fluid]{A moving fluid will drag sound pulses along with it.}
\label{F:moving0}
\end{figure}

\enlargethispage{20pt}
Now the features of general relativity/gravity  that one typically captures in an
``analogue model'' are the \emph{kinematic} features that have to do
with how fields (classical or quantum) are defined on curved spacetime,
and the \emph{sine qua non} of any analogue model is the existence of
some ``effective metric'' that captures the notion of the curved
spacetimes that arise in general relativity. (At the very least, one
might wish to capture the notion of the Minkowski geometry of special
relativity.) Indeed, the verbal description above (and its
generalizations in other physical frameworks) \emph{can} be converted
into a precise mathematical and physical statement, which ultimately was
the reason that analogue models started to attract interest among physicists.
The analogy works at two levels:
\begin{itemize}
\item Geometrical acoustics.
\item Physical acoustics.
\end{itemize}
The advantage of geometrical acoustics is that the derivation of the
precise mathematical form of the analogy is so simple as to be almost
trivial, and that the derivation is extremely general. The disadvantage
is that in the geometrical acoustics limit one can deduce only the
causal structure of the spacetime, and does not obtain a unique
effective metric. The advantage of physical acoustics is that, while the
derivation of the analogy holds in a more restricted regime, the analogy
can do more for you in that it can now uniquely determine a specific effective
metric and accommodate a wave equation for the sound waves.

\subsection{Geometrical acoustics}
 
At the level of geometrical acoustics we need only assume that:
\begin{itemize}
\item The speed of sound $c_s$, relative to the fluid, is well
defined.
\item The velocity of the fluid $\vbf$, relative to the
laboratory, is well defined.
\end{itemize}
Then, relative to the laboratory, the velocity of a sound ray
propagating, with respect to the fluid, along the direction
defined by the unit vector $\n$, is
\begin{equation}
\frac{\d\x}{\d t}=c_s\,\n+\vbf.
\end{equation}
This defines a sound cone in spacetime given by the condition
$\n^2=1$, i.e.,
\begin{equation}
- c_s^2 \; \d t^2 + \left(\d\x - \vbf \,\d t \right)^2 = 0\,.
\end{equation}
That is
\begin{equation}
- [c_s^2-v^2] \;\d t^2 - 2 \vbf \cdot \d\x \;\d t 
+\d\x\cdot\d\x=0\,.
\end{equation}

\begin{figure}[htpb]
    \centerline{\includegraphics[width=0.9\textwidth]{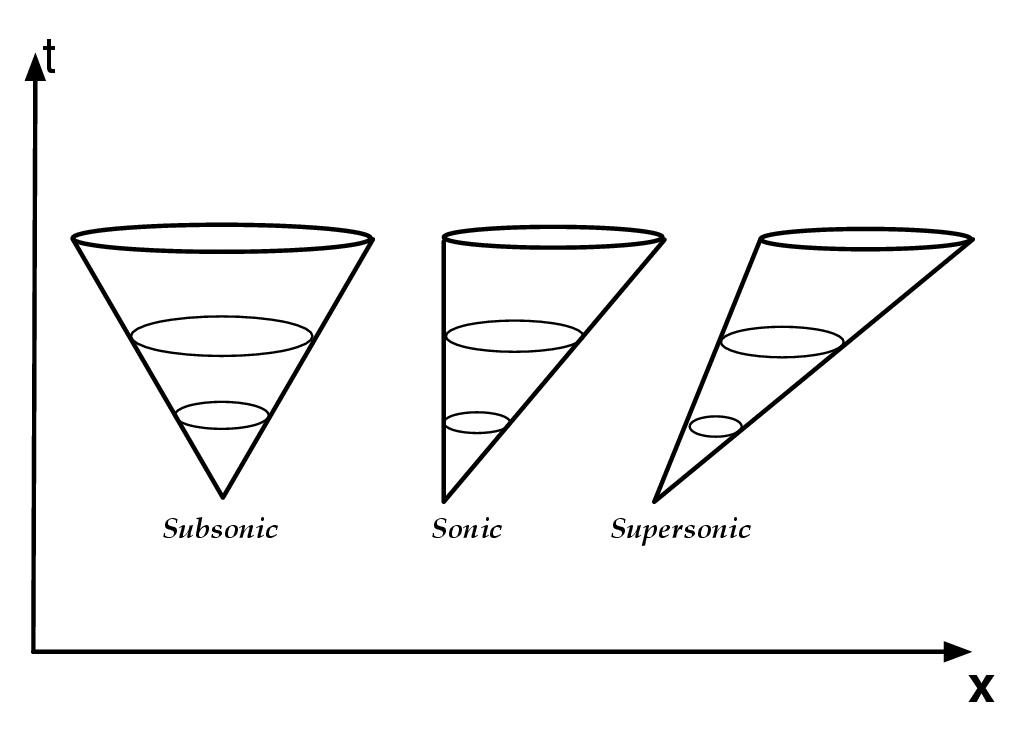}}
\caption[Sound cones]{A moving fluid will tip the ``sound cones''
as it moves. Supersonic flow will tip the sound cones past the
vertical.}
\label{F:cones}
\end{figure}

Solving this quadratic equation for $\d \x$ as a function of $\d t$
provides a double cone associated with each point in space and time. This
is associated with a conformal class of Lorentzian
metrics \citep{Unruh:1981cg, Visser:1993ub, Visser:1998qn, Visser:1997ux, Novello:2002qg}
\begin{equation}
{\mbox{\sl g}}  = \Omega^2\left[ \begin{array}{c|c}
-(c_s^2-v^2) & -{{\vbf}\,}^T\\
\hline -{\vbf} & \mathbf{I}
\end{array}\right]\,,
\label{g-matrix}
\end{equation}
where $\Omega$ is an unspecified but non-vanishing function.

The virtues of the geometric approach are its extreme simplicity and the
fact that the basic structure is dimension-independent. Moreover, this
logic rapidly (and relatively easily) generalises to more complicated
physical situations.\footnote{For instance, whenever one has a system of
PDEs that can be written in first-order quasi-linear symmetric hyperbolic
form, then it is an exact non-perturbative result that the matrix of
coefficients for the first-derivative terms can be used to construct a
conformal class of metrics that encodes the causal structure of the system
of PDEs. For barotropic hydrodynamics this is briefly discussed
in \citet{Piotr}. This analysis is related to the behaviour of
characteristics of the PDEs, and ultimately can be linked back to the
Fresnel equation that appears in the eikonal limit. 
In simple physical situations the causal structure resembles that of the acoustic metric, 
though in more general situations one might encounter bi-refringence or even multi-refringence. 
}

\subsection{Physical acoustics}

It is well known that for a static homogeneous inviscid fluid
the propagation of sound waves is governed by the simple wave
equation \citep{Lamb,Landau-Lifshitz,Milne-Thomson,Skudrzyk}
\begin{equation}
\partial_t^2 \phi = c_s^2 \; \nabla^2 \phi.
\end{equation}
Generalizing this result to a fluid that is non-homogeneous, or to a
fluid that is in motion, possibly even in non-steady motion, is more
subtle than it at first would appear. To derive a wave equation in this
more general situation we shall start by adopting a few simplifying
assumptions to allow us to derive the following theorem.

\begin{theorem}
If a fluid is barotropic and inviscid, and the flow is irrotational
(though possibly time dependent) then the equation of motion for the
velocity potential describing a linearized acoustic disturbance $\phi_1$ around some assumed background flow {$\phi_0$} is identical to
the d'Alembertian equation of motion for a minimally-coupled massless
scalar field propagating in a (3+1)-dimensional Lorentzian
geometry {$[g_0]_{ab}$ specified by the background flow}
\begin{equation}
\Delta \phi_1 \equiv \frac{1}{\sqrt{-g_0}} \partial_\mu 
\left( \sqrt{-g_0} \; [g_0]^{\mu\nu} \; \partial_\nu \phi_1 \right) = 0.
\end{equation}
Under these conditions, the propagation of sound is governed by an
\emph{acoustic metric} -- $[g_0]_{\mu\nu}(t,\x)$. This acoustic
metric describes a (3+1)-dimensional Lorentzian (pseudo--Riemannian)
geometry. The metric depends algebraically on the background flow: the density, velocity
of flow, and local speed of sound in the fluid. Specifically
\begin{equation}
[g_0]_{\mu\nu}(t,\x) 
\equiv 
{\rho_0\over c_s}
\begin{bmatrix}
   -(c_s^2-v_0^2)&\vdots&-{\vbf_0}^T\\
   \cdots\cdots\cdots\cdots&\cdot&\cdots\cdots\\
   -{\vbf_0}&\vdots& \mathbf{I}\\ 
\end{bmatrix}.
\end{equation}
(Here $\mathbf{I}$ is the $3\times3$ identity matrix.) In general, when
the fluid is non-homogeneous and flowing, the \emph{acoustic Riemann
tensor} associated with this Lorentzian metric will be nonzero.
\end{theorem}

\begin{remark}
It is quite remarkable that even though the underlying fluid dynamics is
Newtonian, nonrelativistic, and takes place in flat space-plus-time, the
fluctuations (sound waves) are governed by a curved (3+1)-dimensional
Lorentzian (pseudo-Riemannian) spacetime geometry. For practitioners of
general relativity this observation describes a very simple and concrete
physical model for certain classes of Lorentzian spacetimes, including
(as we shall later see) black holes. On the other hand, this discussion
is also potentially of interest to practitioners of continuum mechanics
and fluid dynamics in that it provides a simple concrete introduction to
Lorentzian differential geometric techniques.
\end{remark}

\begin{proof}
The fundamental equations of fluid
dynamics \citep{Lamb,Landau-Lifshitz,Milne-Thomson,Skudrzyk} are the
equation of continuity 
\begin{equation}
\partial_t \rho + \bnabla\cdot(\rho \; \vbf) = 0,
\label{E-continuity}
\end{equation}
and Euler's equation (equivalent to $\F= m\mathbf{a}$ 
applied to small lumps of fluid)
\begin{equation}
\rho \frac{\d \vbf}{\d t} \equiv 
\rho \left[ \partial_t \vbf + (\vbf \cdot \bnabla) \vbf \right] =
\f.
\label{E-euler}
\end{equation}
Start the analysis by assuming the fluid to be inviscid (zero
viscosity), with the only forces present being those due to
pressure.\footnote{It is straightforward to add external forces, at least
conservative body forces such as Newtonian gravity.} Then, for the force
density, we have
\begin{equation}
\f = - \bnabla p.
\label{E-force}
\end{equation}
Via standard manipulations the Euler equation can be rewritten as
\begin{equation}
\partial_t {\vbf} = 
\vbf \times ( \bnabla \times \vbf)  - \frac{1}{\rho} \;\bnabla p
- \bnabla\left( {\half} \; v^2\right).
\label{E-euler-2}
\end{equation}
Now take the flow to be \emph{vorticity free}, that is, \emph{locally
irrotational}. We can then introduce the
velocity potential $\phi$ such that $\vbf = -\bnabla \phi$, at
least locally. If one further takes
the fluid to be \emph{barotropic} (this
means that $\rho$ is a function of $p$ only), it becomes possible
to define
\begin{equation}
h(p) = \int_0^p \frac{\d p'}{\rho(p')}; 
\qquad \hbox{so that} \qquad
\bnabla h = \frac{1}{\rho} \; \bnabla p.
\end{equation}
Thus, the specific enthalpy, $h(p)$, is a function of $p$ only.
Euler's equation now reduces to
\begin{equation}
-\partial_t \phi + h + \half (\bnabla\phi)^2  = 0.
\label{E-bernoulli}
\end{equation}
This is a version of Bernoulli's equation.

Now linearize these equations of motion around some (possibly time-dependent and/or inhomogeneous) background
$(\rho_0,p_0,\phi_0)$.  By introducing a small expansion parameter $\epsilon\ll1$ we can write
\begin{eqnarray}
\rho &=& \rho_0 + \epsilon \rho_1 +\O(\epsilon^2),
\\
p &=& p_0 + \epsilon p_1 +\O(\epsilon^2),
\\
\phi &=&
\phi_0 + \epsilon \phi_1 + \O(\epsilon^2).   
\end{eqnarray}
Sound is \emph{defined} to
be these linearized fluctuations in the dynamical quantities.  Note that
this is the \emph{standard definition} of (linear) sound and more
generally of acoustical disturbances. In principle, of course, a fluid
mechanic might really be interested in solving the complete equations of
motion for the fluid variables $(\rho, p, \phi)$. In practice, it is
both traditional and extremely useful to separate the exact motion,
described by the exact variables, $(\rho, p, \phi)$, into some average
bulk motion, $(\rho_0,p_0,\phi_0)$, plus low amplitude acoustic
disturbances, $(\epsilon \rho_1,\epsilon p_1,\epsilon \phi_1)$. See, for
example, \citet{Lamb, Landau-Lifshitz, Milne-Thomson, Skudrzyk}. 

Since this is a subtle issue which has caused considerable confusion in the
past, let us be even more explicit by asking the rhetorical question:
\emph{``How can we tell the difference between a wind gust and a sound
wave?''} The answer is that the difference is to some extent a matter of
convention -- sufficiently low-frequency long-wavelength disturbances
(wind gusts) are conventionally lumped in with the average bulk motion.
Higher-frequency, shorter-wavelength disturbances are conventionally
described as acoustic disturbances. If you wish to be hyper-technical,
we can introduce a high-pass filter function to define the bulk motion
by suitably averaging the exact fluid motion. There are no deep physical
principles at stake here -- merely an issue of convention. The place
where we are making a specific physical assumption that restricts the
validity of our analysis is in the requirement that the amplitude of the
high-frequency short-wavelength disturbances be small. This is the
assumption underlying the linearization programme, and this is why
sufficiently high-amplitude sound waves must be treated by direct
solution of the full equations of fluid dynamics.

Linearizing the continuity equation results in the pair of equations
\begin{eqnarray}
&&
\partial_t \rho_0 + 
\bnabla\cdot(\rho_0 \; \vbf_0) = 0,
\\
&&
\partial_t \rho_1 + 
\bnabla\cdot(\rho_1 \; \vbf_0 + \rho_0 \; \vbf_1) = 0.
\end{eqnarray}
Now, the barotropic condition implies
\begin{equation}
h(p) = 
h\left(p_0 + \epsilon p_1 +  O(\epsilon^2)\right) = 
h_0 + \epsilon \; \frac{p_1}{\rho_0} + \O(\epsilon^2).
\end{equation}
Use this result in linearizing the Euler equation. We obtain the pair
\begin{eqnarray}
&&-\partial_t \phi_0 + h_0 + \half (\bnabla\phi_0)^2  = 0.
\\
&&-\partial_t \phi_1 + \frac{p_1}{\rho_0} - \vbf_0 \cdot \bnabla\phi_1 = 0.
\end{eqnarray}
This last equation may be rearranged to yield
\begin{equation}
p_1 =  \rho_0\; \left( \partial_t \phi_1 + \vbf_0 \cdot \bnabla\phi_1\right).
\label{E-linear-euler}
\end{equation}
Use the barotropic assumption to relate
\begin{equation}
\rho_1 = 
\frac{\partial \rho}{\partial p} \; p_1 = 
\frac{\partial \rho}{\partial p} \; \rho_0  \;
( \partial_t \phi_1 + \vbf_0 \cdot \bnabla\phi_1 ). 
\label{E-linear-barotropic}
\end{equation}
Now substitute this consequence of the linearized Euler equation into
the linearized equation of continuity. We finally obtain, up to an
overall sign, the wave equation:
\begin{equation}
{}-\partial_t  
     \left( \frac{\partial\rho}{\partial p}\rho_0 \; 
            ( \partial_t \phi_1 + \vbf_0 \cdot \bnabla\phi_1 ) 
     \right)
+ \bnabla \cdot 
     \left( \rho_0 \bnabla\phi_1 
            - \frac{\partial\rho}{\partial p} \rho_0 \vbf_0 \,
              ( \partial_t \phi_1 + \vbf_0 \cdot \bnabla\phi_1 )      
     \right)
=0.
\label{E-wave-physical}
\end{equation}
This wave equation describes the propagation of the linearized
scalar potential $\phi_1$. Once $\phi_1$ is determined,
Eq.~(\ref{E-linear-euler}) determines $p_1$, and
Eq.~(\ref{E-linear-barotropic}) then determines $\rho_1$.  Thus this 
wave equation completely determines the propagation of acoustic
disturbances.  The background fields $p_0$, $\rho_0$ and $\vbf_0
= - \bnabla \phi_0$, which appear as time-dependent and
position-dependent coefficients in this wave equation, are
constrained to solve the equations of fluid motion for a
barotropic, inviscid, and irrotational flow.
Apart from these constraints, they are otherwise permitted to have
\emph{arbitrary} temporal and spatial dependencies.

Now, written in this form, the physical import of this wave equation
is somewhat less than pellucid. To simplify things algebraically,
observe that the local speed of sound is defined by
\begin{equation}
{c^{-2}_s} \equiv \frac{\partial\rho}{\partial p}. 
\end{equation}
Now construct the symmetric $4\times4$ matrix
\begin{equation}
f^{\mu\nu}(t,\x) \equiv \frac{\rho_0}{c^2_s}
\begin{bmatrix}
   -1&\vdots&-v_0^j\\
   \cdots\cdots&\cdot&\cdots\cdots\cdots\cdots\\
   -v_0^i&\vdots&( c^2_s\;\delta^{ij} - v_0^i v_0^j )\\
\end{bmatrix}.
\label{E-explicit}             
\end{equation}
(Greek indices run from 0\,--\,3, while Roman indices run from
1\,--\,3.)  Then, introducing (3+1)-dimensional space-time
coordinates, which we write as $x^\mu \equiv (t; x^i)$ ,  
the above wave Eq.~(\ref{E-wave-physical}) is easily rewritten as 
\begin{equation}
\partial_\mu ( f^{\mu\nu} \; \partial_\nu \phi_1) = 0.
\end{equation}
This remarkably compact formulation is completely equivalent to
Eq.~(\ref{E-wave-physical}) and is a much more promising
stepping-stone for further manipulations. The remaining steps are
a straightforward application of the techniques of curved space
(3+1)-dimensional Lorentzian geometry.

Now in any Lorentzian (i.e., pseudo--Riemannian) manifold the curved
space scalar d'Alembertian is given in terms of the metric
$g_{\mu\nu}(t,\x)$ by 
\begin{equation}
\Delta \phi \equiv \frac{1}{\sqrt{-g}} 
\partial_\mu \left( \sqrt{-g} \; g^{\mu\nu} \; \partial_\nu \phi \right).
\end{equation}
(See, for example, \citealt{Fock, Moller, Synge, MTW, Hawking-Ellis, Wald}.) 
The inverse metric, $g^{\mu\nu}(t,\x)$, is pointwise the matrix
inverse of $g_{\mu\nu}(t,\x)$, while $g \equiv \det(g_{\mu\nu})$.
Thus one can rewrite the physically derived wave
Eq.~(\ref{E-wave-physical}) in terms of the d'Alembertian provided one 
identifies
\begin{equation}
\sqrt{-g} \; g^{\mu\nu} = f^{\mu\nu}.
\end{equation}
This implies, on the one hand, 
\begin{equation}
\det(f^{\mu\nu}) = (\sqrt{-g})^4 \; g^{-1} = g.
\end{equation} 
On the other hand, from the explicit expression (\ref{E-explicit}),
expanding the determinant in minors (in 3+1 dimensions) yields
\begin{equation}
\det(f^{\mu\nu}) 
= 
\left(\frac{\rho_0}{c^2_s}\right)^4 \cdot
\left[(-1) \cdot (c^2_s - v_0^2) - (-v_0)^2\right] \cdot
\left[c^2_s\right] \cdot
 \left[c^2_s\right]
=
- \frac{\rho_0^4}{c^2_s}.
\end{equation} 
Thus,
\begin{equation}
g = - \frac{\rho_0^4}{c^2_s}; \qquad \sqrt{-g} = \frac{\rho_0^2}{c_s}.
\end{equation}
Therefore, we can pick off the coefficients of the \emph{inverse}
(contravariant) acoustic metric
\begin{equation}
g^{\mu\nu}(t,\x) \equiv 
\frac{1}{\rho_0 c_s}
\begin{bmatrix}
   -1&\vdots&-v_0^j\\
   \cdots\cdots&\cdot&\cdots\cdots\cdots\cdots\\
   -v_0^i&\vdots&(c^2_s \; \delta^{ij} - v_0^i \; v_0^j )\\
\end{bmatrix}.         
\end{equation}
We could now determine the metric itself simply by inverting this
$4\times4$ matrix (and if the reader is not a general relativist,
proceeding in this direct manner is definitely the preferred
option). On the other hand, for general relativists it is even easier
to recognise that one has in front of one a specific example of the
Arnowitt--Deser--Misner split of a (3+1)-dimensional  Lorentzian
spacetime metric into space + time, more commonly used in discussing
initial value data in general relativity. (See, for example,
\citealt[pp.~505--508]{MTW}.) The (covariant) acoustic metric is then
read off by inspection
\begin{equation}
g_{\mu\nu} \equiv 
\frac{\rho_0}{c_s}
\begin{bmatrix}
   -(c^2_s-v_0^2)&\vdots&-v_0^j\\
   \cdots\cdots\cdots\cdots&\cdot&\cdots\cdots\\
   -v_0^i&\vdots&\delta_{ij}\\
\end{bmatrix}.         
\end{equation}
Equivalently, the acoustic interval (acoustic line-element) can be expressed as
\begin{equation}
\label{E:acoustic-line-element}
\d s^2 \equiv g_{\mu\nu} \; \d x^\mu \; \d x^\nu =
\frac{\rho_0}{c_s} 
\left[
- c^2_s \; \d t^2 + (\d x^i - v_0^i \; \d t) \; \delta_{ij} \; (\d x^j - v_0^j \; \d t )
\right].
\end{equation}
This completes the proof of the theorem. 
\end{proof}

We have presented the theorem and
proof, which closely follows the discussion in \citet{Visser:1997ux}, in
considerable detail because it is a standard template that can be readily
generalised in many ways.  This discussion can then be used as a starting
point to initiate the analysis of numerous and diverse physical models.

\subsection{General features of the acoustic metric}
\label{S:general-features}

A few brief comments should be made before proceeding further:
\begin{itemize}
\item
  Observe that the signature of this effective metric is indeed $(-,+,+,+)$, as
  it should be to be regarded as Lorentzian.
\item
  Observe that in physical acoustics it is the inverse metric density,
  \begin{equation}
    f^{\mu\nu} = \sqrt{-g} \; g^{\mu\nu}
  \end{equation}
  that is of more fundamental significance for deriving the wave equation
  than is the metric $g_{\mu\nu}$ itself.  (This observation continues to hold
  in more general situations where it is often significantly easier to calculate
  the tensor density $f^{\mu\nu}$ than it is to calculate the effective metric
  $g_{\mu\nu}$.) 
\item
  It should be emphasised that there are two distinct metrics relevant
  to the current discussion:
  \begin{itemize}
  \item
    The \emph{physical spacetime metric} is, in this case, just the usual flat
    metric of Minkowski space:
    \begin{equation}
      \eta_{\mu\nu} \equiv (\diag[-c_\mathrm{light}^2,1,1,1])_{\mu\nu}. 
    \end{equation}
    (Here $c_\mathrm{light}$ is the speed of light in vacuum.) The fluid particles
    couple only to the physical metric $\eta_{\mu\nu}$. In fact the fluid
    motion is completely non-relativistic, so that $||v_0||   \ll
    c_\mathrm{light}$, and it is quite sufficient to consider Galilean relativity
    for the underlying fluid mechanics. 
  \item
    Sound waves on the other hand, do not ``see'' the physical metric
    at all. Acoustic perturbations couple only to the effective \emph{acoustic
      metric} $g_{\mu\nu}$.
  \end{itemize}
  
\item
  It is quite remarkable that (to the best of our knowledge) a version
  of this acoustic metric was first derived and used in Moncrief's
  studies of the relativistic hydrodynamics of accretion flows
  surrounding black holes \citep{Moncrief}. Indeed, Moncrief was working
  in the more general case of a curved background ``physical'' metric,
  in addition to a curved ``effective'' metric. We shall come back to
  this work later on, in our historical section. (See also
  Sect.~\ref{S:relativistic-acoustic}.)
  
\item However, the geometry determined by the acoustic metric does
  inherit some key properties from the existence of the underlying flat
  physical metric. For instance, the \emph{topology} of the manifold does
  not depend on the particular metric considered.  The acoustic geometry
  inherits the underlying topology of the physical metric -- ordinary
  $\Re^4$ -- with possibly a few regions excised (due to whatever
  hard-wall boundary conditions one might wish to impose on the fluid). In
  systems constrained to have effectively less than 3 space-like dimensions
  one can reproduce more complicated topologies (consider, for example, an
  effectively one-dimensional flow in a tubular ring). 
  
\item
  Furthermore, the acoustic geometry automatically inherits from the underlying
  Newtonian time parameter, the important  
  property of ``stable causality'' \citep{Hawking-Ellis,Wald}. Note that
  \begin{equation}
    g^{\mu\nu} \, (\nabla_\mu t) \, (\nabla_\nu t) = -\frac{1}{\rho_0 \; c_s} < 0.
  \end{equation}
  The fact that $\nabla t$, the gradient of laboratory time, is always timelike as viewed by the analogue metric automatically would seem to preclude some of the more entertaining causality-related
  pathologies that sometimes arise in general relativity. 
  (For a general discussion of causal pathologies in general relativity, see, for
  example, \citealt{Hawking-Ellis, Hawking:1991nk, Hawking:1991pk, Cassidy:1997kv, Hawking:2002yr, Visser:2002ua})

    It is however interesting to note that there are situations in which the above conclusion has to be somewhat refined
    --- if the ``analogue time" is physically radically different from laboratory time. Indeed, it was shown in \cite{Barcelo:2022jbr} that certain configurations with closed time curves (CTCs) can be reproduced in an analogue system.
    
    However, such spacetimes contain what one might call ``trivial CTCs", as they are associated to the signature perceived by observers internal to the analogue system being different from the laboratory signature, and characterized by the fact that the one dimension with odd sign with respect to the others is a periodic angular coordinate. Physically relevant situations correspond instead to analogue spacetimes with chronological horizons separating regions with CTCs from regions without CTCs. It is this category of spacetimes that was found to lead to pathologies when attempting to simulate them~\citep{Barcelo:2022jbr}.
    
    It is worth noticing that the obstructions so found resonate with Hawking’s idea of a Chronology Protection mechanism. Analogue gravity in this case seems to suggest that such a mechanism might be rooted in the emergent nature of spacetime, and in  the causal structure of its underlying microphysics. We shall see, later on, that this might be just one of the many lessons derivable from analogue gravity when used as a toy model for Emergent gravity scenarios.\\
  
\item
  Other concepts that translate immediately are those of ``ergo-region'',
  ``trapped surface'', ``apparent horizon'', and ``event horizon''.
  These notions will be developed more fully in the following subsection.
\item
  The properly normalised four-velocity of the fluid is
  \begin{equation}
    V^\mu = \frac{(1; \; v^i_0)}{\sqrt{\rho_0 \; c_{s}}},
  \end{equation}
  so that
  \begin{equation}
    g_{\mu\nu}\, V^\mu\, V^\nu = g(V,V) = -1.
  \end{equation}
  This four-velocity is related to the gradient of the natural time parameter by
  \begin{equation}
    \nabla_\mu t = (1,0,0,0); \qquad \qquad 
    \nabla^\mu t =  -\frac{(1; \; v^i_0)}{\rho_0 \; c_{s}} 
    = -\frac{V^\mu}{\sqrt{\rho_0 \; c_{s}}}.
  \end{equation}
  Thus the integral curves of the fluid velocity field are orthogonal
  (in the Lorentzian metric) to the constant time surfaces. The
  acoustic proper time along the fluid flow lines (streamlines) is
  \begin{equation}
    \tau = \int \sqrt{\rho_0 \; c_{s}} \; \d t,
  \end{equation}
  and the integral curves are geodesics of the acoustic metric if
  and only if $\rho_0\,c_{s}$ is position independent.
  
\item Note that $V_a = -\sqrt{\rho_0 c_{s}} \; \nabla_a t$ which in the language of differential forms becomes $\V^\flat = \alpha \; \d t $ implying $(\d \V^\flat) \wedge \V^\flat =0$; so in the current framework, where the 3-velocity $\vbf=\bnabla\phi$ is taken to be irrotational, the 4-velocity is automatically  hyper-surface orthogonal.

\item Note that the 4-acceleration of the fluid is explicitly given by
\begin{equation}
A_a = V^b \nabla_b V_a = V^b (\nabla_b V_a - \nabla_a V_b) +  V^b \nabla_a V_b  = V^b (\nabla_b V_a - \nabla_a V_b) .
\end{equation}
But 
\begin{equation}
(\nabla_b V_a - \nabla_a V_b) = {1\over2} (\nabla_b \ln[\rho_0 c_{s}] \; V_a - \nabla_a \ln[\rho_0 c_{s}] \; V_b)
\end{equation}
thereby implying
\begin{equation}
A_a = {1\over2} (\delta_a{}^b + V_a V^b)\;  \nabla_b \ln[\rho_0 c_{s}],
\end{equation}
or alternatively
\begin{equation}
A^a = {1\over2} (g^{ab} + V^a V^b) \; \nabla_b \ln[\rho_0 c_{s}].
\end{equation}
This verifies that, (as required), the 4-acceleration of the fluid is indeed \break \mbox{4-orthogonal} to the 4-velocity.

\item
  Observe that in a completely general (3+1)-dimensional Lorentzian
  geometry the metric has 6 degrees of freedom per point in spacetime 
  ($4\times4$ symmetric matrix $\Longrightarrow$ 10 independent
  components; then subtract 4 coordinate conditions). 

  In contrast,
  the acoustic metric is more constrained. Being specified completely by
  the three scalars $\phi_0(t, \x)$, $\rho_0(t, \x)$, and $c_{s}(t,
  \x)$, the acoustic metric has, at most, 3 degrees of freedom per
  point in spacetime. The equation of continuity actually reduces this to
  2 degrees of freedom, which can be taken to be $\phi_0(t, \x)$
  and $c_{s}(t, \x)$. 
  
  \emph{Thus, the simple acoustic metric of this section can, at best, reproduce
    some subset of the generic metrics of interest in general relativity.} 
\item
  A point of notation: Where the general relativist uses the word
  ``stationary'' the fluid dynamicist uses the phrase ``steady flow''.
  The general-relativistic word ``static'' translates to a rather messy
  constraint on the fluid flow (to be discussed more fully below).\footnote{In GR ``stationary'' merely requires a timelike Killing vector, whereas ``static'' requires a timelike hypersurface-orthogonal Killing vector.}
\item
  Finally, we should emphasise that in Einstein gravity the spacetime metric is
  related to the distribution of matter by the nonlinear
  Einstein--Hilbert differential equations. In contrast, in the present
  context, the acoustic metric is related to the distribution of matter
  in a simple algebraic fashion.
\end{itemize}

\subsubsection{Horizons and ergo-regions}

In the next two subsections we shall undertake to more fully explain
some of the technical details underlying the acoustic
analogy. Concepts and quantities such as horizons, ergo-regions and
``surface gravity'' are  important features of standard general
relativity, and analogies are useful only insofar as they adequately
preserve these notions. 

Let us start with the notion of an ergo-region: Consider integral
curves of the vector 
\begin{equation}
K^\mu \equiv (\partial/\partial t)^\mu = (1,0,0,0)^\mu. 
\end{equation}
If the flow is steady, then this is the time
translation Killing vector.  Even if the flow is not steady the
background Minkowski metric provides us with a natural definition of
``at rest''.  Then\footnote{Henceforth, in the interests of
notational simplicity, we shall drop the explicit subscript 0 on
background field quantities unless there is specific risk of confusion.}
\begin{equation}
g_{\mu\nu} \; (\partial/\partial t)^\mu \;(\partial/\partial t)^\nu = 
g_{tt} = 
-[c_{s}^2 - v^2]. 
\end{equation}
This quantity changes sign when $ ||\vbf|| > c_{s} $.  Thus, any region of
supersonic flow is an ergo-region.  (And the boundary of the
ergo-region may be deemed to be the ergo-surface.) The analogue of this
behaviour in general relativity is the ergosphere surrounding any
spinning black hole -- it is a region where space ``moves'' with
superluminal velocity relative to the fixed
stars \citep{MTW,Hawking-Ellis,Wald}.

A trapped surface in acoustics is defined as follows: Take any closed
two-surface. If the fluid velocity is everywhere inward-pointing and the
normal component of the fluid velocity is everywhere greater than the
local speed of sound, then no matter what direction a sound wave
propagates, it will be swept inward by the fluid flow and be trapped
inside the surface. The surface is then said to be outer-trapped. (For
comparison with the usual situation in general relativity see
\citealt[pp.\ 319--323]{Hawking-Ellis}, or \citealt[pp.\ 310--311]{Wald}.)
Inner-trapped surfaces (anti-trapped surfaces) can be defined by
demanding that the fluid flow is everywhere outward-pointing with
supersonic normal component. 

It is only because of the fact that the
background Minkowski metric provides a natural definition of ``at rest''
that we can adopt such a simple and straightforward definition. In
ordinary general relativity we need to develop considerable additional
technical machinery, such as the notion of the ``expansion'' of bundles
of ingoing and outgoing null geodesics, before defining trapped
surfaces. That the above definition for acoustic geometries is a
specialization of the usual one can be seen from the discussion
in \citet[pp.\ 262--263]{Hawking-Ellis}. The acoustic trapped region
is now defined as the region containing outer trapped surfaces, and
the acoustic (future) apparent horizon as the boundary of the trapped
region. That is, the acoustic apparent horizon is the two-surface for
which the normal component of the fluid velocity is everywhere equal
to the local speed of sound. 
We can also define anti-trapped regions and past apparent horizons though these notions might seem
of limited utility in general relativity\footnote{This discussion naturally leads us to what is perhaps one of the central questions of analogue gravity -- just how much of the standard ``laws of black hole mechanics'' \citep{Bardeen:1973gs, Wald:1999vt} carry over into these analogue models? } 

The fact that the apparent horizon seems to be fixed in a
foliation-independent manner is only an illusion due to the way in which
the analogies work. A particular fluid flow reproduces a specific
``metric'', (a matrix of coefficients in a specific coordinate system,
not a ``geometry''), and, in particular, a specific foliation of
spacetime. (Only ``internal'' observers see a ``geometry''; see the
discussion in Sect.~\ref{S:internal}). The same ``geometry'' written
in different coordinates would give rise to a different fluid flow (if
at all possible, as not all coordinate representations of a fixed
geometry give rise to acoustic metrics) and, therefore, to a
different apparent horizon.

\begin{figure}[htpb]
    \centerline{\includegraphics[width=0.6\textwidth]{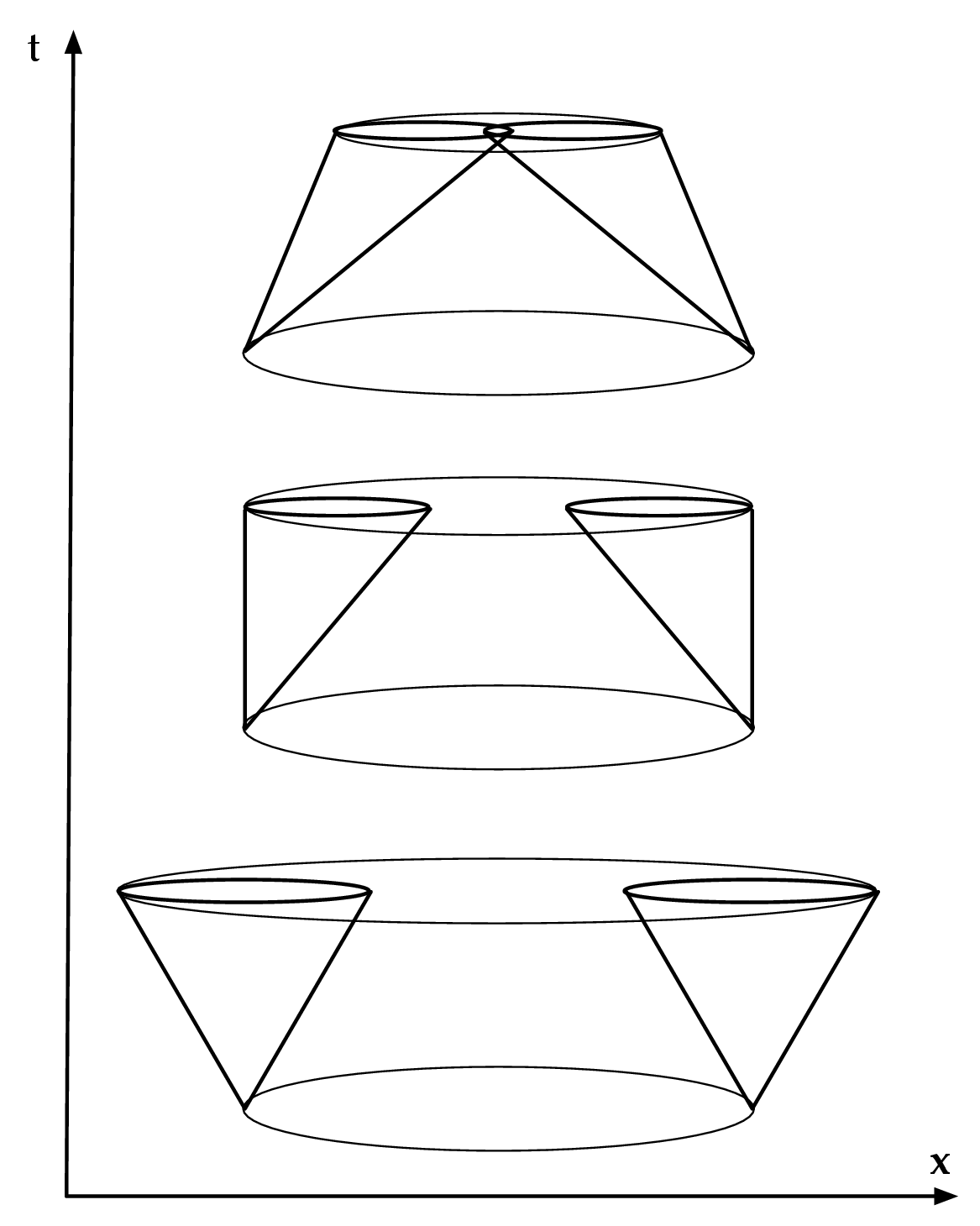}}
\caption[Trapped surfaces]{A moving fluid can form ``trapped
surfaces'' when supersonic flow tips the sound cones past the
vertical.}
\label{F:trapped}
\end{figure}

\begin{figure}[htpb]
    \centerline{\includegraphics[width=\textwidth]{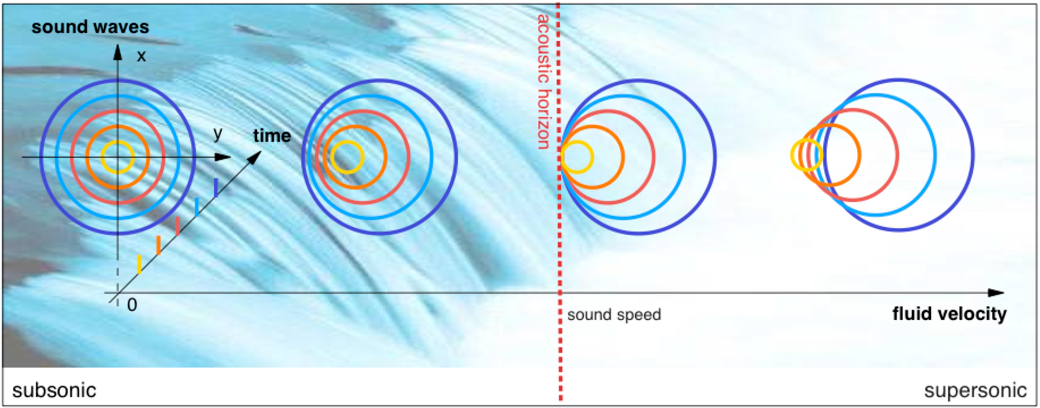}}
\caption[Acoustic horizon]{A moving fluid can form an ``acoustic
  horizon'' when supersonic flow prevents upstream motion of sound
  waves.}
\label{F:horizon}
\end{figure}

The event horizon (absolute horizon) is defined, as in general
relativity, by demanding that it be the boundary of the region from
which null geodesics (now phonons rather than the more usual photons) cannot escape. This is actually
the future event horizon. A past event horizon can be defined in
terms of the boundary of the region that cannot be reached by
incoming phonons -- strictly speaking this requires us to define
notions of past and future null infinities, but we will simply take
all relevant incantations as understood. In particular, the event
horizon is a null surface, the generators of which are null geodesics.

In all stationary geometries the apparent and event horizons coincide,
and the distinction is immaterial. In time-dependent geometries the
distinction is often important. When computing the surface gravity, we
shall restrict attention to stationary geometries (steady flow). In
fluid flows of high symmetry (spherical symmetry, plane symmetry), the
ergosphere may coincide with the acoustic apparent horizon, or even the
acoustic event horizon. This is the analogue of the result in general
relativity that for static (as opposed to stationary) black holes the
ergosphere and event horizon coincide. For many more details, including
appropriate null coordinates and Carter--Penrose diagrams, both in
stationary and time-dependent situations, see \citet{Barcelo:2004wz}.

\subsubsection{Surface gravity} 

Because of the definition of event horizon in terms of phonons (null
geodesics) that cannot escape the acoustic black hole, the event horizon
is automatically a null surface, and the generators of the event horizon
are automatically null geodesics. In the case of acoustics, there is one
particular parameterization of these null geodesics that is ``most
natural'', which is the parameterization in terms of the Newtonian time
coordinate of the underlying physical metric. This allows us to
unambiguously define a ``surface gravity'' even for non-stationary
(time-dependent) acoustic event horizons, by calculating the extent to
which this natural time parameter fails to be an affine parameter for
the null generators of the horizon. (This part of the construction fails
in general relativity where there is no universal natural
time-coordinate unless there is a time-like Killing vector -- this is
why extending the notion of surface gravity to non-stationary geometries
in general relativity is so difficult.)

When it comes to explicitly calculating the surface gravity in terms of
suitable gradients of the fluid flow, it is nevertheless very useful to
limit attention to situations of steady flow (so that the acoustic
metric is stationary). This has the added bonus that for stationary
geometries the notion of ``acoustic surface gravity'' in acoustics is
unambiguously equivalent to the general relativity definition. It is
also useful to take cognizance of the fact that the situation simplifies
considerably for static (as opposed to merely stationary) acoustic
metrics.

To set up the appropriate framework, write the general stationary
acoustic metric in the form
\begin{equation}
\d s^2 = {\rho\over c_s} 
\left[ - c_s^2 \; \d t^2 + (\d\x - \vbf \; \d t)^2 \right].
\end{equation}
The time translation Killing vector is simply 
$ K^\mu = (1;\vec 0\,)$, with 
\begin{equation}
K^2 \equiv g_{\mu\nu} K^\mu K^\nu \equiv - ||\K||^2 = -
{\rho\over c_s}[ c_s^2 - v^2]\,,
\end{equation}
where $v=||\mathbf{v}||$.

The metric can also be written as
\begin{equation}
\d s^2 = {\rho\over c_s} 
\left[ 
-(c_s^2-v^2) \; \d t^2 - 2 \vbf\cdot \d\x \; \d t + (\d\x\,)^2 
\right].
\end{equation}
%

\paragraph*{Static acoustic spacetimes:\;} 
Now suppose that the vector $\vbf/(c_s^2 - v^2)$ is integrable, (the
gradient of some scalar), then we can define a new time coordinate by
\begin{equation}
\d\tau = \d t + \frac{\vbf \cdot \d\x}{c_s^2 - v^2 }.
\end{equation}
Substituting this back into the acoustic line element gives
%
\begin{equation}
\d s^2 = {\rho\over c_s} 
\left[ 
-(c_s^2-v^2) \; \d\tau^2 + 
\left\{ \delta_{ij} + \frac{v^i v^j}{c_s^2 - v^2} \right \}
\d x^i \; \d x^j
\right].
\end{equation}
In this coordinate system the absence of the time-space cross-terms
makes manifest that the acoustic geometry is in fact static (there exists a
family of space-like hypersurfaces orthogonal to the time-like Killing
vector). The condition that an acoustic geometry be static, rather
than merely stationary, is thus seen to be
\begin{equation}
\bnabla \times \left\{ \frac{\vbf}{(c_s^2 - v^2)} \right\} = 0.
\end{equation}
That is, (since in deriving the existence of the effective 
metric we have already assumed the fluid to be irrotational $\vbf = \bnabla\phi$),
\begin{equation}
\vbf \times \bnabla (c_s^2 - v^2) = 0.
\label{Eq:stcond}
\end{equation}
This requires the fluid flow to be parallel to another vector that is
not quite the 3-acceleration of the flow but is closely related to it. (Note that,
because of the assumptions of steady flow and zero vorticity, $\half \bnabla v^2$ is
just the three-acceleration of the fluid. In complete generality 
$\mathbf{a}=\d\mathbf{v}/\d t = \left[ \partial_t \vbf + (\vbf \cdot \bnabla) \vbf \right]$, but steady flow reduces this to 
$\mathbf{a} \to  (\vbf \cdot \bnabla) \vbf$, and zero vorticity then implies 
$\mathbf{a} \to  \half \bnabla v^2$.
It is the occurrence of a
possibly position-dependent speed of sound that complicates the
discussion.) Note that because of the barotropic assumption we have
%
\begin{equation}
\bnabla c_s^2 = {\partial c_s^2\over\partial \rho}\; {\partial\rho\over\partial p} \; \bnabla p
 = -{\partial^2 p\over\partial \rho^2}\; {\partial\rho\over\partial p} \; \rho \; \mathbf{a}.
\end{equation}
where in the last step we have used the Euler equation in the form  $\rho \,\mathbf a = - \bnabla p$. 
That is \\
\begin{equation}
\bnabla (c_s^2 - v^2) = - \left( {\partial^2 p\over\partial \rho^2}\; {\partial\rho\over\partial p} \; \rho + 2\right) \mathbf{a}.
\end{equation}
So (given that the geometry is already {at the very least} stationary) the condition \eqref{Eq:stcond} for a
static acoustic geometry reduces to
\begin{equation}
\left( {\partial^2 p\over\partial \rho^2}\; {\partial\rho\over\partial p} \; \rho + 2\right) \vbf \times \mathbf{a} = 0.
\end{equation}
This condition can be satisfied in two ways, either by having $\vbf
\parallel \mathbf{a}$, or by having the very specific (and not
particularly realistic) equation of state
\begin{equation}
p = - k \rho^{-1} + C; \qquad c_s^2 = k \rho^{-2}.
\end{equation}
Note that  this particular barotropic equation of state represents Chaplygin gas.

Once we have a static geometry, we can of course directly apply
all of the standard tricks \citep{Membrane} for calculating the
surface gravity developed in general relativity. We set up a system
of fiducial observers (FIDOS) by properly normalizing the
time-translation Killing vector
\begin{equation}
\V_\mathrm{FIDO} \equiv 
\frac{\K}{||\K|| } = \frac{\K}{\sqrt{(\rho/c_s)\;[c_s^2-v^2]}}.
\end{equation}
The four-acceleration of the FIDOS is defined as 
\begin{equation}
\A_\mathrm{FIDO} \equiv
(\V_\mathrm{FIDO} \cdot \bnabla) \V_\mathrm{FIDO}, 
\end{equation}
and using the fact that $\K$ is
a Killing vector, it may be computed in the standard manner
\begin{equation}
\A_\mathrm{FIDO} = +\half \frac{\bnabla ||\K||^2}{||\K||^2 }.
\end{equation}
That is 
\begin{equation}
\A_\mathrm{FIDO} = \half
\left[ 
\frac{\bnabla (c_s^2-v^2)}{(c_s^2-v^2)} + \frac{\bnabla (\rho/c_s)}{(\rho/c_s)}
\right].
\end{equation}
The surface gravity is now defined by taking the norm $||\A_\mathrm{FIDO}||$,
multiplying by the lapse function, $ ||\K|| = \sqrt{(\rho/c_s)\;[c_s^2-v^2]}$,
and taking the limit as one approaches the horizon: $|v|\to c$ (remember 
that we are currently dealing with the static case). The net result is
\begin{equation}
||\A_\mathrm{FIDO}|| \; ||\K|| = 
\half \n \cdot \bnabla (c_s^2-v^2) + \O(c_s^2-v^2),
\end{equation}
so that the surface gravity is given in terms of a normal derivative
by\footnote{Because of the background Minkowski metric there can be no
possible confusion as to the definition of this normal derivative.
Sign conventions are to be chosen such that the surface gravity of outer horizons is positive, 
while the surface gravity of inner horizons is negative. }
%
\begin{equation}
\label{acbh:surfgrav}
g_{H} = 
\left. \half \frac{\partial (c_s^2-v^2)}{\partial n}\right|_H = 
{\left. c_s \right|_H } \; \left. \frac{\partial |c_s-v|}{\partial n}\right|_H.
\end{equation}
This is not quite Unruh's result \citep{Unruh:1981cg, Unruh:1995je}
since he implicitly took the speed of sound to be a
position-independent constant. (This is of course a completely
appropriate approximation for water, which was the working fluid he
was considering.) The fact that prefactor $\rho/c_s$ drops out of the
final result for the surface gravity can be justified by appeal to the
known conformal invariance of the surface
gravity \citep{Jacobson:1993pf}. Though derived in a totally different
manner, this result is also compatible with the expression for the
``surface-gravity'' obtained in the solid-state black holes of
\citet{Reznik:1997ag}, wherein a position dependent (and
singular) refractive index plays a role analogous to the acoustic
metric. As a further consistency check, one can go to the spherically
symmetric case and check that this reproduces the results for ``dirty
black holes'' enunciated in \citet{Visser:1992qh}. Finally, note that
we can also write the expression for surface gravity as
\begin{equation}
\label{acbh:surfgrav2}
g_{H} = \left(1+\half {\partial^2 p\over\partial \rho^2}\; {\partial\rho\over\partial p} \; \rho\right) \; \mathbf{a}\cdot \mathbf{n},
\end{equation}
demonstrating that (in a static acoustic spacetime) the surface
gravity is (up to a dimensionless factor depending on the equation of
state) directly related to the acceleration of the fluid as it crosses
the horizon. For water $p=k\rho+C$, $c_s^2=k$, and $g_H =
\mathbf{a}\cdot \mathbf{n}$; for a Bose--Einstein condensate (BEC) we shall later
on see that $p = \half k \rho^2 +C$, implying $c_s^2=k\rho$, which then
leads to the simple result $g_H={3\over2}\mathbf{a}\cdot \mathbf{n} $. 
Indeed for any power law equation of state, $p = {1\over n} k \rho^n +C$, one has $c_s^2 = k \rho^{n-1}$, and so
 $g_H={n+1\over2}\;\mathbf{a}\cdot \mathbf{n}$. 

Since this is a static geometry, the relationship between the Hawking
temperature and surface gravity may be verified in the usual fast-track
manner -- using the Wick rotation trick to analytically continue to
Euclidean space \citep{Gibbons:1976ue}. If you don't like Euclidean
signature techniques (which are in any case only applicable to
equilibrium situations) you should go back to the original Hawking
derivations \citep{Hawking:1974rv, Hawking:1974sw}.\footnote{There are a
few potential subtleties in the derivation of the existence of Hawking
radiation, which we are, for the time being, glossing over; see
Sect.~\ref{S:Hawking-radiation} for details.} 

We should emphasize that the formula for the Hawking temperature
contains both the surface gravity $g_H$ and the speed of sound $c_H$
at the horizon \citep{Visser:1997ux}. Specifically
\begin{equation}
k T_H = {\hbar g_H \over 2\pi c_H}.
\end{equation}
In view of the explicit formula for $g_H$ above, this can also be written as
\begin{equation}
k T_H = {\hbar \over 2\pi} \left. \frac{\partial |c_s-v|}{\partial n}\right|_H,
\end{equation}
which is closer to the original form provided by
\citet{Unruh:1981cg} (which corresponds to $c_s$ being
constant). Purely on dimensional grounds it is a spatial derivative of
velocity (which has the same engineering dimension as frequency) that
is the determining factor in specifying the physically-normalised
Hawking temperature. Since there is a strong tendency in classical
general relativity to adopt units such that $c\to 1$, and even in
these analogue models it is common to adopt units such that $c_H\to1$,
this has the potential to lead to some confusion. If you choose units
to measure the surface gravity as a physical acceleration, then it is
the quantity $g_H/c_H$, which has the dimensions of frequency, that
governs the Hawking flux \citep{Visser:1997ux}.

One final comment to wrap up this section: The coordinate transform we
used to put the acoustic metric into the explicitly static form is
perfectly good mathematics, and from the general relativity point of
view is even a simplification. However, from the point of view of the
underlying Newtonian physics of the fluid, this is a rather bizarre
way of deliberately de-synchronizing your clocks to take a perfectly
reasonable region -- the boundary of the region of supersonic flow
-- and push it out to ``time'' plus infinity. From the fluid dynamics
point of view this coordinate transformation is correct but perverse,
and it is easier to keep a good grasp on the physics by staying with
the original Newtonian time coordinate.

\paragraph*{Stationary (non-static) acoustic spacetimes:\;}

If the fluid flow does not satisfy the integrability condition, which
allows us to introduce an explicitly static coordinate system, then
defining the surface gravity is a little trickier. 

Recall that by construction the acoustic apparent horizon is, in general,
defined to be a two-surface for which the normal component of the fluid
velocity is everywhere equal to the local speed of sound, whereas the
acoustic event horizon (absolute horizon) is characterised by the
boundary of those null geodesics (phonons) that do not escape to
infinity. In the stationary case these notions coincide, and it is still
true that the horizon is a null surface, and that the horizon can be
ruled by an appropriate set of null curves. Suppose we have somehow
isolated the location of the acoustic horizon, then, in the vicinity of
the horizon, we can split up the fluid flow into normal and tangential
components
\begin{equation}
\vbf = \vbf_\perp+ \vbf_\parallel; 
\qquad \hbox{where} \qquad
\vbf_\perp = v_\perp \; \hat \n .
\end{equation}
Here (and for the rest of this particular section) it is essential
that we use the natural Newtonian time coordinate inherited from
the background Newtonian physics of the fluid. In addition $\hat
\n$ is a unit vector field that, at the horizon,
 is perpendicular to
it, and away from the horizon is some suitable smooth extension.
(For example, take the geodesic distance to the horizon and consider
its gradient.) We only need this decomposition to hold in some open
set encompassing the horizon and do not need to have a global
decomposition of this type available. Furthermore, by definition
we know that $v_\perp = c_s$ at the horizon. Now consider the vector
field
\begin{equation}
L^\mu = (\,1; \; v_\parallel^i \,).
\end{equation}
Since the spatial components of this vector field are by definition
tangent to the horizon, the
integral curves of this vector field will be generators for the
horizon. Furthermore, the norm of this vector (in the acoustic
metric) is
\begin{equation}
|| {\mathbf L} ||^2 = 
-{\rho\over c_s} 
\left[ 
\vphantom{\Big|}
-(c_s^2 - v^2) 
- 2\vbf_\parallel \cdot \vbf 
+ \vbf_\parallel \cdot \vbf_\parallel 
\right]
=
{\rho\over c_s} \; (c_s^2 - v_\perp^2).
\end{equation}
In particular, on the acoustic horizon $L^\mu $ defines a null
vector field, the integral curves of which are generators for the
acoustic horizon. We shall now verify that these generators are
geodesics, though the vector field ${\mathbf L}$ is not normalised with an
affine parameter, and in this way shall calculate the surface
gravity. 

Consider the quantity $({\mathbf L}\cdot\bnabla){\mathbf L}$ and calculate
\begin{equation}
L^\alpha \nabla_\alpha L^\mu 
= 
L^\alpha (\nabla_\alpha L_\beta - \nabla_\beta L_\alpha) g^{\beta \mu}
+ \half \nabla_\beta (L^2) g^{\beta \mu}.
\end{equation}
To calculate the first term note that
\begin{equation}
L_\mu = {\rho\over c_s} \; (-[c_s^2-v_\perp^2]; \vbf_\perp).
\end{equation}
Thus, 
\begin{equation}
L_{[\alpha,\beta]} = - 
\begin{bmatrix}
   0&\vdots&-\nabla_i\left[{\rho\over c_s}(c_s^2-v_\perp^2)\right]\\
   \cdots\cdots\cdots\cdots&\cdot&\cdots\cdots\\
   +\nabla_j\left[{\rho\over c_s}(c_s^2-v_\perp^2)\right]&\vdots&
\left({\rho \over c_s} \; v^\perp\right){}_{[i,j]}\\
\end{bmatrix}.     
\end{equation}
And so:
\begin{equation}
L^\alpha L_{[\beta,\alpha]} = 
\left(
\vbf_\parallel \cdot \bnabla\left[{\rho\over c_s} (c_s^2-v_\perp^2)\right]; \;
\nabla_j\left[{\rho\over c_s}(c_s^2-v_\perp^2)\right] 
+ v_\parallel^i \left({\rho\over c_s} \; |v^\perp| \hat n \right)_{[j,i]} 
\right).
\end{equation}
On the horizon, where $c_s=v_\perp$, and additionally assuming
$\vbf_\parallel \cdot \bnabla \rho=0$ so that the density is constant
over the horizon, this simplifies tremendously
\begin{equation}
(L^\alpha L_{[\beta,\alpha]})|_\mathrm{horizon} = 
-{\rho\over c_s} \; \left(0; \nabla_j(c_s^2-v_\perp^2)\right) = 
-{\rho\over c_s} \; \frac{\partial(c_s^2-v_\perp^2)}{\partial n} \; \left(0;\hat n_j\right).
\end{equation}
Similarly, for the second term we have
\begin{equation}
\nabla_\beta (L^2) = 
\left(0; \nabla_j\left[{\rho\over c_s}(c_s^2-v_\perp^2)\right] \right).
\end{equation}
On the horizon this again simplifies
\begin{equation}
\nabla_\beta (L^2)|_\mathrm{horizon} = 
+{\rho\over c_s} \; \left(0; \nabla_j(c_s^2-v_\perp^2) \right) 
= +
{\rho\over c_s} \; \frac{\partial(c_s^2-v_\perp^2)}{\partial n} \; \left(0;\hat n_j\right).
\end{equation}
There is partial cancellation between the two terms, and so
\begin{equation}
(L^\alpha \nabla_\alpha L_\mu)_\mathrm{horizon}
= 
+\half \; {\rho\over c_s} \; \frac{\partial(c_s^2-v_\perp^2)}{\partial n} \; \left(0;\hat n_j\right),
\end{equation}
while
\begin{equation}
(L_\mu)_\mathrm{horizon} = {\rho\over c_s} \; \left(0; c_s\;\hat n_j\right).
\end{equation}
Comparing this with the standard definition of surface
gravity \citep{Wald}\footnote{There is an issue of normalization
here. On the one hand we want to be as close as possible to general
relativistic conventions. On the other hand, we would like the
surface gravity to really have the dimensions of an acceleration. The
convention adopted here, with one explicit factor of $c_s$, is the best
compromise we have come up with. (Note that in an acoustic setting, where the
speed of sound is not necessarily a constant, we cannot simply set $c_s\to1$ by
a choice of units.)} 
\begin{equation}
(L^\alpha \nabla_\alpha L_\mu)_\mathrm{horizon} 
= + \frac{g_H}{c_s} \; (L_\mu)_\mathrm{horizon}\,, 
\end{equation}
we finally have
\begin{equation}
g_H = 
\half \frac{\partial(c_s^2-v_\perp^2)}{\partial n} =
 c_s \; \frac{\partial(c_s-v_\perp)}{\partial n}\,.
\end{equation}

This is in agreement with the previous calculation for static acoustic
black holes, and insofar as there is overlap, is also consistent with
results of \citet{Unruh:1981cg, Unruh:1995je},
\citet{Reznik:1997ag}, and the results for ``dirty black
holes'' \citep{Visser:1992qh}. From the construction it is clear that the
surface gravity is a measure of the extent to which the Newtonian time
parameter inherited from the underlying fluid dynamics fails to be an
affine parameter for the null geodesics on the horizon.\footnote{There
are situations in which this surface gravity is a lot larger than one
might naively expect \citep{Liberati:2000pt}.}

Again, the justification for going into so much detail on this specific
model is that this style of argument can be viewed as a template -- it
will (with suitable modifications) easily generalise to more complicated
analogue models.

\subsection{Regaining geometric acoustics}

Up to now, we have been developing general machinery to force acoustics
into Lorentzian form. This can be justified either with a view to using
fluid mechanics to teach us more about general relativity, or to using
the techniques of Lorentzian geometry to teach us more about fluid
mechanics.

For example, given the machinery developed so far, taking the short
wavelength/high frequency limit to obtain geometrical acoustics is now
easy. Sound rays (phonons) follow the \emph{null geodesics} of the
acoustic metric. Compare this to general relativity, where in the
geometrical optics approximation, light rays (photons) follow \emph{null
geodesics} of the physical spacetime metric. Since null geodesics are
insensitive to any overall conformal factor in the
metric \citep{MTW, Hawking-Ellis, Wald}, one might as well simplify life by
considering a modified conformally-related metric
\begin{equation}
h_{\mu\nu} \equiv 
\begin{bmatrix}
   -(c_s^2-v_0^2)&\vdots&-v_0^j\\
   \cdots\cdots\cdots\cdots&\cdot&\cdots\cdots\\
   -v_0^i&\vdots&\delta^{ij}\\
\end{bmatrix}.     
\end{equation}
This immediately implies that, in the geometric acoustics limit,
sound propagation is insensitive to the density of the fluid. In
this limit, acoustic propagation depends only on the local speed
of sound and the velocity of the fluid. It is only for specifically
wave-related properties that the density of the medium becomes
important.

We can rephrase this in a language more familiar to the acoustics
community by invoking the eikonal approximation. Express the
linearized velocity potential, $\phi_1$, in terms of an amplitude,
$a$, and phase, $\varphi$, by $\phi_1 \sim a e^{i\varphi}$. Then,
neglecting variations in the amplitude $a$, the wave equation
reduces to the \emph{eikonal equation}
\begin{equation}
h^{\mu\nu} \; \partial_\mu \varphi \; \partial_\nu \varphi = 0.
\end{equation}
This eikonal equation is blatantly insensitive to any overall
multiplicative prefactor (conformal factor).
As a sanity check on the formalism, it is useful to re-derive some
standard results. For example, let the null geodesic be parameterised
by $x^\mu(t) \equiv (t; \x(t))$. Then the null condition implies
\begin{eqnarray}
&& h_{\mu\nu} \frac{\d x^\mu}{\d t} \frac{\d x^\nu}{\d t} = 0
\nonumber\\
&&\iff
-(c_s^2 - v_0^2) - 2 v_0^i \frac{\d x^i}{\d t} 
+ \frac{\d x^i}{\d t} \frac{\d x^i}{\d t} = 0
\nonumber\\
&&\iff
\left\Vert \frac{\d{\x}}{\d t} - \vbf_0 \right\Vert = c_s.
\end{eqnarray}
Here the norm is taken in the flat physical metric. This has the
obvious interpretation that the ray travels at the speed of sound,
$c_s$, relative to the moving medium.

Furthermore, if the geometry is stationary (steady flow) one can do slightly
better. Let $x^\mu(s) \equiv (t(s); \x(s))$ be some null path
from $\x_1$ to $\x_2$, parameterised in terms of physical
arc length (i.e., $|| \d{\x}/\d s || \equiv 1$). Then the
tangent vector to the path is
\begin{equation}
\frac{\d x^\mu}{\d s} = \left( \frac{\d t}{\d s}; \frac{\d x^i}{\d s} \right).
\end{equation}
The condition for the path to be null (though not yet necessarily
a null geodesic) is
\begin{equation}
g_{\mu\nu} \frac{\d x^\mu}{\d s} \frac{\d x^\nu}{\d s} = 0.
\end{equation}
Using the explicit algebraic form for the metric, this can be
expanded to show
\begin{equation}
-(c_s^2 - v_0^2) \left(\frac{\d t}{\d s}\right)^2 
- 2 v_0^i \left(\frac{\d x^i}{\d s}\right) \left(\frac{\d t}{\d s}\right)
+1 = 0.
\end{equation}
Solving this quadratic 
\begin{equation}
\left(\frac{\d t}{\d s}\right) 
= \frac{- v_0^i \left(\frac{\d x^i}{\d s}\right)
    + \sqrt{ c_s^2 - v_0^2 + \left(v_0^i \; \frac{\d x^i}{\d s}\right)^2 }}
  {c_s^2 - v_0^2}.
\end{equation}
Therefore, the total time taken to traverse the path is  
\begin{equation}
T[\gamma] = \int_{\x_1}^{\x_2} (\d t/\d s)\,\d s 
= \int_\gamma \frac{1}{c_s^2-v_0^2} 
\Big\{ 
\sqrt{(c_s^2 - v_0^2)\d s^2 + (v_0^i \, \d x^i)^2 } - v_0^i \, \d x^i 
\Big\}.
\end{equation}
If we now recall that extremising the total time taken is Fermat's
principle for sound rays, we see that we have checked the formalism
for stationary geometries (steady flow) by reproducing the discussion
of \citet[p.~262]{Landau-Lifshitz}.\footnote{Mathematically,
  one can view the time taken to traverse such a path as a particular
  instance of Finsler distance -- it is, in fact, the distance function
  associated with a Randers metric. See \citet{Gibbons:2008zi}, and
  brief discussion in Sect.~\ref{S:Randers}.}

\subsection{Example: Vortex geometry}

As an example of a fluid flow where the distinction between ergosphere
and acoustic event horizon is critical, consider the ``draining bathtub''
fluid flow. We shall model a draining bathtub by a (3+1) dimensional
flow with a linear sink along the z-axis. Let us start with the
simplifying assumption that the background density $\rho$ is a
position-independent constant throughout the flow (which automatically
implies that the background pressure $p$ and speed of sound $c$ are also
constant throughout the fluid flow). The equation of continuity then
implies that for the radial component of the fluid velocity we must have
\begin{equation}
v^{\hat r} \propto \frac{1}{r}\,.
\end{equation}
In the tangential direction, the requirement that the flow be
vorticity free (apart from a possible delta-function contribution at
the vortex core) implies, via Stokes' theorem, that 
\begin{equation}
v^{\hat t} \propto \frac{1}{r}\,.
\end{equation}

(If these flow velocities are nonzero, then following the discussion
of \citealt{Visser:2004zs} there must be some external force present to
set up and maintain the background flow. Fortunately it is easy to see
that this external force affects only the background flow and does not
influence the linearized fluctuations we are interested in.)

For the background velocity potential we must then have
\begin{equation}
\phi(r,\theta) = -A\; \ln(r/a) - B \;\theta.
\end{equation}
Note that, as we have previously hinted, the velocity potential is not a
true function (because it has a discontinuity on going through $2\pi$
radians). The velocity potential must be interpreted as being defined
patch-wise on overlapping regions surrounding the vortex core at
$r=0$. The velocity of the fluid flow is
\begin{equation}
\vbf = - \bnabla \phi = \frac{(A \; \hat r + B \; \hat\theta)}{r}.
\end{equation}

Dropping a position-independent prefactor, the acoustic metric for a
draining bathtub is explicitly given by
\begin{equation}
\d s^2 = 
- c^2 \; \d t^2 
+ \left(\d r - \frac{A}{r} \d t\right)^2 
+ \left(r \, \d\theta - \frac{B}{r} \d t\right)^2 + \d z^2.
\end{equation}
Equivalently
\begin{equation}
\d s^2 = 
- \left(c^2 -\frac{A^2+B^2}{r^2}\right) \d t^2  
- 2\frac{A}{r} \, \d r \, \d t - 2 B \, \d\theta \, \d t + \d r^2 + r^2 \d\theta^2 + \d z^2.
\end{equation}
A similar metric, restricted to $A=0$ (no radial flow), and
generalised to an anisotropic speed of sound, has been exhibited by
\citet{Volovik:1996qw}, that metric being a model for the acoustic
geometry surrounding physical vortices in superfluid $^{3}$He. (For a
survey of the many analogies and similarities between the physics of
superfluid $^{3}$He, see Sect.~\ref{S:helium} and references
therein. For issues specifically connected to the Standard Electroweak
Model see \citealt{Volovik:1995yn}.)
Note that the metric given above is \emph{not}
identical to the metric of a spinning cosmic string, which would
instead take the form \citep{Visser:1995cc}
\begin{equation}
\d s^2 = - c^2 (\d t - {\tilde A} \, \d\theta)^2 + \d r^2 + (1-{\tilde B}) r^2 \d\theta^2 +\d z^2.
\end{equation}
%

\begin{figure}[htpb]
\centerline{\includegraphics[width=0.6\textwidth]{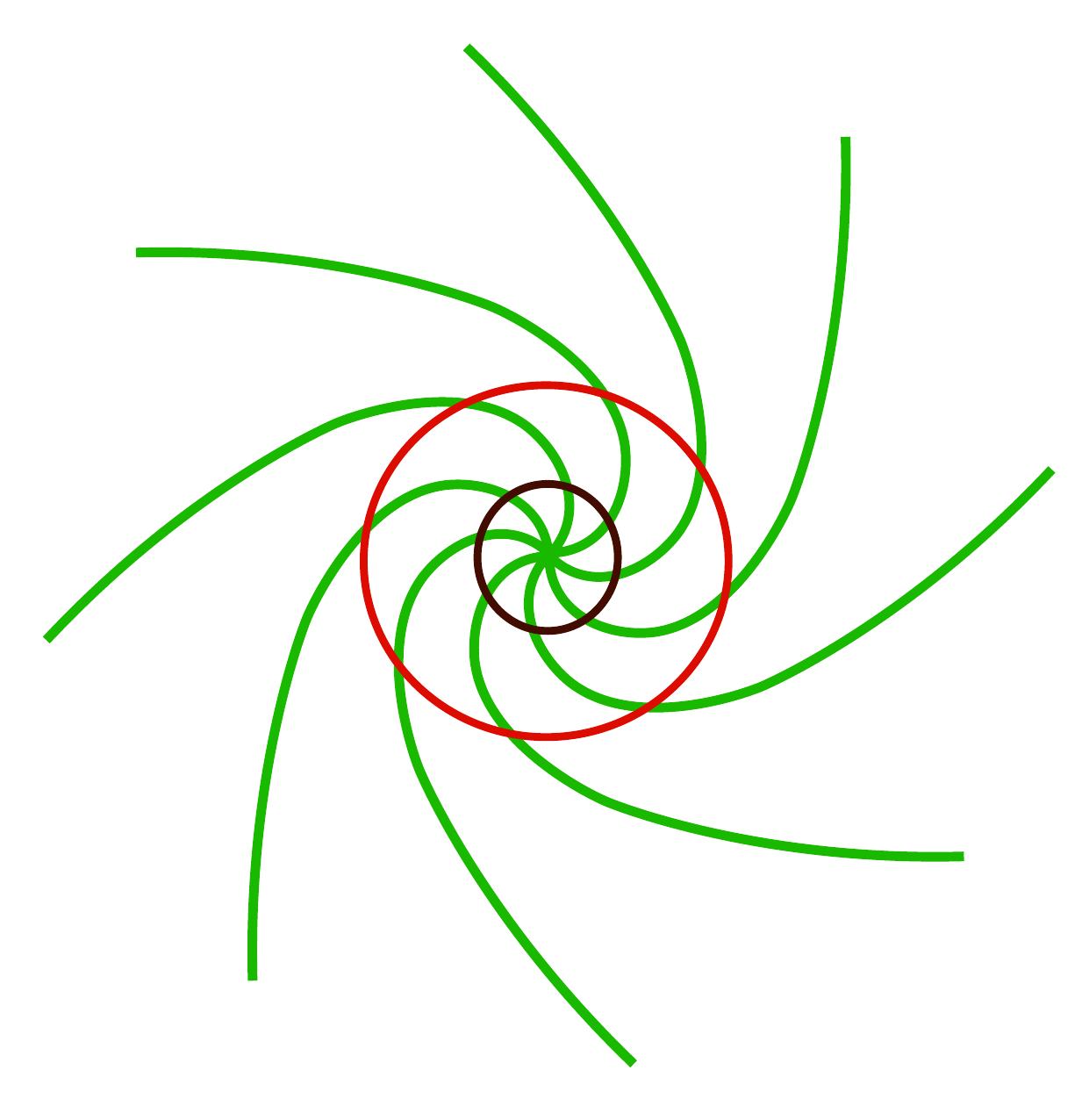}}
\caption[Vortex geometry]{A collapsing vortex geometry (draining
bathtub): The green spirals denote streamlines of the fluid flow. The
outer circle represents the ergo-surface (ergo-circle) while the inner circle
represents the [outer] event horizon.
}
\label{F:vortex}
\end{figure}

In conformity with previous comments, the vortex fluid flow is seen to
possess an acoustic metric that is stably causal and which does not
involve closed time-like curves. At large distances it is possible to
\emph{approximate} the vortex geometry by a spinning cosmic
string \citep{Volovik:1996qw}, but this approximation becomes progressively
worse as the core is approached. Trying to force the existence of
closed time-like curves leads to the existence of evanescent waves in
what would be the achronal region, and therefore to the breakdown of
the analogue model description \citep{Anderson:2010yd}.

The ergo-surface (ergo-sphere, in the current context an ergo-circle) forms at
\begin{equation}
r_\mathrm{{ergo-surface}} = \frac{\sqrt{A^2 + B^2}}{c_s}.
\end{equation}
Note that the sign of $A$ is irrelevant in defining the ergo-surface and
ergo-region: It does not matter if the vortex core is a source or a
sink.

The acoustic event horizon forms once the radial component of the
fluid velocity exceeds the speed of sound, that is, at
\begin{equation}
r_\mathrm{horizon} = \frac{|A|}{c_s}.
\end{equation}
The sign of $A$ now makes a difference. For $A<0$ we are dealing with
a future acoustic horizon (acoustic black hole), while for $A>0$ we
are dealing with a past event horizon (acoustic white hole).
This vortex geometry was the theoretical basis for a number of experiments carried out at Nottingham~\citep{Torres:2016iee}. For more on this point see section~\ref{S:experiments} on experimental investigations.

\subsection{Example: Slab geometry}

A popular model for the investigation of event horizons in the
acoustic analogy is the one-dimensional slab geometry where the
velocity is always along the $z$ direction and the velocity profile
depends only on $z$. The continuity equation then implies that
$\rho(z) \; v(z)$ is a constant, and the acoustic metric becomes
\begin{equation}
\d s^2 \propto \frac{1}{v(z) \; c_s(z) }
\left[ 
- c_s(z)^2 \; \d t^2 + \left\{ \d z - v(z) \; \d t \right\}^2 
+ \d x^2 + \d y^2 
\right].
\end{equation}
That is
\begin{equation}
\d s^2 \propto \frac{1}{v(z) \; c_s(z) }
\left[ 
- \left\{c_s(z)^2-v(z)^2\right\} \d t^2 - 2 v(z) \; \d z\; \d t+ \d x^2 + \d y^2 + \d z^2
\right].
\end{equation}

If we (somewhat unphysically) set $c_s(z)=1$, and ignore the $x, y$ coordinates, and the conformal
factor, then we have the toy model acoustic geometry discussed in many
papers (see, for instance, the early papers by \citealt[p.~2828,
  Eq.~(8)]{Unruh:1995je}, \citealt[p.~7085,
  Eq.~(4)]{Jacobson:1996zs}, \citealt{Corley:1996ar},
and \citealt{Corley:1997nw}):
\begin{equation}
\d s^2 \to  
- \left\{1-v(z)^2\right\} \d t^2 - 2 v(z) \; \d z\; \d t + \d z^2.
\end{equation}
Depending on the velocity profile
one can simulate black holes, white holes or black hole-white hole
pairs, interesting for analyzing the black hole laser
effect \citep{Corley:1998rk} or aspects of the physics of
warp drives \citep{Finazzi:2009jb}.

In this situation one must again invoke an external force to set up
and maintain the fluid flow. Since the conformal factor is regular at
the event horizon, we know that the surface gravity and Hawking
temperature are independent of this conformal
factor \citep{Jacobson:1993pf}. In the general case it is important to
realise that the flow can go supersonic for either of two reasons: The
fluid could speed up, or the speed of sound could decrease. When it
comes to calculating the ``surface gravity'' both of these effects
will have to be taken into account.

\subsection{Example: Conformal Schwarzschild geometry}

To see how close the acoustic metric can get to reproducing the
Schwarzschild geometry it is first useful to introduce one of the
more exotic representations of the Schwarzschild geometry: the
\Painleve--Gullstrand line element, which is simply an unusual
choice of coordinates on the Schwarzschild spacetime.\footnote{The
\Painleve--Gullstrand line element is sometimes called the \Lemaitre 
line element.} In modern notation the Schwarzschild geometry in
outgoing (+) and ingoing (--) \Painleve--Gullstrand coordinates
may be written as: 
\begin{equation}
\d s^2 = 
- \d t^2 +
\left( \d r \pm \sqrt{\frac{2GM}{r}} \d t \right)^2 
+ r^2\left( \d\theta^2 + \sin^2\theta \; \d\phi^2 \right).
\end{equation}
Equivalently
\begin{equation}
\d s^2 = 
- \left(1-\frac{2GM}{r}\right) \d t^2 
\pm \sqrt{\frac{2GM}{r}}\, \d r \, \d t 
+ \d r^2 + r^2\left( \d\theta^2 + \sin^2\theta \; \d\phi^2 \right).
\end{equation}
This representation of the Schwarzschild geometry was not (until the
advent of the analogue models) particularly well-known, and it has been
independently rediscovered several times during the 20th century. See,
for instance, \citet{Painleve}, \citet{Gullstrand},
\citet{Lemaitre}, the related discussion by
\citet{Hawking:300}, and more recently, the paper by \citet{Kraus:1994fh}.
The \Painleve--Gullstrand coordinates are
related to the more usual Schwarzschild coordinates by
\begin{equation}
t_\mathrm{PG} = t_\mathrm{S} \pm 
\left[ 
4 M \; \arctanh\left(\sqrt{\frac{2GM}{r}}\right) - 2 \; \sqrt{2GMr} 
\right].
\end{equation}
Or equivalently
\begin{equation}
\d t_\mathrm{PG} = \d t_\mathrm{S} \pm \frac{\sqrt{2GM/r}}{1-2GM/r}\; \d r.
\end{equation}
With these explicit forms in hand, it becomes an easy exercise to
check the equivalence between the \Painleve--Gullstrand line element
and the more usual Schwarzschild form of the line element. It should
be noted that the $+$~sign corresponds to a coordinate patch that
covers the usual asymptotic region plus the region containing the
future singularity of the maximally-extended Schwarzschild spacetime.
Thus, it covers the future horizon and the black hole singularity. On
the other hand the $-$~sign corresponds to a coordinate patch that
covers the usual asymptotic region plus the region containing the past
singularity. Thus it covers the past horizon and the white hole
singularity.

As emphasised by Kraus and Wilczek, the \Painleve--Gullstrand line
element exhibits a number of features of pedagogical interest. In
particular the constant-time spatial slices are completely flat. That is, 
the curvature of space is zero, and all the spacetime curvature of the
Schwarzschild geometry has been pushed into the time--time and
time--space components of the metric.

Given the \Painleve--Gullstrand line element, it might seem trivial to
force the acoustic metric into this form: Simply take $\rho$ and $c$
to be constants, and set $v=\sqrt{2GM/r}$. While this certainly forces
the acoustic metric into the \Painleve--Gullstrand form, the problem
with this is that this assignment is incompatible with the continuity
equation $\bnabla\cdot(\rho\vbf)\neq 0$ that was used in deriving
the acoustic equations.

The best we can actually do is this: Pick the speed of sound $c_s$
to be a position-independent constant, which we normalise to unity
($c_s=1$). Now set $v=\sqrt{2GM/r}$, and use the continuity equation
$\bnabla\cdot(\rho\vbf)= 0$ plus spherical symmetry to deduce $\rho|\vbf| \propto 1/r^2$ so that $\rho \propto r^{-3/2}$. 
Since the speed of sound
is taken to be constant, we can integrate the relation $c_s^2 =
\d p/\d\rho$ to deduce that the equation of state must be $p = p_\infty +
c_s^2\, \rho$, and that the background pressure satisfies $p-p_\infty
\propto c_s^2 \, r^{-3/2}$. \footnote{Of course,  in order to realise such flow one must also satisfy the Euler equation. 
It is often assumed in the literature that this can be done by a suitable choice of external potential. However, sometimes, as in this particular case, the required external potential is rather non-trivial and difficult to realise in practice. See e.g.~\cite{Liberati:2000ag}.}
Overall, the acoustic metric is now
\begin{equation}
\d s^2 \propto r^{-3/2}
\left[ - \d t^2 +
\left( \d r \pm \sqrt{\frac{2GM}{r}} \d t \right)^2 
+ r^2\left( \d\theta^2 + \sin^2\theta \; \d\phi^2 \right) 
\right].
\end{equation}

So we see that the net result is conformal to the
\Painleve--Gullstrand form of the Schwarzschild geometry but not
identical to it. For many purposes this is good enough. We have
an event horizon; we can define surface gravity; we can analyse
Hawking radiation.\footnote{Similar constructions work for the
  Reissner--Nordstr\"om geometry \citep{Liberati:2000pt}, as long as
  one does not get too close to the singularity. (With $c=1$ one needs
  $r>Q^2/(2m)$ to avoid an imaginary fluid velocity.) Likewise, certain
  aspects of the Kerr geometry can be emulated in this
  way \citep{Visser:2004zs}. (One needs $r>0$ in the Doran
  coordinates \citep{Doran:1999gb, Visser:2007fj} to avoid closed
  timelike curves.) As a final remark, let us note that de~Sitter space
  corresponds to $v \propto r$ and $\rho\propto 1/r^3$. For further
  details see Sect.~\ref{SS:cosmological}.} Since surface gravity and
Hawking temperature are conformal invariants \citep{Jacobson:1993pf}
this is sufficient for analysing basic features of the Hawking
radiation process. The only way in which the conformal factor can
influence the Hawking radiation is through backscattering off the
acoustic metric. (The phonons are minimally-coupled scalars, not
conformally-coupled scalars, so there will in general be effects on
the frequency-dependent greybody factors.) If we focus attention on
the region near the event horizon, the conformal factor can simply be
taken to be a constant, and we can ignore all these complications. 

In closing this section, it is perhaps interesting to add that the \Painleve--Gullstrand coordinates have shown themselves useful, 
not only in the above\-mentioned model, but also in the study of cosmological spacetimes~\citep{Gaur:2022hap}. 
Nonetheless, the extension of this coordinate system to rotating black holes has proven much more difficult. 
Indeed, this difficulty is related to the impossibility of  casting the Kerr geometry in acoustic form, or even (apparently) in the Gordon form~\citep{Visser:2022fwx}, albeit one can successfully mimic the equatorial slice of a Kerr black hole~\citep{Visser:2004zs}, and close approximations to the full Kerr metric are possible when using relativistic fluids~\citep{Giacomelli:2017eze,Liberati:2018osj}.

\subsection{Example: Canonical acoustic black hole}
Given the not particularly practical form of the external potential necessary to simulate a Schwarzschild black hole, one might wonder what is the simplest realizable geometry within an incompressible fluid flow able to simulate the salient features of a spherically symmetric black hole. This is indeed the so called canonical acoustic black hole geometry.

Let us start by by assuming incompressibility and spherical symmetry then, since the density $\rho$ is position independent, the continuity equation implies $v\propto 1/r^2$. 
Similarly, given the position independence of $\rho$ and the barotropic assumption, also the pressure will be the same all over the space, and hence the speed of sound as well. 
So if define a normalization constant $r_0$ and set

\begin{equation}
  \label{eq:vsc}
  {\rm v}=c_s \,\frac{r^{2}_{0}}{r^2}
\end{equation}
The acoustic metric than takes the form:
\begin{equation}
  \label{eq:metrsc}
  \d s^2=-c_s^2 \d t^2
  + \left( \d r \pm c_s \,\frac{r^{2}_{0}}{r^2}\, \d t \right)^2 
  + r^2 \left( \d \theta^2+\sin^2\theta\d \phi^2\right) 
\end{equation}
While this does not fall into the category of any standard geometries typically considered in general relativity, it can be viewed, as explained earlier, as the canonical acoustic black hole.
It is also worth noticing that that by promoting $r_0$ to be time-dependent the above metric can describe the acoustic geometries associated to cavitating bubbles, such as those producing sonoluminescence (see e.g.~\cite{Visser:1997ux} and \cite{Liberati:2000ag}).

\subsubsection{Causal structure of acoustic black-hole spacetimes}

We can now turn to another aspect of acoustic black holes, i.e., their
global causal structure, which we shall illustrate making use of the
\emph{Carter--Penrose} conformal
diagrams \citep{MTW,Hawking-Ellis}. A systematic study in this sense
was performed in \citet{Barcelo:2004wz} for 1+1 geometries (viewed
either as a dimensional reduction of a physical 3+1 system, or
directly as geometrical acoustic metrics). The basic idea underlying
the conformal diagram of any non-compact 1+1 manifold is that its
metric can always be conformally mapped to the metric of a compact
geometry, with a boundary added to represent events at infinity. Since
compact spacetimes are in some sense ``finite'', they can then
properly be drawn on a sheet of paper, something that is sometimes
very useful in capturing the essential features of the geometry at
hand.

The basic steps in the acoustic case are the same as in standard
general relativity: Starting from the coordinates ($t, x$) as in
Eq.~(\ref{E:acoustic-line-element}), one has to introduce
appropriate null coordinates (analogous to the Eddington--Finkelstein
coordinates) ($u, v$), then, by exponentiation, null Kruskal-like
coordinates ($U, W$), and finally compactify by means of a new coordinate
pair (${\cal{U}}, {\cal{W}}$) involving a suitable function mapping
an infinite range to a finite one (typically the arctan function). We
shall explicitly present only the conformal diagrams for an acoustic
black hole, and a black hole-white hole pair, as these particular
spacetimes will be of some relevance in what follows. We redirect the
reader to \citet{Barcelo:2004wz} for other geometries and technical
details.

\paragraph*{Acoustic black hole:\;} 

\begin{figure}[!htpb]
    \centerline{\includegraphics[width=0.75\textwidth]{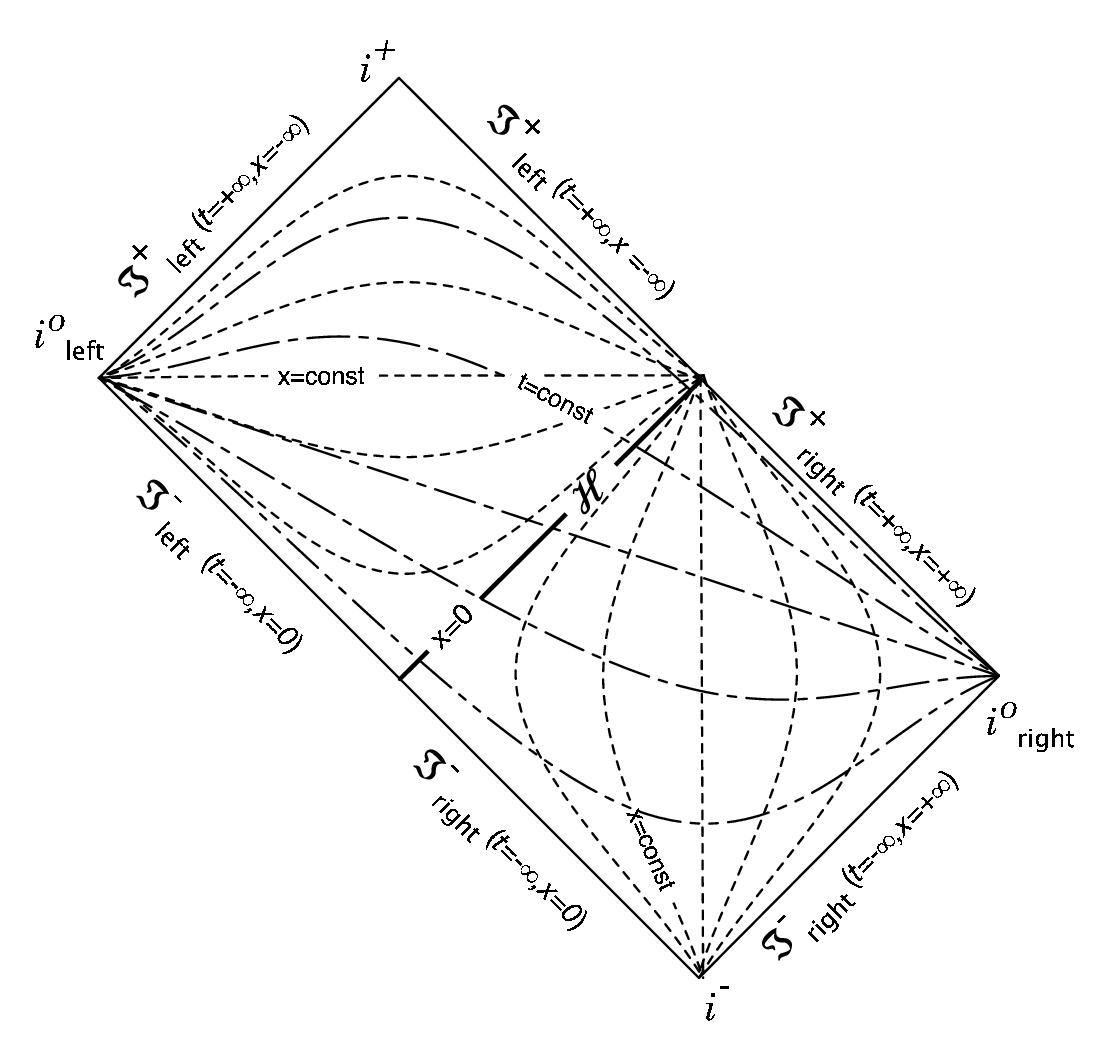}}
\caption[Acoustic Black Hole conformal structure]{Conformal diagram of an acoustic black hole.}
\label{F:conf-bh}
\end{figure}

For the case of a single isolated black-hole horizon we find the
Carter--Penrose diagram of Fig.~\ref{F:conf-bh}. (In the
figure we have introduced an aspect ratio different from unity for the
coordinates $\cal U$ and $\cal W$, in order to make the various
regions of interest graphically more clear.) As we have already
commented, in the acoustic spacetimes, with no periodic identifications,
there are two clearly-differentiated notions of asymptotia, ``right''
and ``left''. In all our figures we have used subscripts ``right'' and
``left'' to label the different null and spacelike infinities.
In addition, we have denoted the different sonic-point boundaries
with ${\Im}^{\pm}_\mathrm{right}$ or ${\Im}^{\pm}_\mathrm{left}$
depending on whether they are the starting point (--~sign) or the ending
point (+~sign) of the null geodesics in the right or left parts of the 
diagram.
  
In contradistinction to the Carter--Penrose diagram for the
Schwarzschild black hole (which in the current context would have to
be an eternal black hole, not one formed via astrophysical stellar
collapse) there is no singularity. On reflection, this feature of the
conformal diagram should be obvious, since the fluid flow underlying
the acoustic geometry is nowhere singular.

Note that the event horizon $\cal{H}$ is the boundary of the
causal past of future right null infinity; that is, ${\cal
H} = \dot{J}^-({\Im}^+_\mathrm{right})$, with standard
notations \citep{MTW}.
Notice that from the point of  view of general relativity this geometry can be (analytically) extended through the past null boundaries 
${\Im}^-_i(t=-\infty,x=0)$. From the analogue gravity perspective this extension would not be part of our universe.

\begin{figure}[!htpb]
    \centerline{\includegraphics[width=0.75\textwidth]{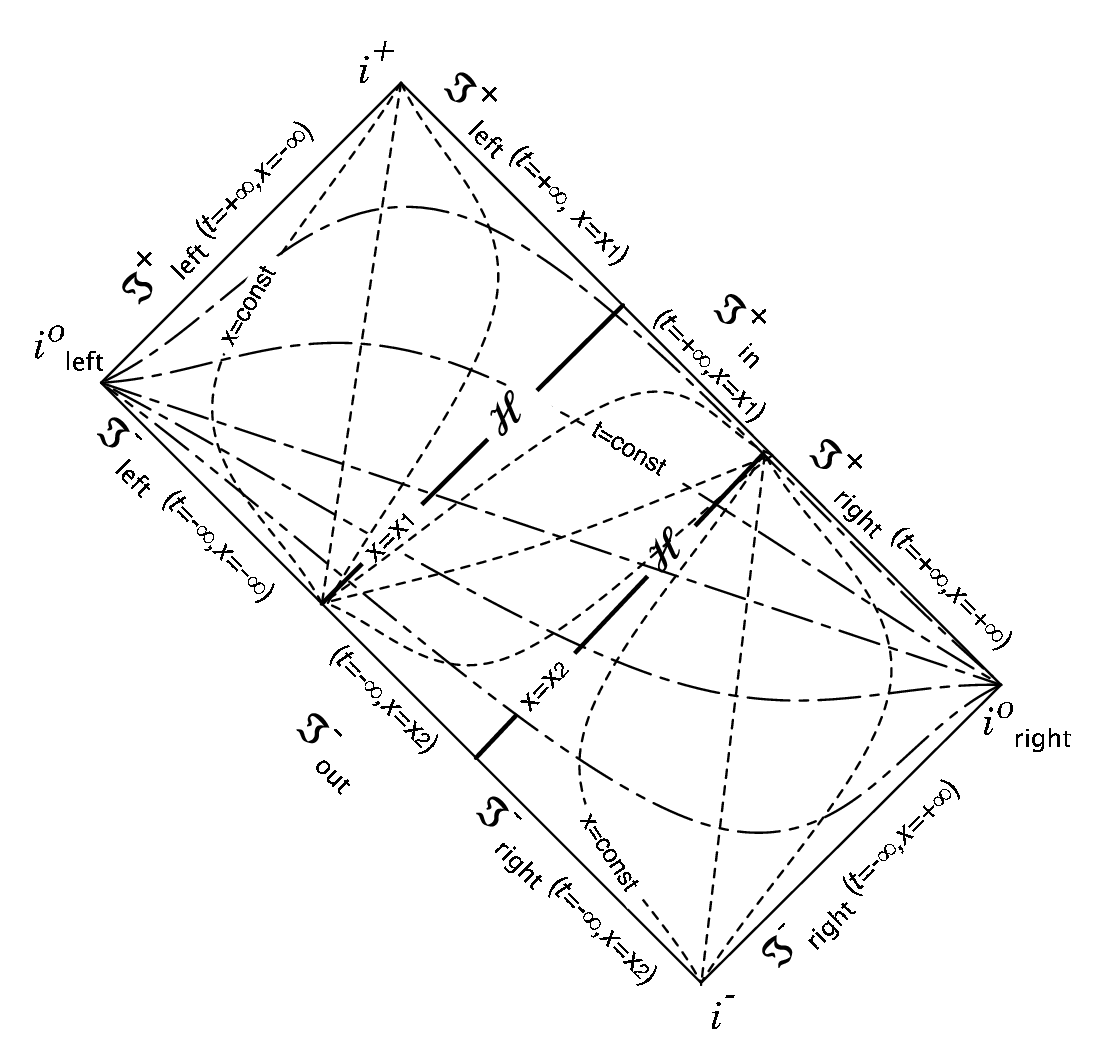}}
\caption[Acoustic Black-Hole--White-Hole conformal structure]{Conformal
  diagram of an acoustic black-hole--white-hole pair. Note the complete absence of singularities.}
\label{F:conf-bh-wh}
\end{figure}

\paragraph*{Acoustic black-hole--white-hole pair:\;}
The acoustic geometry for a black-hole--white-hole
combination again has no singularities in the fluid flow, and no
singularities in the spacetime curvature. In particular, from
Fig.~\ref{F:conf-bh-wh}, we note the complete absence of
singularities. Further discussion about several other acoustic black holes maximal analytic extensions can be found in~\citet{Barcelo:2004wz}.

\subsection{Example: Cosmological metrics}
\label{SS:cosmological}

In a cosmological framework the key items of interest are the so-called
Friedmann--Robertson--Walker (FRW) geometries, though they should more properly be called the
Friedmann--\Lemaitre--Robertson--Walker (FLRW) geometries. The
simulation of such geometries has been considered in various works such
as \citet{Barcelo:2003et, Barcelo:2003wu, Calzetta:2002jz,
  Calzetta:2002zz, Fedichev:2003id, Fedichev:2003bv, Fedichev:2003dj,
  Jain:2007gg, Lidsey:2003ze, Weinfurtner:2004in, Weinfurtner:2004mu,
  Weinfurtner:2007br, Weinfurtner:2008ns, Weinfurtner:2008if} with a
specific view to enhancing our understanding of ``cosmological
particle production'' driven by the expansion of the universe. 

Essentially there are two ways to use the acoustic metric, written as 
\begin{eqnarray}
\d s^2={\rho \over \csound}
\left[-(\csound^2- v^2)\,
\d t^2-2\vbf \cdot \mathbf{dx} \, \d t +{\mathbf{dx}}^2 \right],
\label{fluid-geometry}
\end{eqnarray}
to reproduce
cosmological spacetimes. One is based on physical explosion, the other
on rapid variations in the ``effective speed of light''.

\subsubsection{Explosion}

We can either let the explosion take place more or less spherically
symmetrically, or through a pancake-like configuration, or through a
cigar-like configuration. 

\paragraph*{Three-dimensional explosion:\;}

Following the cosmological ideas implemented by  \citet{Barcelo:2003et,
  Barcelo:2003wu, Calzetta:2002jz, Calzetta:2002zz, Fedichev:2003bv,
  Weinfurtner:2004in, Weinfurtner:2004mu, Weinfurtner:2007br}, and the
BEC technologies described in \citet{Castin:1996, Kagan:1996,
  Kagan:1997tc}, one can take a homogeneous system $\rho(t)$,
$\csound(t)$ and a radial profile for the velocity $\vbf=(\dot b/b)
{\mathbf r}$, with $b$ a scale factor depending only on $t$. (This is
actually very similar to the situation in models for Newtonian
cosmology, where position is simply related to velocity via ``time of
flight''.) Then, defining a new radial coordinate as $r_b=r/b$ the
metric can be expressed as
\begin{eqnarray}
\d s^2={\rho \over \csound}
\left[-\csound^2\,\d t^2+b^2(\d r_b^2+r_b^2\; \d\Omega_2^2)\right].
\label{fluid-geometry2}
\end{eqnarray}
Introducing a Hubble-like parameter,
\begin{equation}
H_b(t) = {\dot b(t)\over b(t)}, 
\end{equation}
the equation of continuity can be written as 
\begin{equation}
\dot \rho + 3 H_b(t) \; \rho = 0; 
\qquad \Rightarrow \qquad 
\rho(t)= {\rho_0\over b^3(t)},
\label{eq:rho}
\end{equation}
with $\rho_0$ constant. Finally, we arrive at the metric of a spatially-flat FLRW
geometry
\begin{eqnarray}
\d s^2=-T^2(t)\; \d t^2+a_s^2(t) \; (\d r_b^2+r_b^2\;\d\Omega_2^2),
\end{eqnarray}
with
\begin{eqnarray}
T(t)\equiv \sqrt{\rho \, \csound};\qquad a_s(t) \equiv \sqrt{\rho \over \csound} \;b. 
\end{eqnarray}
The proper Friedmann time, $\tau$, is related to the laboratory time,
$t$, by 
\begin{equation}
\tau=\int T(t) \;\d t.
\end{equation}
Then, 
\begin{eqnarray}
\d s^2=-\d \tau^2+a_s^2(\tau) \; (\d r_b^2+r_b^2\;\d\Omega_2^2).
\end{eqnarray}
The ``physical'' Hubble parameter is
\begin{equation}
H = {1\over a_s} \, {d a_s\over d\tau}. 
\end{equation}
If one now wishes to specifically mimic de Sitter
expansion, then we would make
\begin{equation}
a_s(\tau) = a_0 \; \exp( H_0 \tau).
\end{equation}
Whether or not this can be arranged (in this explosive model with
comoving coordinates) depends on the specific equation of state (which
is implicitly hidden in $c_s(t)$) and the dynamics of the explosion
(encoded in $b(t)$).

\paragraph*{Two-dimensional explosion:\;}
By holding the trap constant in the $z$ direction, and allowing the
BEC to expand in a pancake in the $x$ and $y$ directions (now best
relabeled as $r$ and $\phi$) one can in principle arrange
\begin{eqnarray}
\d s^2={\rho(x,t) \over \csound(x,t)}
\left[-\{\csound^2(x,t)- v(r,t)^2\}\,
\d t^2-2v(r,t) \d r \, \d t +\d r^2 + r^2 \d\phi^2 + \d z^2 \right].
\label{flrw-2D}
\end{eqnarray}

\paragraph*{One-dimensional explosion:\;}
An alternative ``explosive'' route to FLRW cosmology is to take a long
thin cigar-shaped BEC and let it expand along its axis, while keeping
it trapped in the transverse directions \citep{Fedichev:2003dj,
  Fedichev:2003id}. The relevant acoustic metric is now
\begin{eqnarray}
\d s^2={\rho(x,t) \over \csound(x,t)}
\left[-\{\csound^2(x,t)- v(x,t)^2\}\,
\d t^2-2v(x,t) \d x \, \d t +\d x^2 + \{ \d y^2 + \d z^2\} \right].
\label{flrw-1D}
\end{eqnarray}
The virtue of this situation is that one is keeping the condensate
under much better control and has a simpler dimensionally-reduced
problem to analyze. (Note that the true physics is 3+1 dimensional,
albeit squeezed along two directions, so the conformal factor
multiplying the acoustic metric is that appropriate to 3+1 dimensions.
See also Sect.~\ref{S:dimensionality}.)

\subsubsection{Varying the effective speed of light}

The other avenue starts from a fluid at rest $v=0$ with respect to the 
laboratory at all times:
\begin{eqnarray}
\d s^2=-\rho \,\csound \; \d t^2+{\rho \over \csound} \; 
{\mathbf{dx}}^2.
\end{eqnarray}
Now it is not difficult to imagine a situation in which $\rho$ remains
spatially and temporally constant, in a sufficiently large region of
space, while the speed of sound decreases with time (e.g., we shall
see that this can be made in analogue models based on Bose--Einstein
condensates by changing with time the value of the scattering
length \citep{Barcelo:2003et, Barcelo:2003wu, Jain:2007gg,
  Weinfurtner:2007br, Weinfurtner:2008ns, Weinfurtner:2008if}). This
again reproduces an expanding spatially-flat FLRW Universe. The proper
Friedmann time, $\tau$, is again related to the laboratory time, $t$,
by 
\begin{equation}
\tau=\int \sqrt{\rho \, \csound} \;\d t.
\end{equation}
Then, defining $a_s(t) \equiv \sqrt{\rho/ \csound}$, we have
\begin{eqnarray}
\d s^2=-\d \tau^2+a_s^2(\tau) \; (\d r_b^2+r_b^2\;\d\Omega_2^2).
\end{eqnarray}
If one now specifically wishes to specifically mimic de~Sitter
expansion then we would wish
\begin{equation}
a_s(\tau) = a_0 \; \exp( H_0 \tau).
\end{equation}
Whether or not this can be arranged (now in this non-explosive model
with tuneable speed of sound) depends on the specific manner in which
one tunes the speed of sound as a function of laboratory time.

\subsection{Generalizing the physical model}

There are a large number of ways in which the present particularly-simple analogue model can be generalised. Obvious issues within the
current physical framework are:
\begin{itemize}
\item Adding external forces.
\item Working in truly (1+1) or (2+1) dimensional systems.
\item Adding vorticity, to go beyond the irrotational constraint.
\end{itemize}
Beyond these immediate questions, we could also seek similar effects in other 
physical or mathematical frameworks. 

\subsubsection{External forces}

Adding external forces is (relatively) easy, an early discussion can be found
in \citet{Visser:1997ux} and more details are available
in \citet{Visser:2004zs}. The key point is that with an external force
one can to some extent shape the background flow (see for example the
discussion in \citet{Giovanazzi:2004pm}). However, upon linearization, the
fluctuations are insensitive to any external force.

\subsubsection{The role of dimension}
\label{S:dimensionality}

The role of spacetime dimension in these acoustic geometries is
sometimes a bit surprising and potentially confusing. This is important
because there is a real physical distinction, for instance, between
truly (2+1)-dimensional systems and effectively (2+1)-dimensional
systems in the form of (3+1)-dimensional systems with cylindrical (pancake-like)
symmetry. Similarly, there is a real physical distinction between a
truly (1+1)-dimensional system and a (3+1)-dimensional system with
transverse (cigar-like) symmetry. We emphasise that in Cartesian
coordinates the wave equation
\begin{equation} 
\label{wavef_eq}
\frac{\partial}{\partial {x^{\mu}}} 
\left( f^{\mu \nu} \frac{\partial}{\partial {x^{\nu}}}\; \phi \right)=0,
\end{equation}
where
\begin{equation}
f^{\mu \nu}=\left[
\begin{array}{c|c}
-{\rho}/{c_s^2}          & -{\rho}\;{v^j} /{c_s^2}\\
\hline
-{\rho}\; {v^i}/{c_s^2}  & \rho \;\{\delta^{ij} - v^i v^j /{c_s^2}\}
\end{array}
\right],
\end{equation}
holds \emph{independent} of the dimensionality of spacetime. It
depends only on the Euler equation, the continuity equation, a
barotropic equation of state, and the assumption of irrotational
flow \citep{Unruh:1981cg, Visser:1993ub, Visser:1998qn, Visser:1997ux}.

Introducing the inverse acoustic metric $g^{\mu \nu}$, defined by
\begin{equation}
f^{\mu \nu}=\sqrt{-g} \, g^{\mu \nu}; 
\qquad
g=\frac{1}{\det(g^{\mu \nu})},
\end{equation} 
the wave Eq.~(\ref{wavef_eq}) corresponds to the d'Alembertian wave
equation in a curved space-time with contravariant metric tensor:
\begin{equation}
g^{\mu \nu}=\left(\rho\over c_s\right)^{-2/(d-1)}
\left[
\begin{array}{c|c}
-1/c_s^2            & -\vbf^{\,T}/c_s^2 \\
\hline
\vphantom{\Big|}
- \vbf/c_s^2  & 
\mathbf{I}_{d\times d} - \vbf\otimes \vbf^{\,T}/c_s^2
\end{array}
\right],
\end{equation}
where $d$ is the dimension of \emph{space} (not spacetime).
In $d$ space dimensions the covariant acoustic metric is then
\begin{equation}
g_{\mu \nu}=\left( \rho\over c_s \right)^{2/(d-1)}
\left[
\begin{array}{c|c}
-\left( c_s^2-v^2 \right) & -\vbf^{\,T} \\
\hline
-\vbf                     & \mathbf{I}_{d\times d}
\end{array}
\right]. 
\end{equation}
The physically interesting specific cases are:
\begin{description}
\item[\textit{d = 3:}]
The acoustic line element for three space and one time dimension reads
\begin{equation} 
g_{\mu \nu}=\left( \rho\over c_s \right)
\left[
\begin{array}{c|c}
-\left( c_s^2-v^2 \right) & -\vbf^{\,T} \\
\hline
-\vbf                     & \mathbf{I}_{3\times 3}
\end{array}
\right]. 
\end{equation}

\item[\textit{d = 2:}]
The acoustic line element for two space and one time dimension reads
\begin{equation} 
g_{\mu \nu}=\left( \rho\over c_s \right)^2
\left[
\begin{array}{c|c}
-\left( c_s^2-v^2 \right) & - \vbf^{\,T} \\
\hline
-\vbf                     & \mathbf{I}_{2 \times 2}
\end{array}
\right]. 
\end{equation}
This situation would be appropriate, for instance, when dealing with
surface waves or excitations confined to a particular substrate.

\item[\textit{d = 1:}]
The naive form of the acoustic metric in (1+1) dimensions is
ill-defined, because the conformal factor is raised to a formally
infinite power. This is a side effect of the well-known conformal
invariance of the Laplacian in 2 dimensions. The wave equation in terms
of the densitised inverse metric $f^{\mu\nu}$ continues to make good
sense; it is only the step from $f^{\mu\nu}$ to the effective metric
that breaks down. Acoustics in intrinsically (1+1) dimensional systems
does not reproduce the conformally-invariant wave equation in (1+1)
dimensions.  

Note that this issue only presents a difficulty for physical systems
that are \emph{intrinsically} one-dimensional. A three-dimensional system with
plane symmetry, or a two-dimensional system with line symmetry, provides
a perfectly well-behaved model for (1+1) dimensions, as in the cases
$d=3$ and $d=2$ above.

\end{description}

\subsubsection{Adding vorticity}

For the preceding analysis to hold, it is necessary and sufficient that
the flow locally be vorticity free, $\bnabla \times \vbf = 0$, so
that velocity potentials exist on an atlas of open patches. Note that
the irrotational condition is automatically satisfied for the
superfluid component of physical superfluids. (This point has been
emphasised by \citealt{Comer}, who has also pointed out that in
superfluids there will be multiple acoustic metrics -- and multiple
acoustic horizons -- corresponding to first and second sound.) Even for
normal fluids, vorticity-free flows are common, especially in situations
of high symmetry. Furthermore, the previous condition enables us to
handle vortex filaments, where the vorticity is concentrated into a thin
vortex core, provided we do not attempt to probe the vortex core itself.
It is not necessary for the velocity potential $\phi$ to be globally
defined.

Though physically important, dealing with situations of distributed vorticity
is much more difficult, 
and the relevant wave equation is more complicated in that the velocity
scalar is now insufficient to completely characterise the fluid
flow.\footnote{Vorticity is automatically generated, for instance,
whenever the background fluid is non-barotropic, and, in particular, when
$\nabla \rho \times \nabla p \neq 0$. Furthermore, it has been argued
in \citet{Schutzhold:2005ex} that quantum backreaction can also act as a
source for vorticity.} 
An approach similar to the spirit of the present discussion, but in terms
of Clebsch potentials, can be found in \citet{PerezBergliaffa:2001nd}.
The eikonal approximation (geometrical acoustics) leads to the same
conformal class of metrics previously discussed, but in the realm of
physical acoustics the wave equation is considerably more complicated
than a simple d'Alembertian. More recently, see~\citet{Liberati:2018uev}. (Roughly speaking, 
in a fluid mechanics context the vorticity becomes a
source for the d'Alembertian, while the vorticity evolves in response to
gradients in a generalised scalar potential. This seems to take us
outside the realm of models of direct interest to the general relativity
community.)

However, there are good general relativistic reasons for wanting to introduce at least some form of vorticity into the discussion. All astrophysical black holes rotate to some extent, and the resulting Kerr spacetime exhibits a non-zero shift vector with non-zero vorticity~\citep{Visser:2007fj,Liberati:2018uev,Cropp:2015tua}.

\subsubsection{Simple Lagrangian meta-model}

As a first (and rather broad) example of the very abstract ways in which
the notion of an acoustic metric can be generalised, we start from the
simple observation that irrotational barotropic fluid mechanics can be
described by a Lagrangian, and ask if we can extend the notion of an
acoustic metric to all (or at least some wide class of) Lagrangian
systems?

Indeed, suppose we have a single scalar field $\phi$ whose dynamics is
governed by some generic Lagrangian $\L(\partial_\mu\phi, \phi)$,
which is some arbitrary function of the field and its first
derivatives (here we will follow the notation and ideas
of \citealt{Barcelo:2001ah}). In the general analysis that follows the
previous irrotational and inviscid fluid system is included as a
particular case; the dynamics of the scalar field $\phi$ is now much
more general. We want to consider linearized fluctuations around some
background solution $\phi_0(t,\x)$ of the equations of motion, and to
this end we write
\begin{equation}
\phi(t,\x) = \phi_0(t,\x) + \epsilon \phi_1(t,\x) + 
\frac{\epsilon^2}{2} \phi_2(t,\x) + \O(\epsilon^3).
\end{equation}
Now use this to expand the Lagrangian around the classical solution
$\phi_0(t,\x)$:
\begin{eqnarray}
{\L}(\partial_\mu \phi,\phi) 
&=& 
{\L}(\partial_\mu \phi_0,\phi_0)
\nonumber\\
&+&
\epsilon \left[ 
\frac{\partial \L}{\partial(\partial_\mu \phi)} \; \partial_\mu \phi_1
+
\frac{\partial \L}{\partial\phi} \; \phi_1
\right]
\nonumber\\
&+&
{\epsilon^2\over2} \left[ 
\frac{\partial \L}{\partial(\partial_\mu \phi)} \; \partial_\mu \phi_2
+
\frac{\partial \L}{\partial\phi} \; \phi_2
\right]
\nonumber\\
&+&
\frac{\epsilon^2}{2} \Bigg[ 
\frac{\partial^2 \L}{\partial(\partial_\mu \phi) \; \partial(\partial_\nu \phi)} 
\; \partial_\mu \phi_1 \; \partial_\nu \phi_1
+
2 \frac{\partial^2 \L}{\partial(\partial_\mu \phi)\; \partial \phi} 
\;\partial_\mu \phi_1 \; \phi_1
+
\frac{\partial^2 \L}{\partial\phi\; \partial\phi}\; \phi_1 \; \phi_1
\Bigg]
\nonumber\\
&+&
\O(\epsilon^3).
\nonumber\\
&&
\end{eqnarray}
It is particularly useful to consider the action
\begin{equation}
\label{E:bare-action}
S[\phi] = \int \d^{d+1} x \; \L(\partial_\mu\phi,\phi),
\end{equation}
since doing so allows us to integrate by parts. (Note that the
Lagrangian $\L$ is taken to be a scalar density, not a true scalar.) We
can now use the Euler--Lagrange equations for the background field
\begin{equation}
\partial_\mu \left( \frac{\partial \L}{\partial(\partial_\mu \phi)} \right) 
- \frac{\partial \L}{\partial \phi} = 0,
\end{equation}
to discard the linear terms (remember we are linearizing around a
solution of the equations of motion) and so we get
\begin{eqnarray}
S[\phi] &=& S[\phi_0] 
+
{\epsilon^2\over2}
\int \d^{d+1} x \Bigg[
\left\{
  \frac{\partial^2 \L}{\partial(\partial_\mu \phi) \; \partial(\partial_\nu \phi)} 
   \right\} 
\; \partial_\mu \phi_1 
\; \partial_\nu \phi_1 
\nonumber\\
&& \qquad\qquad\qquad\qquad
+
\left(
\frac{\partial^2 \L}{\partial\phi\; \partial \phi} - 
\partial_\mu \left\{
\frac{\partial^2 \L}{\partial(\partial_\mu \phi)\; \partial \phi} 
\right\}
\right) 
\;
\phi_1 \; \phi_1
\Bigg]
+
\O(\epsilon^3).\;\;\;\;\;\;
\end{eqnarray}
Having set things up this way, the equation of motion for the
linearized fluctuation is now easily read off as
\begin{equation}
\partial_\mu \left(
  \left\{
    \frac{\partial^2 \L}{\partial(\partial_\mu \phi) \; \partial(\partial_\nu \phi)} 
  \right\}
  \partial_\nu \phi_1 \right)
- \left(
  \frac{\partial^2 \L}{\partial\phi\; \partial \phi} - 
  \partial_\mu 
  \left\{
    \frac{\partial^2 \L}{\partial(\partial_\mu \phi)\; \partial \phi} 
  \right\}
\right)
\phi_1
= 0. 
\end{equation}
This is a second-order differential equation with position-dependent
coefficients (these coefficients all being implicit functions of the
background field $\phi_0$). 
This can
be given a nice clean geometrical interpretation in terms of a
d'Alembertian wave equation -- provided we \emph{define} the
effective spacetime metric by
\begin{equation}
\sqrt{-g} \; g^{\mu\nu} \equiv f^{\mu\nu} \equiv 
\left.\left\{
{\partial^2 \L\over\partial(\partial_\mu \phi) \; \partial(\partial_\nu \phi)} 
\right\}\right|_{\phi_0}.
\end{equation}
Note that this is another example of a situation in which calculating 
the inverse metric density is easier than calculating the metric itself.

Suppressing the $\phi_0$ except when necessary for clarity, this
implies [in (d+1) dimensions, d space dimensions plus 1 time
dimension]
\begin{equation}
(-g)^{(d-1)/2} = -\det  
\left\{ 
{\partial^2 \L\over\partial(\partial_\mu \phi) \; \partial(\partial_\nu \phi)} 
\right\}.
\end{equation}
Therefore,
\begin{equation}
 g^{\mu\nu}(\phi_0) = 
\left.
\left(
- \det  
\left\{
{\partial^2 \L\over\partial(\partial_\mu \phi) \; \partial(\partial_\nu \phi)} 
\right\}
\right)^{-1/(d-1)}
\right|_{\phi_0}
\;
\left.
\left\{
{\partial^2 \L\over\partial(\partial_\mu \phi) \; \partial(\partial_\nu \phi)} 
\right\}
\right|_{\phi_0}.
\end{equation}
And, taking the inverse,
\begin{equation}
g_{\mu\nu}(\phi_0) =  
\left.
\left( - 
\det\left\{
{\partial^2 \L\over\partial(\partial_\mu \phi) \; \partial(\partial_\nu \phi)} 
\right\}
\right)^{1/(d-1)}
\right|_{\phi_0}
 \; \; 
\left.
\left\{
{\partial^2 \L\over\partial(\partial_\mu \phi) \; \partial(\partial_\nu \phi)} 
\right\}^{-1}
\right|_{\phi_0}.
\end{equation}
We can now write the equation of motion for the linearized
fluctuations in the geometrical form
\begin{equation}
\label{E:geometrical}
\left[\Delta(g(\phi_0)) - V(\phi_0)\right] \phi_1 = 0,
\end{equation}
where $\Delta$ is the d'Alembertian operator associated with the
effective metric $g(\phi_0)$, and $V(\phi_0)$ is the
background-field-dependent (and so, in general, position-dependent) ``mass term'':
\begin{equation}
V (\phi_0)=
{1\over\sqrt{-g}} \;
\left(
{\partial^2 \L\over \partial\phi\; \partial \phi} - 
\partial_\mu 
\left\{
{\partial^2 \L\over \partial(\partial_\mu \phi)\; \partial \phi} 
\right\}
\right)
\end{equation}
That is,
\begin{equation}
V(\phi_0) =
 \left(
- \det  
\left\{
{\partial^2 \L\over\partial(\partial_\mu \phi) \; \partial(\partial_\nu \phi)} 
\right\}
\right)^{-1/(d-1)}
\times
\left(
{\partial^2 \L\over \partial\phi\; \partial \phi} - 
\partial_\mu 
\left\{
{\partial^2 \L \over \partial(\partial_\mu \phi)\; \partial \phi} 
\right\}
\right). 
\end{equation}
Thus, $V(\phi_0)$ is a true scalar (not a density). Note that the
differential Eq.~(\ref{E:geometrical}) is automatically formally
self-adjoint (with respect to the measure $\sqrt{-g} \; \d^{d+1}x$).

It is important to realise just how general the result is (and where the
limitations are): It works for \emph{any} Lagrangian depending only on a
single scalar field and its first derivatives. The linearized PDE will
be \emph{hyperbolic} (and so the linearized equations will have
wave-like solutions) if and only if the effective metric $g_{\mu\nu}$
has Lorentzian signature $\pm[-,+^d]$. Observe that if the Lagrangian
contains nontrivial second derivatives you should not be too surprised
to see terms beyond the d'Alembertian showing up in the linearized
equations of motion.

As a specific example of the appearance of effective metrics due to
Lagrangian dynamics we reiterate the fact that inviscid irrotational
barotropic hydrodynamics naturally falls into this scheme (which is why,
with hindsight, the derivation of the acoustic metric presented earlier
in this review was so relatively straightforward). In inviscid
irrotational barotropic hydrodynamics the lack of viscosity
(dissipation) guarantees the existence of a Lagrangian; which a priori
could depend on several fields. Since the flow is irrotational $\vbf =
 -\bnabla \phi$ is a function only of the velocity
potential, and the Lagrangian is a function only of this potential and
the density. Finally, the equation of state can be used to eliminate the
density leading to a Lagrangian that is a function only of the single
field $\phi$ and its derivatives \citep{Barcelo:2001ah}.

\subsection{Going further}
\label{SS:going-further-2}

The class of analogue models based on fluid mechanics is now quite large
and the literature is extensive. Most of the relevant detailed discussion will be
deferred until subsequent sections, so for the time being we shall just
mention reasonably immediate generalizations such as:

\begin{itemize}
\item Working with specific fluids.
  \begin{itemize}
  \item Superfluids.
  \item Bose--Einstein condensates.
  \end{itemize}
\item Abstract generalizations.
  \begin{itemize}
  \item Normal modes in generic systems.
  \item Multiple signal speeds.
  \end{itemize}
\end{itemize}
We next turn to a brief historical discussion, seeking to place the 
work of the last two decades into its proper historical perspective.

\clearpage
\section{History}
\label{S:history}

From the perspective of the general relativity community, the history of analogue models can be conveniently (though somewhat superficially) divided into several historical periods:
\begin{itemize}
\item Historical (pre-1981).
\item Classical (1981--2002).
\item Early modern (2003--2008).
\item Late modern (2009--2016).
\item Contemporary (2017--2024).
\end{itemize}
In the next section (Sect.~\ref{sec:historical-period}), we will focus more precisely on the early history of analogue models, specifically on those works that appear to have a direct historical connection with the significant developments of the last 40 years.

We shall then divide the post-1981 period into ``classical'' (1981-2002), ``early modern'' (2003-2008),  ``late modern'' (2009--2016) and ``contemporary'' (2017-2024), indicating a number of key theoretical and experimental milestones that can be used to justify these sub-divisions.
The large number of works, especially in the more recent of these periods, shows the current vitality of the field. 
Herein, the reader will just find a rapid introduction to the literature divided by topics. 
Many references will be mentioned again in more detail in other sections.

\subsection{Historical period (pre-1981)}
\label{sec:historical-period}

Of course, the division into pre-1981 and post-1981 articles is at a
deeper level somewhat deceptive. There have been several analogue models
investigated over the years, with different aims, different levels of
sophistication, and ultimately different levels of development. Armed
with a good library and some hindsight it is possible to find
interesting analogues in a number of places.\footnote{Indeed,
historically, though not of direct relevance to general relativity, analogue
models played a key role in the development of electromagnetism -- Maxwell's
derivation of his equations for the electromagnetic field was guided by a
rather complicated ``analogue model'' in terms of spinning vortices of
aether. Of course, once you have the equations in hand you can treat them in
their own right and forget the model that guided you -- which is exactly
what happened in this particular case.}

\subsubsection{Optics -- the Gordon metric} 

The first paper to seriously discuss analogue models and effective metric techniques was likely written by Gordon (\citeyear{Gordon}).  Yes, this is the same Gordon as of the Klein-Gordon equation. Gordon appeared to be primarily interested in describing dielectric media through an ``effective metric". In other words, he aimed to use a gravitational field to mimic the properties of a dielectric medium. The expression now commonly known as the Gordon metric is
\begin{equation}
[g_\mathrm{effective}]_{\mu\nu}=\eta_{\mu \nu} +\left[1- n^{-2}\right] V_{\mu} V_{\nu}, 
\end{equation}
where $\eta_{\mu\nu}$ is the flat Minkowski metric, $n$ is the
position-dependent refractive index, and $V_\mu$ is the  (constant)
4-velocity of the medium. 

After Gordon's contribution, there was sporadic interest in effective metric techniques.
An historically-important contribution was one of the \emph{problems}
in the well-known book ``The Classical Theory of Fields'' by
\citet{Landau:Lifshitz}. See the end of Chapt.~10, paragraph~90, and the
problem immediately thereafter: ``Equations of
electrodynamics in the presence of a gravitational field''. Note that in
contrast to Gordon, here the interest is in using dielectric media to
mimic a gravitational field.

In France, the idea was taken up by \citet{Pham-Mau-Quan},
who showed that (under certain conditions) Maxwell's equations can be
expressed directly in terms of the effective metric specified by the
coefficients
\begin{equation}
[g_\mathrm{effective}]_{\mu\nu}=g_{\mu \nu} +\left[1- {1\over\epsilon\mu}\right]
V_{\mu} V_{\nu}, 
\end{equation}
where $g_{\mu\nu}$ is the ordinary spacetime metric, $\epsilon$ and $\mu$ are 
the permeability and permittivity, and $V_\mu$ is the 4-velocity of the medium.
The trajectories of the electromagnetic rays are interpreted in this case as 
geodesics of null length of this new effective metric.

Three articles that directly used the dielectric analogy to analyse
specific physics problems are those of \citet{Skrotskii},
\citet{Balazs}, and \citet{Winterberg}. The general
formalism was more fully developed in articles such as those by
\citet{Plebanski, Plebanski2}, and a good summary of this
classical period can be found in the article by \citet{deFelice}.
In summary and with the benefit of hindsight:
An arbitrary gravitational field can always be represented as an equivalent
optical medium, but subject to the somewhat unphysical restriction that 
\begin{equation}
\hbox{ [magnetic permitivity]} \propto \hbox{[electric permeability].}
\end{equation}
If an optical medium does not satisfy this constraint (with a position
independent proportionality constant) then it is not completely equivalent
to a gravitational field. 
For a position dependent proportionality constant complete equivalence can be
established in the geometric optics limit, but for wave optics the equivalence
is not complete. 

The Gordon metric, and its generalizations, have now led to a major surge of theoretical and experimental interest in meta-materials, 
to the extent that relevant research within the condensed-matter/optics community and within the general relativity community is now largely pursued independently. 
Researchers in this specific area would be well advised to check out both sets of (largely independent) literature.

Subsequently, \citet{1975ApJ...202..454A}
extended and modified the notion of the Gordon metric to allow the
medium to be a flowing fluid -- so the 4-vector $V^a$ is no longer
assumed to be a constant. In hindsight, using modern notation and
working in full generality, the Gordon metric can be justified as
follows: Consider an arbitrary curved spacetime with physical metric
$g_{ab}$ containing a fluid of 4-velocity $V^a$ and refractive index
$n$. Pick a point in the manifold and adopt Gaussian normal
coordinates around that point so that $g_{ab} \to \eta_{ab}$. Now
perform a Lorentz transformation to go to to a local inertial frame
co-moving with the fluid, so that $V^a \to (1;\vec 0)$ and one is now
in Gaussian normal coordinates co-moving with the fluid. In this
coordinate patch, the light rays, by definition of the refractive
index, locally propagate along the light cones
\begin{equation}
- {1\over n^2} \; \d t^2 + ||\d \vec x||^2 = 0,
\end{equation}
implying in these special coordinates the existence of an optical
metric
\begin{equation}
{\cal G}_{\mu\nu} \propto \left[\begin{array}{cccc}-{1/ n^2}&0&0&0\\0&1&0&0\\0&0&1&0\\0&0&0&1\end{array} \right].
\end{equation}
That is, transforming back to arbitrary coordinates:
\begin{equation}
{\cal G}_{\mu\nu} \propto -{1\over n^2} V_\mu\, V_\nu + \left\{ g_{\mu\nu} + V_\mu\, V_\nu\right\} 
\propto g_{\mu\nu} + \left\{1-{1\over n^2}\right\} V_\mu\, V_\nu.
\end{equation}
Note that in the ray optics limit, because one only has the light
cones to work with, one can neither derive (nor is it even meaningful
to specify) the overall conformal factor.

\subsubsection{Acoustics} 

After the pioneering hydrodynamical paper by \citet{white:1700}, which studied
acoustic ray tracing in non-relativistic moving fluids, there were several papers
in the 1980s using an acoustic analogy to investigate the propagation of shockwaves
in astrophysical situations, most notably those of
\citet{Moncrief} and \citet{Matarrese, Matarrese2, Matarrese3}.  
In particular, in \citet{Moncrief} the linear perturbations of a
relativistic perfect fluid on a Schwarzschild background were
studied, and it was shown that the wave equation for such perturbations can be
expressed as a relativistic wave equation on some effective (acoustic) metric
(which can admit acoustic horizons). In this sense, \citet{Moncrief} can be
seen as a precursor to the later works on acoustic geometries and acoustic
horizons. Indeed, because they additionally permit a general
relativistic Schwarzschild background, the results of
\citet{Moncrief} are, in some sense, more general than those
considered in the mainstream acoustic gravity papers that followed.

In spite of these impressive results, we consider these papers to be part of
the ``historical period'' for the main reason that such works are
philosophically orthogonal to modern developments in analogue gravity. 
Indeed the main motivation for such works was the study of perfect fluid
dynamics in accretion flows around black holes, or in cosmological expansion,
and in this context the description via an acoustic effective background was
just a tool in order to derive results concerning conservation laws and
stability. This is probably why, even if temporally, \citet{Moncrief}
pre-dates Unruh's 1981 paper by one year, and while \citet{Matarrese,
  Matarrese2, Matarrese3} post-date Unruh's 1981 paper by a few years,
there seems to have not been any cross-connection. 

\subsubsection{Surface waves} 

Somewhat ironically, 1983 marked the appearance of some purely
experimental results on surface waves in water obtained by 
\citet{badulin:782}. At the time these results passed unremarked
by the relativity community, but they are now of increasing interest,
and are seen to be precursors of the theoretical work reported
in \citet{Schutzhold:2002rf,Rousseaux:2010md} and the modern
experimental work reported in \citet{Rousseaux:2007is, Weinfurtner:2010nu}.
 
\clearpage
\subsection{Classical period (1981--2002)}

We shall denote the ``classical'' period as that period between the 1981 publication of Unruh's paper ``Experimental black hole evaporation'' \citep{Unruh:1981cg}, and the 2002 publication of  the book ``Artificial Black Holes'' \citep{Novello:2002qg}.
While progress in the early part of this period was relatively sedate, by the year 2000, articles on one or another aspect of analogue gravity
were appearing at the rate of over 20 per year, and it becomes
impractical to summarise more than a selection of them. Note that some of the individual sub-sections in the discussion below will of necessity overlap.

\subsubsection{Dumb holes}

The key event in the ``Early Modern'' period (though largely unrecognised at
the time) was the 1981 publication of Unruh's paper ``Experimental black
hole evaporation'' \citep{Unruh:1981cg}, which implemented an analogue
model based on fluid flow, and then used the power of that analogy to
probe fundamental issues regarding Hawking radiation from ``real''
general-relativistic black holes.

We believe that Unruh's 1981 article represents the first observation
of the now widely established fact that Hawking radiation has nothing to
do with general relativity \emph{per se}, but that Hawking radiation is
instead a fundamental curved-space quantum field theory phenomenon that
occurs whenever a horizon is present in an effective
geometry.\footnote{We emphasise: To get Hawking radiation you need an
effective geometry, a horizon, and a suitable quantum field theory on
that geometry.} Though Unruh's 1981 paper was seminal in this regard, it
lay largely unnoticed for many years.

Some 10 years later Jacobson's article ``Black-hole evaporation and
ultra-short distances'' \citep{Jacobson:1991gr} used Unruh's analogy to
build a physical model for the ``trans-Planckian modes'' believed to be
relevant to the Hawking radiation process. Progress then sped up with
the relatively rapid appearance of \citet{Jacobson:1993hn}
and \citet{Unruh:1995je}. (This period also saw the
independent rediscovery of the fluid analogue model by one of the
present authors \citep{Visser:1993ub}.)
More precise formulations of the notions of horizon, ergosphere, and surface
gravity in analogue models were developed in \citet{Visser:1998qn,Visser:1997ux}, and
discussions of the implications of analogue models regarding
Bekenstein--Hawking entropy appeared in \citet{Visser:1998qn,Visser:1997yu}.
Finally, analogue spacetimes based on \emph{special} relativistic
acoustics (not non-relativistic acoustics) were considered in \citet{Bilic:1999sq}.
Work on ``near-horizon'' physics appeared in \citet{Fischer:2000hg}.
Vorticity was discussed in \citet{PerezBergliaffa:2001nd}.  
Acoustics in
an irrotational vortex were investigated in \citet{Fischer:2001jz}.
The stability of an acoustic white hole was investigated
in \citet{Leonhardt:2002yu}.

\subsubsection{Superfluids}

The first explicit consideration of superfluids in this regard was due to \citet{Comer}. 
This ``classical'' period also saw the
introduction of the more general class of superfluid models considered
by Volovik and his collaborators \citep{Volovik:1995ja, Volovik:1996he,
Kopnin:1997jy, Eltsov:1998fv, Volovik:1998de, Volovik:1998pf2,
Volovik:1998pf, Jacobson:1998he, Volovik:1999fc, Volovik:1999cn}.
2001  also marked the appearance of a review article on superfluid
analogues \citep{Volovik:2000ua}.
The excitation spectrum in superfluids, specifically the fermion zero
modes, were investigated in \citet{Volovik:2001rx,Huhtala:2001cj}, while
the relationship between rotational friction in superfluids and
super-radiance in rotating spacetimes was discussed
in \citet{Calogeracos:2001pw}.
 Specifically $^{3}$He-A based models were considered
in \citet{Jacobson:2000gs,Volovik:1999mj}. 
The most radical proposal to appear in 2000 was that of 
\citet{Chapline:2000en}. Based on taking a superfluid
analogy rather literally, they mooted an actual physical breakdown of
general relativity at the horizon of a black
hole \citep{Chapline:2000en}.
2001 saw further developments regarding the use of
$^{3}$He-A \citep{Dziarmaga:2001st} as an analogue for electromagnetism.

\subsubsection{Bose--Einstein condensates}

Key developments in 2000 were the introduction, by Garay and
collaborators, of the use of Bose--Einstein condensates as a working
fluid \citep{Garay:1999sk,Garay:2000jj}, and the extension of those ideas
by the present authors \citep{Barcelo:2000tg}. BEC related analogues would prove to become a recurring and increasingly important  theme over subsequent decades. 

\subsubsection{Surface waves in shallow water}

In 2002 Sch\"utzhold and Unruh developed a rather different fluid-based analogue
based on gravity waves in shallow water \citep{Schutzhold:2002rf}. These shallow-water gravity waves will also become a recurring and increasingly important theme in future developments, both theoretical and experimental. 

\subsubsection{Non-linear electrodynamics}

Analogue models based on nonlinear electrodynamics were developed and investigated
in \citet{Baldovin:2000xy} and  \citet{DeLorenci:2001ch}. 
Further developments along these lines  were presented by
Novello and collaborators in \citet{DeLorenci:2002mi, DeLorenci:2002ws,
Novello:2002ed, Novello:2002iq, Fonseca-Barbatti:2002hq}. 
Non-linear electrodynamics has now become a major research area in its own right. 

\subsubsection{Multiple effective metrics and multi-refringence}

The transference of the idea of analogue-inspired ``multiple metric'' theories into cosmology, 
where they can be used as the basis for a precise definition of what is meant by a
VSL (``variable speed of light'') cosmology, was addressed in \citet{Bassett:2000wj}.
Multi-metric theories, in their various incarnations, have now become a major research area 
in their own right.

\subsubsection{Normal modes}

Closer to the heart of the analogue programme were the development of a
``normal mode'' analysis in \citet{Barcelo:2001ah, Barcelo:2001cp} and \citet{Visser:2001fe}. 
This ``normal mode'' analysis sought to establish a general framework for analogue spacetimes whenever one can meaningfully linearize a classical field theory around a well-defined background.

\subsubsection{Optical analogues and slow light}

 Optical models were considered in \citet{Leonhardt:1999fe},  and ``slow light'' models in
quantum dielectrics were considered in \citet{Leonhardt:2000aa,
Leonhardt:2000hf, Leonhardt:2000zp}. More work on ``slow light'' appeared in  \citet{Brevik:2001nf}
 and  in \citet{Fiurasek:2002ay,
  Piwnicki:2002xw}.

\subsubsection{Solid state}

New and rather different contributions arose in the form of the solid-state models
considered by \citet{Reznik:1997ag,Reznik:1997xt}. Somewhat related are the dielectric analogues discussed 
in \citet{Schutzhold:2001fw}.

\subsubsection{Possible Lorentz violations.}

The possible role of Lorentz violations at
ultra-high energy was emphasised in \citet{Jacobson:2001kz}.
Though analogue models lead naturally to the idea of high-energy
violations of Lorentz invariance, it must be stressed that definite
observational evidence for violations of Lorentz invariance is lacking
-- in fact, there are rather strong constraints on how strong any
possible Lorentz violating effect might be \citep{Jacobson:2002hd,
Jacobson:2002ye}.
The use of BECs as a model for the breakdown of Lorentz invariance is discussed  
in \citet{Visser:2001jd}.

\subsubsection{Other analogue models}

The later 1990s saw continued work by Jacobson and his
group \citep{Jacobson:1995ak, Jacobson:1996zs, Corley:1996ar,
  Corley:1997bx, Corley:1998rk, Jacobson:1999ay}. In counterpoint, \citet{Hochberg} is an attempt at
connecting Hawking evaporation with the physics of collapsing
bubbles. This was part of a more general programme aimed at connecting
black-hole thermodynamics with perfect-fluid
thermodynamics \citep{Hochberg:1996mb}. 
Further afield, the
trans-Planckian problem also reared its head in the context of
cosmological inflation, and analogue model ideas previously applied to
Hawking radiation were reused in that context \citep{Kempf:2000ac,
Niemeyer:2000eh}.

Subsequently, 2001 saw more applications of analogue-inspired ideas to
cosmological inflation \citep{Easther:2001fi, Mersini:2001su,
Mersini:2001jt, Kempf:2001fa, Niemeyer:2001qe}, to neutron star
cores \citep{Carter:2001tg},  to the cosmological
constant \citep{Volovik:2001fm, Volovik:2001jw},
and to speculations regarding the possibly
emergent nature of Einstein gravity \citep{Barcelo:2001tb,
Visser:2001fe}.
Some experimental proposals were considered in \citet{Barcelo:2001ca,
Visser:2001fe, Sakagami:2001ph}.

In 2002 ``What did we learn from studying acoustic black holes?'' was the title
and theme of article by \citet{Parentani:2002bd}, while
super-radiance was investigated
in \citet{Basak:2002aw}. The propagation of phonons and
quasiparticles was discussed in \citet{Fischer:2002jn, Fischer:2002ma}.
 Applications to inflationary cosmology were
developed in \citet{Niemeyer:2002kh}, while analogue spacetimes
relevant to braneworld cosmologies were considered
by \citet{Barcelo:2002xw}.

\subsubsection{The Rio de Janeiro workshop}

The workshop on ``Analogue models of general relativity'',
held at CBPF (Rio de Janeiro) in October 2000, gathered some 20 international
participants and greatly stimulated the field, ultimately leading to the
publication (in 2002) of the book ``Artificial Black Holes'' \citep{Novello:2002qg}.
Somewhat arbitrarily we shall take the publication of this book 
as the demarcation point for the end of the classical period.

\clearpage
\subsection{Early Modern period (2003--2008)}

\bigskip

What we shall refer to as the ``Early Modern'' period covers the six years 
dating from the publication of the book ``Artificial Black Holes''~\citep{Novello:2002qg} 
to the appearance in late in late 2007/early 2008 of Rousseaux's water tank experiments~\citep{Rousseaux:2007is}. 

Some of the subsections below describe further developments of work initiated in the ``classical'' period, but other subsections describe significant new developments and new approaches.
During this period one often encounters some 30 or more articles per year appearing on (or closely related to) the topic of analogue models.
The original version of this Living Review appeared during this Early Modern period in May of
2005 \citep{Barcelo:2005fc}, and since then activity has, if anything, increased.

\subsubsection{Acoustics} 

The lecture notes by \citet{Jacobson:2003vx}
give a nice introduction to Hawking radiation and its connection to
analogue spacetimes,
while the
physical realizability of acoustic Hawking radiation was addressed
by \citet{Das:2004cs} and \citet{Vachaspati:2004wn}. 
The causal structure of analogue
spacetimes was considered in \citet{Barcelo:2004wz}, while quasinormal
modes attracted attention in \citet{Berti:2004ju, Lepe:2004kv,
Cardoso:2004fi, Nakano:2004ha}. 
A mini-survey was presented in \citet{Cardoso:2005ij}.

Singularities in the acoustic geometry are
considered in \citet{Cadoni:2005jg}, while backreaction has received
more attention in \citet{Schutzhold:2005ex}. 
 A nice survey
of analogue ideas and backreaction effects was presented
in \citet{Balbinot:2006ua}, and related articles \citep{Balbinot:2004dc, Balbinot:2004da}. 
The background fluid flow was discussed in \citet{Cadoni:2005nh}.
Quasinormal modes again received attention in \citet{Cherubini:2005dh} and \citep{Saavedra:2005ug}.

Back-reaction effects were again considered in \citet{Balbinot:2006cy}.
Quasinormal frequencies were considered
in \citet{Chen:2006zy}. Finally, we emphasise particularly the
realization that the occurrence of Hawking-like radiation does not
require the presence of an event horizon or even a trapped
region \citep{Barcelo:2006uw, Barcelo:2006np}.

Acoustic cross-sections were considered
in \citet{Crispino:2007zz}. ``Rimfall'' was discussed
in \citet{Unruh:2007zza}. The specific shape of the de
Laval nozzle needed to acoustically reproduce linearized perturbations
of the Schwarzschild geometry was discussed in \citet{Abdalla:2007dz},
and quasinormal frequencies were investigated
in \citet{Chen:2006zy} and \citet{Xi:2007yb}.

\subsubsection{Bose--Einstein condensates}
\enlargethispage{20pt}
BEC-based horizons were again considered in \citet{Giovanazzi:2004pm, 
Giovanazzi:2004zv}, while backreaction effects were the focus of
attention in \citet{Balbinot:2004da, Balbinot:2004dc, Kim:2004sf}. 
A proposal for using analogue models to generate massive phonon modes in
BECs was advanced in \citet{Visser:2004qp, Visser:2005ss}.
Further work on BEC-related models appearing in 2005 included 
\citep{Liberati:2005id,
  Liberati:2005pr, Weinfurtner:2005by, Visser:2005ss,
  Weinfurtner:2005jd, Uhlmann:2005hf}.

More BEC-related developments appeared in 2006: 
\citet{Weinfurtner:2006wt, Barcelo:2006uw, Barcelo:2006np,
  Barcelo:2006yi, Balbinot:2006qe, Fagnocchi:2006gn, Fagnocchi:2006cz}. 
 The book \citet{, Fischer:2005iy} first appeared as an e-print in 2005 but was published in 2007.
 
\subsubsection{Phase horizons \emph{versus} group horizons}
In dispersive media one can distinguish horizons based on phase velocity from horizons based on group velocity --- both are interesting, albeit for different reasons. 
Vachaspati argued for an analogy between phase
boundaries and acoustic horizons in \citet{Vachaspati:2003rb}. 
Analogue-inspired ideas regarding the
possible ``localization'' of the origin of the Hawking flux were
investigated in \citet{Unruh:2007zz}. 

\subsubsection{Possible Lorentz violations}

We mention the development of yet more strong observational
bounds on possible ultra-high-energy Lorentz
violation \citep{Jacobson:2004rj, Jacobson:2004qt}.
We also mention the appearance in 2005 of
another Living Review, one that summarises and systematises the very
stringent bounds that have been developed on possible
ultra-high-energy Lorentz violation \citep{Mattingly:2005re}.
 Dissipation-induced breakdown of Lorentz invariance was
considered by Parentani in \citet{Parentani:2007uq, Parentani:2007dw},
and BEC-based models continued to attract
attention \citep{Weinfurtner:2007br, Barcelo:2007ru, Barcelo:2007zz,
Jannes:2007zz}. 

\subsubsection{Super-radiance}

Super-radiance was
further investigated in \citet{Basak:2003uj, Basak:2003yt}, and
in 2005 yet more studies of the super-resonance phenomenon
appeared in references \citep{Basak:2005fv, Federici:2005ty, Kim:2005pb, Savage},
and \citep{Choy:2005qx}.
Superradiance and disclinations were subsequently considered
in \citet{deA.Marques:2007qc}.

\subsubsection{FLRW cosmology and related issues}

2003 saw further discussion of analogue-inspired models for the cosmological constant \citep{Volovik:2003kt}, 
and the related development of analogue models for FLRW
geometries \citep{Fedichev:2003bv, Fedichev:2003id, Barcelo:2003et,
Dumin:2003gs, Lidsey:2003ze}. 
Effective geometries in astrophysics were discussed by \citet{PerezBergliaffa:2004xg}.
More cosmological issues were
raised in \citet{Vachaspati:2004wn,Weinfurtner:2004mu}, while a
specifically astrophysical use of the acoustic analogy was invoked
in \citet{Das:2004gn,Das:2004zm,Das:2004wf}.
More issues
relating to the simulation of FLRW cosmologies were raised
in \citet{Fischer:2004hr,Fischer:2004bf}.

Astrophysical applications to accretion flow were discussed in \citet{Abraham:2005ah,
  Dasgupta:2005hh}.  More analogue-inspired work on
black-hole accretion appeared in \citet{Das:2006an}. 
Applications to the cosmological constant
were considered in \citet{Samuel:2006ga}. 

Analogue-inspired ideas were adapted to the study of gravitational
collapse in \citet{Barcelo:2007yk}, while the importance of nonlocal
correlations in the Hawking flux was emphasized
in \citet{Balbinot:2007de}. 
Additionally, an
analogue inspired analysis of accretion appeared in \citet{Das:2007ix},
while astrophysical constraints on modified dispersion relations were
improved and extended in \citet{Maccione:2007yc, Liberati:2007vx}.

\subsubsection{Effective geometries and emergence}

Effective geometry was the theme in \citet{Novello:2003je}, while
applications of nonlinear electrodynamics (and its effective metric) to
cosmology were presented in \citet{Novello:2003kh}.
Emergent relativity was again addressed in \citet{Laughlin:2003yh}.
The review article by \citet{Burgess:2003jk} emphasised the role
of general relativity as an effective field theory -- the \emph{sine
qua non} for any attempt at interpreting general relativity as an
emergent theory.

Emergent geometry \citep{Babichev:2007wg,
  Babichev:2007zz, Babichev:2007dw, Vikman:2007sj} proved an important theme in 2007, as
were efforts at moving beyond the semiclassical
description \citep{Parentani:2007mb}. 
The use of analogue spacetimes as ``toy models'' for quantum/emergent gravity
was emphasized in \citet{Visser:2007du, Visser:2007nx, Girelli:2008gc, Girelli:2008qp}.

\subsubsection{Bose--Einstein condensates and other superfluids}

There were several further developments
regarding the foundations of BEC-based models in \citet{Barcelo:2003wu,
Fedichev:2003dj}, while analogue spacetimes in superfluid neutron stars
were further investigated in \citet{Carter:2003im}.
In 2006 BEC-based analogue models were
adapted to investigating ``signature change events''
in \citet{Weinfurtner:2007dq}.

\subsubsection{Slow light and optical analogues}

Limitations of the ``slow light'' analogue were explained
in \citet{Unruh:2003ss}. Within the
optics community Philbin, Leonhardt, and co-workers initiated the
study of ``fibre-optic black holes'' \citep{Philbin:2007jj,Philbin:2007ji}.

\subsubsection{Quark matter}

A key article, which appeared in 2006, involved the ``inverse'' use of
the acoustic metric to help understand hydrodynamic fluid flow in
quark-gluon plasma \citep{CasalderreySolana:2006sq}. 
While analyzing quark matter, acoustic metrics were
found to be useful in \citet{Manuel:2007pz} (see
also \citealt{Manuel:2008ri}).

\subsubsection{Finsler spacetime}
\label{S:Randers}

The relationship
between modified dispersion relations and Finsler spacetimes was
discussed in \citet{Girelli:2006fw}.
Gibbons and co-workers used analogue-based ideas in their analysis of
general stationary spacetimes, demonstrating that the spatial slices
of stationary spacetimes are best thought of as a special class of
Finsler spaces, in particular, Randers
spaces \citep{Gibbons:2008zi}. 
Attempts at developing a generally
useful notion of Finsler \emph{spacetime} were discussed
in \citet{Skakala:2008kf, Skakala:2008jp}, with Finslerian applications
to the Higgs mechanism being investigated
in \citet{Sindoni:2007rh}.

\subsubsection{Analogue entropy?}

2003 saw further discussion of analogue-inspired models for black-hole 
entropy \citep{Volovik:2003ga}.
Entanglement entropy was investigated in \citet{Jacobson:2007jx}. 
Analogue implications \emph{vis-\'a-vis}
entanglement entropy were discussed in \citet{Fursaev:2006ng,Fursaev:2006ja}.

\subsubsection{Surface waves in shallow water}

 The connection between white hole horizon and the
classical notion of a ``hydraulic jump'' was explored
in \citet{Volovik:2005ga} and in \citet{2005EPJB...48..417S}, 
while ``ripplons'' (quantized surface capillary waves) were discussed
in \citet{Volovik:2006cz}. Theoretical aspects of the circular
hydraulic jump were investigated in \citet{2007PhLA..371..241R}. 
Within the fluid dynamics community, wave-tank
experiments were initiated \citep{Rousseaux:2007is} by Rousseaux and
co-workers, who demonstrated the presence of negative phase-velocity
waves.

\subsubsection{Quantum graphity }

Using analogue ideas as backdrop, Markopoulou developed a
pre-geometric model for quantum gravity
in \citet{Markopoulou:2006qh}.

\enlargethispage{25pt}
\subsubsection{Other analogue models}

Unruh and Sch\"utzhold discussed the universality of the Hawking
effect \citep{Unruh:2004zk}, and a new proposal for possibly detecting
Hawking radiation in an electromagnetic wave
guide \citep{Schutzhold:2004tv}. Two dimensional analogue models were
considered in \citet{Cadoni:2004my}.
There were attempts at modeling the Kerr geometry \citep{Visser:2004zs}, 
and (e-printed in 2004 but published a decade later) generic ``rotating'' spacetimes \citep{Chapline:2004mu},
and an extension of the usual formalism for
representing weak-field gravitational lensing in terms of an analogue
refractive index \citep{Boonserm:2004wp}.

The Magnus force was reanalysed in terms of the acoustic geometry
in \citet{Zhang:2005va}. 
Dynamical phase transitions were considered in \citet{Schutzhold:2005fh}.
 A ``spacetime condensate'' point of view was advocated
in \citet{Hu:2005ub}, and analogue applications to ``quantum
teleportation'' were considered in \citet{Ge:2005xe}. 

 A microscopic analysis of the micro-theory
underlying acoustic Hawking radiation in a ``piston'' geometry
appeared in \citet{Giovanazzi:2006nd}. 
Volovik extended and explained
his views on quantum hydrodynamics as a model for quantum gravity
in \citet{Volovik:2006ft}. 
 An analogue model based on a suspended
``shoestring'' was explored in \citet{Heyl:2006dv}. 
Quantum field theoretic anomalies were
considered in \citet{Kim:2007ge}, 

Analogues based on ion traps were considered
in \citet{Schutzhold:2007mx}, while a toy model for backreaction was
explored in \citet{Maia:2007vy}. Further afield, analogue models were
used to motivate a ``Abraham--Lorentz'' interpretation of relativity
in terms of a physically-real aether and physically-real
Lorentz--FitzGerald contraction \citep{Barcelo:2007iu}. In a similar
vein analogue models were used to motivate a counter-factual
counter-historical approach to the Bohm versus Copenhagen
interpretations of quantum
physics \citep{Nikolic:2007su}. 
Gravitational collapse was again discussed in \citet{Barcelo:2007yk}. 

Quantum field theoretic anomalies in an
acoustic geometry were considered in \citet{Becar:2008yn} and \citet{Wu:2008yx},
while (2+1) acoustic black-hole thermodynamics were investigated
in \citet{Kim:2008aa}. 

\subsubsection{End of the Early Modern period} 

We define the end of the ``Early Modern'' period by the appearance in late 
in late 2007/early 2008 of Rousseaux's water tank experiments~\citep{Rousseaux:2007is},
{and Philbin et al experiments with fiber optics~\citep{Philbin:2007ji}. }
Prior to this point, the analogue spacetime programme was largely theory driven --- subsequently experimental efforts came to play an increasingly important role.


\clearpage
\subsection{Late Modern period (2008--2016)}

What we shall denote as the Late Modern period beings with the
Rousseaux {and Philbin \emph{et al.}  experiments~\citep{Rousseaux:2007is,Philbin:2007ji}, }
and ends with Steinhauer's BEC-based observation of Hawking radiation and entanglement~\citep{Steinhauer:2015saa}.
Though bracketed by these two experiments, this period also saw an explosion in theoretical (and other experimental) developments.
The second edition of this Living Review \citep{Barcelo:2005fc} appeared in 2010/2011 in the middle of this Late Modern period.

\subsubsection{Dumb holes and acoustics}

Quasinormal modes of dumb holes  were again considered
in \citet{Dolan:2009nk, Zhidenko:2009zx}, with a survey appearing
in \citet{Berti:2009kk}. 
Acoustic scattering was considered
in \citet{Dolan:2009zza}. 
A mini-review was provided in \citet{Schutzhold:2009fic}.
 Relativistic fluids have been
revisited in \citet{Visser:2010xv}.
The acoustic geometry of polytrope
rotating Newtonian stars was considered in \citet{Bini:2010zz}. Random
fluids were investigated in \citet{Krein:2010ee}. 

2+1 dimensional draining-bathtub
geometries were probed in \citet{Oliveira:2010zzb}.
Various theoretical models were explored in~\citet{Hossenfelder:2010zj, Ge:2010eu, Ge:2010wx, Mayoral:2010ck} 
 and \citet{Menicucci:2010xs}.
Various analogues based on ``draining bathtub. vortices'' were studied in \citet{Dolan:2011zza, Anacleto:2011tr} and \citet{Cherubini:2011zza}.
2011 saw another brief review~\citet{Sindoni:2011ej},  and several conference articles \citep{Visser:2011mf,Liberati:2011bp,Finazzi:2011jd}.
In 2112 several abstract analyses of acoustic horizons appeared in \citet{Fleurov:2011rbl, Leigh:2012jv, Bilic:2012yh, Bilic:2012bn}.

Technical issues regarding the de Laval nozzle were addressed in \citet{Cuyubamba:2013iwa}. 
The classical limit of Hawking radiation was discussed in \citet{Weinfurtner:2013zfa}. 
Acoustic ``clouds'' were the subject of \citet{Benone:2014nla}. 
Wave ``blocking'' was specifically addressed in \citet{Euve:2014aga}. 
Acoustic black holes were again explored in \citet{Vieira:2014rva} and \citet{Ananda:2014gga}. 

Adding vorticity to acoustic models was again addressed in \citet{Cropp:2015tua}, while analogue wormholes were explored in \citet{Peloquin:2015rnl}. Acoustic ``frame dragging'' was the subject of \citet{Chakraborty:2015ioa},
while \citet{Hossenfelder:2015pza} investigated planar acoustic black holes.
Polytropic acoustic configurations were discussed in \citet{Hossenfelder:2015pza}. 
Perturbations of the Bernoulli equation were explored in \citet{Datta:2016bgm} and \citet{Shaikh:2016qko}. 
The acoustic Aharanov-Bohm effect was addressed in \citet{Anacleto:2016ukc}. 
Sub-critical flows were the subject of \citet{Coutant:2016vsf}. 
Dynamical instabilities were explored in \citet{Coutant:2016bgk}.

\clearpage
\subsubsection{Horizons and ergo-regions} 

Various differing types of horizon were discussed in~\citet{Cropp:2013sea} and \citet{Cropp:2013zxi}. 
The interplay between Hawking radiation and the WKB approximation was addressed in \citet{Schutzhold:2013mba}
Surface gravities were addressed in \citet{Tarafdar:2013oqa}. 
Infra-red instabilities were the subject of \citet{Busch:2014hla}. 
The physical observability of horizons was explored in \citet{Visser:2014zqa}. 
Dispersion-induced horizon broadening was discussed in \citet{Coutant:2014cwa}. 
The delicate. issue of what exactly constitutes an ``observation'' of Hawking radiation was the subject of \citet{Unruh:2014hua}. 

Universal horizons were discussed in \citet{Michel:2015rsa}, and ``slow sound'' in \citet{Auregan:2015uva}. 
Non-standard horizons were again explored in \cite{Cropp:2016gkn} and in \citet{Cropp:2016teb}. 
De Sitter and anti-de Sitter analogue models were investigated in \citet{Dey:2016khw}. 
Abstract questions regarding thermality were dealt with in \citet{Visser:2014ypa}. 
Stability issues were addressed in \citet{Coutant:2012vff}.
We emphasize the experimental detection of
photons associated with a phase velocity horizon by \citet{Belgiorno:2010xxx}.

\subsubsection{Bose--Einstein condensates}

The BEC theme continued to generate
attention in 2008 \citep{Kurita:2008fb}, in particular regarding cosmological
particle production \citep{Weinfurtner:2008if, Weinfurtner:2008ns}, and
Hawking radiation \citep{Carusotto:2008ep}. A mini-review appeared
in \citet{Schutzhold:2010yf}. 
{The extension of the analogue gravity framework to
relativistic BECs was reported in \citet{Fagnocchi:2010sn} where it was shown that a Gordon-like metric replaces the standard acoustic one for the phonons of the system (as expected in general for relativistic inviscid, irrotational, flows~\citep{Visser:2010xv}). }

BEC based models for signature change events
were again considered in \citet{Weinfurtner:2009cj}.
The BEC paradigm for acoustic geometry is
again discussed in \citet{Jannes:2009yr} and \citet{Recati:2009ya}.
Step-function discontinuities in BECs were considered
in \citet{Fabbri:2010tj,Mayoral:2010un}. (Signature change can be
viewed as an extreme case of step-function
discontinuity \citep{Weinfurtner:2007dq, Weinfurtner:2009cj,
  White:2008xr}.) 
Signature change, now not in a BEC context,
was again addressed in \citet{White:2008xr}. 

Quantum sound in BECs was again investigated in \citet{Barcelo:2010bq,Barcelo:2010zz},
BEC-based particle creation in \citet{Kurita:2010bu}, and BEC-based
black hole lasers in \citet{Finazzi:2010nc}.
Most remarkably, a BEC-based black hole
analogue was experimentally realised in \citet{Lahav:1183480}.  
Then, Steinhauer's programme continued with \citep{Shammass:2012,Schley:2013}
and \citep{Steinhauer:2014dra} where he reported the observation of self-amplified Hawking radiation.
(See Sections~\ref{S:Lahav} and \ref{SS:going-further-7} of this review.)
A BEC based acoustic analogue of the dynamical Casimir effect was realized in \cite{Jaskula:2012ab}.

BECs were again discussed in~\citet{Kuhnel:2013uxa}, \citet{Castellanos:2013ena},
 \citet{Anderson:2013ux},  and \citet{Belenchia:2014hga}. 
One dimensional BECs were specifically addressed in \citet{Anderson:2014jua}. 
Amplification issues were the subject of \citet{Balbinot:2014cfa}. 
Dipolar condensates were explored in \cite{Cha:2016esj}. 
A synthetic Unruh effect was analyzed in \citet{Rodriguez-Laguna:2016kri}.
Trans-sonic flow was again investigated in \citet{Michel:2016tog}. 
Non-classical excitations were the subject of \citet{Finke:2016wcn}.

\subsubsection{Hawking radiation --- generic theoretical aspects}

  The use of correlations as a potential experimental probe has been theoretically
investigated in \citet{Balbinot:2010my, Carusotto:2009re,
  Fagnocchi:2010zz, Parentani:2010bn, Prain:2010zq, Schutzhold:2010ig,
  2010arXiv1005.2645U,Finazzi:2010yq}, while an analysis of optimality conditions for
the detection of Hawking--Unruh radiation appeared
in \citet{Aspachs:2010hh}. 

Possible measurement protocols for Hawking radiation in ionic systems were
discussed in \citet{Horstmann:2010xd}.

\enlargethispage{35pt}
\subsubsection{Possible Lorentz violations}

The theme of nontrivial dispersion relations was revisited in references \citep{
Macher:2009nz, Macher:2009tw, Maccione:2009ju, Liberati:2009pf, Mattingly:2009jf}. 
Modified dispersion relations again
attracted attention \citep{Girelli:2006sc}, 
and analogue inspired ideas
concerning constrained systems were explored in \citet{Konopka:2006sa}.

Localization of the Hawking radiation was again addressed
in \citet{Schutzhold:2008tx}, while sensitivity of the Hawking flux to
the presence of superluminal dispersion was considered
in \citet{Barcelo:2008qe}. 
Astrophysical constraints on modified
dispersion relations were again considered in \citet{Maccione:2008iw, Maccione:2008tq}.
More drastically, the analogue paradigm was used as a framework for studying possible violations of Lorentz invariance~\citep{Maccione:2011fr,Coutant:2011fz,Coutant:2011in, Carmona:2011wc, Baccetti:2011aa,Baccetti:2011us,Skakala:2011aw}.
Possible analogue-inspired violations of Lorentz invariance were also explored in~\citet{Liberati:2012th, Anacleto:2012ba,Tate:2012xd}. 

Foundational aspects of Lorentz invariance were addressed in \citet{Garcia-Meca:2013soa,Garcia-Meca:2013zia}, 
and experimental/observational  bounds were updated in \citet{Liberati:2013xla}.
Analogue-inspired causality constraints were explored in \citet{Liberati:2016brg}. 

\subsubsection{Super-radiance}

Universal aspects of super-radiant scattering are considered
in \citet{Richartz:2009mi}. 
Super-resonance was
again discussed in \citet{Kobes:2006zz}.
In 2012 super-radiance was addressed in~\citet{Richartz:2012bd}.

\subsubsection{FLRW cosmology and related issues}

Attempts at including
backreaction within a cosmological fluid context were investigated
in \citet{Naddeo:2009em, Naddeo:2009en}. 
Analogue models were again used to provide new insight into the cosmological constant in~\citet{Finazzi:2011zw}.

The simulation of FLRW universes was again the subject of investigation 
in~\citet{Fagnocchi:2010sn} and \citet{Bilic:2013qpa}, as well as 
the issue of cosmological particle creation~\citep{Prain:2010zq}.
Horizon area products were the subject of \citet{Anacleto:2013esa}. 
Experimental quantum cosmology was explored in \citet{Westerberg:2014lra}. 

\subsubsection{Effective geometries and emergence}

The theme of ``emergence'' was
represented in articles such as \citet{Girelli:2008qp, Girelli:2008gc,
  Jannes:2008nm}. The more specific theme of emergent gravity continues to play a
role \citep{Hu:2009jd, Liberati:2009uq, Sindoni:2009fc}. 
Being necessarily
very selective, we first mention work related to ``entropic'' attempts
at understanding the ``emergence'' of general relativity and the
spacetime ``degrees of freedom'' from the quantum
regime \citep{Padmanabhan:2009vy, Padmanabhan:2010xh,
  ChangYoung:2010rz, Steinacker:2010rh, Kolekar:2010dm}. 

More work on ``emergent horizons''
has appeared in \citet{Schutzhold:2008zzb,
  Schutzhold:2010rm, Schutzhold:2010cr}. 
In \citet{Barcelo:2010vc} analogue spacetimes were used to carefully separate the notion of
``emergent manifold'' from that of ``emergent curvature''.
Padmanabhan then used the analogue paradigm as background in several articles~\citep{Padmanabhan:2012ik, Padmanabhan:2012gx,Padmanabhan:2012qz}.
More abstract investigations include those of \citet{Nielsen:2012xu, Busch:2012ne, Baccetti:2012ge,Ford:2012xk, Finazzi:2012iu}.
Analogue gravity inspired reasoning led to conjectured dissipative effects at the Planck scale; the implied constraints were analysed in~\citet{Liberati:2013usa}.

\subsubsection{Slow light and optical analogues}

Models based on ultra-short laser pulses are considered in \citet{Faccio:2009yw}. 
 Black holes induced by dielectric effects, and their
associated Hawking radiation, were considered
in \citet{Belgiorno:2010iz}. 
Optical effective
geometries in Kerr media were discussed
in \citet{Cacciatori:2010vr}. 
Experimental applications to optical glass were discussed in \citet{Liberati:2011ep}. 
Laser pulse analogues in optical fibres were studied in \citet{Rubino:2011zq, Faccio:2012uy, Unruh:2012tz, Finazzi:2012bn}. 
Optical media were addressed in \citet{Finazzi:2013nla}.

\subsubsection{Analogues based on quantum fluids of light}

A new analogue model based on a ``photon fluid'' was introduced in \citet{Marino:2008kk}.
The model was further developed in~\citep{Marino:2009,Fouxon:2009be,Solnyshkov:2011sf,Gerase:2012,Larre:2013tba}.
A review on quantum fluids of light was put together in~\citet{Carusotto:2013}.
A first experimental realization was reported in~\citep{Nguyen:2015}, with further analyses in~\citep{Grisins:2016gru}.

\subsubsection{Quark matter}

Acoustic analogue applications to quark matter were further
investigated in \citet{Mannarelli:2008jq, Mannarelli:2008je}
and \citet{Mannarelli:2009qt}. 

\subsubsection{Finsler space}

Finsler spacetime geometries were again considered in \citet{Skakala:2010hw,Skakala:2010wz}, 
while the relationship between analogue spacetimes and foundational mathematical
relativity was discussed in \citet{Chrusciel:2010fn}.

\subsubsection{Analogue entropy?}

Entanglement issues were further explored
in \citet{MartinMartinez:2010ur}.  
Further afield, the analogue paradigm was used to study horizon entropies and Hawking radiation~\citep{Barbado:2011dx,Giovanazzi:2011az,Jannes:2011vb, Rinaldi:2011aa,Rinaldi:2011nb}, 
 and address the trans-Planckian problem~\citep{Barbado:2011ai}.
 
 Entanglement entropy was the subject of \citet{Bruschi:2013tza} and \citet{Busch:2014bza}.
Quantifying the entropy one needs to hide in the Hawking flux was addressed in \citet{Alonso-Serrano:2015bcr} and \citet{Alonso-Serrano:2015trn}.  See also \citet{Alonso-Serrano:2017rfv}.

\subsubsection{Surface waves in shallow water}

Theoretical and historical analyses of surface waves in a wave tank 
were presented in \citet{Rousseaux:2010md}. 
Finally, we mention the stunning
experimental verification by \citet{Weinfurtner:2010nu} the existence of
classical stimulated Hawking radiation in a wave
flume.

More work on surface waves appeared in~\citet{Coutant:2012mf, Coutant:2012zh, Dolan:2012yc, Zhao:2012zz}. 
More water tank experiments were reported in \citet{Euve:2015vml}. 
The ``draining bathtub'' was again investigated in \citet{Richartz:2014lda}, and in
\citet{Dempsey:2016wad}.

\subsubsection{\Horava gravity}

2009 saw intriguing and unexpected relations develop between
analogue spacetimes and \Horava gravity \citep{Sotiriou:2009gy,
  Sotiriou:2009bx, Visser:2009fg, Volovik:2009av, Sindoni:2009vj}.
These connections seem primarily related to the way \Horava's
projectability condition interleaves with the ADM decomposition of the
metric, and to the manner in which \Horava's distinguished spacetime
foliation interleaves with the preferred use of Painlev\'e--Gullstrand-like coordinates.

\subsubsection{Quantum graphity }

2008 saw the introduction of ``quantum
graphity'' \citep{Konopka:2008hp, Konopka:2008ds}, an analogue-inspired
model for quantum gravity. 
A variant of quantum graphity was further
developed in \citet{Hamma:2009xb}, and a matrix model implementation of
analogue spacetime was developed
in \citet{Franchini:2009cf}.  Quantum
graphity was again considered in \citet{Caravelli:2010xx}.

\enlargethispage{35pt}
\subsubsection{Other analogue models}

Possible applications to high-temperature superconductivity were reported
in \citet{Minic:2008an}. 
Black-hole lasers were
considered in \citet{Leonhardt:2008js}, and the fluid-gravity
correspondence in \citet{Ambrosetti:2008mt}. 
Backreaction of the
Hawking flux was again considered in \citet{Schutzhold:2008zzc}.
In \citet{Bini:2008zz} analogue
ideas were applied to polytrope models of Newtonian stars, while
super-radiance was considered in \citet{2008JLTP..150..624T}.
In \citet{Katti:2009mi}, the
universe was interpreted as a ``soap film''. 
Nonlinear
electrodynamics was again considered
in \citet{GoulartdeOliveiraCosta:2009pr}. 
A model based on liquid
crystals appeared in \citet{Pereira:2009vb}.

Possible experimental implementations of acoustic black holes 
using circulating ion rings
are discussed in \citet{Horstmann:2009yh}, while 
analogue-inspired lessons regarding the fundamental nature of time
were investigated in \citet{Jannes:2009ns}
and \citet{Girelli:2009ip}. Non-canonical quantum fields were
considered in \citet{Indurain:2009db}.

In \citet{Ford:2009im, Ford:2009zz} analogue ideas are
implemented in an unusual direction: fluid dynamics is used to model
aspects of quantum field theory. That the trans-Planckian and
information loss problems are linked is argued
in \citet{Liberati:2009ak}. 
Further afield, analogue spacetimes were used as an aid to understanding 
``warp drive'' spacetimes \citep{Barcelo:2010pu}.
 More abstract models included~\citet{Goulart:2011rs, Goulart:2011cb, Visser:2011jp}.

A graphene-based analogue was discussed in~\citet{Chen:2012uc}, while the Gordon metric was revisited in~\citet{Novello:2012tk}. 
Nematic liquid crystals were discussed in \citet{Pereira:2013eq}.
Abstract PDE approaches were explored in \citet{Barcelo:2007mga}. 
Gravity duals were the subject of \citet{Hossenfelder:2014gwa}.
Some issues regarding ``emergence'' of gravitational dynamics in analogue-like scenarios  were addressed in \citet{Marolf:2014yga}; see section \ref{S:Marolf-theorem}. Emergent electromagnetism was investigated in \citet{Barcelo:2014yna}. 
Oceanic issues were addressed in \citet{Prain:2014nra}.
The borderlands between general relativity and quantum physics were explored in \citet{Baccetti:2014tja}.
No-hair theorems were explored in \citet{Michel:2015mlr}, and analogue aspects of extreme gravitational collapse were addressed in \citet{Barcelo:2015noa}. 
Super-radiance was again explored in \citet{Brito:2015oca}. Acoustic wormholes were the subject of \citet{Jusufi:2016eav}.

\subsubsection{End of the Late Modern period} 

We define the end of the ``Late Modern'' period by the appearance in late 
 2016 of Steinhauer's BEC-based experiment  probing (phononic) Hawking radiation and entanglement~\citep{Steinhauer:2015saa}.

\clearpage
\subsection{Contemporary period (2017--2024)}

What we shall call the contemporary period covers the last decade or so --- during which there has been a massive outpouring (at least 50 articles per year) of both theoretical and experimental developments. 
We again sub-divide the discussion thematically, noting that there will inevitably be overlaps and redundancies in the classification.

\subsubsection{Acoustics} 

A hydrodynamic wormhole was explored in \citet{Euve:2017gfj}. 
Rotating analogues appeared in \citet{Giacomelli:2017eze} and \citet{Shaikh:2017mfs}. 
The de Laval nozzle again made an appearance in \citet{daRocha:2017lqj}. 
Trapping and frame dragging were considered in \citet{Banerjee:2018wev}. 
Relativistic viscous fluids were analyzed in \citet{Bittencourt:2018yrc}.
Vorticity in analogue models was again studied in \citet{Liberati:2018uev}
The use of non-isentropic fluids was explored in \cite{Bilic:2018fsk} and \citet{Bilic:2021yhz}.
Step-like configurations were addressed in \citet{Curtis:2018qey}.

Particle production was considered in \citet{Balbinot:2019mei}. 
Dumb hole spectroscopy was addressed in \citet{Torres:2019sbr}.
The influence of external pressure was considered in \citet{Bilic:2021fsx}.
Slowly rotating acoustic black holes were studied in \citet{Vieira:2021ozg}.
Dynamical configurations were explored in \citet{Fernandes:2021gkf}. 
The analogue Penrose process was explored in \citet{Ma:2022zow}
Plasma-based acoustics was the topic of \citet{Ditta:2023lny}. 
Potential flow in a tube was investigated in \citet{Tsuda:2023ypv}. 
Shock wave singularities were addressed in \citet{Fischer:2022bkl}. 
Prospects for synergies between quantum-gravity and classical and quantum fluid theories are discussed in~\citet{Braunstein:2023jpo}.

\subsubsection{{Hydrodynamic vortex flow}} 

Vortex flows were again studied in \citet{Torres:2017vaz},
and were subsequently again reconsidered in \citet{Svancara:2023yrf}. 
Rotating sources and observers were explicated in \citet{Ruggiero:2023ker}.
Circular motion was again explored in \citet{Bunney:2023vyj}.

\subsubsection{Horizons and ergo-regions} 

Near-horizon physics was addressed in \citet{Boonserm:2017cql}. 
See also \citet{Boonserm:2018orb}. 
Ergo-region instabilities were probed in \citet{Oliveira:2018ckz}. 
See also \citet{Assumpcao:2018bka}.

Analogue horizons were addressed in \citet{Barcelo:2018ynq},
and the dynamics of horizon formation  was analyzed in \citet{Tettamanti:2020wrp}. 
Ergo-region instabilities were discussed in \citet{Giacomelli:2020jlz}.
Acoustic silhouettes and near horizon physics were presented in \citet{Ling:2021vgk}.
Cross-horizon correlations were investigated in \citet{Balbinot:2021bnp}. 
Rotating black holes were again the subject of \citet{TorresVicente:2019ahs}.

Horizons based on hydrodynamic electron flow were discussed in \citet{Dave:2022wgk}. 
Synthetic horizons were studied in \citet{Mertens:2022jij}.
A lattice model based on Weyl cones was addressed in \citet{Konye:2022wds}.
The BTZ black hole was investigated in \citet{deOliveira:2022csc}. 
Dissipation at the acoustic horizon was explored in \citet{Chiofalo:2022qkp}.
Both black and white horizons were probed in \citet{Bossard:2023zyq,Bossard:2024eys}.

\subsubsection{Hawking radiation} 

A Rindler quench was considered in \citet{Louko:2018pij}.
Hawking radiation was interpreted as an instability in \citet{Bermudez:2018ogk}. 
The experimental situation was summarized in \citet{Weinfurtner:2019zyc}. 

Acoustic Hawking phenomena were considered in \citet{Mannarelli:2020ebs}, 
and back reaction in \citet{Liberati:2020mdr}.
The influence of PT symmetry was the topic of \citet{Stalhammar:2021rks}. 
Unruh and Cerenkov radiation were analyzed in \citet{Svidzinsky:2021zbv}. 
Laser-driven plasmas were studied in \citet{Fiedler:2021vbg} and \citet{Fiedler:2021xjn}. 
The Hawking effect, viewed as a quantum catastrophe, was investigated in \citet{Farrell:2021hib} and \citet{Farrell:2022mzv}.
Hawking-effect-induced destabilization of horizons was explored in \citet{Anacleto:2022lnt}.
A review article dedicated to the analogue Hawking effect and its experimental realization was presented in~\cite{Almeida:2022otk}.

\subsubsection{Causality} 

Chronology protection was studied in \citet{Barcelo:2022jbr}, while 
causal hierarchies were explored in \citet{Carballo-Rubio:2020ttr}. 
The possibility of analogue non-causal curves were investigated in \citet{Sabin:2022mmj}.

\subsubsection{Bose--Einstein condensates} 

Steinhauer's experiments have continued to stimulate many subsequent discussions~\citep{Tettamanti:2016ntx,Wang:2016jaj,Wang:2017wnj, Michel:2016tog,Finke:2016wcn,Leonhardt:2016qdi,Steinhauer:2016hfa, Steinhauer:2016ftg,Steinhauer:2018qzg,Tettamanti:2020rvt}. 
Among other things, we have learned how to distinguish between black hole laser 
and Bogoliubov--Cherenkov--Landau stimulations~\citep{Steinhauer:2021xxj,deNova:2023yyu}.

Correlations were the main topic in \citet{Fabbri:2017ntv},
while ``gravitational waves'' were discussed in \citet{Hartley:2017fdv}. 
A rapidly expanding BEC was investigated in \citet{Eckel:2017uqx}. 
Entanglement issues were addressed in \citet{Robertson:2017ysi}. 
An exactly solvable BEC model appeared in \citet{Parola:2017qpn}. 
A BEC based version of the Penrose process was studied in \citet{Solnyshkov:2018dgq}. 
A BEC based version of the Unruh effect was reported in \citet{Kosior:2018vgx}.
One dimensional BECs were again studied in \citet{Robertson:2018gwi}.

2+1 dimensional BECs were addressed in \cite{Giacomelli:2019tvr}. 
Phonon redshift was studied in \cite{Eckel:2020qee}.
Toroidal BEC configurations were the subject of \citet{Bhardwaj:2020ndh}. 
BEC analog cosmological reheating was discussed in \citet{Chatrchyan:2020cxs}.
{Hawking quanta correlators in 1+1 dimensional BECs  were further investigated in \citet{Dudley:2020toe} while Unruh detectors for BECs were discussed in \citet{Gooding:2020scc}. 
The formation of an acoustic black hole in a Bose-Einstein condensate, and the correspondent emergence of analog Hawking radiation were studied in~\citet{Fabbri:2020unn}.}

Quasi 1+1 dimensional configurations were studied in \citet{Ribeiro:2021fpk}.
A system of fermionic cold atoms to simulate cosmological particle production of Dirac fermions in an expanding universe was considered in \citet{Fulgado-Claudio:2022fut}. 
Density correlations were investigated in \citet{Palan:2022sly}. 
BEC based ``expanding cosmologies'' were considered in references \citep{Tolosa-Simeon:2022umw, Banik:2022fua, Viermann:2022fet, GutierrezGalan:2021tcr,Butera:2017oig}. 
Entanglement in an expanding toroidal Bose-Einstein condensate was the subject of  \citet{Bhardwaj:2023squ}.
Curved-spacetime Gross-Pitaevskii hydrodynamics was investigated \citet{Roitberg:2023rps}.

\subsubsection{Meta-materials} 

The use of meta-materials were specifically addressed in \citet{Thompson:2017tus} and \citet{Thompson:2017pdl}.
See also \citep{Schuster:2017mdx,Schuster:2018cwt, Schuster:2018bqx, Schuster:2018ycr}. 
Wave propagation in meta-materials was studied in \citet{FernandezNunez:2018yug}, 
while analogues based on wave guides were considered in \cite{Lang:2018tpj}. 
Black and white horizons at material junctions was the topic of \citet{Kedem:2020aqz}.
Dielectric models were reconsidered in \citet{Bittencourt:2021jki} and in \citet{Guerrieri:2022ymn}.
Meta-materials were again analyzed in \citet{FernandezNunez:2018yug}. 
2+1 dimensional meta-materials were addressed in \citet{Goulart:2022tfv}. 
Magneto-electric models were studied in \citet{DeLorenci:2022pdw}.
Moir\'e meta-materials were explored in \citet{Tolosa-Simeon:2023bqc}. 
General aspects of meta-material models were the focus of \citet{Schuster:2023jfa}.

\subsubsection{Optics} 

Results of an optical experiment are reported in \citet{Drori:2018ivu}.
Dielectric media were considered in \citet{Lang:2019avs}, and
moving dielectric media were explored in \citet{Guo:2019tmr}. 
More work on dielectric media appeared in \citet{Belgiorno:2020zzy}. 
Hawking radiation in optical analogues was addressed in \citet{Aguero-Santacruz:2020krw},
optical horizons were considered in \citet{Rosenberg:2020jde},
and electromagnetic analogies were explicated in \citet{Bera:2020doh}.
A fibre-optic analogue was studied in \citet{Felipe-Elizarraras:2022ifv}. 
Optical media were again addressed in \citet{Westerberg:2019wtk}.

Disformal electro-dynamics was explored in \citet{Bittencourt:2023ikm}. 
Optical analogues for BTZ black holes were investigated in \citet{Chen:2023xsz}. 
Penrose super-radiance was explored in \citet{Zhang:2023kyr}. 
Electromagnetic analogues were again considered in \citet{Falcon-Gomez:2023vmm}. 

A spiralling light front has been designed to simulate rotational effects~\citep{Vocke:2015,Vocke:2016}, with experimental results in~\citep{Vocke:2017tif}. After several theoretical studies~\citep{Prain:2019jqk,Braidotti:2020ize,Solnyshkov:2018dgq} an experimental observation of superradiance has been reported in~\citep{Braidotti:2021nhw}.

\subsubsection{Quantum fluids of light} 

Work on analogues based on quantum fluids of light has been strengthened.  New experimental results appeared
in~\citep{Jacquet:2020znq} and~\citep{Falque:2023ctx}, with further analyses and discussion of prospects in~\citep{Jacquet:2021scv,Claude:2021rkt,Claude:2022wfm,Jacquet:2022vak}. 
A different experiment, now probing cosmological particle production, has also been conducted~\citep{Steinhauer:2021fhb}.
(See section \ref{S:6.5} below).

\subsubsection{Shallow water surface waves} 

Analogue models based on the K-dV and B-dG equations were explored in \citet{Coutant:2017nea} and  \citet{Coutant:2017qnz}.
Experimental QNMs were studied in \citet{Torres:2020tzs}. 
Co-current surface waves were addressed in \cite{Euve:2018uyo}. 
The draining bathtub model continued to attract  attention in \citet{Patrick:2018orp}, while 
modulated flows were the subject of \citet{Michel:2018oyz}. 

Back-reaction was considered in \citet{Patrick:2019kis}. 
Circular motion was analyzed in \citet{Biermann:2020bjh}.
Dispersive phenomena were addressed in \citet{Patrick:2020yyy} and \citet{Patrick:2020baa}. 
Surface wave analogues of the Unruh effect were studied in \citet{Barros:2020zkt}.

More work on bathtub vortices was presented in \citet{Patrick:2019prh} and in \citet{Patrick:2021oqk}.
Froude horizons were discussed in \citet{Porporato:2021rmm}. 
The Stokes and Darwin drifts were explicated in \citet{Sheikh-Jabbari:2023eba}.

\subsubsection{Astrophysical topics} 

Attempts at constructing a Gordon form of the Kerr spacetime are reported in \citet{Liberati:2018osj}.
Kerr-like configurations were again studied in references \citep{Baines:2020dqm,Baines:2020unr,Baines:2020egv,Baines:2021jbj}  
Cosmological analogues were the subject of \citet{Faraoni:2020swm}. 
Potential connections with the cosmological constant were explored in \citet{Padmanabhan:2021uim}.
Applications to the Hawking--Ellis classification appeared in \citet{Martin-Moruno:2021niw}.
Kerr-like geometries again reappeared in \citet{Visser:2022fwx}.
Analogue inspired applications to cosmology appeared in \citet{Gaur:2022hap},
and numerical issues were explored in \citet{Butera:2022kwi}.

Draining vortices as analogues for astrophysical compact objects were studied in \citet{Torres:2022bto}. 
Inner horizon instabilities were probed in \citet{DiFilippo:2022qkl}. 
2+1 dimensional cosmologies were explored in \citet{Sanchez-Kuntz:2022gds}.
Kerr-like geometries were again reconsidered in \citet{Schmidt:2023mjg}.
 Black hole mergers were addressed in \citet{Solnyshkov:2023sjl}. 
Instabilities were the focus of \citet{Geelmuyden:2023cow}. 

Kerr-like configurations were revisited in \citet{Baines:2023cac} and in \citet{Baines:2023dhq}.
Cosmological analogues were again studied in \citet{Faraoni:2023kzf}.

\subsubsection{Entropy} 

An analogue perspective on information loss was presented in \citet{Liberati:2019fse}. 
Entangling q-bits were discussed in \cite{Sabin:2020ans}. 
More discussion of information loss appeared in \citet{Buoninfante:2021ijy}.
Entanglement in an expanding universe was studied in both \citet{Sanchez-Kuntz:2023ssh} and \citet{Kranas:2023aph}.

\subsubsection{Lorentz invariance} 

Multiple horizons in Lorentz violating systems were considered in \citet{DelPorro:2023lbv}. 

\subsubsection{Emergence} 

Emergent gauge symmetries were addressed in \citet{Barcelo:2016xhp},
while emergent gravity was explored in \citet{Linnemann:2017hdo}.
Oriti investigated spacetime viewed as a many-body system \citep{Oriti:2017twl}. 
Other contributions on the theme of emergence include \citet{Mielczarek:2017cdp} and \citet{Liberati:2017jnr}. 
More ideas on emergence were reported in \citet{Hossenfelder:2017eoh}, and in \citet{CarballoRubio:2016tok}. 

The topic of emergence was revisited in both \citet{Barcelo:2021ryh} and \citet{Tricella:2020rzl}.
Emergent gravity in particle traps was discussed in \citet{Penin:2023szu}.
Emergent symmetries were probed in \citet{Torrieri:2023udd}.

\subsubsection{Historical, foundational, and philosophical issues} 
\nopagebreak

Historical and philosophical issues were considered in references \citep{pittphilsci20365,Almeida:2022otk, ElSkaf:2022qmg, Shahbazi:2023wok}.
The inverse problem was addressed in \citet{Albuquerque:2023lzw}.

\subsubsection{Exotic analogues} 

Two particularly exotic analogue models are these:
A vehicular traffic analogue was investigated in \citet{deSouza:2022fgg}.
An analogue based on tsunami focussing was explored in \cite{Torres:2022zua}.
Rather unusual was the use of tachyonic media in \citet{Bilson-Thompson:2023zqj}. 
Finally, an unexpected formal analogy between cosmology and geophysical flows was found in \citet{Faraoni:2023kzf}, with the term ``geophysical flows" encompassing many natural phenomena, including lava flows, the creep of glacier ice, avalanches, and mud slides.

\subsubsection{Other analogue models} 

The classical analogue of the Unruh effect was investigated in \citet{Leonhardt:2017lwm}. 
A 1+1 dimensional analogue appeared in \citet{Pedernales:2017eue}.
Nonlinear effects were considered in \citet{Michel:2017joy}
Conformal models were the subject of \citet{Hossenfelder:2017iom}.
Philosophical issues were explored in \citet{Dardashti:2017kfw}. 
An analogue based on a superconducting circuit was investigated in \citet{Tian:2017wij}, 
while quantum Hall systems were considered in \citet{Hegde:2018xub}. 
Holographic issues were considered in \citet{Hossenfelder:2018lhy}.
Tolman temperature gradients were addressed in \citet{Santiago:2018lcy}. 
See also \citet{Santiago:2019aem}. 

Many-body analogues appeared in \cite{Yang:2019kbb}. 
Pedagogical models were addressed in \citet{Possel:2018mnd}. 
Lessons for quantum gravity were explored in \citet{Versteegen:2019qvz},
as were black bounces in \citet{Simpson:2019oft}. 
Particle scattering in analogue models was the subject of \citet{Todd:2020ztj}. 
Undoped graphene was discussed in \citet{Sorge:2020vou}.
The next generation of experimental proposals was explicated in \citet{Jacquet:2020bar}.

Specific analogues for Schwarzschild and Reissner--Nordstr\"om were considered in \citet{deOliveira:2021edr}. 
Analogue wormholes appeared in \citet{Churilov:2021olu}. 
Relations with ``warp drives'' were addressed in \citet{Santiago:2021aup}
Analogue quantum superpositions were investigated in \citet{Barcelo:2021nhs}. 
Graphene was again studied in \citet{Gallerati:2021rtp}, and in \citet{Gallerati:2022egf}.

Symplectic circuits were analyzed in \citet{Brady:2022ffk}.
Super-radiance made yet another appearance in \citet{Cardoso:2022yin}. 
Nonlinear perturbations were addressed in \citet{Fernandes:2022bwo}. 
Computational complexity was explored in \citet{Parvizi:2022lbv}.
Analogue black bounces were discussed in \citet{Pal:2022ahd}, and in \citet{GarciaMartin-Caro:2023jjq}. 
A ``Higgs phonon'' was studied in \citet{Yan:2023hut}. 
The Petrov classification was considered in \citet{Baak:2023zjf}. 
Dilaton models were investigated in \citet{Moss:2023qzn}.

\subsection{Theses --- PhD and MSc} 

We should explicitly mention the recent appearance of many analogue-inspired PhD and MSc theses, both theoretical and experimental.
(In the lists below we are largely constrained by whether or not the theses were entered into the appropriate international bibliographic databases. We are sure there must be more that we have missed.)

For some relevant MSc theses see for instance \citet{Weinfurtner:2004in,Kolishetty:2014evq, Simpson:2019oft,Mukherjee:2020, daSilva:2020hjf, Baines:2021jbj,Simons:2022}.

For relevant PhD theses up to 2010 one might peruse \citet{Corley:1997bx,Liberati:2000ag, Weinfurtner:2007br, Vikman:2007sj, Sindoni:PhD, Jannes:2009yr, Zhidenko:2009zx}.
For PhD theses from 2011 to 2020 one should check \citet{Skakala:2011aw, 
Finazzi:2011yig, Prain:PhD, Cook:2012jqn, Coutant:2012vff, 
Baccetti:2014tja, Tolic:2015oki, Lawrence:2015ytm, Cropp:2016gkn, CarballoRubio:2016tok, 
Jacquet:2017nxs, Michel:2017joy, Vocke:2017lsk,  Butera:2017oig,
FernandezNunez:2018yug, Vieira:2018xjy, Schuster:2018lmz, 
Fitzgerald:2019hze, Hartley:2019ybu, Westerberg:2019wtk, Santiago:2019aem, 
Versteegen:2019qvz, Patrick:2019prh, TorresVicente:2019ahs}.
More recently (2020 to present) consider \citet{Tricella:2020rzl, 
Fiedler:2021xjn, Farrell:2021hib, Banik:2022fua, Viermann:2022fet, Fifer:2022zfh, 
GutierrezGalan:2021tcr, Prizia:2022tad, Katayama:2022yey,
Geelmuyden:2023cow,Roitberg:2023rps,Busnaina:2023jdt,Sanchez-Kuntz:2023ssh,
Kranas:2023aph, BarrosoSilveira:2023tcx, Cominotti:2023hvf, Baak:2023vru}.
Overall, well over 50 theses have now been written on analogue and analogue-adjacent topics.

\subsection{Survey articles}

Various survey articles, both full surveys and focussed micro-surveys,  have appeared over the decades. We should particularly mention
\citet{Maynard:2001zz,Cardoso:2005ij,Fischer:2005iy,Balbinot:2006ua,Das:2007ix,Berti:2009kk, Nation:2011dka, Jacobson:2012ei, Carlip:2012wa, Balbinot:2012xw, Robertson:2012ku, Visser:2012rct, Balbinot:2014cfa, Liberati:2017jnr, Rousseaux:2020cjo, Jacquet:2020bar,Castorina:2022gar}
For book-length expositions see~\citet{30-years} and \citet{Belgiorno:2019ofm}, and of course~\citet{Novello:2002qg} and \citet{Volovik:2003fe}. Over 20 reviews/surveys/books on this topic have now appeared.

\subsection{The future?}

The last decade (2015-2024) has seen a veritable explosion of experimental efforts aimed at testing the existence and robustness of of (analogue) Hawking radiation. As of November 2024 significant experimental progress has been made. 
Experimental probes of the closely related Unruh
radiation are also under consideration. 

In short, interest in analogue models, analogue spacetimes, and analogue gravity
is intense and shows no signs of abating. Interest in these ideas now
extends far beyond the general relativity community, and there is
significant promise for considerably more direct laboratory-based experimental input.

We particularly wish to encourage the reader
to keep an eye out for future developments regarding the experimental
study of Hawking radiation and/or the closely related Unruh
radiation. 

\subsection{ Going further } 
\label{SS:going-further-3}

To further complicate the history, there is large body of work for which
analogue spacetime ideas provide part of the background gestalt, even if
the specific connection may sometimes be somewhat tenuous. Among such
articles, we mention:

\begin{itemize}

\item Analogue-based ``geometrical'' interpretations of pseudo-momentum,
  Iordanskii forces, Magnus forces, and the acoustic Aharanov--Bohm
  effect \citep{Furtado:2004ei, Stone:1997nj, Stone:1999gi, 
Stone:2000vortex, Stone:2000gb, Volovik:1998pc}.

\item An analogue-inspired interpretation of the Kerr spacetime \citep{Hamilton:2004au}.
More recently, see also \citet{Liberati:2018osj} and \citet{Baines:2022xnh}. 

\item The use of analogies to clarify the Newtonian limit of general
  relativity \citep{Trautman}, to provide heuristics for motivating interest in
  specific spacetimes \citep{Rosquist:2003qs, Visser:2003tt}, and to discuss a
  simple interpretation of the notion of a horizon \citep{Obukhov:2003vv}. 

\item Discrete \citep{Sorkin:2003bx, Henson:2006kf, Mattingly:2007be, Loll:2019rdj, Ambjorn:2024pyv} and non-commutative \citep{Chu:2000ww} spacetimes partially influenced and flavoured by analogue ideas. 

\item Analogue-based hints on how to implement ``doubly special relativity''
  (DSR) \citep{Kowalski-Glikman:2000dz, Kowalski-Glikman:2002ft,
    Kowalski-Glikman:2002be, Tamaki:2002iz}, and a cautionary analysis of why
  this might be difficult \citep{Schutzhold:2003yp}. 

\item Possible black-hole phase transitions placed in an analogue
  context \citep{Stephens:2001sd}. 
  
\item Attempts at deriving inertia and passive gravity, (though not
  active gravity), from analogue ideas \citep{Milgrom:2006wf}. 

\item Applications of analogue ideas in braneworld \citep{Ge:2007tr}
  and Kaluza--Klein settings \citep{Ge:2007ts}.

\item Analogue inspired views on the ``mathematical universe'' \citep{Jannes:2009bb}. 

\item Cosmological structure formation viewed as noise
  amplification \citep{Siemienic-Ozieblo:2001ef}. 

\item Modified inflationary scenarios \citep{Chu:2000ww,Collins:2003ze}.

\item Discussions of unusual topology, ``acoustic wormholes'', and unusual
  temporal structure \citep{Nandi:2004rr, Nandi:2004nw, Radu:1998sk,
    Smolyaninov:2003wb, Smolyaninov:2003qt, Zloshchastiev:1998uz,
    Zloshchastiev:1999rr}. 
    
\item Magnon contributions to BEC physics \citep{Bunkov:2010fq,
  Volovik:2007dj}. 

\item Analogue models based on plasmon
  physics \citep{Smolyaninov:2003wb, Smolyaninov:2003qt}.

\item Abstract quantum field theoretic considerations of the Unruh
  effect \citep{Zaslavsky:1998in}. 
  
\item The interpretation of black-hole entropy in terms of universal
  ``near horizon'' behaviour of quantum fields living on
  spacetime \citep{Carlip:2006fm, Carlip:2008wv}. 

\item Numerous suggestions regarding possible trans-Planckian
  physics \citep{Arteaga:2003we, Bastero-Gil:2001js, Casadio:2003vk,
    Chang:2003he, Chang:2003sa, Ho:2003zm, Mersini:2001jt,
    Rosu:1996pw, Tanaka:2000jw}.

\item Numerous suggestions regarding a minimum length in quantum
  gravity \citep{Bhattacharyya:2004dy, Blaut:2001fy, Brout:1998ei,
    Consoli:2003up, Garay:1994en, Hossenfelder:2004gj,
    Hossenfelder:2004up, Hossenfelder:2005ed, Kowalski-Glikman:2000dz,
    Kowalski-Glikman:2002ft, Kowalski-Glikman:2002be, Lubo:1999xg,
    Lubo:2003mr, Lubo:2003rs, Smolin:1995ai}. 

\item Standard quantum field theory physics reformulated in the light of
  analogue models \citep{Anglin:1992ur, Antunes:2003jr, Fischer:1999cp,
    Ford:1997hb, Liberati:2000ag, Liberati:2000pt, Liberati:2000mp,
    Liberati:2001sd, Martin:2004um, Obadia:2001hj, Obadia:2002qe,
    Parentani:1995ts, Parentani:1995iw, Parentani:1996sz,
    Renaud:2004ta, Raval:1996vt, Zaslavsky:1998in}. 

\item Standard general relativity supplemented with analogue viewpoints and
  insights \citep{Kokkotas:1999bd, Larsen:1996ah, Martin:2004um,
    Nielsen:2005af, Jacobson:2008cx, Okninski:2005gz}. 

\item The discussion of, and argument for, a possible reassessment of
  fundamental features of quantum physics and general
  relativity \citep{Arbona:2003re, Gorski:1997re, Kempf:2003qu, Laschkarew,
    Liberati:2001sd, Parentani:1997ua, Rovelli:1994ge, Saul:2004ch}. 

\item Non-standard viewpoints on quantum physics and general
  relativity \citep{Czerniawski:2002sv, Horwitz:2004fm, Oron:2002xk,
    Rosu:1994je, Rosu:1994ky, Schmelzer:2000da, Schmelzer:2000db,
    Schmelzer:2001kx, Schmelzer:2002es}. 

\item Soliton physics \citep{Pashaev:1998sk}, defect
  physics \citep{MdeCarvalho:2004ic}, and the Fizeau
  effect \citep{Nandi:2002fe}, presented with an analogue flavour. 

\item Analogue-inspired models of black-hole
  accretion \citep{Ray:2002gd, Ray:2003gb, Ray:2007up}. 

\item Cosmological horizons from an analogue spacetime
  perspective \citep{Ghafarnejad:1997ps}. 

\item Analogue-inspired insights into renormalization group
  flow \citep{Calzetta:1999zr}. 

\item An analysis of ``wave catastrophes'' inspired by analogue
  models \citep{Kiss:2003kp}. 

\item Improved numerical techniques for handling wave
  equations \citep{Winicour:2000zj}, and analytic techniques for handling wave
  tails \citep{Bombelli:1994rh}, partially based on analogue ideas. 

\item {Analogue gravity inspired emergent gravity scenarios in Group Field Theory, both in regards to cosmology \citep{Oriti:2017twl, Gielen:2013kla, Gielen:2013naa},  as well as providing an operative framework for emergence, see~\citep{Oriti:2016acw} or more recently~\citep{Oriti:2024qav} and references therein.}
\end{itemize}
From the above the reader can easily appreciate the broad interest in, and
wide applicability of, analogue spacetime models. 

There is not much more that we can usefully say here. We have doubtless
missed some articles of historical importance, but with a good library
or a fast Internet connection the reader will be in as good a position
as we are to find any additional historical articles.


\clearpage
\section{Catalogue of Models}
\label{S:catalogue}

In this section, without being completely exhaustive, we attempt to categorize the very many analogue models researchers have investigated. Perhaps the most basic subdivision is into classical models and quantum models, but even then many other levels of refinement are possible. 
We have retained most of the material in the previous version of the LRR, but given the experimental relevance they are acquiring, have added an explicit item on quantum fluids of light.
Consider for instance the following list:
\begin{itemize}
\item Classical models:
  \begin{itemize}
  \item Classical sound.
  \item Sound in relativistic hydrodynamics.
  \item Shallow water waves (gravity waves).
  \item Classical refractive index.
  \item Normal modes.
  \end{itemize}
\item Quantum models:
  \begin{itemize}
  \item Bose--Einstein condensates (BECs). 
  \item The heliocentric universe. \\ 
    (Helium as an exemplar for just about anything.)
  \item Slow light.
  \item Quantum fluids of light.
  \item Lattice models
  \item Graphene.
  \end{itemize}
\end{itemize}
We will now provide a few words on each of these topics.

\subsection{Classical models}
\label{S:classical-models}

\subsubsection{Classical sound}

Sound in a non-relativistic moving fluid has already been extensively discussed in
Sect.~\ref{S:simple}, and we will not repeat such discussion here.
In contrast, sound in a solid exhibits its own distinct and interesting
features, notably in the existence of a generalization of the normal
notion of birefringence -- longitudinal modes travel at a different
speed (typically faster) than do transverse modes. This may be viewed as
an example of an analogue model which breaks the ``light cone'' into two
at the classical level; as such this model is not particularly useful if
one is trying to simulate special relativistic kinematics with its
universal speed of light, though it may be used to gain insight into yet
another way of ``breaking'' Lorentz invariance (and the equivalence principle).

\subsubsection{Classical sound in relativistic hydrodynamics} 
\label{S:relativistic-acoustic}

When dealing with relativistic sound, key historical papers are those
of \citet{Moncrief} and \citet{Bilic:1999sq}, with
astrophysical applications being more fully explored
in \citet{Das:2004wf, Das:2004zm, Das:2004gn}, and with a more recent
pedagogical follow-up in \citet{Visser:2010xv}. It is convenient to
first quickly \emph{motivate} the result by working in the limit of
relativistic ray acoustics where we can safely ignore the wave
properties of sound. In this limit we are interested only in the
``sound cones''. Let us pick a curved manifold with physical spacetime
metric $g_{\mu\nu}$, and a point in spacetime where the background fluid
4-velocity is $V^\mu$ while the speed of sound is $c_s$. Now
(in complete direct conformity with our discussion of the
generalised optical Gordon metric) adopt Gaussian normal coordinates
so that $g_{\mu\nu}\to \eta_{\mu\nu}$, and go to the local rest frame
of the fluid, so that $V^\mu \to (1; \vec 0)$ and 
\begin{equation}
g_{\mu\nu} \to \left[\begin{array}{cccc}-1&0&0&0\\0&1&0&0\\0&0&1&0\\0&0&0&1\end{array} \right];
\qquad
h_{\mu\nu} = g_{\mu\nu} + V_\mu\, V_\nu \to \left[\begin{array}{cccc}0&0&0&0\\0&1&0&0\\0&0&1&0\\0&0&0&1\end{array} \right].
\end{equation}
In the rest frame of the fluid the sound cones are (locally) given by
\begin{equation}
- c_s^2 \; \d t^2 + ||\d \vec x||^2 = 0,
\end{equation}
implying in these special coordinates the existence of an acoustic
metric
\begin{equation}
{\cal G}_{\mu\nu} \propto \left[\begin{array}{cccc}-c_s^2&0&0&0\\0&1&0&0\\0&0&1&0\\0&0&0&1\end{array} \right].
\end{equation}
That is, transforming back to arbitrary coordinates:
\begin{equation}
{\cal G}_{\mu\nu} \propto -c_s^2 V_\mu\, V_\nu + \left\{ g_{\mu\nu} + V_\mu\, V_\nu\right\} \propto g_{\mu\nu} + \left\{1-c_s^2\right\} V_\mu \,V_\nu.
\end{equation}
Note again that in the ray acoustics limit, because one only has the
sound cones to work with, one can neither derive (nor is it even
meaningful to specify) the overall conformal factor. When going beyond
the ray acoustics limit, seeking to obtain a relativistic wave
equation suitable for describing physical acoustics, all the ``fuss''
is simply over how to determine the overall conformal factor (and to
verify that one truly does obtain a d'Alembertian equation using the
conformally-fixed acoustic metric).

One proceeds by combining the relativistic Euler equation, the
relativistic energy equation, an assumed barotropic equation of state,
and assuming a relativistic irrotational flow of the
form \citep{Visser:2010xv}
\begin{equation}
V^\mu = {g^{\mu\nu} \; \nabla_\nu \Theta\over ||\nabla \Theta||}.
\end{equation}
In this situation the relativistic Bernoulli equation can be shown to be
\begin{equation}
 \ln||\nabla \Theta|| =
 + \int_0^p {\d p\over\varrho(p)+p},
\end{equation}
where we emphasize that $\varrho$ is now the energy density (not the
mass density), and the total particle number density can be shown to
be
\begin{equation}
n(p) = n_{(p=0)} \; \exp\left[ \int_{\varrho_{(p=0)}}^{\varrho(p)} {\d\varrho\over\varrho+p(\varrho)} \right].
\end{equation}
After linearization around some suitable background \citep{Moncrief,
  Bilic:1999sq, Visser:2010xv}, the perturbations in the scalar
velocity potential $\Theta$ can be shown to satisfy a
dimension-independent d'Alembertian equation
\begin{equation}
\nabla_\mu \left\{  {n_0^2\over\varrho_0+p_0} 
\;\;
 \left[  -{1\over c_s^2} V_0^\mu V_0^\nu +h^{\mu\nu} \right] \; \nabla_\nu \Theta_1 
   \right\} = 0,
\end{equation}
which leads to the identification of the relativistic acoustic metric
as 
\def\G{{\mathcal{G}}}
\begin{equation}
\sqrt{-\G}\; \G^{\mu\nu} = {n_0^2\over\varrho_0+p_0} 
\;\;
 \left[  -{1\over c_s^2} V_0^\mu V_0^\nu +h^{\mu\nu} \right].
 \end{equation}
The dimension-dependence now comes from solving this equation for
$\G^{\mu\nu}$. Therefore, we finally have the (contravariant) acoustic
metric
\begin{equation}
\G^{\mu\nu} = \left( {n_0^2\; c_s^{-1} \over\varrho_0+p_0} \right)^{-2/(d-1)} \\
 \left\{  -{1\over c_s^2} V_0^\mu V_0^\nu +h^{\mu\nu} \right\},
\end{equation}
and (covariant) acoustic metric
\begin{equation}
\G_{\mu\nu} = \;\;\left( {n_0^2\; c_s^{-1}\over\varrho_0+p_0} \right)^{2/(d-1)} \\
 \left\{ -{ c_s^2} \; [V_0]_\mu [V_0]_\nu +h_{\mu\nu} \right\}.
 \label{eq:relhydroacmet}
\end{equation}
In the non-relativistic limit $p_0 \ll \varrho_0$ and $\varrho_0
\approx \bar m\; n_0$, where $\bar m$ is the average mass of the
particles making up the fluid (which by the barotropic assumption is a
time-independent and position-independent constant). So in the
non-relativistic limit we recover the standard result for the
conformal factor \citep{Visser:2010xv}
\begin{equation}
 {n_0^2\; c_s^{-1}\over\varrho_0+p_0} \to { n_0\over \bar m c_s} = {1\over \bar m^2} \; {\rho_0\over c_s} \propto {\rho_0\over c_s}.
\end{equation}
Under what conditions is the fully general relativistic discussion of
this section necessary? (The non-relativistic analysis is, after all,
the basis of the bulk of the work in ``analogue spacetimes'', and is
perfectly adequate for many purposes.) The current analysis will be
needed in three separate situations:
\begin{itemize}
\item when working in a nontrivial curved general relativistic background;
\item whenever the fluid is flowing at relativistic speeds;
\item less obviously, when the internal degrees of freedom of the
  fluid are relativistic, even if the overall fluid flow is
  non-relativistic. (That is, in situations where it is necessary to
  distinguish the energy density $\varrho$ from the mass density
  $\rho$; this typically happens in situations where the fluid is
  strongly self-coupled -- for example in neutron star cores or in
  relativistic BECs \citep{Fagnocchi:2010sn}. See
  Sect.~\ref{S:quantum-models}.)
\end{itemize}

\subsubsection{Shallow water waves (gravity waves)} 
\label{S:WW}

\begin{figure}[htpb]
    \centerline{\includegraphics[width=0.8\textwidth]{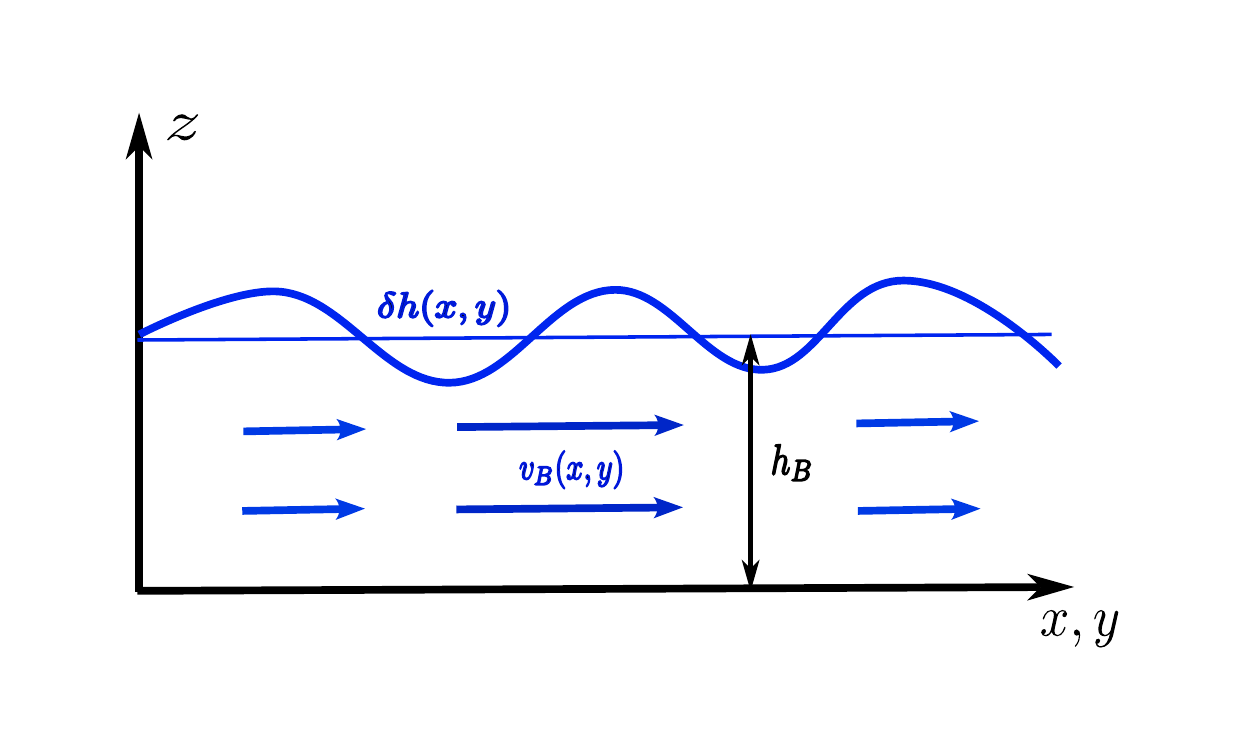}}
  \caption[Gravity waves in a shallow fluid basin]{Gravity waves in a 
  shallow fluid basin with a background horizontal flow.}
  \label{F:fig1}
\end{figure}

A wonderful example of the occurrence of an effective metric in nature
is that provided by gravity waves in a shallow basin filled with
liquid \citep{Schutzhold:2002rf}. (See Fig.~\ref{F:fig1}.)%
\footnote{Of course, we now mean
  ``gravity wave'' in the 
  traditional fluid mechanics sense of a water wave whose restoring force
  is given by ordinary Newtonian gravity. Waves in the fabric of spacetime
  are more properly called ``gravitational waves'', though this usage
  seems to be in decline within the general relativity community. Be very
  careful in any situation where there is even a possibility of confusing
  the two concepts.} 
If one neglects the
viscosity and considers an irrotational flow,
$\vbf=\bnabla\phi$, one can write Bernoulli's equation in
the presence of Earth's gravity as %
\begin{eqnarray}
\partial_t \phi + \half (\bnabla\phi)^2 = -\frac{p}{\rho}-gz-V_{\parallel}.
\label{bernoulli-gravity}
\end{eqnarray}
Here $\rho$ is the density of the fluid, $p$ its pressure, $g$ the
gravitational acceleration and $V_{\parallel}$ a potential associated
with some external force necessary to establish an horizontal flow in
the fluid. We denote that flow by $\vbf_\mathrm{B}^{\parallel}$. We must
also impose the boundary conditions that the pressure at the surface,
and the vertical velocity at the bottom, both vanish. That is,
$p(z=h_\mathrm{B})=0$, and $v_\perp(z=0)=0$.

Once a horizontal background flow is established, one can see that the
perturbations of the velocity potential satisfy
\begin{eqnarray}
\partial_t \,\delta \phi + 
\vbf_\mathrm{B}^{\parallel} \cdot \bnabla_{\parallel} \delta \phi
=- {\delta p \over \rho}.
\end{eqnarray}
If we now expand this perturbation potential in a Taylor series 
\begin{eqnarray}
\delta \phi =
\sum_{n=0}^{\infty} {z^n \over n!} \; \delta \phi_n(x,y),
\end{eqnarray}
it is not difficult to prove \citep{Schutzhold:2002rf} that surface
waves with long wavelengths (long compared with the depth of the basin,
$\lambda \gg h_\mathrm{B}$), can be described to a good approximation by $\delta
\phi_0(x,y)$ and that this field ``sees'' an effective metric of the
form
\begin{eqnarray}
\d s^2={1 \over c^2} \left[-(c^2 - v_\mathrm{B}^{\parallel 2})\; \d t^2
-2\vbf_\mathrm{B}^{\parallel} \cdot \mathbf{dx} \; \d t
+\mathbf{dx} \cdot \mathbf{dx}\right],
\end{eqnarray}
where $c \equiv \sqrt{gh_\mathrm{B}}$. The link between small variations of the
potential field and small variations of the position of the surface is
provided by the following equation
\begin{eqnarray}
\delta v_{\perp}=-h_\mathrm{B} \;\nabla_{\parallel}^2 \delta \phi_0=
\partial_t \,\delta h + 
\vbf_\mathrm{B}^{\parallel} \cdot \bnabla_{\parallel} \delta h
=
{\d\over\d t} \,\delta h.
\end{eqnarray}
The entire previous analysis can be generalised to the case in which
the bottom of the basin is not flat, and the background flow not purely
horizontal \citep{Schutzhold:2002rf}. Therefore, one can create
different effective metrics for gravity waves in a shallow fluid basin
by changing (from point to point) the background flow velocity and the
depth, $h_\mathrm{B}(x,y)$.

The main advantage of this model is that the velocity of the surface
waves can very easily be modified by changing the depth of the basin. 
This velocity can be made very slow, and consequently, the creation of
ergoregions should be relatively easier than in other models. 
As described here, this model is completely classical given that the analogy
requires long wavelengths and slow propagation speeds for the gravity
waves. Although the latter feature is convenient for the practical realization
of analogue horizons, it is a disadvantage in trying to detect analogue
Hawking radiation as the relative temperature will necessarily be very
low. (This is why, in order to have a possibility of experimentally observing (spontaneous)
Hawking evaporation and other quantum phenomena, one would need to use ultra-cold quantum fluids.) However, the gravity wave analogue can certainly serve
to investigate the classical phenomena of mode mixing that underlies the
quantum processes. 

It is this particular analogue model (and its extensions to finite depth and surface tension) that underlies the historically-important experimental work of \citet{badulin:782}, and the modern experimental works of Rousseaux et~al. and Weinfurtner et~al. (See section on experimental developments.)

\subsubsection{More general water waves} 
\label{S:WW-gen}

If one moves beyond shallow-water surface waves the physics becomes
more complicated. In the shallow-water regime (wavelength $\lambda$
much greater than water depth $d$) the co-moving dispersion relation
is a simple linear one $\omega = c_s k$, where the speed of sound can
depend on both position and time. Once one moves to finite-depth
($\lambda\sim d$) or deep ($\lambda \ll d$) water, it is a standard
result that the co-moving dispersion relation becomes
\begin{equation}
\omega = \sqrt{gk \tanh(kd)} = c_s k \; \sqrt{ \tanh(kd)\over kd}.
\end{equation}
See, for instance, \citet{Lamb} \S228, p. 354, Eq.~(5). A more
modern discussion in an analogue spacetime context is available
in \citet{Visser:2007du}. Adding surface tension requires a brief
computation based on \citet{Lamb} \S267 p. 459, details can be
found in \citet{Visser:2007du}. The net result is
\begin{equation}
\omega = c_s k \; \sqrt{1+k^2/K^2} \; \sqrt{ \tanh(kd)\over kd}. \qquad 
\end{equation}
Here $K^2 = {g\rho/\sigma}$ is a constant depending on the
acceleration due to gravity, the density, and the surface
tension \citep{Visser:2007du}. Once one adds the effects of fluid
motion, one obtains
\begin{equation}
\omega = \vbf \cdot \k + c_s k \; \sqrt{1+k^2/K^2} \; \sqrt{ \tanh(kd)\over kd}. \qquad 
\end{equation}
All of these features, fluid motion, finite depth, and surface tension
(capillarity), seem to be present in the experimental
investigations by \citet{badulin:782}. All of these
features should be kept in mind when interpreting the
experimental \citep{Rousseaux:2007is} and
theoretical \citep{Rousseaux:2010md} work of Rousseaux et~al., and the
experimental work by \citet{Weinfurtner:2010nu}. 

A feature that is sometimes not remarked on is that the careful
derivation we have previously presented of the acoustic metric, or in
this particular situation the derivation of the shallow-water-wave
effective metric \citep{Schutzhold:2002rf}, makes technical assumptions
tantamount to asserting that one is in the regime where the co-moving
dispersion relation takes the linear form $\omega\approx c_s k$. Once
the co-moving dispersion relation becomes nonlinear, the situation is
more subtle, and based on a geometric acoustics approximation to the
propagation of signal waves one can introduce \emph{several} notions
of conformal ``rainbow'' metric (momentum-dependent metric). Consider
\begin{equation}
g_{ab}(k^2) \propto \left[ \begin{array}{c|c}
\vphantom{\Big|}
- \left\{c^2(k^2) - \delta_{ij} \; v^i\; v^j \right\} & +v_j \\
\hline
\vphantom{\Big|}
+v_i & + h_{ij} \\
\end{array}
\right],
\end{equation}
and the inverse
\begin{equation}
g^{ab}(k^2) \propto \left[ \begin{array}{c|c}
\vphantom{\Big|}
- 1 & +v^j \\
\hline
\vphantom{\Big|}
+v^i & c^2 (k^2) \; h^{ij} - v^i \; v^j \\
\end{array}
\right].
\end{equation}
At a minimum we could think of using the following notions of
propagation speed
\begin{equation}
c(k^2) \to \left\{ \begin{array}{l}
c_\mathrm{phase}(k^2);\\
c_\mathrm{group}(k^2);\\
c_\mathrm{sound} = \lim\limits_{k\to0} c_\mathrm{phase}(k^2) \hbox{ provided this equals } \lim\limits_{k\to0} c_\mathrm{group}(k^2) ;\\
c_\mathrm{signal} = \lim\limits_{k\to\infty} c_\mathrm{phase}(k^2).\\
\end{array}
\right. 
\end{equation}
Brillouin, in his classic book \citep{Brillouin}, identified at
least six useful notions of propagation speed, and many would argue
that the list can be further refined. Each one of these choices for
the rainbow metric encodes different physics, and is useful for
different purposes. It is still somewhat unclear as to which of these
rainbow metrics is ``best'' for interpreting the experimental results
reported in \citet{badulin:782, Rousseaux:2007is, Weinfurtner:2010nu}.

\subsubsection{Classical refractive index}
\label{sec:cri}

The macroscopic Maxwell equations inside a dielectric take the well-known form
\begin{eqnarray}
&\bnabla \cdot \mathbf{B}=0,\quad
&\bnabla \times \mathbf{E} +\partial_t \mathbf{B}=0, \\ 
&\bnabla \cdot \mathbf{D}=0,\quad
&\bnabla \times \mathbf{H} - \partial_t \mathbf{D}=0,
\end{eqnarray}
with the constitutive relations 
$\mathbf{H}=\mathbold{\mu}^{-1}\cdot \mathbf{B}$ and 
$\mathbf{D}=\mathbold{\epsilon}\cdot \mathbf{E}$. Here, 
$\mathbold{\epsilon}$ is the $3\times 3$ permittivity tensor and $\mathbold{\mu}$ 
the $3\times 3$ permeability tensor of the medium. 
These equations can be written in a condensed way as 
\begin{equation}
\partial_\alpha \left( Z^{\mu\alpha\nu\beta} \; F_{\nu\beta}\right)=0
\label{elec-equation}
\end{equation}
where $F_{\nu\beta}= A_{[\nu,\beta]}$ is the electromagnetic tensor, 
\begin{equation}
F_{0i}=-F_{i0}=-E_i,\quad F_{ij}=\varepsilon_{ijk}B^k,
\end{equation}
and (assuming the medium is at rest) the non-vanishing components of
the 4th-rank tensor $Z$ are given by
\begin{eqnarray}
&&Z^{0i0j}=-Z^{0ij0}=Z^{i0j0}=-Z^{i00j}=
-\half\epsilon^{ij}; \\
&&Z^{ijkl}=\half\varepsilon^{ijm}\;\varepsilon^{kln}\; \mu^{-1}_{mn};
\end{eqnarray}
supplemented by the conditions that $Z$ is antisymmetric on its first
pair of indices and antisymmetric on its second pair of indices. Without
significant loss of generality we can ask that $Z$ also be symmetric
under pairwise interchange of the first pair of indices with the second
pair -- thus $Z$ exhibits most of the algebraic symmetries of the
Riemann tensor, though this appears to merely be accidental, and not
fundamental in any way.

If we compare this to the Lagrangian for electromagnetism in curved
spacetime
\begin{equation}
\L = \sqrt{-g} \; g^{\mu\alpha} \; g^{\nu\beta}\; F_{\mu\nu} \; F_{\alpha\beta}
\end{equation}
we see that in curved spacetime we can also write the electromagnetic 
equations of motion in the form~(\ref{elec-equation}) where now (for 
some constant $K$, often chosen to be $K\to{1\over2}$):
\begin{equation}
Z^{\mu\nu\alpha\beta} = K\; \sqrt{-g} 
\;\left\{ g^{\mu\alpha} \; g^{\nu\beta} 
- g^{\mu\beta} \; g^{\nu\alpha}\right\}. 
\end{equation}
If we consider a static gravitational field we can always re-write 
it as a conformal factor multiplying an ultra-static metric 
\begin{eqnarray}
g_{\mu\nu}=\Omega^2\; \left\{-1\oplus g_{ij} \right\}
\end{eqnarray}
then
\begin{eqnarray}
&&
Z^{0i0j}=-Z^{0ij0}=Z^{i0j0}=-Z^{i00j}=
-K\;\sqrt{-g} \; g^{ij}; \\
&&Z^{ijkl}= K\; \sqrt{-g} \left\{ g^{ik} \; g^{jl} 
- g^{il} \; g^{jk}\right\}. 
\end{eqnarray}

The fact that $Z$ is independent of the conformal factor $\Omega$ is
simply the reflection of the well-known fact that the Maxwell equations
are conformally invariant in (3+1) dimensions. Thus, if we wish to have
the analogy (between a static gravitational field and a dielectric
medium at rest) hold \emph{at the level of the wave equation} (physical
optics) we must satisfy the two stringent constraints

\begin{eqnarray}
&&
K\;\sqrt{-g} \; g^{ij} = {1 \over 2}\;\epsilon^{ij};
\\
&&
K\; \sqrt{-g} \left\{ g^{ik} \; g^{jl} - g^{il} \; g^{jk}\right\} = 
\half\;\varepsilon^{ijm}\;\varepsilon^{kln}\; \mu^{-1}_{mn}.
\end{eqnarray}
The second of these constraints can be written as
\begin{equation}
K\; \sqrt{-g} \; \varepsilon_{ijm}\;\varepsilon_{kln} 
\left\{ g^{ik} \; g^{jl} \right\} = \mu^{-1}_{mn}.
\end{equation}
In view of the standard formula for $3\times3$ determinants
\begin{equation}
\varepsilon_{ijm}\;\varepsilon_{kln} \left\{ X^{ik} \; X^{jl} \right\} 
= 2 \;\det{X}\; X^{-1}_{mn},
\end{equation}
this now implies
\begin{equation}
2 K {g_{ij}\over\sqrt{-g} } = \mu^{-1}_{ij},
\end{equation}
whence
\begin{equation}
{1\over2 K} \; \sqrt{-g} \; g^{ij} = \mu^{ij}.
\end{equation}
Comparing this with
\begin{equation}
{2 K} \; \sqrt{-g} \; g^{ij} = \epsilon^{ij},
\end{equation}
we now have:
\begin{eqnarray}
\epsilon^{ij} &=& 4 \; K^2 \; \mu^{ij};
\\
g^{ij} &=& {4\;K^2\over\det\mathbold{\epsilon}}\; \epsilon^{ij}
= {1\over4\;K^2\;\det\mathbold{\mu}}\; \mu^{ij}.
\label{dielectric-metric-physical}
\end{eqnarray}
To rearrange this, introduce the matrix square root
$[\mathbold{\mu}^{1/2}]^{ij}$, which
always exists because $\mathbold{\mu}$ is real positive definite and
symmetric. Then
\begin{equation}
g^{ij} = 
\left[\left\{ {
\mathbold{\mu}^{1/2} \; \mathbold{\epsilon} \; \mathbold{\mu}^{1/2}
\over \det(\mathbold{\mu}\;\mathbold{\epsilon}) } \right\}^{1/2}
\right]^{ij}.
\end{equation}
Note that if you are given the static gravitational field (in the form
$\Omega$, $g_{ij}$) you can always solve it to find an equivalent analogue
in terms of permittivity/permeability (albeit an analogue that satisfies
the mildly unphysical constraint
$\mathbold{\epsilon}\propto\mathbold{\mu}$).%
\footnote{The existence of
  this constraint has been independently re-derived several times in the
  literature. In contrast, other segments of the literature seem blithely
  unaware of this important restriction on just when permittivity and
  permeability are truly equivalent to an effective metric.} 
On the other hand, if you are given permeability and permittivity tensors 
$\mathbold{\epsilon}$ and $\mathbold{\mu}$, then it is \emph{only} for
that subclass of media that satisfy
$\mathbold{\epsilon}\propto\mathbold{\mu}$ that one can perfectly mimic
\emph{all} of the electromagnetic effects by an equivalent gravitational
field. Of course, this can be done provided one only considers
wavelengths that are sufficiently long for the macroscopic description
of the medium to be valid. In this respect it is interesting to note
that the behaviour of the refractive medium at high frequencies has been
used to introduce an effective cutoff for the modes involved in Hawking
radiation \citep{Reznik:1997ag}. We shall encounter this model (which 
is also known in the literature as a solid state analogue model) later on
when we consider the trans-Planckian problem for Hawking
radiation. Let us stress that if one were able to directly
probe the quantum effective photons over a dielectric medium, then one
would be dealing with a quantum analogue model instead of a classical
one.

\paragraph*{Eikonal approximation:\;}

With a bit more work this discussion can be extended to a medium in
motion, leading to an extension of the Gordon metric. Alternatively, one
can agree to ask more limited questions by working at the level of
geometrical optics (adopting the eikonal approximation), in which case
there is no longer any restriction on the permeability and permittivity
tensors. To see this, construct the matrix

\begin{eqnarray}
C^{\mu\nu}=Z^{\mu\alpha\nu\beta} \; k_\alpha \, k_\beta.
\end{eqnarray}
The dispersion relations for the propagation of photons (and therefore
the sought for geometrical properties) can be obtained from the reduced
determinant of $C$ (notice that the [full] determinant of $C$ is identically
zero as $C^{\mu\nu}k_\nu=0$; the reduced determinant is that
associated with the three directions orthogonal to $k_\nu=0$). By
choosing the gauge $A_0=0$ one can see that this reduced determinant
can be obtained from the determinant of the $3\times3$ sub-matrix $C^{ij}$.
This determinant is
\begin{eqnarray}
\det( C^{ij})={1 \over 8} \det 
\left(-\omega^2 \epsilon^{ij} + 
\varepsilon^{ikm}\;\varepsilon^{jln}\; \mu^{-1}_{mn} k_k k_l\right)
\end{eqnarray}
or, after making some manipulations,
\begin{eqnarray}
\det( C^{ij})= {1 \over 8} \det 
\left[-\omega^2 \epsilon^{ij} + (\det \mathbold{\mu})^{-1} 
(\mu^{ij} \mu^{kl} k_k k_l -\mu^{im} k_m \mu^{jl} k_l) \right].
\end{eqnarray}
To simplify this, again introduce the matrix square roots
$[\mathbold{\mu}^{1/2}]^{ij}$ and $[\mathbold{\mu}^{-1/2}]_{ij}$, which
always exist because the relevant matrices are real positive definite and
symmetric.
Then define
\begin{equation}
\tilde k^i = [\mathbold{\mu}^{1/2}]^{ij} \; k_j
\end{equation}
and
\begin{equation}
[\mathbold{\tilde\epsilon}]^{ij} = \det(\mathbold{\mu}) \; 
[\mathbold{\mu}^{-1/2} \; \mathbold{\epsilon} \;\mathbold{\mu}^{-1/2}]_{ij} 
\end{equation}
so that
\begin{eqnarray}
&&\det( C^{ij}) \propto 
\det \left\{
-\omega^2 \;
[\mathbold{\tilde\epsilon}]^{ij} 
+ \delta^{ij} \; [\delta_{mn} \; \tilde k^m \; \tilde k^n]
-\tilde k^i \; \tilde k^j 
\right\}.
\end{eqnarray}
The behaviour of this dispersion relation now depends critically on the
way that the eigenvalues of $\mathbold{\tilde\epsilon}$ are distributed.

\paragraph*{3 degenerate eigenvalues:\;} 

If all eigenvalues are degenerate then $\mathbold{\tilde \epsilon} =
\tilde \epsilon \;\mathbf{I}$, implying
$\mathbold{\epsilon}\propto\mathbold{\mu}$ but now with the possibility of
a position-dependent proportionality factor (in the case of physical
optics the proportionality factor was constrained to be a
position-independent constant). In this case we now easily evaluate
\begin{equation}
\mathbold{\epsilon} = 
{\tr(\mathbold{\epsilon})\over\tr(\mathbold{\mu})} \; \mathbold{\mu}
\qquad
\hbox{and}
\qquad
\tilde\epsilon = \det\mathbold{\mu} \; 
{\tr(\mathbold{\epsilon})\over\tr(\mathbold{\mu})},
\end{equation}
while
\begin{equation}
\det( C^{ij}) \propto \omega^2 \;
\left\{ \omega^2 - [ {\tilde\epsilon}^{-1}\; \delta_{mn} \; 
\tilde k^m\;\tilde k^n]  \right\}^2.
\end{equation}
That is
\begin{equation}
\det( C^{ij}) \propto \omega^2 
\left\{ \omega^2 - [g^{ij} \; k_i\; k_j]  \right\}^2,
\end{equation}
with
\begin{equation}
\label{dielectric-metric}
g^{ij} = {1\over\tilde\epsilon}\; [\mathbold{\mu}]^{ij} = 
{ \tr(\mathbold{\mu})\;[\mathbold{\mu}]^{ij}\over \tr(\mathbold{\epsilon}) \; 
\det\mathbold{\mu} }
= 
{ \tr(\mathbold{\epsilon})\;[\mathbold{\epsilon}]^{ij}\over \tr(\mathbold{\mu}) \; 
\det\mathbold{\epsilon} }.
\end{equation}
This last result is compatible with but more general than the result
obtained under the more restrictive conditions of physical optics. In
the situation where both permittivity and permeability are isotropic,
($\epsilon^{ij} =\epsilon\;\delta^{ij}$ and $\mu^{ij} =
\mu\;\delta^{ij}$) this reduces to the perhaps more expected result

\begin{equation}
g^{ij} = {\delta^{ij}\over \epsilon\;\mu}.
\end{equation}

\paragraph*{2 distinct eigenvalues:\;} 

If $\mathbold{\tilde\epsilon}$ has two distinct eigenvalues then the
determinant $\det(C^{ij})$ factorises into a trivial factor of
$\omega^2$ and two quadratics. Each quadratic corresponds to a distinct
effective metric. This is the physical situation encountered in uni-axial
crystals, where the \emph{ordinary} and \emph{extraordinary} rays each
obey distinct quadratic dispersion relations \citep{Born-Wolf}. From
the point of view of analogue models this corresponds to a two-metric
theory.

\paragraph*{3 distinct eigenvalues:\;} 

If $\mathbold{\tilde\epsilon}$ has three distinct eigenvalues then the
determinant $\det(C^{ij})$ is the product of a trivial factor of
$\omega^2$ and a \emph{non-factorizable quartic}. This is the physical
situation encountered in bi-axial crystals \citep{Born-Wolf,
Visser:2002sf}, and it seems that no meaningful notion of the effective
Riemannian metric can be assigned to this case. (The use of Finsler
geometries in this situation is an avenue that may be worth
pursuing \citep{EDM}. But note some of the negative results obtained
in \citet{Skakala:2008kf, Skakala:2008jp, Skakala:2010hw}.)

\paragraph*{Abstract linear electrodynamics:\;}

Hehl and co-workers have championed the idea of using the linear constitutive
relations of electrodynamics as the primary quantities, and then treating the
spacetime metric (even for flat space) as a \emph{derived}
concept. 
See \citet{Obukhov:1999ug, Hehl:2004yk, Lammerzahl:2004ww, Hehl:2004tk}.

\paragraph*{Nonlinear electrodynamics:\;}

In general, the permittivity and permeability tensors can be modified by
applying strong electromagnetic fields (this produces an effectively
nonlinear electrodynamics). The entire previous discussion still
applies if one considers the photon as the linear perturbation of the
electromagnetic field over a background configuration
\begin{equation}
F_{\mu\nu}=F_{\mu\nu}^\mathrm{bg}+f_{\mu\nu}^\mathrm{ph}. 
\end{equation}
The background
field $F_{\mu\nu}^\mathrm{bg}$ sets the value of 
$\epsilon^{ij}(F^\mathrm{bg})$, and $\mu^{ij}(F^\mathrm{bg})$. 
Eq.~(\ref{elec-equation}) then
becomes an equation for $f_{\mu\nu}^\mathrm{ph}$. This approach has been
extensively investigated by Novello and co-workers \citep{Novello:1999pg,
  Novello:2001gk, DeLorenci:2002mi, Novello:2002ed, Novello:2003je,
  Novello:2003kh, Novello:2002iq, Fonseca-Barbatti:2002hq}. 


\paragraph*{Summary:\;}

The propagation of photons in a dielectric medium characterised by
$3\times3$ permeability and permittivity tensors constrained by
$\mathbold{\epsilon} \propto \mathbold{\mu}$ is equivalent (at the level
of geometric optics) to the propagation of photons in a curved spacetime
manifold characterised by the ultra-static metric
(\ref{dielectric-metric}), provided one only considers wavelengths that
are sufficiently long for the macroscopic description of the medium to
be valid. If, in addition, one takes a fluid dielectric, by controlling
its flow one can generalise the Gordon metric and again reproduce
metrics of the Painlev\'e--Gullstrand type, and therefore geometries
with ergo-regions. If the proportionality constant relating
$\mathbold{\epsilon} \propto \mathbold{\mu}$ is position independent, one
can make the stronger statement (\ref{dielectric-metric-physical}) which
holds true at the level of physical optics. Recently this topic has
been revitalised by the increasing interest in (classical)
meta-materials. See for instance~\citep{Schuster:2017mdx, Schuster:2018ycr,Schuster:2018cwt,Schuster:2018bqx}.

\subsubsection{Normal mode meta-models}

We have already seen how linearizing the Euler--Lagrange equations for a
single scalar field naturally leads to the notion of an effective
spacetime metric. If more than one field is involved the situation
becomes more complicated, in a manner similar to that of geometrical
optics in uni-axial and bi-axial crystals. (This should, with hindsight,
not be too surprising since electromagnetism, even in the presence of a
medium, is definitely a Lagrangian system and definitely involves more
than one single scalar field.) A normal mode analysis based on a general
Lagrangian (many fields but still first order in derivatives of those
fields) leads to a concept of \emph{refringence}, or more specifically
\emph{multi-refringence}, a generalization of the birefringence of
geometrical optics. To see how this comes about, consider a
straightforward generalization of the one-field case.

We want to consider linearized fluctuations around some background
solution of the equations of motion. As in the single-field case we
write (here we will follow the notation and conventions of
\citealt{Barcelo:2001cp})
\begin{equation}
\phi^A(t,\vec x) = \phi_0^A(t,\vec x) + \epsilon \;\phi_1^A(t,\vec x) + 
{\epsilon^2\over2} \; \phi_2^A(t,\vec x) + \O(\epsilon^3).
\end{equation}
Now use this to expand the Lagrangian
\begin{eqnarray}
{\L}(\partial_\mu \phi^A,\phi^A) 
&=& 
{\L}(\partial_\mu \phi^A_0,\phi^A_0)
+\epsilon \left[ 
{\partial \L\over\partial(\partial_\mu \phi^A)} \; \partial_\mu \phi^A_1
+
{\partial \L\over\partial\phi^A} \; \phi^A_1
\right]
\nonumber
\\
&+&
{\epsilon^2\over2} \left[ 
{\partial \L\over\partial(\partial_\mu \phi^A)} \; \partial_\mu \phi^A_2
+
{\partial \L\over\partial\phi^A} \; \phi^A_2
\right]
\nonumber
\\
&+&
{\epsilon^2\over2} \Bigg[ 
{\partial^2 \L\over\partial(\partial_\mu \phi^A) \; \partial(\partial_\nu \phi^B)} 
\; \partial_\mu \phi^A_1 \; \partial_\nu \phi^B_1
\nonumber
\\
&&
+
2 {\partial^2 \L\over \partial(\partial_\mu \phi^A)\; \partial \phi^B} 
\; \partial_\mu \phi^A_1 \; \phi^B_1
+
{\partial^2 \L\over \partial\phi^A\; \partial\phi^B} 
\; \phi^A_1 \; \phi^B_1
\Bigg]
\nonumber
\\
&&
+
\O(\epsilon^3).
\end{eqnarray}
Consider the action
\begin{equation}
S[\phi^A] = \int \d^{d+1} x \; \L(\partial_\mu\phi^A,\phi^A).
\end{equation}
Doing so allows us to integrate by parts. As in the single-field
case we can use the Euler--Lagrange equations to
discard the linear terms (since we are linearizing around a solution
of the equations of motion) and so get
\begin{eqnarray}
S[\phi^A] &=& S[\phi_0^A] 
\nonumber
\\
&+&
{\epsilon^2\over2}
\int \d^{d+1} x \Bigg[
\left\{
  {\partial^2 \L\over
   \partial(\partial_\mu \phi^A) \; \partial(\partial_\nu \phi^B)} 
\right\}
\; \partial_\mu \phi^A_1
\; \partial_\nu \phi^B_1
\nonumber
\\
&&
\quad
+
2 
\left\{
{\partial^2 \L\over \partial(\partial_\mu \phi^A)\; \partial \phi^B} 
\right\}
\; \partial_\mu \phi^A_1 
\; \phi^B_1
+
\left\{
{\partial^2 \L\over \partial\phi^A\; \partial \phi^B} 
\right\}
\; \phi^A_1 \; \phi^B_1
\Bigg]
\nonumber
\\
&+&
\O(\epsilon^3).
\end{eqnarray}
Because the fields now carry indices ($AB$\/) we cannot cast the
action into quite as simple a form as was possible in the single-field
case. The equation of motion for the linearized fluctuations are now
read off as
\begin{eqnarray}
&&  \partial_\mu \left(\left\{
  {\partial^2 \L\over\partial(\partial_\mu \phi^A) \; 
  \partial(\partial_\nu \phi^B)} 
   \right\}
 \partial_\nu \phi^B_1 \right)
+  \partial_\mu \left(
{\partial^2 \L\over \partial(\partial_\mu \phi^A)\; \partial \phi^B} 
\; \phi^B_1 \right)
\nonumber
\\
&&
\qquad
- \partial_\mu \phi^B_1 \; 
{\partial^2 \L\over \partial(\partial_\mu \phi^B)\; \partial \phi^A} 
- \left(
{\partial^2 \L\over \partial\phi^A\; \partial \phi^B} 
\right)
\phi^B_1
= 0. 
\label{E:unmaked-up}
\end{eqnarray}
This is a linear second-order {\emph{system}} of partial differential
equations with position-dependent coefficients. This system of PDEs is
automatically self-adjoint (with respect to the trivial ``flat''
measure $\d^{d+1} x$).

To simplify the notation we introduce a number of definitions. First
\begin{equation}
f^{\mu\nu}{}_{AB} \equiv
{1 \over 2}\left(
{\partial^2 {\cal L}\over\partial(\partial_\mu \phi^A) \; 
\partial(\partial_\nu \phi^B)}+
{\partial^2 {\cal L}\over\partial(\partial_\nu \phi^A) \; 
\partial(\partial_\mu \phi^B)} 
\right).
\end{equation}
This quantity is independently symmetric under interchange of $\mu$,
$\nu$ and $A$, $B$. We will want to interpret this as a generalization
of the ``densitised metric'', $f^{\mu\nu}$, but the interpretation is
not as straightforward as for the single-field case. Next, define
\begin{eqnarray}
\Gamma^\mu{}_{AB} &\equiv&
+
{\partial^2 {\cal L}\over \partial(\partial_\mu \phi^A)\; \partial \phi^B}
-
{\partial^2 {\cal L}\over \partial(\partial_\mu \phi^B)\; \partial \phi^A} 
\nonumber\\
&&
+
{1 \over 2}\partial_{\nu}\left(
{\partial^2 {\cal L}\over\partial(\partial_\nu \phi^A) \; 
\partial(\partial_\mu \phi^B)}-
{\partial^2 {\cal L}\over\partial(\partial_\mu \phi^A) \; 
\partial(\partial_\nu \phi^B)} 
\right).
\end{eqnarray}
This quantity is anti-symmetric in $A$, $B$. One might want to interpret
this as some sort of ``spin connection'', or possibly as some
generalization of the notion of ``Dirac matrices''. Finally, define
\begin{equation}
K_{AB} =
-{\partial^2 \L\over \partial\phi^A\; \partial \phi^B} 
+
{1\over2}
\partial_\mu \left(
{\partial^2 \L\over \partial(\partial_\mu \phi^A)\; \partial \phi^B} 
\right)
+
{1\over2}
\partial_\mu \left(
{\partial^2 \L\over \partial(\partial_\mu \phi^B)\; \partial \phi^A} 
\right).
\end{equation}
This quantity is by construction symmetric in $(AB)$. We will want
to interpret this as some sort of ``potential'' or ``mass matrix''.
Then the crucial point for the following discussion is to realise that
Eq.~(\ref{E:unmaked-up}) can be written in the compact form
\begin{equation}
\partial_\mu \left( f^{\mu\nu}{}_{AB}\; \partial_\nu \phi^B_1 \right)
+ {1\over2} 
\left[ \Gamma^\mu_{AB} \; \partial_\mu \phi^B_1 
+ \partial_\mu (\Gamma^\mu_{AB} \; \phi^B_1) \right]
+ K_{AB} \; \phi^B_1 
= 0.
\label{E:system}
\end{equation}
Now it is more transparent that this is a formally self-adjoint
second-order linear \emph{system} of PDEs. Similar considerations can be 
applied to the linearization of any hyperbolic system of second-order PDEs. 

Consider an eikonal approximation for an arbitrary direction
in field space; that is, take
\begin{equation}
\phi^A(x) = \epsilon^A(x) \; \exp[-i\varphi(x)],
\end{equation}
with $\epsilon^A(x)$ a slowly varying amplitude, and $\varphi(x)$ a
rapidly varying phase. In this eikonal approximation (where we neglect
gradients in the amplitude, and gradients in the coefficients of the
PDEs, retaining only the gradients of the phase) the linearized system
of PDEs (\ref{E:system}) becomes
\begin{equation}
\left\{
f^{\mu\nu}{}_{AB}\; \partial_\mu \varphi(x) \; \partial_\nu \varphi(x) 
+
\Gamma^\mu{}_{AB}\;  \partial_\mu \varphi(x)
+
K_{AB} 
\right\} \; 
\epsilon^B(x) = 0.
\label{E:eikonal}
\end{equation}
This has a nontrivial solution if and only if $\epsilon^A(x)$ is a
null eigenvector of the matrix 
\begin{equation}
f^{\mu\nu}{}_{AB}\; k_\mu \; k_\nu
+ \Gamma^\mu{}_{AB}\;k_\mu 
+ K_{AB},
\end{equation}
where $k_\mu=\partial_\mu\varphi(x)$.
Now, the condition for such a null eigenvector to exist is that
\begin{equation}
F(x,k) \equiv 
\det \left\{ f^{\mu\nu}{}_{AB}\; 
k_\mu \; k_\nu 
+
\Gamma^\mu{}_{AB}\;  k_\mu 
+
K_{AB}
\right\} = 0,
\label{E:fresnel}
\end{equation}
with the determinant to be taken on the field space indices $AB$. This
is the natural generalization to the current situation of the Fresnel
equation of birefringent optics \citep{Born-Wolf,Landau}. Following the
analogy with the situation in electrodynamics (either nonlinear
electrodynamics, or more prosaically propagation in a birefringent
crystal), the null eigenvector $\epsilon^A(x)$ would correspond to a
specific ``polarization''. The Fresnel equation then describes how
different polarizations can propagate at different velocities (or in more
geometrical language, can see different metric
structures). In the language of particle physics, this determinant condition
$F(x,k)=0$ is the natural generalization of the ``mass shell''
constraint. Indeed, it is useful to define the mass shell as a subset of
the cotangent space by
\begin{equation}
{\cal F}(x) \equiv
\left\{ 
k_\mu \; \bigg| \; F(x,k) = 0
\right\}.
\label{E:mass-shell}
\end{equation}
In more mathematical language we are looking at the null space of the
determinant of the ``symbol'' of the system of PDEs. By investigating
$F(x,k)$ one can recover part (not all) of the information encoded in
the matrices $f^{\mu\nu}{}_{AB}$, $\Gamma^\mu{}_{AB}$, and $K_{AB}$, or
equivalently in the ``generalised Fresnel equation'' (\ref{E:fresnel}).
(Note that for the determinant equation to be useful it should be
non-vacuous; in particular one should carefully eliminate all gauge and
spurious degrees of freedom before constructing this ``generalised
Fresnel equation'', since otherwise the determinant will be identically
zero.) We now want to make this analogy with optics more precise, by
carefully considering the notion of characteristics and characteristic
surfaces. We will see how to extract from the the high-frequency
high-momentum regime described by the eikonal approximation all the
information concerning the causal structure of the theory.

One of the key structures that a Lorentzian spacetime metric provides is
the notion of causal relationships. This suggests that it may be
profitable to try to work backwards from the causal structure to
determine a Lorentzian metric. Now the causal structure implicit in the
system of second-order PDEs given in Eq.~(\ref{E:system}) is
described in terms of the characteristic surfaces, and it is for this
reason that we now focus on characteristics as a way of encoding causal
structure, and as a surrogate for some notion of a Lorentzian metric. Note
that, via the Hadamard theory of surfaces of discontinuity, the
characteristics can be identified with the infinite-momentum limit of
the eikonal approximation \citep{Hadamard}. That is, when extracting the
characteristic surfaces we neglect subdominant terms in the generalised
Fresnel equation and focus only on the leading term in the symbol
($f^{\mu\nu}{}_{AB}$). In the language of particle physics, going to the
infinite-momentum limit puts us on the light cone instead of the mass
shell; and it is the light cone that is more useful in determining
causal structure. The ``normal cone'' at some specified point $x$,
consisting of the locus of normals to the characteristic surfaces, is
defined by
\begin{equation}
{\cal N}(x) \equiv
\left\{ 
k_\mu \; \bigg| \; 
\det\left(f^{\mu\nu}{}_{AB} \;\; k_\mu \; k_\mu\right) = 0
\right\}.
\label{E:normal}
\end{equation}

As was the case for the Fresnel Eq.~(\ref{E:fresnel}), the
determinant is to be taken on the field indices $AB$. (Remember to
eliminate spurious and gauge degrees of freedom so that this determinant
is not identically zero.) We emphasise that the algebraic equation
defining the normal cone is the leading term in the Fresnel equation
encountered in discussing the eikonal approximation. If there are $N$
fields in total then this ``normal cone'' will generally consist of
$N$ nested sheets each with the topology (not necessarily the geometry)
of a cone. Often several of these cones will coincide, which is not
particularly troublesome, but unfortunately it is also common for some
of these cones to be degenerate, which is more problematic.

It is convenient to define a function $Q(x,k)$ on the co-tangent bundle
\begin{equation}
Q(x,k) \equiv \det\left(f^{\mu\nu}{}_{AB}(x) \; k_\mu \; k_\mu \right).
\end{equation}
The function $Q(x,k)$ defines a completely-symmetric spacetime tensor
(actually, a tensor density) with $2N$ indices
\begin{equation}
Q(x,k) = Q^{\mu_1\nu_1\mu_2\nu_2\cdots\mu_N\nu_N}(x) 
\;\; k_{\mu_1} \; k_{\nu_1} \; k_{\mu_2} \; k_{\nu_2} \cdots \; 
k_{\mu_N} \; k_{\nu_N}. 
\end{equation}
(Remember that $f^{\mu\nu}{}_{AB}$ is symmetric in both $\mu\nu$ and
$AB$ independently.) Explicitly, using the expansion of the
determinant in terms of completely antisymmetric field-space
Levi--Civita tensors
\begin{equation}
Q^{\mu_1\nu_1\mu_2\nu_2\cdots\mu_N\nu_N} = 
\frac{1}{N!}\,
\epsilon^{A_1 A_2 \cdots A_N} \; \epsilon^{B_1 B_2 \cdots B_N} \,
f^{\mu_1\nu_1}{}_{A_1 B_1} \; f^{\mu_2\nu_2}{}_{A_2 B_2} \; 
\cdots f^{\mu_N\nu_N}{}_{A_N B_N}.
\end{equation}
In terms of this $Q(x,k)$ function, the normal cone is
\begin{equation}
{\cal N}(x) \equiv
\left\{ 
k_\mu \; \bigg| \; Q(x,k) = 0
\right\}.
\end{equation}
In contrast, the ``Monge cone'' (\emph{aka} ``ray cone'', \emph{aka}
``characteristic cone'', \emph{aka} ``null cone'') is the envelope of the
set of characteristic surfaces through the point $x$. Thus the
``Monge cone'' is dual to the ``normal cone'', its explicit
construction is given by (Courant and Hilbert~\cite[vol.~2, p.~583]{Courant}):
\begin{equation}
{\cal M}(x) = 
\left\{ 
t^\mu = {\partial Q(x,k)\over\partial k_\mu} 
\;
\bigg| \; k_\mu \in {\cal N}(x)
\right\}.
\label{E:null}
\end{equation}

The structure of the normal and Monge cones encode all the information
related with the causal propagation of signals associated with the
system of PDEs. We will now see how to relate this causal structure with
the existence of effective spacetime metrics, from the experimentally
favoured single-metric theory compatible with the Einstein equivalence
principle to the most complicated case of pseudo-Finsler
geometries \citep{EDM}.

\begin{itemize}
\item Suppose that $f^{\mu\nu}{}_{AB}$ factorises
  \begin{equation}
    f^{\mu\nu}{}_{AB} = h_{AB} \;  f^{\mu\nu}.
  \end{equation}
  Then 
  \begin{equation}
    Q(x,k) = \det(h_{AB}) \; [f^{\mu\nu}\; k_\mu\; k_\nu]^N.
  \end{equation}

  The Monge cones and normal cones are then true geometrical cones (with
  the $N$ sheets lying directly on top of one another). The normal modes
  all see the same spacetime metric, defined up to an unspecified
  conformal factor by $g^{\mu\nu}\propto f^{\mu\nu}$. This situation is
  the most interesting from the point of view of general relativity.
  Physically, it corresponds to a single-metric theory, and mathematically
  it corresponds to a strict algebraic condition on the
  $f^{\mu\nu}{}_{AB}$.
  
\item The next most useful situation corresponds to the commutativity
  condition: 
  \begin{equation}
    f^{\mu\nu}{}_{AB}\; f^{\alpha\beta}{}_{BC} = 
    f^{\alpha\beta}{}_{AB}\; f^{\mu\nu}{}_{BC};
    \qquad
    \hbox{that is}
    \qquad
    [\; \bm{f}^{\mu\nu}, \bm{f}^{\alpha\beta} \,] = 0.
  \end{equation}
  If this algebraic condition is satisfied, then for all spacetime indices
  $\mu\nu$ and $\alpha\beta$ the $f^{\mu\nu}{}_{AB}$ can be simultaneously
  diagonalised in field space leading to
  \begin{equation}
    \bar f^{\mu\nu}{}_{AB} = 
    \diag\{\bar f^{\mu\nu}_1, \bar f^{\mu\nu}_2, \bar f^{\mu\nu}_3, 
    \dots, \bar f^{\mu\nu}_N\}
  \end{equation}
  and then
  \begin{equation}
    Q(x,k) = \prod_{A=1}^N \; [\bar f^{\mu\nu}_A \; k_\mu\; k_\nu].
  \end{equation}
  This case corresponds to an $N$-metric theory, where up to an 
  unspecified conformal factor
  $g^{\mu\nu}_A \propto \bar f^{\mu\nu}_A$. This is the natural 
  generalization of the two-metric situation in bi-axial crystals. 
  
\item If $f^{\mu\nu}{}_{AB}$ is completely general, satisfying no
  special algebraic condition, then $Q(x,k)$ does not factorise and is, in
  general, a polynomial of degree $2N$ in the wave vector $k_\mu$. This is
  the natural generalization of the situation in bi-axial crystals. (And
  for any deeper analysis of this situation one will almost certainly need
  to adopt pseudo-Finsler techniques \citep{EDM}. But note some of the
  negative results obtained in \citet{Skakala:2008kf, Skakala:2008jp,
    Skakala:2010hw}.)
  
\end{itemize}

The message to be extracted from this rather formal discussion is that
effective metrics are rather general and mathematically robust objects
that can arise in quite abstract settings -- in the abstract setting
discussed here it is the algebraic properties of the object
$f^{\mu\nu}{}_{AB}$ that eventually leads to mono-metricity,
multi-metricity, or worse. The current abstract discussion also serves
to illustrate, yet again, 

\begin{enumerate}
\item that there is a significant difference between the levels of
  physical normal modes (wave equations), and geometrical normal modes
  (dispersion relations), and
\item that the densitised inverse metric is in many ways more
  fundamental than the metric itself.
\end{enumerate}

\subsection{Quantum models}
\label{S:quantum-models}
\subsubsection{Bose--Einstein condensates} 

We have seen that one of the main aims of research in analogue models of
gravity is the possibility of simulating semiclassical gravity
phenomena, such as the Hawking radiation effect or cosmological particle
production. In this sense systems characterised by a high degree of
quantum coherence, very cold temperatures, and low speeds of sound offer
the best test field. One could reasonably hope to manipulate these
systems to have Hawking temperatures on the order of the environment
temperature ($\sim$~100~nK) \citep{Barcelo:2001ca}. Hence it is not
surprising that in recent years Bose--Einstein condensates (BECs) have
become the subject of extensive study as possible analogue models of
general relativity \citep{Garay:1999sk, Garay:2000jj, Barcelo:2001cp,
  Barcelo:2001ca, Barcelo:2003wu, Fedichev:2003bv, Fedichev:2003id}.

Let us start by very briefly reviewing the derivation of the acoustic
metric for a BEC system, and show that the equations for the phonons
of the condensate closely mimic the dynamics of a scalar field in a
curved spacetime. In the dilute gas approximation, one can describe a
Bose gas by a quantum field ${\widehat \Psi}$ satisfying 
\begin{eqnarray} 
\im\hbar \; \frac{\partial }{\partial t} {\widehat \Psi} = 
\left( 
  - {\hbar^2 \over 2m} \nabla^2 + \Vext(\x) 
  +\g(a)\;{\widehat \Psi}^{\dagger}{\widehat \Psi} 
\right){\widehat \Psi}. 
\end{eqnarray}
Here $\g$ parameterises the strength of the interactions between the
different bosons in the gas. It can be re-expressed in terms of the
scattering length $a$ as 
\begin{equation} 
\g(a) = \frac{4\pi a \hbar^2}{m}. 
\end{equation} 
As usual, the quantum field can be separated into
a macroscopic (classical) condensate and a fluctuation: ${\widehat
\Psi}=\psi+{\widehat \varphi}$, with $\langle {\widehat \Psi}
\rangle=\psi $. Then, by adopting the self-consistent mean-field
approximation (see, for example, \citealt{Griffin}) 
\begin{eqnarray}
{\widehat \varphi}^{\dagger}{\widehat \varphi}{\widehat \varphi}
\simeq 
2\langle {\widehat \varphi}^{\dagger}{\widehat \varphi} \rangle \;
{\widehat \varphi} 
+ \langle {\widehat \varphi} {\widehat \varphi}
\rangle \; {\widehat \varphi}^{\dagger}, 
\end{eqnarray} 
one can arrive
at the set of coupled equations: 
\begin{eqnarray} 
\im \hbar\;\frac{\partial }{\partial t} \psi(t,\x)
&=& 
\left ( - {\hbar^2 \over
2m} \nabla^2 + \Vext(\x) + \g \; n_c \right) \psi(t,\x)
\nonumber\\
&&
\qquad + \g \left\{2\tilde n \psi(t,\x)+ \tilde m \psi^*(t,\x) \right\};
\label{bec-self-consistent1}
\\
&& \nonumber\\ 
\im \hbar \; \frac{\partial }{\partial t} {\widehat \varphi}(t,\x) 
&=& 
\left( - {\hbar^2 \over 2m} \nabla^2 +  
\Vext(\x)  +\g \;2 n_T \right){\widehat \varphi} (t,\x)
\nonumber
\\
&&
\qquad + \g \; m_T \; {\widehat \varphi}^{\dagger}(t,\x). 
\label{quantum-field} 
\end{eqnarray} 
Here 
\begin{eqnarray} 
&& n_c \equiv \left| \psi(t,\x) \right|^2; 
\quad 
m_c \equiv \psi^2(t,\x); 
\\ 
&& \tilde n \equiv \langle 
{\widehat \varphi}^{\dagger}\,{\widehat \varphi} \rangle; 
\quad \quad \quad 
\tilde m \equiv \langle {\widehat \varphi}\, {\widehat \varphi} \rangle; 
\\ 
&& n_T=n_c+\tilde n; 
\quad \quad 
m_T=m_c+\tilde m. 
\end{eqnarray} 
The equation for the classical wave function of the condensate is
closed only when the backreaction effect due to the fluctuations is
neglected. (This backreaction is hiding in the parameters $\tilde m$
and $\tilde n$.) This is the approximation contemplated by the
Gross--Pitaevskii equation. In general, one will have to solve both
equations simultaneously. Adopting the Madelung representation for
the wave function of the condensate
\begin{equation} 
\psi(t,\x)=\sqrt{n_c(t,\x)} \; \exp[-\im\theta(t,\x)/\hbar], 
\end{equation} 
and defining an irrotational ``velocity field'' by $\vbf\equiv
{\bnabla\theta}/{m}$, the Gross--Pitaevskii equation can be rewritten
as a continuity equation plus an Euler equation:
\begin{eqnarray} 
&& \frac{\partial}{\partial t}n_c+\bnabla\cdot({n_c \vbf})=0, 
\label{E:continuity}\\ 
&& m\frac{\partial}{\partial t}\vbf+\bnabla\left(\frac{mv^2}{2}+ 
V_\mathrm{ext}(t,\x)+\g n_c- \frac{\hbar^2}{2m}
\frac{\nabla^{2}\sqrt{n_c}}{\sqrt{n_c}} \right)=0. 
\label{E:Euler1} 
\end{eqnarray} 
These equations are completely equivalent to those of an irrotational
and inviscid fluid apart from the existence of the \emph{quantum
potential} 
\begin{equation}
V_\mathrm{quantum}=
-{\hbar^2\over 2 m} \; {\nabla^{2}\sqrt{n_c}\over \sqrt{n_c}},
\end{equation}
which has the dimensions of an energy. Note that 
\begin{equation} 
n_c \; \nabla_i
V_\mathrm{quantum} \equiv n_c \; \nabla_i \left[ -{\hbar^2\over2m}
{\nabla^{2}\sqrt{n_c}\over\sqrt{n_c}} \right] = \nabla_j \left[
-{\hbar^2\over4m} \; n_c \; \nabla_i \nabla_j \ln n_c \right],
\end{equation} 
which justifies the introduction of the 
\emph{quantum stress tensor} 
\begin{equation} 
\sigma_{ij}^\mathrm{quantum} =
-{\hbar^2\over4m} \; n_c \; \nabla_i \nabla_j \ln n_c. 
\end{equation}
This tensor has the dimensions of pressure, and may be viewed as an
intrinsically quantum anisotropic pressure contributing to the Euler
equation. If we write the mass density of the Madelung fluid as $\rho
= m \; n_c$, and use the fact that the flow is irrotational, then the
Euler equation takes the form 
\begin{equation} \rho \left[
\frac{\partial} {\partial t}\vbf+ (\vbf\cdot\bnabla) \vbf \right] + \rho \;
\bnabla \left[\frac{V_\mathrm{ext}(t,\x)}{m} \right] + \bnabla
\left[{\g \rho^2\over 2 m^2}\right] + \bnabla \cdot \sigma^\mathrm{quantum} 
=0. \label{E:Euler2} 
\end{equation} 
Note that the term
$V_\mathrm{ext}/m$ has the dimensions of specific enthalpy, while $\g
\rho^2/(2m)$ represents a bulk pressure. When the gradients in the
density of the condensate are small one can neglect the quantum stress
term leading to the standard hydrodynamic approximation. Because the
flow is irrotational, the Euler equation is often more conveniently
written in Hamilton--Jacobi form: 
\begin{equation} 
m \frac{\partial}{\partial t}\theta+ \left( \frac{[\bnabla\theta]^2}{2m}
+V_\mathrm{ext}(t,\x)+\g n_c-
\frac{\hbar^2}{2m}\frac{\nabla^{2}\sqrt{n_c}}{\sqrt{n_c}}
\right)=0. \label{E:HJ} \end{equation} 
Apart from the wave function
of the condensate itself, we also have to account for the (typically
small) quantum perturbations of the system
(\ref{quantum-field}). These quantum perturbations can be described in
several different ways, here we are interested in the ``quantum
acoustic representation'' 
\begin{equation} 
\widehat
\varphi(t,\x)= e^{-\im \theta/\hbar} \left({1 \over 2
\sqrt{n_c}} \; \widehat n_1 - \im \; {\sqrt{n_c} \over \hbar} \;\widehat
\theta_1\right), 
\label{representation-change} 
\end{equation} 
where
$\widehat n_1,\widehat\theta_1$ are real quantum fields. By using
this representation, Eq.~(\ref{quantum-field}) can be rewritten as
\begin{eqnarray} 
&&\partial_t \widehat n_1 + {1\over m}
\bnabla\cdot\left( n_1 \; \bnabla \theta + n_c \; \bnabla \widehat
\theta_1 \right) = 0, \label{pt1}
\\ &&\partial_t \widehat \theta_1  + 
{1\over m} \bnabla \theta \cdot \bnabla \widehat \theta_1 
+ \g(a) \; n_1 - {\hbar^2\over2 m}\; D_2 \widehat n_1 = 0. 
\label{pt2} 
\end{eqnarray} 
Here $D_2$ represents a second-order differential operator obtained
by linearizing the quantum potential. Explicitly: 
\begin{eqnarray}
D_2\, \widehat n_1 &\equiv& 
-\half n_c^{-3/2} \;[\nabla^2
(n_c^{+1/2})]\; \widehat n_1 
+\half n_c^{-1/2} \;\nabla^2
(n_c^{-1/2}\; \widehat n_1). 
\end{eqnarray} 
The equations we have just written can be obtained easily by
linearizing the Gross--Pitaevskii equation around a classical
solution: $n_c \rightarrow n_c + \widehat n_1$, $\phi \rightarrow \phi
+ \widehat \phi_1$. It is important to realise that in those
equations the backreaction of the quantum fluctuations on the
background solution has been assumed negligible. We also see in 
Eqs.~(\ref{pt1}) and (\ref{pt2}) that time variations of
$V_\mathrm{ext}$ and time variations of the scattering length $a$
appear to act in very different ways. Whereas the external potential
only influences the background Eq.~(\ref{E:HJ}) (and hence the
acoustic metric in the analogue description), the scattering length
directly influences both the perturbation and background equations.
{From} the previous equations for the linearized perturbations it is
possible to derive a wave equation for $\widehat \theta_{1}$ (or
alternatively, for $\widehat n_{1}$). All we need is to substitute in
Eq.~(\ref{pt1}) the $\widehat n_{1}$ obtained from
Eq.~(\ref{pt2}). This leads to a PDE that is second-order in
time derivatives but infinite-order in space derivatives -- to
simplify things we can construct the symmetric $4 \times 4$ matrix of differential operators
\begin{equation} 
f^{\mu\nu}(t,\x) \equiv
\begin{bmatrix}
   f^{00}&\vdots&f^{0j}\\
   \cdots\cdots&\cdot&\cdots\cdots\cdots\cdots\cr
   f^{i0}&\vdots&f^{ij}\\
\end{bmatrix}.
\label{E:explicit} 
\end{equation} 
(Greek indices run from
0\,--\,3, while Roman indices run from 1\,--\,3.)  Then, introducing
(3+1)-dimensional space-time coordinates, 
\begin{equation}
x^\mu \equiv (t;\, x^i)
\end{equation}
the wave equation for $\theta_{1}$ is easily rewritten as 
\begin{equation}
\partial_\mu ( f^{\mu\nu} \;
\partial_\nu \widehat \theta_1) = 0. \label{weq-phys} 
\end{equation} 
Where the $f^{\mu\nu}$ are now \emph{differential operators} acting on space
only: 
\begin{eqnarray} 
f^{00} &=& - \left[ \g(a) - {\hbar^2\over 2m}\;
D_2 \right]^{-1} \\ f^{0j} &=& -\left[ \g(a) - {\hbar^2\over 2m}\; D_2
\right]^{-1}\; {\nabla^j \theta_0\over m} \\ f^{i0} &=& - {\nabla^{i}
\theta_0\over m} \; \left[ \g(a) - {\hbar^2\over2 m}\; D_2
\right]^{-1} \\ f^{ij} &=& {n_c \; \delta^{ij}\over m} - {\nabla^{i}
\theta_0\over m} \; \left[ \g(a) - {\hbar^2\over2 m}\; D_2
\right]^{-1}\; {\nabla^{j} \theta_0\over m}. 
\end{eqnarray} 
Now, if we make a spectral decomposition of the field $\widehat
\theta_1$ we can see that for wavelengths larger than $\hbar /m\csound$
(this corresponds to the ``healing length'', as we will explain
below), the terms coming from the linearization of the quantum
potential (the $D_2$) can be neglected in the previous expressions, in
which case the $f^{\mu\nu}$ can be approximated by (momentum
independent) numbers, instead of differential operators. (This is the
heart of the acoustic approximation.) Then, by identifying 
\begin{equation} 
\sqrt{-g} \; g^{\mu\nu}=f^{\mu\nu},
\end{equation} 
the equation for the field $\widehat \theta_1$
becomes that of a (massless minimally coupled) quantum scalar field
over a curved background 
\begin{equation}
\Delta\theta_{1}\equiv\frac{1}{\sqrt{-g}}\;
\partial_{\mu}\left(\sqrt{-g}\; g^{\mu\nu}\; \partial_{\nu}\right)
\widehat\theta_{1}=0, 
\end{equation} 
with an effective metric of the form 
\begin{equation} g_{\mu\nu}(t,\x) \equiv {n_c\over m\;
\csound(a,n_c)}
\begin{bmatrix}
   -\{\csound(a,n_c)^2-v^2\}&\vdots& - v_j \\
   \cdots\cdots\cdots\cdots&\cdot&\cdots\cdots\\
   -v_i&\vdots&\delta_{ij}\\
\end{bmatrix}. 
\end{equation} 
Here, the magnitude
$\csound(n_c,a)$ represents the speed of the phonons in the medium: 
\begin{equation} 
\csound(a,n_c)^2={\g(a) \; n_c \over m}. 
\end{equation} 
With this effective metric now in hand, the analogy is fully
established, and one is now in a position to start asking more specific
physics questions.

\paragraph*{Lorentz breaking in BEC models --- the eikonal approximation:\;}

It is interesting to consider the case in which the above
``hydrodynamical'' approximation for BECs does not hold. In order to
explore a regime where the contribution of the quantum potential
cannot be neglected we can use the \emph{eikonal}
approximation, a high-momentum approximation where the phase
fluctuation $\widehat \theta_1$ is itself treated as a slowly-varying
amplitude times a rapidly varying phase. This phase will be taken to
be the same for both $\widehat n_1$ and $\widehat \theta_1$
fluctuations. In fact, if one discards the unphysical possibility that
the respective phases differ by a time-varying quantity, any
time-constant difference can be safely reabsorbed in the definition of
the (complex) amplitudes. Specifically, we shall write 
\begin{eqnarray} 
{\widehat\theta}_1(t,\x) 
&=& \mathrm{Re}\left\{\mathcal{A}_\theta \; \exp(-i\phi) \right\},\\ 
{\widehat n}_1(t,\x)
&=& \mathrm{Re}\left\{\mathcal{A}_\rho \; \exp(-i\phi)\right\}. 
\end{eqnarray} 
As a consequence of our starting assumptions, gradients of the
amplitude, and gradients of the background fields, are systematically
ignored relative to gradients of $\phi$. (Warning: What we are doing
here is not quite a ``standard'' eikonal approximation, in the sense
that it is not applied directly on the fluctuations of the field
$\psi(t,\x)$ but separately on their amplitudes and phases $\rho_{1}$
and $\phi_{1}$.) We adopt the notation 
\begin{equation} 
\omega ={\partial\phi\over\partial t};\qquad k_i = \nabla_i \phi. 
\end{equation} 
Then the operator $D_2$ can be approximated as 
\begin{eqnarray} 
D_2 \;{\widehat n}_1
&\equiv& -\half n_c^{-3/2} \;[\Delta (n_c^{+1/2})]\; {\widehat n}_1
+\half n_c^{-1/2} \;\Delta (n_c^{-1/2} {\widehat n}_1) 
\\ 
&\approx&
+\half n_c^{-1} \;[\Delta {\widehat n}_1] 
\\ 
&=& -\half n_c^{-1}
\;k^2 \;{\widehat n}_1. 
\end{eqnarray} 
A similar result holds for
$D_2$ acting on ${\widehat \theta}_1$. That is, under the eikonal
approximation we effectively replace the {\emph{operator}} $D_2$ by
the {\emph{function}} 
\begin{equation} 
D_2 \to -\half n_c^{-1}k^2. 
\end{equation} 
For the matrix $f^{\mu\nu}$ this effectively
results in the (explicitly momentum dependent) replacement 
\begin{eqnarray} 
f^{00} 
&\to& - \left[\kappa(a) + {\hbar^2 \; k^2\over4m\;n_c} \right]^{-1} \\ 
f^{0j} 
&\to& -\left[ \kappa(a) + {\hbar^2 \; k^2\over4m\;n_c}\right]^{-1}\;
{\nabla^j \theta_0\over m} \\ 
f^{i0} 
&\to& - {\nabla^i \theta_0\over m} \; 
\left[ \kappa(a) + {\hbar^2 \; k^2\over4m\;n_c} \right]^{-1} \\
f^{ij} 
&\to& {n_c \; \delta^{ij}\over m} - {\nabla^i \theta_0\over m}\; 
\left[ \kappa(a) + {\hbar^2 \; k^2\over4m\;n_c} \right]^{-1}\;
{\nabla^j \theta_0\over m}\,.
\end{eqnarray} 
As desired, this has the net effect of making $f^{\mu\nu}$ a matrix
of (explicitly momentum dependent) numbers, not operators. The
physical wave equation~(\ref{weq-phys}) now becomes a nonlinear
dispersion relation 
\begin{equation} 
f^{00} \;\omega^2 + (f^{0i} +f^{i0}) \;\omega \;k_i +
f^{ij} \;k_i \;k_j = 0. 
\end{equation} 
After substituting the approximate $D_2$ into this dispersion relation
and rearranging, we see (remember: $k^2 = ||k||^2 = \delta^{ij}
\;k_i \;k_j$) 
\begin{equation} 
-\omega^2 + 2 \; v_0^i \; \omega k_i
+ {n_c k^2\over m}\left[\kappa(a)+{\hbar^2\over4m n_c} k^2\right] -
(v_0^i \; k_i)^2 = 0. 
\end{equation} 
That is: 
\begin{equation}
\left(\omega - v_0^i \; k_i\right)^2 = {n_c k^2\over
m}\left[\kappa(a)+{\hbar^2\over4m n_c} k^2\right]\,.
\end{equation} 
Introducing the speed of sound $\csound$, this takes the form: 
\begin{equation} \omega= v_0^i \; k_i \pm \sqrt{\csound^2 k^2+\left({\hbar
\over 2 m}\;k^2\right)^2}. 
\label{eq:disprel} 
\end{equation} 
At this stage some observations are in order: 

\begin{enumerate}

  \item It is interesting to recognize that the dispersion
    relation~(\ref{eq:disprel}) is exactly in agreement with that
    found by \citet{Bogoliubov:1947} (reprinted
    in \citealt{Pines:1962}; see also \citealt{LLP}) for the collective
    excitations of a homogeneous Bose gas in the limit $T\to 0$
    (almost complete condensation). In his derivation Bogoliubov
    applied a diagonalization procedure for the Hamiltonian describing
    the system of bosons.

  \item Coincidentally, this is the same dispersion relation that one
    encounters for shallow-water surface waves in the presence of
    surface tension. See Sect.~\ref{S:WW-gen}. 

  \item Because of the explicit momentum dependence of the co-moving
    phase velocity and co-moving group velocity, once one goes to high
    momentum the associated effective metric should be thought of as
    one of many possible ``rainbow metrics'' as in
    Sect.~\ref{S:WW-gen}. See also \citet{Visser:2007du}. (At low
    momentum one, of course, recovers the hydrodynamic limit with its
    uniquely specified standard metric.) 

  \item It is easy to see that Eq.~(\ref{eq:disprel}) actually
    interpolates between two different regimes depending on the value
    of the wavelength $\lambda= 2\pi / ||k||$ with respect to the
    ``acoustic Compton wavelength'' $\lambda_c=h/(m
    \csound)$. (Remember that $\csound$ is the speed of sound; this is
    not a standard particle physics Compton wavelength.) In
    particular, if we assume $v_{0}=0$ (no background velocity), then,
    for large wavelengths $\lambda \gg \lambda_c$, one gets a standard
    phonon dispersion relation $\omega \approx c ||k||$. For
    wavelengths $\lambda \ll \lambda_c$ the quasi-particle energy
    tends to the kinetic energy of an individual gas particle and, in
    fact, $\omega \approx \hbar^2 k^2/(2 m)$. 

    We would also like to highlight that in relative terms, the
    approximation by which one neglects the quartic terms in the
    dispersion relation gets worse as one moves closer to a horizon
    where $v_0=-\csound$. The non-dimensional parameter that provides
    this information is defined by
    \begin{equation} 
      \delta \equiv {\sqrt{1+{\lambda_c^2 \over 4\lambda^2}}-1 \over (1-v_0/\csound)}
      \simeq {1 \over (1-v_0/\csound)}{\lambda_c^2 \over 8\lambda^2}.
    \end{equation}
    As we will discuss in Sect.~\ref{S:Horizon-stability}, this is
    the reason why sonic horizons in a BEC can exhibit different
    features from those in standard general relativity.
  
  \item The dispersion relation~(\ref{eq:disprel}) exhibits a
    contribution due to the background flow $ v_0^i \; k_i$, plus a
    quartic dispersion at high momenta. The group velocity is 
    \begin{equation} 
      v_g^i = {\partial\omega\over\partial k_i} = v_0^i \pm
      { \left(c^2+{\hbar^2 \over 2 m^2}k^2\right) \over \sqrt{c^2
          k^2+\left({\hbar \over 2 m}\;k^2\right)^2} } \; k^i\,. 
    \end{equation}

  Indeed, with hindsight, the fact that the group velocity goes to
  infinity for large $k$ was pre-ordained: After all, we started from the
  generalised nonlinear Schr\"odinger equation, and we know what its
  characteristic curves are. Like the diffusion equation the
  characteristic curves of the Schr\"odinger equation (linear or
  nonlinear) move at infinite speed. If we then approximate this
  generalised nonlinear Schr\"odinger equation in any manner, for
  instance by linearization, we cannot change the characteristic curves:
  For any well-behaved approximation technique, at high frequency and
  momentum we should recover the characteristic curves of the system we
  started with. However, what we certainly do see in this analysis is a
  suitably large region of momentum space for which the concept of the 
  effective metric both makes sense, and leads to finite propagation speed
  for medium-frequency oscillations.
  
\end{enumerate}

\paragraph*{Relativistic BEC extension:\;}

Bose--Einstein condensation can occur not only for non-relativistic bosons
but for relativistic ones as well. The main differences between the
thermodynamical properties of these condensates at finite temperature
are due both to the different energy spectra and also to the presence,
for relativistic bosons, of anti-bosons. These differences result in
different conditions for the occurrence of Bose--Einstein
condensation, which is possible, e.g., in two spatial dimensions for a
homogeneous relativistic Bose gas, but not for its non-relativistic
counterpart -- and also, more importantly for our purposes, in the
different structure of their excitation spectra. 

in \citet{Fagnocchi:2010sn} an analogue model based on a relativistic
BEC was studied. We summarise here the main results. The Lagrangian
density for an interacting relativistic scalar Bose field
$\hat{\phi}(\mathbf{x},t)$ may be written as
\begin{equation}
\hat{{\cal L}}=\frac{1}{c^2}\frac{\partial\hat\phi^\dagger}{\partial t}\frac{\partial\hat\phi}{\partial t}-
\mathbf{\nabla}\hat\phi^\dagger \cdot \mathbf{\nabla}\hat\phi
-\left(\frac{m^2c^2}{\hbar^2} + V(t,\mathbf{x})\right) \hat{\phi}^\dagger \hat{\phi}-U(\hat\phi^\dagger\hat\phi;\lambda_i)~,
\label{Lagrangian}
\end{equation}
where $V(t,\mathbf{x})$ is an external potential depending both on
time $t$ and position $\mathbf{x}$, $m$ is the mass of the bosons and
$c$ is the light velocity. $U$ is an interaction term and the coupling
constant $\lambda_i(t,\mathbf x)$ can depend on time and position too
(this is possible, for example, by changing the scattering length via
a Feshbach resonance; \citealt{Cornish:2000zz, Donley}).
$U$ can be expanded as
\begin{equation}
 U(\hat\phi^\dagger\hat\phi;\lambda_i) = \frac{\lambda_2}{2}\hat{\rho}^2 + \frac{\lambda_3}{6}\hat{\rho}^3 + \cdots
\end{equation}
where $\hat{\rho}=\hat\phi^\dagger\hat\phi$. The usual two-particle
$\lambda_2 \hat{\phi}^4$-interaction corresponds to the first term
$(\lambda_2/2)\hat{\rho}^2$, while the second term represents the
three-particle interaction and so on.

The field $\hat\phi$ can be written as a classical field 
(the condensate) plus perturbation:
\begin{equation}\label{eq:exp}
 \hat\phi = \phi(1+\hat\psi)\,.
\end{equation}
It is worth noticing now that the expansion in Eq.~(\ref{eq:exp})
can be linked straightforwardly to the previously discussed expansion
in phase and density perturbations $\hat\theta_1$, $\hat\rho_1$, by
noting that
$$
\frac{\hat\rho_1}{\rho}=\frac{\hat\psi+\hat\psi^\dagger}2,\quad
\hat\theta_1=\frac{\hat\psi-\hat\psi^\dagger}{2i}\,.
$$
Setting $\psi \propto \mathrm{exp}[i \left( \mathbf{k} \cdot \mathbf{x}
 -\omega t\right)]$ one then gets from the equation of
motion \citep{Fagnocchi:2010sn}
\begin{eqnarray}
\label{eq:fourierexp_gen2}
&&\left(-\frac{\hbar}{m} \mathbf{q} \cdot \mathbf{k}+
\frac{u^0}{c}\omega-\frac{\hbar }{2 mc^2}\omega^2+\frac{\hbar}{2m}k^2\right) 
\left(\frac{\hbar}{m} \mathbf{q} \cdot \mathbf{k}
-\frac{u^0}{c}\omega-\frac{\hbar }{2 mc^2}\omega^2+\frac{\hbar}{2m}k^2\right)
\nonumber\\
&&
\qquad\qquad\qquad\qquad 
 -\left(\frac{c_0}{c}\right)^2\omega^2+c_0^2 k^2=0~,
\end{eqnarray}
where, for convenience, we have defined the following quantities as
\begin{eqnarray}
 u^\mu \equiv \frac{\hbar}{m}\eta^{\mu\nu}\partial_\nu\theta~,\\
 c_{0}^2\equiv\frac{\hbar^2}{2m^2}U''(\rho;\lambda_i)\rho~,\label{eq:defc0}\\
 \mathbf{q}\equiv m \mathbf{u}/\hbar\,.
\end{eqnarray}
Here $\mathbf{q}$ is the speed of the condensate flow and $c$ is the speed of light.
For a condensate at rest ($\mathbf{q}=0$) one then obtains the following dispersion relation
\begin{equation}\label{eq:dispersion}
 \omega^2_\pm=c^2
 \left\{k^2+2\left(\frac{mu^0}{\hbar}\right)^2 \left[1+\left(\frac{c_0}{u^0}\right)^2\right]
 \pm 2\left(\frac{mu^0}{\hbar}\right)
 \sqrt{k^2+\left(\frac{mu^0}{\hbar}\right)^2\left[1+\left(\frac{c_0}{u^0}\right)^2\right]^2}
 \right\}\,.
\end{equation}
The dispersion relation~(\ref{eq:dispersion}) is sufficiently
complicated to prevent any obvious understanding of the regimes
allowed for the excitation of the system. It is much richer than the
non-relativistic case. For example, it allows for both a
massless/gapless (phononic) and massive/gapped mode, respectively for
the $\omega_{-}$ and $\omega_{+}$ branches of Eq.~(\ref{eq:dispersion}). Nonetheless, it should be evident that
different regimes are determined by the relative strength of the the
first two terms on the right-hand side of Eq.~(\ref{eq:dispersion}) (note that the same terms enter in the
square root). This can be summarised, in low and high momentum limits
respectively, for $k$ much less or much greater than 
\begin{equation}
\label{eq:momcond}
\frac{m u^0}{\hbar }\left[1+\left(\frac{c_0}{u^0}\right)^2\right] \equiv \frac{m u^0}{\hbar }(1+b),
\end{equation}
where $b$ encodes the relativistic nature of the condensate
(the larger $b$ the more the condensate is relativistic).

A detailed discussion of the different regimes would be
inappropriately long for this review; it can be found
in \citet{Fagnocchi:2010sn}. The results are summarised in
Table~\ref{tab:regimes}. Note that $\mu \equiv mcu^0$ plays the role
of the chemical potential for the relativistic BEC. One of the most
remarkable features of this model is that it is a condensed matter
system that interpolates between two different Lorentz symmetries, one
at low energy and a different Lorentz symmetry at high energy. 

\begin{table}[!htb]
 \caption[Dispersion relation of gapless and gapped modes in different
   regimes.]{Dispersion relation of gapless and gapped modes in
   different regimes. Note that we have $c_s^2 = c^2 b/(1+b)$, and
   $c_{s,\mathrm{gap}}^2 = c^2(2+b)/(1+b)$, while $m_{\mathrm{eff}} =
   2(\mu/c^2)(1+b)^{3/2}/(2+b)$.}
\label{tab:regimes}
\bigskip
\centering
\small
\begin{tabular}[c]{c c c c c}
\toprule\toprule
~			&	~		 & \multicolumn{2}{c}{Gapless}		& Gapped\\
\cline{3-4}	& ~          & \vphantom{\Big|}$b\ll1$	& $b\gg1$	 &	\\
\midrule
\multirow{2}{*}{$|k|\ll {mu^0(1+b)\over \hbar}$} & \vphantom{\Big|}$|k|\ll{2mc^0\over\hbar}$ &	$\omega^2 = c_s^2 k^2$	&	\multirow{2}{*}{$\omega^2 = c_s^2 k^2$}	&	\multirow{2}{*}{$\omega^2={m_{\mathrm{eff}}^2\,c_{s,\mathrm{gap}}^4\over\hbar^2}+c_{s,\mathrm{gap}}^2 k^2 $}	\\
  						&
\vphantom{\Big|} ${2mc^0\over\hbar}\ll|k|\ll {mu^0\over\hbar}$	&
$\hbar\omega = {(\hbar c k)^2\over2\mu}$	& ~ 			&	\\
\midrule
\vphantom{\Bigg|} $|k|\gg {mu^0(1+b)\over\hbar}$ & ~		&	 \multicolumn{3}{c}{$\omega^2 = c^2 k^2$}\\
\bottomrule  \bottomrule
\end{tabular}
\end{table}

Finally, it is also possible to recover an acoustic metric for the
massless (phononic) perturbations of the condensate in the low
momentum limit ($k\ll m u^0(1+b)/\hbar$):
\begin{equation}\label{eq:becmetric}
 g_{\mu\nu}=\frac{\rho}{\sqrt{1-u_\sigma u^\sigma/c_o^2}}\left[\eta_{\mu\nu}\left(1-\frac{u_\sigma u^\sigma}{c_0^2}\right)+\frac{u_\mu u_\nu}{c_0^2}\right]~.
\end{equation}

As should be expected, it is just a version of the acoustic geometry
for a relativistic fluid previously discussed, and in fact can be cast
in the form of Eq.~(\ref{eq:relhydroacmet}) by suitable variable
redefinitions \citep{Fagnocchi:2010sn}.

\subsubsection{The heliocentric universe} 
\label{S:helium}

Helium is one of the most fascinating elements provided by nature. Its
structural richness confers on helium a paradigmatic character regarding
the emergence of many and varied macroscopic properties from the
microscopic world (see \citealt{Volovik:2003fe} and references therein). 
Here, we are interested in the emergence of effective geometries in
helium, and their potential use in testing aspects of semiclassical
gravity.

Helium four, a bosonic system, becomes superfluid at low temperatures
(2.17~K at vapour pressure).  This superfluid behaviour is associated
with condensation in the vacuum state of a macroscopically large
number of atoms. A superfluid is automatically an irrotational and
inviscid fluid, so, in particular, one can apply to it the ideas worked
out in Sect.~\ref{S:simple}. The propagation of classical acoustic
waves (scalar waves) over a background fluid flow can be described in
terms of an effective Lorentzian geometry: the acoustic
geometry. However, in this system one can naturally go considerably
further, into the quantum domain. For long wavelengths, the
quasiparticles in this system are quantum phonons. One can separate
the classical behaviour of a background flow (the effective geometry)
from the behaviour of the quantum phonons over this background. In
this way one can reproduce, in laboratory settings, different aspects
of quantum field theory over curved backgrounds. The speed of sound in
the superfluid phase is typically on the order of cm/sec. Therefore,
at least in principle, it should not be too difficult to establish
configurations with supersonic flows and their associated ergoregions.

Helium three, the fermionic isotope of helium, in contrast, becomes
superfluid at much lower temperatures (below 2.5 milli-K). The
reason behind this rather different behaviour is the pairing of fermions
to form effective bosons (Cooper pairing), which are then able to
condense. In the $^{3}$He-A phase, the structure of the
fermionic vacuum is such that it possesses two Fermi points, instead of
the more typical Fermi surface. In an equilibrium configuration one can
choose the two Fermi points to be located at $\{p_x=0,p_y=0,p_z=\pm
p_F\}$ (in this way, the z-axis signals the direction of the angular
momentum of the pairs).  Close to either Fermi point the spectrum of
quasiparticles becomes equivalent to that of Weyl fermions. From the
point of view of the laboratory, the system is not isotropic, it is
axisymmetric. There is a speed for the propagation of quasiparticles
along the z-axis, $c_\parallel \simeq \mathrm{cm}/\mathrm{sec}$, and a
different speed, 
$c_\perp\simeq 10^{-5}\,c_\parallel$, for propagation perpendicular to the
symmetry axis.  However, from an internal observer's point of view this
anisotropy is not ``real'', but can be made to disappear by an
appropriate rescaling of the coordinates. Therefore, in the equilibrium
case, we are reproducing the behaviour of Weyl fermions over Minkowski
spacetime. Additionally, the vacuum can suffer collective excitations.
These collective excitations will be experienced by the Weyl
quasiparticles as the introduction of an effective electromagnetic field
and a curved Lorentzian geometry. The control of the form of this
geometry provides the sought for gravitational analogy.

\begin{figure}[htpb]
    \centerline{\includegraphics[width=0.75\textwidth]{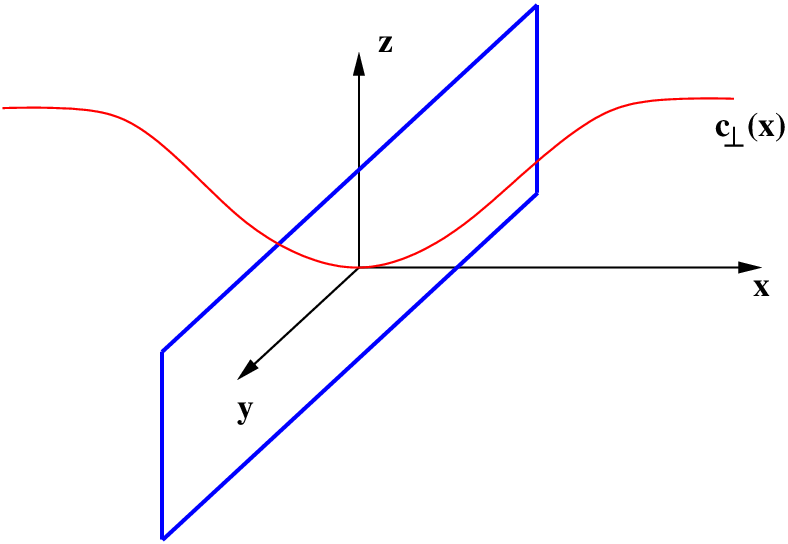}}
  \caption[Domain wall configuration in $^{3}$He]{
Domain wall configuration in $^{3}$He.}
  \label{F:soliton}
\end{figure}

Apart from the standard way to provide a curved geometry based on
producing nontrivial flows, there is also the possibility of creating
topologically nontrivial configurations with a built-in nontrivial
geometry. For example, it is possible to create a domain-wall
configuration \citep{Jacobson:1998ms, Jacobson:1998he} (the wall contains
the z-axis) such that the transverse velocity $c_\perp$ acquires a
profile in the perpendicular direction (say along the x-axis) with
$c_\perp$ passing through zero at the wall (see Fig.~\ref{F:soliton}).
This particular arrangement could be used to reproduce a
black-hole--white-hole configuration only if the soliton is set up to
move with a certain velocity along the x-axis. This configuration has
the advantage that it is dynamically stable, for topological reasons,
even when some supersonic regions are created.

A third way in which superfluid helium can be used to create analogues
of gravitational configurations is the study of surface waves (or
ripplons) on the interface between two different phases of $^{3}$He
\citep{Volovik:2002ci, Volovik:2001yh}.  In particular, if we have a
thin layer of $^{3}$He-A in contact with another thin layer of
$^{3}$He-B, the oscillations of the contact surface ``see'' an
effective metric of the form \citep{Volovik:2002ci, Volovik:2001yh}
\begin{equation}
\d s^2 = \frac{1}{(1- \alpha_A\alpha_B U^2)}
\Bigg[-\left(1-W^2-\alpha_A\alpha_B U^2\right)\; \d t^2
 - 2 \mathbf{W} \cdot \mathbf{dx}\; \d t +\mathbf{dx}\cdot \mathbf{dx} \Bigg],
\end{equation} 
where 
\begin{eqnarray} 
\mathbf{W}\equiv \alpha_A \mathbf{v_A} +\alpha_B \mathbf{v_B}, 
\qquad
\mathbf{U}\equiv \mathbf{v_A} - \mathbf{v_B},
\end{eqnarray} 
and 
\begin{eqnarray} 
\alpha_A \equiv {h_B \;\rho_A \over h_A \;\rho_B + h_B \;\rho_A};
\qquad
\alpha_B \equiv {h_A \;\rho_B \over h_A \;\rho_B + h_B \; \rho_A}. 
\end{eqnarray} 
(All of this provided that we are looking at wavelengths larger than
the layer thickness, $k\,h_A\ll 1$ and $k\,h_B\ll 1$.)
 
\begin{figure}[htpb]
    \centerline{\includegraphics[width=0.75\textwidth]{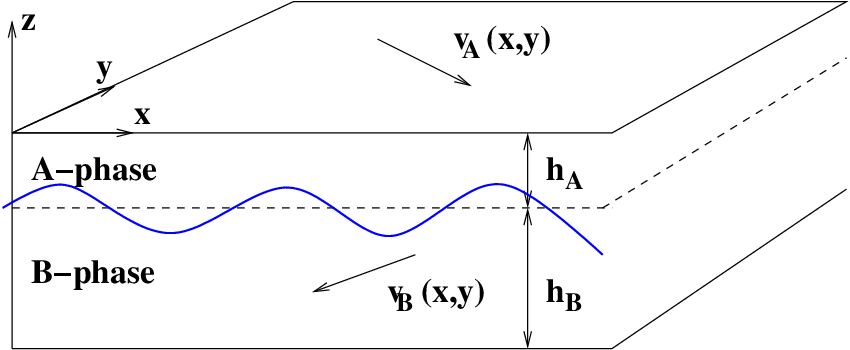}}
  \caption[Interface between two sliding superfluids]{Ripplons in the 
  interface between two sliding superfluids.}
  \label{F:ripplon}
\end{figure}

The advantage of using surface waves instead of bulk waves in
superfluids is that one could create horizons without reaching
supersonic speeds in the bulk fluid. This could alleviate the appearance
of dynamical instabilities in the system, that in this case are
controlled by the strength of the interaction of the ripplons with bulk
degrees of freedom \citep{Volovik:2002ci, Volovik:2001yh}.

\subsubsection{Slow light in fluids}

The geometrical interpretation of the motion of light in dielectric
media leads naturally to conjecture that the use of flowing dielectrics
might be useful for simulating general relativity metrics with
ergoregions and black holes. Unfortunately, these types of geometry
require flow speeds comparable to the group velocity of the light. Since
typical refractive indexes in non-dispersive media are quite close to
unity, it is then clear that it is practically impossible to use them to
simulate such general relativistic phenomena. However recent
technological advances have radically changed this state of affairs. In
particular the achievement of controlled slowdown of light, down to
velocities of a few meters per second (or even down to complete rest;
\citealt{Hau, Kash, Budker, Koch, Phil, Turu, SZ}), has opened a whole new set of
possibilities regarding the simulation of curved-space metrics via
flowing dielectrics. 

But how can light be slowed down to these ``snail-like'' velocities?
The key effect used to achieve this takes the name of
Electromagnetically Induced Transparency (EIT). A laser beam is coupled
to the excited levels of some atom and used to strongly modify its
optical properties. In particular one generally chooses an atom with two
long-lived metastable (or stable) states, plus a higher energy state
that has some decay channels into these two lower states. The coupling of the
excited states induced by the laser light can affect the transition from
a lower energy state to the higher one, and hence the capability of the
atom to absorb light with the required transition energy. The system can
then be driven into a state where the transitions between each of the
lower energy states and the higher energy state exactly cancel out, due
to quantum interference, at some specific resonant frequency. In this
way the higher-energy level has null averaged occupation number. This
state is hence called a ``dark state''. EIT is characterised by a
transparency window, centered around the resonance frequency, where the
medium is \emph{both} almost transparent \emph{and} extremely dispersive
(strong dependence on frequency of the refractive index). This in turn
implies that the group velocity of any light probe would be
characterised by very low real group velocities (with almost vanishing
imaginary part) in proximity to the resonant frequency.

Let us review the most common setup envisaged for this kind of analogue
model. A more detailed analysis can be found in \citet{Leonhardt:2001ye}.
One can start by considering a medium in which an EIT window is opened
via some control laser beam which is oriented perpendicular to the
direction of the flow. One then illuminates this medium, now along the
flow direction, with some probe light (which is hence perpendicular to
the control beam). This probe beam is usually chosen to be weak with
respect to the control beam, so that it does not modify the optical
properties of the medium. In the case in which the optical properties of
the medium do not vary significantly over several wavelengths of the
probe light, one can neglect the polarization and can hence describe the
propagation of the latter with a simple scalar dispersion
relation \citep{Leonhardt:2000aa,Fiurasek:2002ay}
\begin{equation}
k^2=\frac{\omega^2}{c^2}\left[1+\chi(\omega)\right],
\label{eq:sldr}
\end{equation}
where $\chi$ is the susceptibility of the medium, related to the refractive
index $n$ via the simple relation $n=\sqrt{1+\chi}$.

It is easy to see that in this case the group and phase velocities
differ
\begin{equation}
v_\mathrm{g}=\frac{\partial \omega}{\partial k}
=\frac{c}{\sqrt{1+\chi}+\frac{\textstyle \omega}{\textstyle 2n}
\frac{\textstyle \partial\chi}{\textstyle \partial\omega}};
\qquad 
v_\mathrm{ph}=\frac{\omega}{k}=\frac{c}{\sqrt{1+\chi}}.
\end{equation}
So even for small refractive indexes one can get very low group
velocities, due to the large dispersion in the transparency window, and
in spite of the fact that the phase velocity remains very near to $c$.
(The phase velocity is exactly $c$ at the resonance frequency
$\omega_0$). In an ideal EIT regime the probe light experiences a
vanishing susceptibility $\chi$ near the the critical frequency
$\omega_0$, this allows us to express the susceptibility near the
critical frequency via the expansion
\begin{equation}
\chi(\omega)=\frac{2\alpha}{\omega_0}\left(\omega-\omega_0\right)
+\O\left[\left(\omega-\omega_0\right)^3\right], 
\label{eq:grp-ind}
\end{equation}
where $\alpha$ is sometimes called the ``group refractive index''. The
parameter $\alpha$ depends on the dipole moments for the transition from
the metastable states to the high energy one, and most importantly
depends on the ratio between the probe-light energy per photon,
$\hbar\omega_0$, and the control-light energy per
atom \citep{Leonhardt:2001ye}. This might appear paradoxical because it
seems to suggest that for a dimmer control light the probe light would
be further slowed down. However this is just an artificial feature due
to the extension of the EIT regime beyond its range of applicability. In
particular in order to be effective the EIT requires the control beam
energy to dominate all processes and hence it cannot be dimmed at
will.

At resonance we have
\begin{equation}
v_\mathrm{g}=\frac{\partial \omega}{\partial k}
\to\frac{c}{1+\alpha} \approx \frac{c}{\alpha};
\qquad 
v_\mathrm{ph}=\frac{\omega}{k}\to c.
\end{equation}
We can now generalise the above discussion to the case in which our
highly dispersive medium flows with a characteristic velocity profile
$\mathbf{u}(\mathbf{x},t)$. In order to find the dispersion relation of the
probe light in this case we just need to transform the dispersion
relation~(\ref{eq:sldr}) from the comoving frame of the medium to the
laboratory frame. Let us consider for simplicity a monochromatic probe
light (more realistically a pulse with a very narrow range of
frequencies $\omega$ near $\omega_0$). The motion of the dielectric
medium creates a local Doppler shift of the frequency
\begin{equation}
\omega\rightarrow \gamma\left(\omega_0-{\mathbf u}\cdot {\mathbf k}\right),
\label{eq:tilt}
\end{equation}
where $\gamma$ is the usual relativistic factor. Given that
$k^2-\omega^2/c^2$ is a Lorentz invariant, it is then easy to see
that this Doppler detuning affects the dispersion
relation~(\ref{eq:sldr}) only via the susceptibility dependent
term. Given further that in any realistic case one would deal with
non-relativistic fluid velocities $\mathbf{u}\ll c$ we can then
perform an expansion of the dispersion relation up to second order in
$u/c$. Expressing the susceptibility via Eq.~(\ref{eq:grp-ind}) we can
then rewrite the dispersion relation in the
form \citep{Leonhardt:2000aa}
\begin{equation}
g^{\mu\nu}k_\mu k_\nu=0,
\end{equation}
where 
\begin{equation}
k_\nu=\left(\frac{\omega_0}{c},-\mathbf{k}\right),
\end{equation}
and 
\begin{equation}
g^{\mu \nu}=
\left[
\begin{array}{c|c}
-(1+\alpha{u^2}/{c^2})            & -\alpha \mathbf{u}^{T} /c^2 \\
\hline
\vphantom{\Big|}
 -\alpha {\mathbf u}/c^2  & 
\mathbf{I}_{3\times 3} -4\alpha \mathbf{u}\otimes \mathbf{u}^T /c^2
\end{array}
\right].
\label{eq:sl-cont-metr}
\end{equation}
(Note that most of the original articles on this topic adopt the opposite signature $(+---)$.)
The inverse of this tensor will be the covariant effective metric
experienced by the probe light, whose rays would then be null geodesics
of the line element $\d s^2=g_{\mu\nu} \d x^\mu \d x^\nu$. In this sense the
probe light will propagate as in a curved background. Explicitly one
finds the covariant metric to be
\begin{equation}
g_{\mu \nu}=
\left[
\begin{array}{c|c}
-A            & -B \mathbf{u}^T \\
\hline
\vphantom{\Big|}
 -B \mathbf{u}  & 
\mathbf{I}_{3\times 3} -C\mathbf{u}\otimes \mathbf{u}^T
\end{array}
\right],
\label{eq:sl-cov-metr}
\end{equation}
where
\begin{eqnarray}
A&=&\frac{1-4\alpha u^2/c^2}{1+(\alpha^2-3\alpha)u^2/c^2-4\alpha^2u^4/c^4};
\\
B&=&\frac{1}{1+(\alpha^2-3\alpha)u^2/c^2-4\alpha^2u^4/c^4};
\\
C&=&\frac{1-(4/\alpha+4 u^2/c^2)}{1+(\alpha^2-3\alpha)u^2/c^2-4\alpha^2u^4/c^4}.
\end{eqnarray}

Several comments are in order concerning the
metric~(\ref{eq:sl-cov-metr}). First of all, it is clear that, although
more complicated than an acoustic metric, it will still be possible to
cast it into the Arnowitt--Deser--Misner-like
form \citep{Visser:2000pk}
\begin{equation}
g_{\mu \nu}=
\left[
\begin{array}{c|c}
-[c_\mathrm{eff}^2-g_{ab}u_\mathrm{eff}^a u_\mathrm{eff}^b]          
& [u_\mathrm{eff}]_i \\
\hline
\vphantom{\Big|}
[u_\mathrm{eff}]_j & [g_\mathrm{eff}]_{ij}
\end{array}
\right],
\label{eq:adm}
\end{equation}
where the effective speed $\mathbf{u}_\mathrm{eff}$ is proportional to the
fluid flow speed $\mathbf{u}$ and the three-space effective metric 
$g_\mathrm{eff}$ is (rather differently from the acoustic case) nontrivial.

In any case, the existence of this ADM form already tells us that an
ergoregion will always appear once the norm of the effective velocity
exceeds the effective speed of light (which for slow light is
approximately $c/\alpha$, where $\alpha$ can be extremely large due to
the huge dispersion in the transparency window around the resonance
frequency $\omega_0$). However, a trapped surface (and hence an optical
black hole) will form only if the \emph{inward normal component} of the
effective flow velocity exceeds the group velocity of light. In the slow
light setup so far considered such a velocity turns out to be
$u=c/(2\sqrt{\alpha})$. 

The realization that ergoregions and event horizons can be simulated via
slow light may lead one to the (erroneous) conclusion that this is an
optimal system for simulating particle creation by gravitational fields.
 However, as pointed out by Unruh in \citet{Novello:2002qg,Unruh:2003ss},
such a conclusion would turn out to be over-enthusiastic. In order to
obtain particle creation through ``mode mixing'', (mixing between the
positive and negative norm modes of the incoming and outgoing states),
an inescapable requirement is that there must be regions where the
frequency of the quanta as seen by a local comoving observer becomes
negative.

In a flowing medium this can, in principle, occur thanks to the tilting
of the dispersion relation due to the Doppler effect caused by the
velocity of the flow Eq.~(\ref{eq:tilt}); but this also tells us
that the negative norm mode must satisfy the condition
$\omega_0-\mathbf{u}\cdot\mathbf{k}<0$, but this can be satisfied only
if the velocity of the medium exceeds $|\omega_0/k|$, which is the
\emph{phase} velocity of the probe light, not its group velocity. This
observation suggests that the existence of a ``phase velocity
horizon'' is an essential ingredient (but not the only essential
ingredient) in obtaining Hawking radiation. A similar argument
indicates the necessity for a specific form of ``group velocity
horizon'', one that lies on the negative norm branch. Since the phase
velocity in the slow light setup we are considering is very close to
$c$, the physical speed of light in vacuum, not very much hope is left
for realizing analogue particle creation in this particular laboratory
setting.

However, it was also noticed by \citet{Unruh:2003ss} that a different setup for slow light
might deal with this and other issues (see \citealt{Unruh:2003ss} for a detailed 
summary). In the setup suggested by these authors there are two strong-background counter-propagating control beams illuminating the atoms. The
field describing the beat fluctuations of this electromagnetic
background can be shown to satisfy, once the dielectric medium is in
motion, the same wave equation as that on a curved background. In this
particular situation the phase velocity and the group velocity are
approximately the same, and both can be made small, so that the
previously discussed obstruction to mode mixing is removed. So in this
new setup it \emph{is} concretely possible to simulate classical
particle creation such as, e.g., super-radiance in the presence
of ergoregions.

Nonetheless, the same authors showed that this does not open the
possibility for a simulation of quantum particle production
(e.g., Hawking radiation). This is because that effect also
requires the commutation relations of the field to generate the
appropriate zero-point energy fluctuations (the vacuum structure)
according to the Heisenberg uncertainty principle. This is not the case
for the effective field describing the beat fluctuations of the system
we have just described, which is equivalent to saying that it does not
have a proper vacuum state (i.e., analogue to any physical field).
Hence, one has to conclude that any simulation of quantum particle
production is precluded.

\subsubsection{Slow light in fibre optics}

In addition to the studies of slow light in fluids, there has now been
a lot of work done on slow light in a fibre-optics
setting \citep{Philbin:2007ji, Philbin:2007jj, Belgiorno:2010zz,
  Belgiorno:2010iz}, culminating in recent experimental detection of
photons apparently associated with a phase-velocity
horizon \citep{Belgiorno:2010xxx}. The key issue here is that the Kerr
effect of nonlinear optics can be used to change the refractive index
of an optical fibre, so that a ``carrier'' pulse of light traveling
down the fibre carries with it a region of high refractive index, which
acts as a barrier to ``probe'' photons (typically at a different
frequency). If the relative velocities of the ``carrier'' pulse and
``probe'' are suitably arranged then the arrangement can be made to
mimic a black-hole--white-hole pair. This system is described more
fully in Sect.~\ref{SS:fibre}.

\subsubsection{Quantum fluids of light}
\label{SS:QFL}

Analogue systems based on quantum fluids of light have proven to be very rich and technologically versatile. A quantum fluid of light appears when photons are somehow made to interact with each other, so that their effective description amounts to a (generalized) Gross-Pitaevskii equation. (See e.g.~\citet{Frisch:1992,Chiao:1999,Kivshar:2003} and~\citet{Carusotto:2013} for appropriate reviews.) These interactions are mediated by some matter degrees of freedom (as classically photons do not interact with each other). In 2008 it was proposed
a specific analogue spacetime model based on an optical fiber implementation of a quantum fluid of light~\citep{Marino:2008kk}. After that initial effort, several variations on this idea have also been proposed~\citep{Marino:2009,Fouxon:2009be,Solnyshkov:2011sf}.

The electromagnetic field of a light signal moving along a self-defocussing fiber, with a non-linear refraction response  
$n = n_0 - n_2 |E|^2$, satisfies
\begin{equation}
\partial_z E(x,y;z) = {i \over 2k} \nabla_\perp^2 E(x,y;z) - i k {n_2 \over n_0} |E|^2 E(x,y;z),
\end{equation}
This corresponds to a plane-wave approximation with $k$ the wave number of the propagating electromagnetic field. If one re-identifies $z$ as a time coordinate, this equation is formally equal to the Gross--Pitaevskii equation for a 2D BEC with repulsive interactions. Thus, one can then apply the same logic we followed in the BEC subsection to identify a (low-wavelength) acoustic metric. 

Instead of considering a signal moving along a fiber, one can \cbs{also} generate a quantum fluid of light by pumping a laser into an optical microcavity enclosed by mirrors (see e.g.~\cite{Jacquet:2020znq}). 

Specifically, exciton-polaritons are quasi-particles created by the interaction of light and matter within a semiconductor microcavity. When photons emitted by a laser are directed into a cavity formed by two Bragg mirrors, they exhibit Fabry--Pérot dispersion, which gives them an effective (and very low) mass. These confined photons generate excitons, or bound electron-hole pairs, in the microcavity. The strong coupling between photons and excitons trapped in quantum wells results in two eigenstates for the total Hamiltonian, known as the lower polariton (LP) and upper polariton (UP) branches, separated by the Rabi splitting. Additionally, the Coulomb interaction between excitons introduces an effective non-linearity for exciton-polaritons (polaritons). 

If all energies involved are small compared to the Rabi splitting, the exciton-polariton system can be described by a mean field approximation. 
The dynamics of the mean field is then that of a generalized (dissipative) Gross-Pitaevskii equation, leading (via the usual Madelung representation) to Euler and continuity equations, so characterizing the system as a quantum fluid.

Indeed, following~\cite{Jacquet:2020znq}, we have
\begin{equation}
    i\partial_t \Psi = F_p(x,t)+\left( \omega_0 -\frac{\hbar}{2m_*}  \partial_x^2 +V(x)+g|\Psi(x,t)|^2-i\frac{\gamma}{2}\right)\Psi(x,t),
\end{equation}
where $F_p$ is the field of the pump laser, $\omega_0$ is the frequency of the lower polaritons at the bottom of the branch, $m_*$ is their effective mass, $V(x)$ is the ``external potential" (that is controlled via the cavity geometry), $g$ is the effective non-linearity in the refractive index (proportional to $n_2$), $\gamma$ is the loss rate. The field of the lower polaritons $\Psi(x,t)$ is here written in the laboratory frame.

This equation is typically post-processed as follows:
Consider the cavity to have a resonance at $\omega_{\rm cav}$, then the slowly varying envelope of the intra-cavity field $E$ can indeed be shown to satisfy an equation of the form
\begin{equation}
\partial_t E = {i \over 2k} \nabla_{\perp}^2 E - i \omega {n_2 \over n_0} |E|^2 E +i \delta E - \Gamma(E-E_d).
\end{equation}
Here we have changed notation so that $\omega$ is the frequency of the pump laser, $k$ its wavenumber, $\delta=\omega-\omega_{\rm cav}$, $n_2$ and $n_0$ characterize the effective nonlinear refractive index, and $E_d$ is the intensity of the coherent driving field, which is proportional to the incident field.  In this setup analogue time is the actual physical time. 
This is a forced-dissipative system in which the losses $-i\,\Gamma E$ are balanced by the coherent injected field $i\Gamma E_d$. 
The coefficient $\Gamma = c T/ 2n_0 L$ depends on the cavity's length $L$ and the mirror transmissivity $T$.

The central point is that the field again satisfies  a generalized Gross-Pitaevskii equation so we can generate background configurations and analyze the behavior of acoustic fluctuations.  For a (locally) homogeneous configuration we have the (damped quartic) dispersion relation
\begin{equation}
(\Omega -v_0 K -i \Gamma)^2 = c_s^2 K^2 + K^4/K_0^2;~~~~~c_s^2= {c^2 n_2^2 |E_d|^2 \over n_0^3};
~~~~K_0^2 = {4 n_0 k^2 \over c^2},
\end{equation}
where $\Omega$ and $K$ are the frequency and wave number of a phononic excitation. 
This is equal to the BEC dispersion relation except for the dissipative term from which the field acquires attenuation in a length scale of the order $\xi_d= \Gamma/c_s$. For long wavelengths $\lambda \ll \xi_0 \ll \xi_d$ we have (approximate) acoustic propagation.

\subsubsection{Lattice models}
\label{SS:lattice}

The quantum analogue models described above all have an underlying
discrete structure: namely the atoms they are made of. In abstract
terms one can also build an analogue model by considering a quantum
field on specific lattice structures representing different
spacetimes. in \citet{Jacobson:1996zs, Corley:1998ef, Jacobson:1999ay}
a falling-lattice black-hole analogue was put forward, with a view to
analyzing the origin of Hawking particles in black-hole
evaporation. The positions of the lattice points in this model change
with time as they follow freely falling trajectories. This causes the
lattice spacing at the horizon to grow approximately linearly with
time. By definition, if there were no horizons, then for long
wavelengths compared with the lattice spacing one would recover a
relativistic quantum field theory over a classical
background. However, the presence of horizons makes it impossible to
analyze the field theory only in the continuum limit, it becomes
necessary to recall the fundamental lattice nature of the model.

\subsubsection{Graphene}
\label{SS:graphene}

\begin{figure}[hptb]
    \centerline{\includegraphics[width=0.6\textwidth]{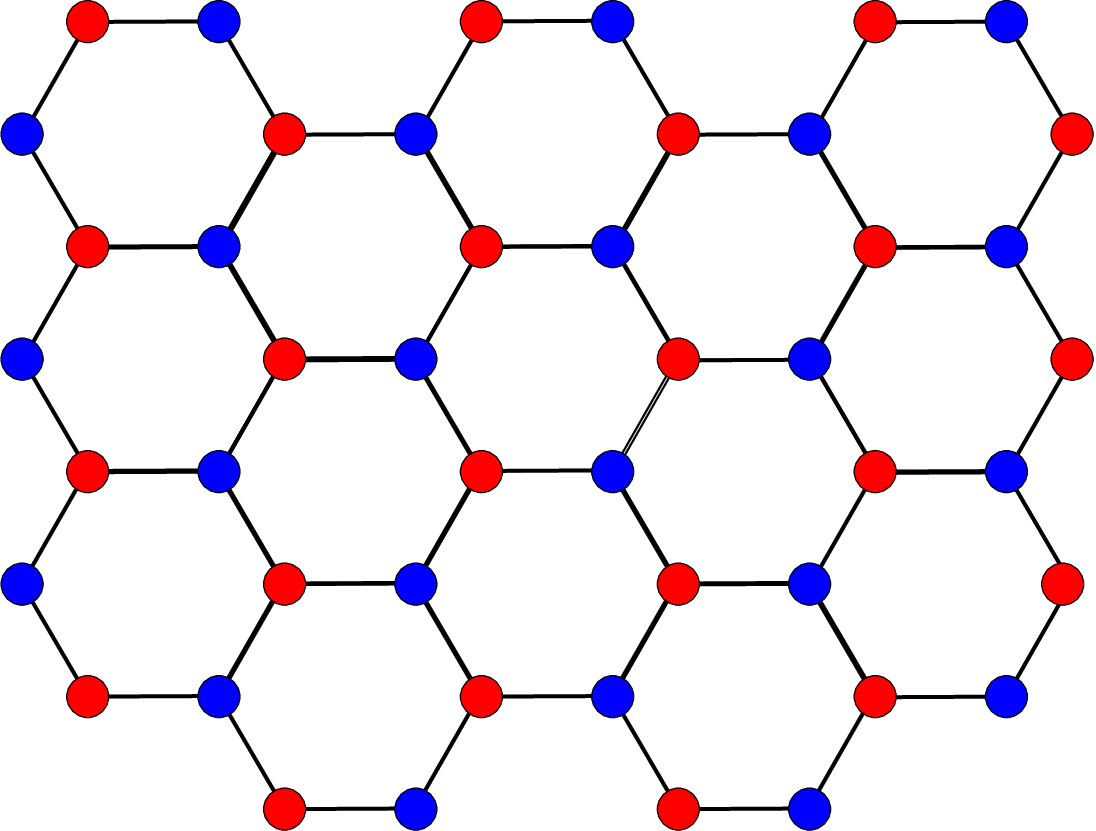}}
\caption[Graphene: hexagonal lattice]{The graphene hexagonal lattice
  is made of two inter-penetrating triangular lattices. Each is
  associated with one Fermi point.}
\label{Fig:graphene}
\end{figure}

A very interesting addition to the catalogue of analogue systems is the
graphene (see, for example, these
reviews by \citealt{RevModPhys.81.109,Katsnelson20073}). Although graphene
and some of its peculiar electronic properties have been known since
the 1940s \citep{PhysRev.71.622}, only recently has it been
specifically proposed as a system with which to probe gravitational
physics \citep{Cortijo:2006xs,Cortijo:2006mh}. Graphene (or mono-layer
graphite) is a two-dimensional lattice of carbon atoms forming a
hexagonal structure (see Fig.~\ref{Fig:graphene}). From the
perspective of this review, its most important property is that its
Fermi surface has two independent Fermi points (see
Sect.~\ref{S:helium} on helium). The low-energy excitations around
these points can be described as massless Dirac fields in which the
light speed is substituted by a ``sound'' speed $c_s$ about 300 times
smaller:
\begin{eqnarray}
\partial_t \psi_j = c_s \sum_{k=1,2} \sigma^k \partial_{x_k} \psi_j\,.
\end{eqnarray}
Here $\psi_j$ with $j=1,2$ represent two types of massless spinors
(one for each Fermi point), the $\sigma^k$ are the Pauli sigma
matrices, and $c_s=3ta/2$, with $a=1.4\mathrm{\ \AA}$ being the interatomic
distance, and $t=2.8\mathrm{\ eV}$ the hopping energy for an electron between
two nearest-neighboring atoms.

From this perspective graphene can be used to investigate
ultra-relativistic phenomena such as the Klein
paradox \citep{Katsnelson20073}. On the other hand, graphene sheets can
also acquire curvature. A nonzero curvature can be produced by adding
strain fields to the sheet, imposing a curved substrate, or by
introducing topological defects (e.g., some pentagons within the
hexagonal structure; \citealt{Vozmediano:2008zz}). It has been suggested
that, regarding the electronic properties of graphene, the sheet
curvature promotes the Dirac equation to its curved space counterpart,
at least on the average \citep{Cortijo:2006mh,Cortijo:2006xs}. If this
proves to be experimentally correct, it will make graphene a good
analogue model for a diverse set of spacetimes. This set, however,
does not include black-hole spacetimes, as the curvatures mentioned
above are purely spatial and do not affect the temporal components of
the metric.

\subsection{Going further}
\label{SS:going-further-5}
\enlargethispage{40pt}
As we said at the beginning of this section, this catalogue is by no means exhaustive, but we feel it covers key items in the history of the subject.
Additional models, like the promising superconducting circuits~\citep{Blencowe:2020ygo} or 
\citep{Wilson:2011rsw} 
can be found in the historical section and the cited reviews.


\clearpage
\section{Phenomenological Applications of Analogue Models}
\label{S:lessons}

One of the first motivations to study analogue models was to see
whether it is possible (both theoretically and in practice) 
to reproduce in the laboratory various gravitational
phenomena whose real existence in nature cannot be currently checked in any direct fashion.
These are phenomena that surpass our present (and foreseeable)
observational capabilities, but yet, we believe in their existence
because it follows from seemingly strong theoretical arguments within
the standard frameworks of general relativity and field theory in curved
space. However, the interest of this approach is not merely to reproduce
these gravitational phenomena in some formal analogue model, but to
see which departures from the ideal case show up in real analogue
models, and to analyse whether similar deviations are likely to appear in
real gravitational systems.

When one thinks about emergent gravitational features in condensed-matter systems, 
one immediately realises that these features only appear
in the low-energy regime of the analogue systems. When these systems are
probed at high energies (short length scales) the effective geometrical
description of the analogue models break down, as one starts to be aware
that the systems are actually composed of discrete pieces (atoms and
molecules). This scenario is quite similar to what one expects to happen
with our geometrical description of the Universe, when explored with
microscopic detail at the Planck scale.

That is, the study of analogue models of general relativity is giving us
insights as to how the standard theoretical picture of different
gravitational phenomena could change when taking into account additional
physical knowledge coming from the existence of an underlying
microphysical structure. 
These studies highlight that an important family of deviations from the standard general relativity picture can be encoded in the form of high-energy violations of Lorentz invariance in particle dispersion relations. 
Beyond these first deviations, the analogue models of general relativity provide well-understood examples (the underlying physics is well known) in which a description in terms of fields in curved spacetimes shows up as a
low-energy-regime emergent phenomena.

We devote this section to {discussing} how analogue models are being used to shed light on phenomenological questions posed within GR at both the classical and semiclassical levels. In this section these questions are discussed theoretically; in the next we will review the experimental developments trying to obtain actual realizations of the phenomena at hand. Section 7, we will go further and discuss the interplay between analogue gravity and the quest for a superseding theory.
Thus, let us now turn to discussing several specific physics issues that are being analysed from this phenomenological perspective.
In particular we shall briefly discuss:
\begin{itemize}
    \item Hawking radiation.
    \begin{itemize}
        \item Robustness.
        \item Necessary and sufficient conditions.
        \item Source of the Hawking quanta.
        \item Which notion of surface gravity?
        \item Correlations.
        \item Solid state and lattice models.
        \item Analogues as background gestalt.
    \end{itemize}
    \item Dynamical stability of horizons.
    \begin{itemize}
        \item Classical.
        \item Semiclassical.
        \item Modifed dispersion relations.
        \item Black/white holes, black/white rings.
    \end{itemize}
    \item Super-radiance.
    \item Cosmological particle production.
    \item Bose-novae.
    \item Romulan cloaking devices.
\end{itemize}

\subsection{Hawking radiation} 
\label{S:Hawking-radiation}
\subsubsection{Basics} 

As is well known, in 1974 Stephen Hawking announced that quantum
mechanically black holes should emit radiation with a spectrum
approximately that of a black body \citep{Hawking:1974rv,
  Hawking:1974sw}. We shall not re-derive the existence of Hawking
radiation from scratch, but will instead assume a certain familiarity
with at least the basics of the phenomenon. (See, for
instance, \citealt{Brout:1995rd} or \citealt{Jacobson:1995ak}.) The collapse
of a distribution of matter will end up forming an evaporating black
hole emitting particles from its horizon toward future null
infinity. Hawking radiation is a quantum-field-in-curved-space effect:
The existence of radiation emission is a kinematic effect that does
not rely on Einstein's equations. Therefore, one can aim to reproduce it
in a condensed-matter system. Within standard field theory, a 
{sufficient}
requirement for having  Hawking radiation, {a strictly quantum spontaneous effect}, is the existence in the
background configuration of an apparent horizon \citep{Visser:2001kq}.
{(Minimal conditions to have a Hawking-like emission have been analyzed in~\cite{Barcelo:2006uw,Barcelo:2010pj},
where it is shown that one does not need the strict formation of neither event nor apparent horizons)}. 
So, in principle, to be able to
reproduce Hawking radiation in a laboratory one would have to fulfill
at least two requirements:

\begin{enumerate}
\item 
To choose an adequate analogue system; it has to be a quantum
 analogue model (see Sect.~\ref{S:catalogue}) such that its description
 could be separated into a classical effective background spacetime plus
 some standard relativistic quantum fields living thereon. (It can happen
 that the quantum fields do not satisfy the appropriate commutation or
 anti-commutation relations \citep{Unruh:2003ss}).

\item 
To configure the analogue geometry such that it includes some sort of
  horizon. That is, within an appropriate quantum analogue model,
 the formation of an apparent horizon for the propagation of the quantum
 fields should excite the fields such as to result in the emission of a
 thermal distribution of field particles.%
 \footnote{One could also imagine
   systems in which the effective metric fails to exist on one side 
   of the horizon (or even more radically, on both sides). The existence of particle
   production in this kind of system will then depend on the specific
   interactions between the sub-systems characterizing each side of the
   horizon. For example, in 
   stationary configurations it will be necessary that these interactions
   allow negative energy modes to disappear beyond the horizon,
   propagating forward in time (as happens in an ergoregion). Whether these
   systems will provide adequate analogue models of Hawking radiation or
   not is an interesting question that deserves future analysis. Certainly systems
   of this type lie well outside the class of usual analogue models. 
 }

\end{enumerate}
This is a straightforward and quite naive translation of the standard Hawking
effect derivation to the condensed matter realm. However, in reality,
this translation process has to take into account additional issues
that, as we are trying to convey, instead of problems, are where the
interesting physics lies.

\begin{enumerate}

\item The effective description of the quantum analogue systems as
 fields over a background geometry breaks down when probed at
 sufficiently short length scales. This could badly influence the main
 features of Hawking radiation. In fact, immediately after the inception
 of the idea that black holes radiate, it was realised that there was a
 potential problem with the calculation \citep{Unruh:1976db}. It strongly
 relies on the validity of quantum field theory on curved backgrounds up
 to arbitrary high energies. Following a wave packet with a certain
 frequency at future infinity backwards in time, we can see that it had
 to contain arbitrarily large frequency components with respect to a
 local free fall observer (well beyond the Planck scale) when it was
 close to the horizon. In principle, any unknown physics at the Planck
 scale could strongly influence the Hawking process so that one should
 view it with caution. 
This is the \emph{trans-Planckian} problem of
 Hawking radiation. To create an analogue model exhibiting Hawking
 radiation will be, therefore, equivalent to providing one
 solution to the trans-Planckian problem.

\item In order to clearly observe Hawking radiation, one should first
 be sure that there is no other source of instabilities in the system
 that could mask the effect. In analogue models such as liquid helium or
 BECs the interaction of a radial flow (with speed on the order of the
 critical Landau speed, which in these cases coincides with the sound
 speed \citep{Kopnin:1997jy}) with the surface of the container (an
 electromagnetic potential in the BECs case) might cause the production
 of rotons and quantized  vortices, respectively. Thus, in order to
 produce an analogue model of Hawking radiation, one has to be somewhat
 ingenious. For example, in the liquid helium case, instead of taking
 acoustic waves in a supersonic flow stream as the analogue model, it is
 preferable to use as analogue model ripplons in the interface between
 two different phases, A and B phases, of liquid helium
 three \citep{Volovik:2002ci}. Another option is to start from a moving
 domain wall configuration. Then, the topological stability of the
 configuration prevents its destruction when creating a
 horizon \citep{Jacobson:1998he, Jacobson:1998ms}. In the case of BECs,
 a way to suppress the formation of quantized  vortices is to take
 effectively one-dimensional configurations. If the transverse
 dimension of the flow is smaller than the healing length, then there
 is no space for the existence of a vortex \citep{Barcelo:2001ca}. In
 either liquid helium or BECs, there is also the possibility of
 creating an apparent horizon by rapidly approaching a critical
 velocity profile (see
 Fig.~\ref{F:critical-dynamical-configuration}), but without
 actually crossing into the supersonic regime \citep{Barcelo:2004wz},
 softening in this way the appearance of dynamical instabilities.

\begin{figure}[htpb]
   \centerline{\includegraphics[width=0.75\textwidth]{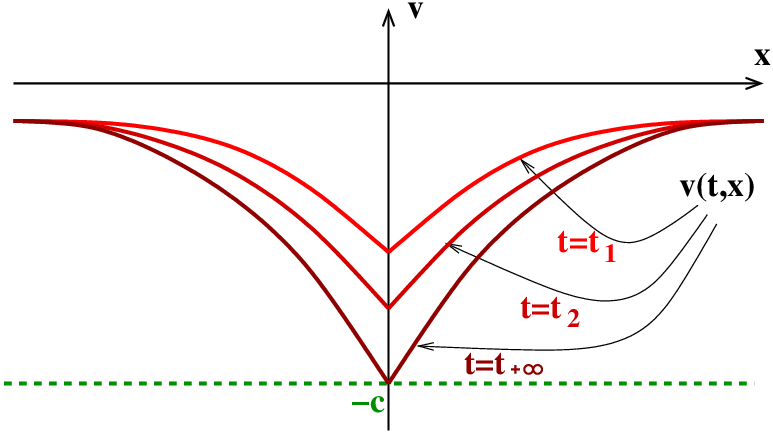}}
\caption[Critical collapse velocity profile]{Velocity profile for a
left going flow; the profile is dynamically modified with time so that
it reaches the profile with a sonic point at some time in the asymptotic future.}
\label{F:critical-dynamical-configuration}
\end{figure}

\item Real analogue models cannot, strictly speaking, reproduce eternal
 black-hole configurations. An analogue model of a black hole has always
 to be created at some finite laboratory time. Therefore, one is forced
 to carefully analyse the creation process, as it can greatly influence the
 Hawking effect. Depending on the procedure of creation, one could end up
 in quite different quantum states for the field and only some of them
 might exhibit Hawking radiation. This becomes more important when
 considering that the analogue models incorporate modified dispersion
 relations. An inappropriate preparation, together with modified
 dispersion relation effects, could completely eliminate Hawking
 radiation \citep{Unruh:2004zk,Barcelo:2010bq}.

\item Another important issue is the need to characterise ``how
 quantum'' a specific analogue model is. Even though, strictly speaking,
 one could say that any system undergoes quantum fluctuations, the point
 is how important they are in its description. In trying to build an
 analogue model of Hawking's quantum effect, the relative value of
 Hawking temperature with respect to the environment 
 is the most basic  quantity that can tell us
 whether the system can be really thought of as a quantum analogue model or
 as effectively classical. For example, in our standard cosmological
 scenario, for a black hole to radiate at temperatures higher than that
 of the Cosmic Microwave Background, $\approx$~3~K, the black hole should
 have a diameter on the order of micrometers or less. We would have to
 say that such black holes are no longer classical, but semiclassical.
 The black holes for which we have some observational evidence are of
 much higher mass and size, so their behaviour can be thought of as
 completely classical. 
 First estimates of the Hawking temperature reachable in
 BECs yielded $T \sim 100\,\mathrm{nK}$ \citep{Barcelo:2001ca}. 
 This has the same order of magnitude as the temperature of the BECs themselves. This was telling
 us that, regarding the Hawking process, BECs can be considered to be
 highly-quantum analogue models. 
 We must advance that a retrieval of the attained temperature in Steinhuaer's actual experiment yields $T_H$ around 1.2nK~\citep{Steinhauer:2016ftg}.
\end{enumerate}

Up to this point, we have talked exclusively about the spontaneous Hawking effect. However, there is also a stimulated version of the effect which does not need a quantum analogue model to be probed~\citep{Unruh:2014hua}. Both effects, spontaneous and stimulated, depend crucially on the form of the high-energy dispersion relations. Thus, let us now review what we know about the effects of high-energy dispersion relations on the Hawking process.

\subsubsection{UV robustness} 

We saw in the introduction to this section that the trans-Planckian
problem of Hawking radiation was one of the strongest motivations for
modern research into analogue models of gravity. In fact, it was
soon realised that such models could provide a physical framework within
which a viable solution of the problem could be found. Let us explain
why and how.

As we have said, the requirement of a reservoir of ultra-high frequency
modes near the horizon seems to indicate a possible (and worrisome)
sensitivity of the black-hole radiation to the microscopic structure of
spacetime. Actually, by assuming exact Lorentz invariance, one could, in
principle, always \emph{locally} transform the problematic ultra-high frequency
modes to low energy ones via some appropriate Lorentz
transformation \citep{Jacobson:1991gr}. However there are (at least)
two problems in doing so:

\begin{enumerate}

\item One has to rely on the physics of reference frames moving ultra
  fast with respect to us, as the reference frame needed would move
  arbitrarily close to the speed of light. Hence, we would have to
  apply Lorentz invariance in a regime of arbitrarily large boosts that is as yet
  untested, and in principle never completely testable given the
  non-compactness of the boost subgroup. The assumption of an exact
  boost symmetry is linked to the scale-free nature of spacetime given
  that unbounded boosts expose ultra-short distances. Hence, the
  assumption of \emph{exact} Lorentz invariance needs, in the end, to
  rely on some ideas regarding the nature of spacetime at ultra-short
  distances.

\item Worse, even given these assumptions, ``one cannot boost away an
  \textit{s}-wave''. That is, given the expected isotropy of Hawking
  radiation, a boost in any given direction could, at most, tame the
  trans-Planckian problem only in that specific direction. Indeed, the
  problem is then not ameliorated in directions orthogonal to the
  boost, and would become even worse on the opposite side of
  the black hole.

\end{enumerate}

It was this type of reasoning that led in the nineties to a careful
reconsideration of the crucial ingredients required for the derivation
of Hawking radiation \citep{Jacobson:1991gr, Jacobson:1993hn,
Unruh:1995je}. In particular investigators explored the possibility that
spacetime microphysics could provide a short-distance, Lorentz-breaking
cutoff, but at the same time leave Hawking's results unaffected at energy
scales well below that set by the cutoff.

Of course, ideas about a possible cutoff imposed by the discreteness of
spacetime at the Planck scale had already been discussed in the
literature well before Unruh's seminal paper \citep{Unruh:1981cg}.
However, such ideas were running into serious difficulties given that a
naive short-distance cutoff posed on the available modes of a free field 
theory results in a complete removal of the evaporation process (see,
e.g., \citealt{Jacobson:1991gr} and references therein.
Indeed 
there are alternative ways through which the effect of the short-scale
physics could be taken into account, and analogue models provide a
physical framework where these ideas could be put to the test. In fact,
analogue models provide explicit examples of emergent spacetime
symmetries; they can be used to simulate black-hole backgrounds; they
may be endowed with quantizable perturbations and, in most of the cases,
they have a well-known microscopic structure. Given that Hawking
radiation can be, at least in principle, simulated in such systems, one
might ask how and if the trans-Planckian problem is resolved in these
cases.

\paragraph*{Modified dispersion relations:\;}

The general feature that most of the work on this subject has focused
on is that in most analogue models --- those formally described as classical or quantum Galilean systems --- the quasi-particles propagating on the
effective geometry are actually collective excitations of atoms. This implies that their dispersion relation will be a
relativistic one only at low energies (large
scales)~\footnote{Actually, even relativistic behaviour at low energy
can be non-generic~\citep{Barcelo:2001cp}, but we assume in this discussion that an analogue
model by definition is a system for which all the linearized perturbations do propagate on the same Lorentzian background at low
energies.}, {while presenting at high energies (short wavelengths) some deviation from the relativistic behaviour. Such deviations are generically characterized by a short length scale
(e.g., intermolecular distance for a fluid, coherence length
for a superfluid, healing length for a BEC) in whose proximity deviations from the relativistic behaviour
will not be negligible anymore.} In general, such microphysics-induced corrections to the dispersion relation take the form
\begin{equation} 
E^2=c^2\left(m^2c^2+k^2+\Delta(k,K)\right),
\end{equation} 
where $K$ is the scale that describes the transition to the full
microscopic system (what we might call -- within this section -- the
``analogue Planck scale'').
Let us stress that the above defined modified dispersion relations are always provided within a preferred reference frame, usually associated to the laboratory frame or to some external field associated to the experimental set up acting as an effective aether.

In general, the best one can do is to expand $\Delta(k,K)$ around $k=0$,
obtaining an infinite power series (of which it will be safe to
retain only the lowest-order terms), although in some special models
(like BEC) the series is automatically finite due to intrinsic
properties of the system. (In any case, one can see that most of the
analogue models so far considered lead to modifications of the form $\pm
k^3/K$ or $\pm k^4/K^2$.) Depending on the sign in front of the
modification, the group velocity at high energy can be larger ($+$) or
smaller ($-$) than the low energy speed of light $c$. These cases are
usually referred to in the literature as ``superluminal'' and
``subluminal'' dispersion relations.

Most of the work on the trans-Planckian problem in the 1990s focused
on studying the effect on Hawking radiation due to such modifications of
the dispersion relations at high energies in the case of acoustic
analogues \citep{Jacobson:1991gr, Jacobson:1993hn, Unruh:1995je,
  Corley:1996ar}, and the question of whether such phenomenology could
be applied to the case of real black holes (see e.g.,
\citealt{Brout:1995wp, Jacobson:1996zs, Corley:1996ar,
  Parentani:2000ts}).\footnote{However, see
  also \citet{Reznik:1997xt, Rosu:1996pw} for a radically different
  alternative approach based on the idea of ``superoscillations''
  where ultra-high frequency modes near the horizon can be mimicked (to
  arbitrary accuracy) by the exponential tail of an exponentially-large
amplitude mostly hidden behind the horizon.} In all the aforementioned
works, Hawking radiation can be recovered under some suitable
assumptions, as long as neither the black-hole temperature nor the
frequency at which the spectrum is considered are too close to the scale
of microphysics $K$. However, the applicability of these assumptions to
the real case of black hole evaporation is an open question. It is also
important to stress that the mechanism by which the Hawking radiation
is recovered is conceptually rather different depending on the type of
dispersion relation considered. We concisely summarise here the main
results (but see, e.g., \citealt{Unruh:2004zk} for further details).

\paragraph*{Subluminal dispersion relations:\;} 

This was the case originally considered by \citet{Unruh:1995je},
\begin{equation} 
\omega=K\left(\tanh(k/K)^n\right)^{1/n}, 
\label{eq:Un}
\end{equation}
and by \citet{Corley:1996ar}
\begin{equation} 
\omega^2=k^2-k^4/K^2,
\label{eq:CJ} 
\end{equation}
where both dispersion relations are given in the co-moving frame.

\begin{figure}[htbp]
    \centerline{\includegraphics[width=4.0in]{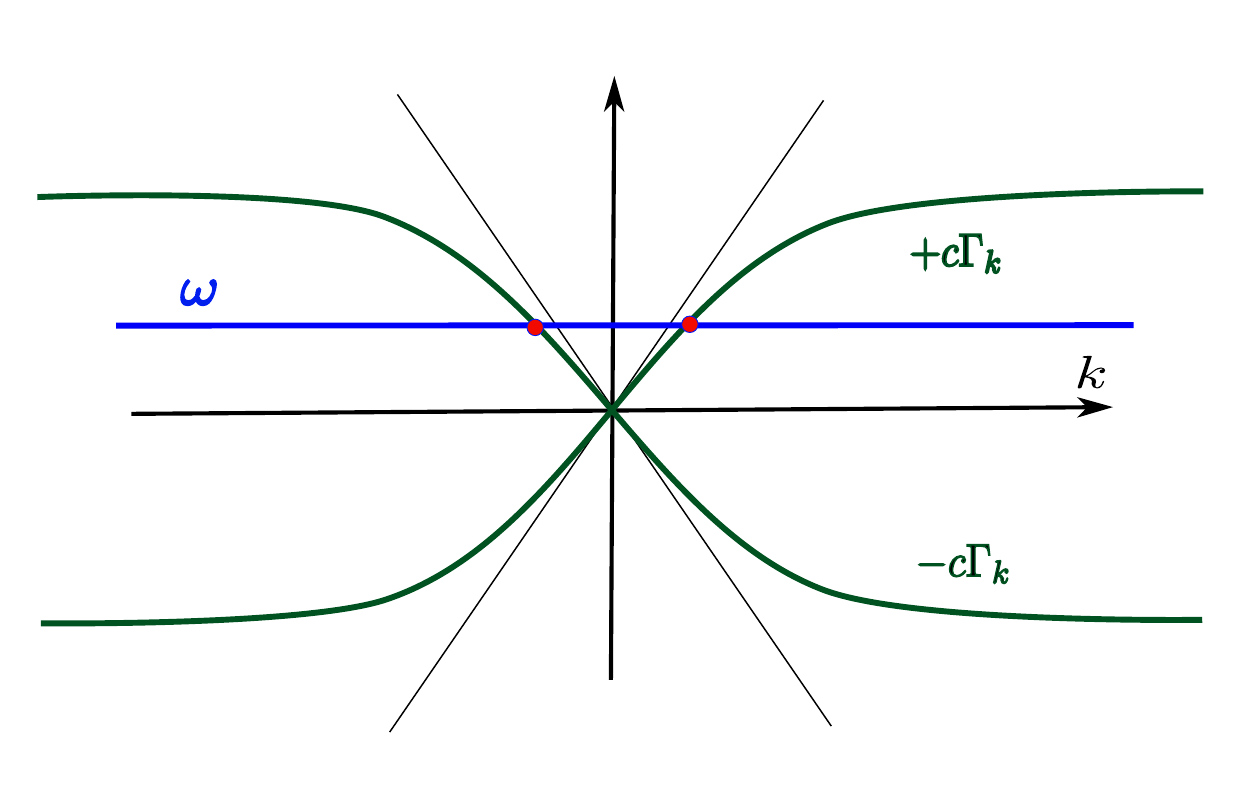}}
    \centerline{\includegraphics[width=4.0in]{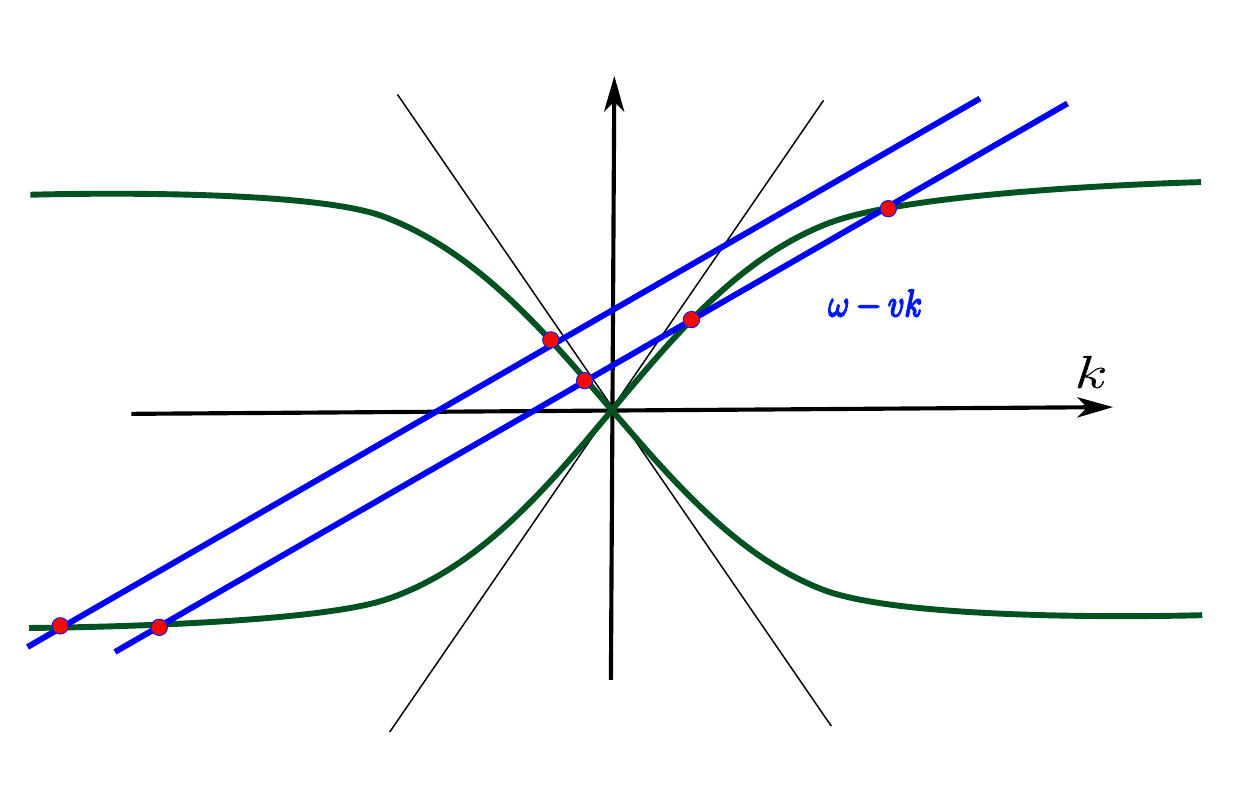}}
    \centerline{\includegraphics[width=4.0in]{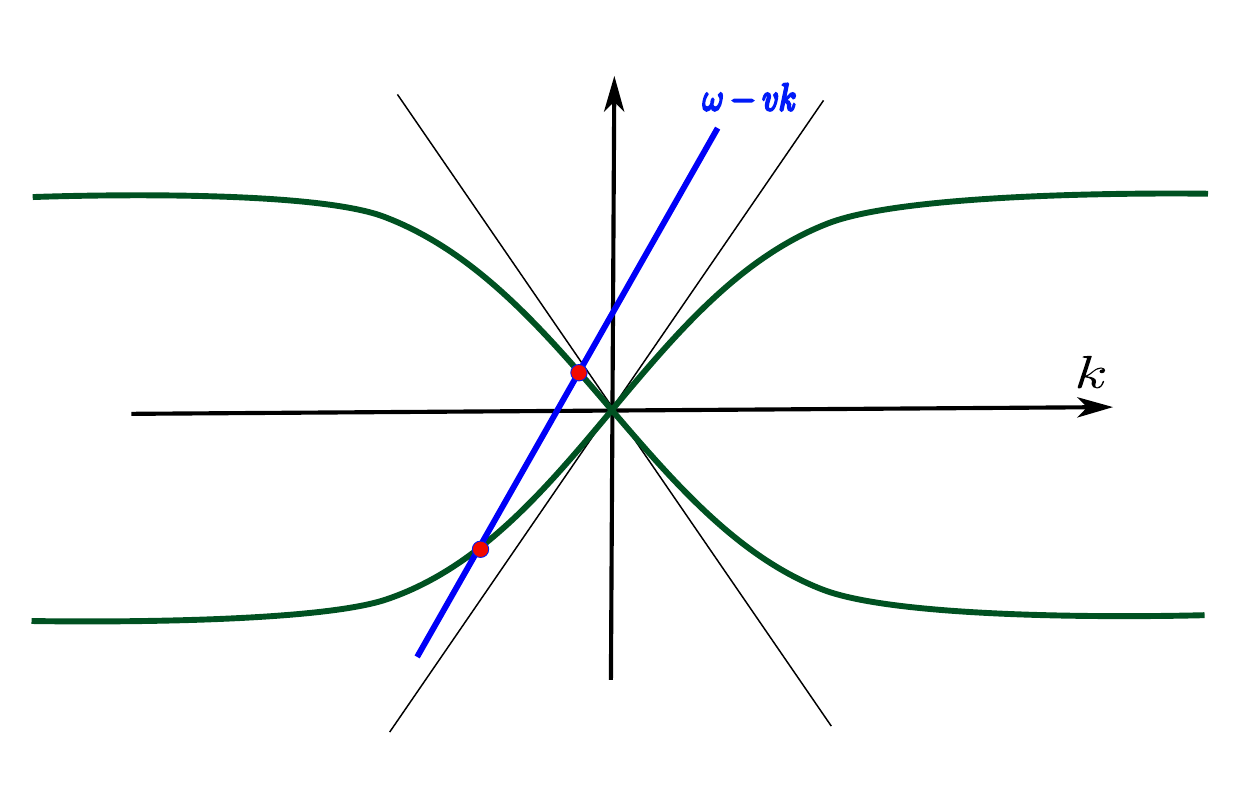}}
  \caption[Subluminal dispersion relation]{The picture shows a
    subsonic dispersion relation as a relation $\omega-vk = \pm c_s
    |k|\Gamma_k$ for the three cases $v=0$, $v<c_s$ and $v>c_s$. 
    In particular we plot a dispersion of the type
    $\Gamma_k=(K/|k|)\sqrt{\tanh(k^2/K^2)}$ originally employed by
    Unruh.  For subsonic velocities, $v<c_s$, and $\omega$ less than a
    critical frequency $\omega_c$, the dispersion relation has four
    real roots, though for $\omega > \omega_c$ one only has two roots.
    For supersonic velocities $v>c_s$, the dispersion relation has two
    real roots.
    }
\label{F:sub-dispersion}
\end{figure}

The key feature is that in the presence of a subluminal modification
the group velocity of the modes increases with $k$ only up to some
turning point (which is equivalent to saying that the group velocity
does not asymptote to $c$, which could be the speed of sound, but
instead is upper bounded). For values of $k$ beyond the turning point
the group velocity decreases to zero (for Eq.~(\ref{eq:Un})) or becomes
imaginary (for Eq.~(\ref{eq:CJ})). In the
latter case, this can be interpreted as signifying the breakdown of the
regime where the dispersion relation~(\ref{eq:CJ}) can be trusted. 
In Fig.~\ref{F:sub-dispersion} one can see the behaviour of the 
dispersion relation for a specific case of Eq.~(\ref{eq:Un})).

For this class of dispersion relations, an accurate analysis of the 
propagating modes leads to a surprising conclusion regarding the 
origin of the outgoing Hawking quanta at infinity. In fact, if one
traces back in time an outgoing mode, as it approaches the horizon, it
decreases its group velocity below the speed of sound. At some point
before reaching the horizon, the outgoing mode will appear as a
combination of ingoing modes dragged into the black hole by the
flow. Stepping further back in time it is seen that such modes are
located at larger and larger distances from the horizon, and tend to
have very high wave numbers far away at early times. The important
point is that one of these modes has ``negative norm''. (That is, the
norm is negative in the appropriate Klein--Gordon-like inner
product \citep{Barcelo:2010bq}.) In this way one finds what might be
called a ``mode conversion'', where the origin of the outgoing Hawking
quanta seems to originate from ingoing modes, which have ``bounced
off'' the horizon without reaching trans-Planckian frequencies in its
vicinities. 

Several detailed analytical and numerical calculations
have shown that such a conversion indeed happens \citep{Unruh:1995je,
  Brout:1995wp, Corley:1996ar, Corley:1997pr, Himemoto:1999kd,
  Saida:1999ap, Unruh:2004zk,  Macher:2009nz, Macher:2009tw} and that
the Hawking result, specifically the presence of a Planckian spectrum
of particles at a temperature $T \simeq \kappa/(2\pi)$ moving outwards
toward the asymptotic region, can be recovered for $K \gg \kappa$
where $\kappa$ is the black-hole surface gravity. 

Actually, the
condition to recover a Planckian spectrum is a bit more subtle. In the
mode conversion process every frequency experiences the horizon as
located at a different position, and so, having a different surface
gravity (a rainbow geometry). To have an approximately Planckian
spectrum one would need this frequency-dependent surface gravity to
stay almost constant for frequencies below and around $\kappa$~(see,
for example, the analysis in \citet{Macher:2009nz, Macher:2009tw, 
  Finazzi:2010nc}). 
This condition appears difficult to attain in current experiments.

In addition, we would like to point out another subtlety of these
analogue black-hole configurations: It is necessary that the constant
flow velocity reached at the asymptotic region is different from
zero. For the particular dispersion relation in
Eq.~(\ref{eq:CJ}), $|v|\geq \kappa/K$ so that the Planckian
form {of the spectrum} is not cut {off} at frequencies around $\kappa$.

Finally, let us mention here that more details on this case {will be provided} in Sec.~\ref{S:water-tank}, given that the classical counterpart of the above-described mode-conversion mechanism has indeed been observed in several water talk experiments.

\paragraph*{Superluminal dispersion relations:\;} 

\begin{figure}[htpb]
  \centerline{
  \includegraphics[width=2.0in]{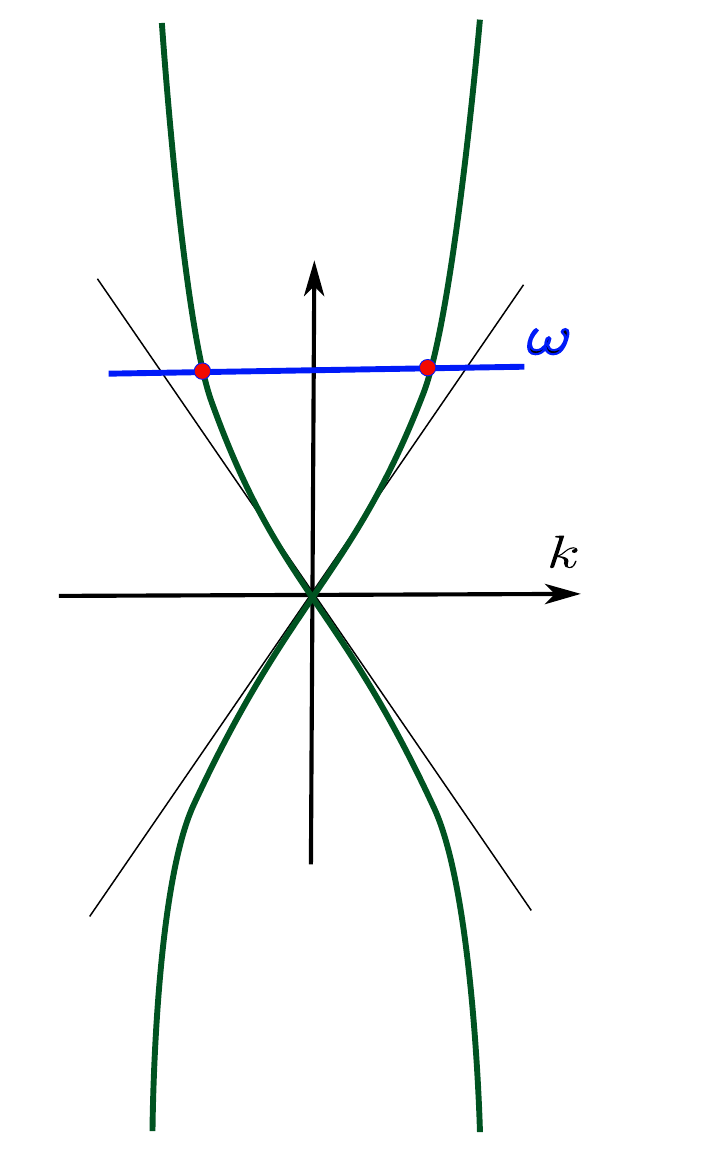}
  \includegraphics[width=2.0in]{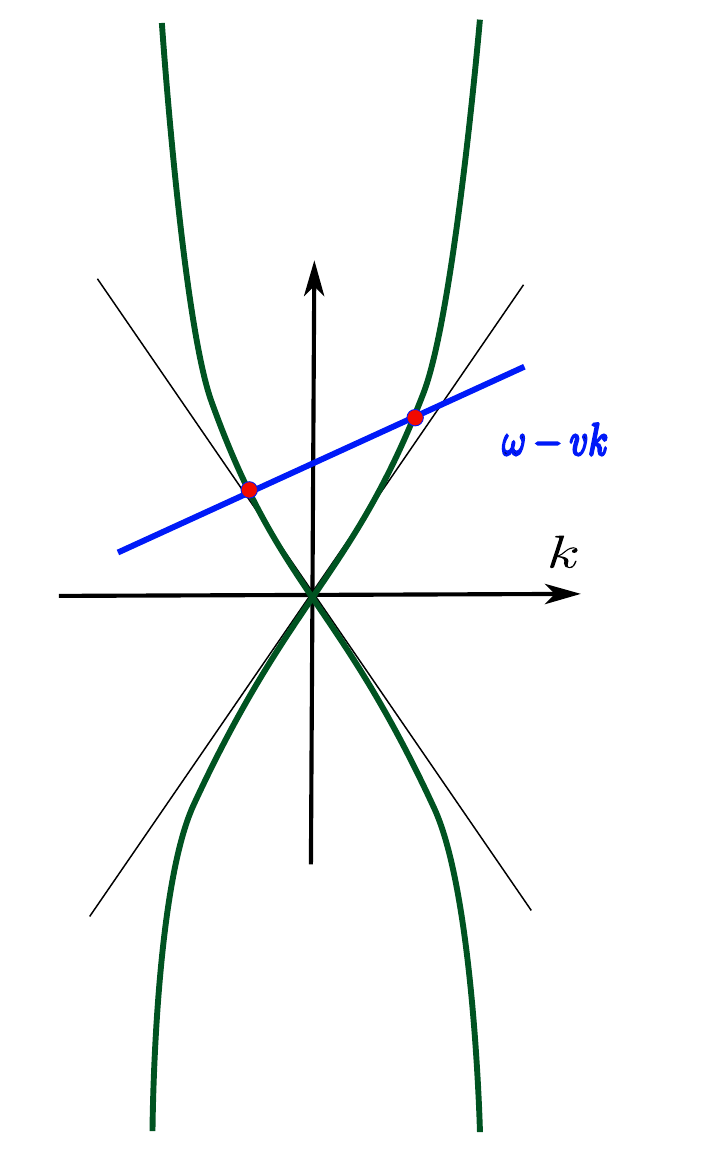}
  \includegraphics[width=2.0in]{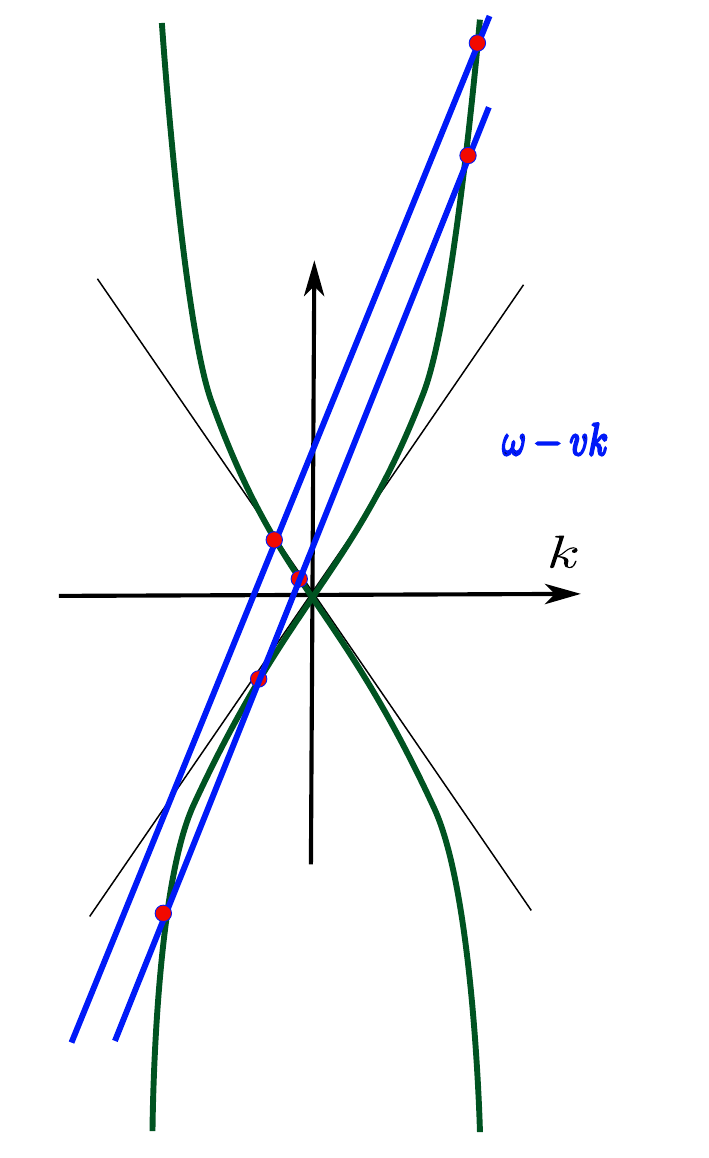}
  }
  \caption[Superluminal dispersion relation]{The picture shows a
    supersonic dispersion relation as a relation $\omega-vk = \pm c_s
    |k|\Gamma_k$, for the three cases $v=0$, $v<c_s$, and $v>c_s$. In particular, we plot a dispersion of the type
    $\Gamma_k=\sqrt{1+k^2/K^2}$. For subsonic velocities $v<c_s$ the
    dispersion relation has two real roots. For supersonic velocities $v>c_s$,
    and $\omega$ less than a critical frequency $\omega_c$, the
    dispersion relation has four real roots, though for $\omega>\omega_c$ one again recovers two real roots.
    }
\label{F:super-dispersion}
\end{figure}

The case of a superluminal dispersion relation is quite different and,
as we have seen, has some experimental interest, given that this is the
kind of dispersion relation that arises, 
{for instance, in BECs and quantum fluids of light.}
In this situation, the outgoing modes are actually
seen as originating from behind the horizon (see
Fig.~\ref{F:super-dispersion}). This implies that these modes
somehow originate from the singularity (which can be a region of high
turbulence in acoustic black-hole analogues), and hence it would seem
that not much can be said in this case. However, it is possible to
show that if one imposes vacuum boundary conditions on these modes
near the singularity, then it is still possible to recover the Hawking
result, i.e., thermal radiation outside the black hole \citep{Corley:1997pr}.

The understanding of the physics behind the presence of Hawking
radiation in superluminal dispersive theories has greatly \blue{improved}
recently through the detailed analysis of 1+1 stationary
configurations possessing one black or white horizon connecting two
asymptotic regions \citep{Macher:2009tw} (for previous work dealing
with mode conversion through a horizon,
see \citealt{Leonhardt:2002yu}). One of the asymptotic regions
corresponds to the asymptotic region of a black-hole spacetime and is
subsonic; the other asymptotic region is supersonic and replaces the
internal singularity. Once the appropriate acoustic geometry is
defined, this analysis considers a Klein--Gordon field equation in
this geometry, modified by a $\partial_x^4$ term that gives rise to
the quartic dispersion relation $\omega^2=k^2+k^4/K^2$. For this setup
it has been shown that the relevant Bogoliubov coefficients have the
form $\beta_{\omega',i;\omega,j}=\delta(\omega-\omega')
\beta_{\omega;ij}$, with $\beta_{\omega;ij}$ a $3\times3$ matrix. 

In order to recover from these coefficients a Planckian spectrum of particles at
the external asymptotic region, the relevant condition happens to be
not $\kappa \ll K$ but $\kappa \ll \omega_c$ where
$\omega_c=Kf(v_{\mathrm{int-asym}})$ and with $f(\cdot)$ a specific
function of $v_{\mathrm{int-asym}}$, the value of the flow velocity at
the internal (or supersonic) asymptotic
region \citep{Macher:2009tw}. In all cases, for $\omega > \omega_c$ the
Bogoliubov coefficients are exactly zero. When $v_{\mathrm{int-asym}}$ is
just above $c=1$, that is, when the flow is only ``slightly
supersonic'', the function $f(\cdot)$ can become very small $f
(v_{\mathrm{int-asym}}) \propto (v_{\mathrm{int-asym}}-1)^{3/2}$, and thus
also the critical frequency $\omega_c$ at which particle production is
cut off. If this happens at frequencies comparable with $\kappa$ the
whole Planckian spectrum will be truncated and distorted. Therefore,
to recover a Planckian spectrum of particles at the external
asymptotic region one needs to have a noticeable supersonic region.

As with subsonic dispersion, the existence of a notion of rainbow
geometry makes modes of different frequency experience different
surface gravities. One can think of the distortion of the Planckian
spectrum as having a running $\kappa(\omega)$ that, for these
configurations, interpolates between its low energy, or geometric
value, $\kappa(\omega=0)=\kappa_0$ and a value of zero for $\omega \to
\omega_c$. This transition is remarkably sudden for the smooth
profiles analyzed. If $\kappa_0 \ll \omega_c$, then $\kappa(\omega)$
will stay constant and equal to $\kappa_0$ throughout the relevant
part of the spectrum reproducing Hawking's result. However, in general 
theoretical terms one can say that the spectrum takes into account the 
characteristic of the profile deep inside the supersonic region (the 
analogue of the black hole interior). The existence of superluminal 
modes makes it possible to obtain information from inside the 
(low energy) horizon.

In these analyses based on stationary configurations, the quantum field
was always assumed to be in the \emph{in}-vacuum state. {It is this vacuum
state which has thermal properties in terms of
\emph{out}-observers. However, it is interesting to note that the dispersive 
character of the theory allows in principle to select for these
systems (albeit most probably through fine-tuning) a different, perfectly 
regular state, which is empty of both incoming and outgoing particles in 
the external asymptotic region \citep{Barcelo:2010bq}. 
This state can be interpreted as the
regular generalization to a dispersive theory of the Boulware state
for a field in a stationary black hole (let us recall that this state
is not regular at the horizons). As such it will be characterized by a very large vacuum polarization which will be cut-off at the characteristic high energy scale $K$. In this sense the realization of such state would require a very fine tuned setup and it will be energetically disfavoured. 
Nonetheless, this implies that, in principle, one
could set up a semi-classically stable acoustic black hole geometry in this setting. Finally, let us mention that another important result of the analysis
carried on in \citet{Macher:2009tw}, is that stationary white holes do also Hawking
radiate and in a very similar way to black holes.}

These analyses have been repeated for the specific case of the
fluctuations of a BEC \citep{Macher:2009nz} with identical qualitative
results. The reasons for this is that, although the Bogoliubov--de
Gennes system of equations is different from the modified scalar field
equation analyzed in previous papers, they share the same quartic
dispersion relation. Apart from the more formal treatments of BECs,
\citet{Carusotto:2008ep, Carusotto:2009re} reported
the numerical observation of the Hawking effect in simulations in
which a black-hole horizon has been created dynamically from an
initially homogeneous flow. The observation of the effect has been
through the calculation of two-point correlation functions (see
Sect.~\ref{S:correlations} below). These simulations strongly
suggest that any non-quasi-static formation of a horizon would give
place to Hawking radiation.

It is particularly interesting to note that this recovery of the
standard result is not always guaranteed in the presence of
superluminal dispersion relations. 
\citet{Corley:1998rk} in fact discovered a very peculiar type
of instability due to such superluminal dispersion in the presence of
black holes with inner horizons. The net result of the investigation
carried out in \citet{Corley:1998rk} is that the compact ergo-region
characterizing such configurations is unstable to self-amplifying
Hawking radiation. The presence of such an instability was also
identified in the dynamical analysis carried on in \citet{Garay:1999sk,
  Garay:2000jj, Barcelo:2006yi} where Bose--Einstein condensate
analogue black holes were considered. 

These configurations, \emph{in addition to} a steady Hawking flux,
produce a self-amplified particle emission; from this feature arises
their name ``black-hole lasers''. In the recent analyses
in \citet{Coutant:2009cu, Finazzi:2010nc} it has been shown that the
complete set of modes to be taken into account in these configurations
is composed of a continuous sector with real frequencies, plus a
discrete sector with complex frequencies of positive imaginary
part. These discrete frequencies encode the unstable behaviour of
these configurations, and are generated as resonant modes inside the
supersonic cavity encompassed between the two horizons.
{One lesson one could take from these analyses is that, under 
superluminal dispersion, regular black holes {endowed with 
inner-outer horizon pairs} might be {generically} unstable.}

Finally, let us mention that a completely analytical derivation of the Hawking temperature associated to acoustic horizons was recently carried out within the so-called tunneling method~\citep{Parikh:1999mf} in \citet{DelPorro:2024tuw}. While less informative about the full spectrum retrieved far away from the horizon, this analysis has the advantage of being equally applicable to subsonic and supersonic flows as well as to superluminal or subluminal dispersion relations.

\subsubsection{General conditions for Hawking radiation}

Is it possible to reduce the rather complex phenomenology just described to a
few basic assumptions that must be satisfied in order 
to recover Hawking radiation in the presence of Lorentz-violating
dispersion relations? A tentative answer is given
in \citet{Unruh:2004zk}, where the robustness of the Hawking result is
considered for general modified (subluminal as well as superluminal)
dispersion relations. The authors of \citet{Unruh:2004zk} assume that the
geometrical optics approximation breaks down only in the proximity of
the event horizon (which is equivalent to saying that the particle
production happens only in such a region). Here, the would-be
trans-Planckian modes are converted into sub-Planckian ones. Then, they
try to identify the minimal set of assumptions that guarantees that such
``converted modes'' are generated in their ground states (with respect
to a freely falling observer), as this is a well-known condition in
order to recover Hawking's result. They end up identifying three basic
assumptions that guarantee such emergence of modes in the ground state
at the horizon. 
\begin{itemize}
\item 
First, the preferred frame selected by the breakdown of
Lorentz invariance must be the freely falling one instead of the rest
frame of the static observer at infinity (which coincides in this limit
with the laboratory observer). 
\item
Second, the Planckian excitations must
start off in the ground state with respect to freely falling observers.
\item
Finally, they must evolve in an adiabatic way (i.e., the Planck
dynamics must be much faster than the external sub-Planckian dynamics).
\end{itemize}
Of course, although several systems can be found in which such
conditions hold, it is also possible to show \citep{Unruh:2004zk} that
realistic situations in which at least one of these assumptions is
violated can be imagined. Hence, it is still an open question whether
real black hole physics does indeed satisfy such conditions, and whether it is
therefore robust against modifications induced by the violation of Lorentz
invariance.

\subsubsection{Origin of the Hawking quanta} 

There is a point of view (not universally shared within the community)
that asserts that the trans-Planckian problem also makes it clear that
the ray-optics limit cannot be the whole story behind Hawking
radiation. Indeed, it is precisely the ray optics approximation that
leads to the trans-Planckian problem. Presumably, once one goes beyond
ray optics, to the wave optics limit, it will be the region within a
wavelength or so of the horizon (possibly the region between the
horizon and the unstable circular photon orbit) that proves to be
quantum-mechanically unstable and will ultimately be the ``origin'' of
the Hawking photons. If this picture is correct, then the black-hole
particle production is a low-frequency and low-wavenumber
process. See, for instance, \citet{Schutzhold:2008tx, Unruh:2007zz,
Unruh:2008zz, Giddings:2015uzr, Dey:2017yez, Dey:2019ugf} or \citet{Barbado:2016nfy} for an alternative perspective.

\subsubsection{Which surface gravity?} 
\label{S:which-kappa}

One issue that has become increasingly important, particularly in view
of recent experimental advances, is the question of exactly which
particular definition of surface gravity is the appropriate one for
controlling the temperature of the Hawking radiation. In standard
general relativity with Killing horizons there is no ambiguity, but
there is already considerable maneuvering room once one goes to
evolving horizons in general relativity, and even more ambiguities
once one adopts modified dispersion relations (as is very common in
analogue spacetimes).

Already at the level of time-dependent systems in standard general
relativity there are two reasonably natural definitions of surface
gravity, one in terms of the inaffinity of null geodesics skimming
along the event horizon, and another in terms of the peeling
properties of those null geodesics that escape the black hole to reach
future null infinity. It is this latter definition that is relevant
for Hawking-like fluxes from non-stationary systems (e.g., evaporating
black holes) and in such systems it never coincides with the
inaffinity-based definition of the surface gravity except possibly at
asymptotic future-timelike infinity $i^+$. Early comments along these
lines can be found in \citet{Brout:1995wp, Brout:1995rd}; more recently
this point was highlighted in \citet{Barcelo:2010pj, Barcelo:2010xk}.

Regarding analogue models of gravity, the conclusions do not
change when working in the hydrodynamic regime (where there is a
strict analogy with GR). This point was implicitly made
in \citet{Barcelo:2004wz, Barcelo:2010xk} and clearly stressed
in \citet{Macher:2009nz}. If we now add modified dispersion relations,
there are additional levels of complication coming from the
distinction between ``group velocity horizons'', and ``phase velocity
horizons'', and the fact that null geodesics have to be replaced by
modified characteristic curves. 
The presence of dispersion also makes explicit that the crucial notion
underlying Hawking emission is the ``peeling'' properties of null ray
characteristics. For instance, the relevant ``peeling'' surface
gravity for determining Hawking fluxes has to be determined locally,
in the vicinity of the Killing horizon, and over a finite frequency
range. (See for instance \citealt{Unruh:2003ss, Philbin:2007ji,
  Philbin:2007jj, Rousseaux:2007is, Belgiorno:2010xxx,
  Weinfurtner:2010nu, Macher:2009nz} for some discussion of this and
related issues.) This ``surface gravity'' is actually an emergent
quantity coming from averaging the naive surface gravity (the slope of
the $c$--$v$ profile) on a finite region around the would-be Killing
horizon associated with the acoustic geometry \citep{Finazzi-Parentani-2010}. 
This quantity is highly idealized and, on a case by case basis, 
it needs to be carefully compared with the actual experimental results.

\subsubsection{How to detect Hawking radiation: Correlations} 
\label{S:correlations}

While the robustness of Hawking radiation against UV violations of the
acoustic Lorentz invariance seems a well-established feature by now (at
least in static or stationary geometries), its strength is indubitably
a main concern for a future detection of this effect in a
laboratory. As we have seen, the Hawking temperature in acoustic
systems is simply related to the gradient of the flow velocity at the
horizon (see Eq.~(\ref{acbh:surfgrav})). This gradient cannot be
made arbitrarily large and, for the hydrodynamic approximation to hold,
one actually needs it to be at least a few times the typical coherence
length (e.g., the healing length for a BEC) of the superfluid used for
the experiment. This implies that in a cold system, with low speed of
sound, like a BEC, the expected power loss due to the Hawking emission
could be estimated to be on the order of $P \approx 10^{-48}
\mathrm{\ W}$ (see, e.g., \citealt{Barcelo:2001ca}): arguably too faint to be
detectable above the thermal phonon background due to the finite
temperature of the condensate (alternatively, one can see that the
Hawking temperature is generally below the typical temperature of the
BEC, albeit they are comparable and both in the nano-Kelvin range).
Despite this, it is still possible that a detection of the spontaneous
quantum particle creation can be obtained via some other feature
rather than the spectrum of the Hawking flux. A remarkable possibility
in this sense is offered by the fact that vacuum particle creation
leads generically to a spectrum, which is (almost) Planckian but not
thermal (in the sense that all the higher order field correlators are
trivial combinations of the two-point one).\footnote{Note
  however, that for a real black hole and Lorentz-invariant physics,
  the spectrum observed at infinity is indistinguishable from thermal
  given that no correlation measurement is allowed, since the Hawking
  partners are space-like separated across the horizon. This fact is
  indeed the root of the \emph{information paradox.}}

Indeed, particles created by the mode mixing (Bogoliubov) mechanism
are generically in a squeezed state (in the sense that the \emph{in}
vacuum appears as a squeezed state when expressed in terms of the
\emph{out} vacuum; \citealt{Jacobson:2003vx}) and such a state can be
distinguished from a real thermal one exactly by the nontrivial
structure of its correlators. This discrimination mechanism was
suggested a decade ago in the context of dynamical Casimir effect
explanations of Sonoluminescence \citep{Belgiorno:1999ha}, and later
envisaged for analogue black holes in \citet{Barcelo:2001ca}, but was
finally investigated and fully exploited only recently in a stream of
papers focussed on the BEC set up \citep{Balbinot:2007de,
  Carusotto:2008ep, Macher:2009nz, Recati:2009ya, Balbinot:2010my,
  Carusotto:2009re, Fagnocchi:2010zz, Parentani:2010bn, Prain:2010zq,
  Schutzhold:2010ig, 2010arXiv1005.2645U}. 
  
The outcome of such
investigations (carried out taking into account the full Bogoliubov
spectrum) is quite remarkable as it implies that indeed, while the
Hawking flux is generically out-powered by the condensate intrinsic
thermal bath, it is, in principle, possible to have a clear cut
signature of the Hawking effect by looking at the density-density
correlator for phonons on both sides of the acoustic horizon. In fact,
the latter will show a definite structure totally absent when the flow
is always subsonic or always supersonic. Even more remarkably, it was
shown, both via numerical simulations as well as via a detailed
analytical investigation, that a realistic finite temperature
background does not spoil the long distance correlations which are
intrinsic to the Hawking effect (and, indeed, for non-excessively large
condensate temperatures the correlations can be amplified).

The current situation is that \blue{correlation} patterns are actually 
being successfully used in analysing the generation of Hawking-like 
fluxes in several experiments (see next section). 
Correlator analyses can be applied to a wide class of analogue
systems, for example, it has been applied to analogue black holes
based on cold atoms in ion rings \citep{Horstmann:2009yh}, or extended
to the study of the particle creation in time-varying external fields
(dynamical Casimir effect) in Bose--Einstein condensates (where the
time-varying quantity is the scattering length, \blue{adjusted} via a Feshbach
resonance) \citep{Carusotto:2009re}.

\subsubsection{Open issues} 

In spite of the remarkable insight given by the models discussed above
(based on modified dispersion relations) it is not possible to consider
them fully satisfactory in addressing the trans-Planckian problem. In
particular, {in the subluminal case} it was soon recognised 
\citep{Corley:1998ef,Jacobson:1999zk}
that in this framework it is not possible to explain the origin of the
short wavelength incoming modes, which are ``progenitors'' of the
outgoing modes after bouncing off in the proximity of the horizon. For
example, in the Unruh model~(\ref{eq:Un}), one can see that if one keeps
tracking a ``progenitor'' incoming mode back in time, then its group velocity
(in the co-moving frame) drops to zero as its frequency becomes more and more
blue shifted (up to arbitrarily large values), just the situation one was
trying to avoid. This is tantamount to saying that the trans-Planckian problem
has been moved from the region near the horizon out to the region near
infinity. In the Corley--Jacobson model~(\ref{eq:CJ}) this 
unphysical behaviour is removed thanks to the presence of the physical
cutoff $K$. However, it is still true that in tracking the incoming modes
back in time one finally sees a wave packet so blue shifted that
$|k|=K$. At this point one can no longer trust the dispersion
relation~(\ref{eq:CJ}) (which in realistic analogue models is
emergent and not fundamental anyway), and hence the model has no predictive power regarding the ultimate origin of the relevant incoming modes.

In the superluminal case the situation is that the high frequency modes {originate in the region} where the classical singularity of a GR black hole would be localized. In this case, the {resolution} of the trans-Planckian problem is intertwined with the very problem of regularization of the internal black hole singularity.

These conclusions regarding the impossibility of clearly predicting the
origin at early times of the modes ultimately to be converted into
Hawking radiation are not specific to the particular dispersion
relations~(\ref{eq:Un}), (\ref{eq:CJ}) {or superluminal} one is using. In fact, the
Killing frequency is conserved on a static background; thus, the
incoming modes must have the same frequency as the outgoing
ones. Hence, in the case of strictly Lorentz invariant dispersion
relations there can be no mode-mixing and particle creation. This is
why one actually has to assume that the WKB approximation fails in the
proximity of the horizon and that the modes are there in the vacuum
state for the co-moving observer. In this sense, the need for these
assumptions can be interpreted as evidence that these models are not
yet fully capable of solving the trans-Planckian problem. Ultimately,
these issues underpin the analysis by Sch\"utzhold and Unruh regarding
the spatial ``origin'' of the Hawking quanta \citep{Schutzhold:2008tx,
  Unruh:2007zz, Unruh:2008zz}.

Finally, an intriguing feature of the acoustic analogies is the possibility of non-thermal particle creation even in the absence of acoustic horizons. 
This phenomenon has been observed in scenarios involving everywhere subcritical flows with subsonic dispersion relations~\citep{Michel:2014zsa,Michel:2015aga,Coutant:2016vsf}, as well as in scenarios with everywhere supercritical flows and supersonic dispersion relations~\citep{Finazzi:2011jd}. Notably, the first scenario is highly relevant because shallow water wave experiments primarily involve ``near-critical" flows~\citep{Weinfurtner:2010nu,Weinfurtner:2013zfa,Euve:2014aga,Euve:2015vml,Euve:2021mnj}, which have nonetheless demonstrated some amount of particle production.

Focusing on the (subcritical flow)-(subsonic dispersion) setting, the underlying mechanism becomes apparent: for flows sufficiently close to forming a horizon, and for modes with sufficiently high energy, the subsonic dispersion relation permits a turning point in the Hawking quanta trajectory. Such modes effectively experience an effective horizon before the flow ceases to increase its speed, thus remaining subcritical. Consequently, these modes are unaffected by the absence of a real acoustic horizon and undergo mode conversion at their own effective horizon. This mode conversion results in a spectrum of excitations that is generally non-thermal, with deviations from thermality increasing as the flow moves further from forming a horizon. 
For flows further and further from criticality, only modes with higher and higher energy (smaller and smaller group velocity) can be excited, but these would pertain to a regime where the dispersion relation untrustworthy (e.g.~$\omega^2$ becomes negative), so for all practical purposes the effect would rapidly shut down. 

Switching from purely subcritical flow to purely supercritical flows and adopting supersonic dispersion relations leads to a similar explanation. This correspondence in behaviour is a manifestation of the duality between (supercritical flow)-(supersonic dispersion) and (subcritical flow)-(subsonic dispersion) settings noted in previous analogue gravity studies. 

This discussion could also be rephrased in terms of the frequency-dependent ``rainbow'' metrics we have previously discussed, with frequency-dependent ``effective horizons'' --- effectively a Finslerian structure.  For another more quantitative (albeit idealized) analysis of these cases, for both subsonic and supersonic dispersion relations via the tunneling method, we refer to the recent analysis presented in~\citep{DelPorro:2024tuw}. 

Furthermore, in spite of the previous discussion, it is important to stress that mode conversion by classical back scattering, especially of the low energy modes, is also present and, depending on the setup, possibly dominant (see references above).

\subsubsection{Solid state and lattice models} 

It was to overcome this type of issue that alternative ways of
introducing an ultra-violet cutoff due to the microphysics were
considered \citep{Reznik:1997xt, Reznik:1997ag, Corley:1998ef}. In
particular, in \citet{Reznik:1997ag} the transparency of the refractive
medium at high frequencies has been used to introduce an effective
cutoff for the modes involved in Hawking radiation in a classical
refractive index analogue model (see Sect.~\ref{sec:cri}). In this
model an event horizon for the electromagnetic field modes can be
simulated by a surface of singular electric and magnetic
permeabilities. This would be enough to recover Hawking radiation but
it would imply the unphysical assumption of a refractive index which
is valid at any frequency. However, it was shown
in \citet{Reznik:1997ag} that the Hawking result can be recovered even
in the case of a dispersive medium, which becomes transparent above
some fixed frequency $K$ (which we can imagine as the plasma frequency
of the medium); the only (crucial) assumption being again that the
``trans-Planckian'' modes with $k>K$ are in their ground state near
the horizon.

An alternative avenue was considered in \citet{Corley:1998ef}.
There, a lattice description of the background was used for imposing a
cutoff in a more physical way with respect to the continuum dispersive
models previously considered. In such a discretised spacetime, the field
takes values only at the lattice points, and wavevectors are identified
modulo $2\pi/\ell$ where $\ell$ is the lattice characteristic spacing;
correspondingly one obtains a sinusoidal dispersion relation for the
propagating modes. Hence, the problem of recovering a smooth evolution
of incoming modes to outgoing ones is resolved by the
intrinsically-regularised behaviour of the wave vectors
field. In \citet{Corley:1998ef} the authors explicitly considered the
Hawking process for a discretised version of a scalar field, where the
lattice is associated with the free-fall coordinate system (taken as
the preferred system). With such a choice, it is possible to preserve
a discrete lattice spacing. Furthermore, the requirement of a fixed
short-distance cutoff leads to the choice of a lattice spacing
constant at infinity, and that the lattice points are at rest at
infinity and fall freely into the black hole.%
\footnote{\citet{Corley:1998ef} also considered the case
  of a lattice with proper distance spacing constant in time but this has
  the problem that the proper spacing of the lattice goes to zero at
  spatial infinity, and hence there is no fixed short-distance cutoff.} 
In this case, the lattice spacing grows in time and the lattice points
spread in space as they fall toward the horizon. However, this time
dependence of the lattice points is found to be of order $1/\kappa$, and
hence unnoticeable to long-wavelength modes and relevant only for those
with wavelengths on the order of the lattice spacing. The net result is
that, on such a lattice, long wavelength outgoing modes are seen to 
originate from short wavelength incoming modes via a process analogous
to the Bloch oscillations of accelerated electrons in
crystals \citep{Corley:1998ef,Jacobson:1999zk}.

\subsubsection{Analogue spacetimes as background gestalt}

In addition, among the many papers using analogue spacetimes as part of their
background mindset when addressing these issues we mention some historically relevant papers: 

\begin{itemize}

\item ``Top down'' calculations of Hawking radiation starting from some
  idealised model of quantum gravity \citep{Amati:1996pu, Jevicki:2002fq,
    Nikolic:2003xs, Nikolic:2004wu}. 

\item ``Bottom up'' calculations of Hawking radiation starting from curved
  space quantum field theory \citep{Barrabes:1998iw, Barrabes:2000fr,
    Canfora:2003zx, Canfora:2003cc, Casher:1996ct, Englert:1994sm,
    Englert:1994qe, Ford:1997zb, Massar:1994iy, Massar:1996tx,
    Massar:1995im, Massar:1999wg, Parentani:1999qv, Rosu:1989cb}.

\item Trans-Hawking versions of Hawking radiation, either as reformulations
  of the physics, or as alternative scenarios \citep{Belgiorno:2002iv,
    Casadio:2001hp, Frolov:1995pt, Frolov:1995qp, Glass:1999yt, Hambli:1995pp,
    Helfer:2000wh, Helfer:2003va, Helfer:2004jx, Leonhardt:2003mp,
    Oppenheim:2002kx, Padmanabhan:2003gd, Rosu:1999sa, Russo:1996qa,
    Salehi:1994em, Salehi:2003pj, Schutzhold:2000ju, Schutzhold:2000jn,
    Paddy2, Stephens:2001sd, DelPorro:2023lbv}. 
    
\item Black hole entropy viewed in the light of analogue spacetimes \citep{Damour:2004kw}. 
  
\item Hawking radiation interpreted as a statement about particles traveling
  along complex spacetime trajectories \citep{Paddy3, Paddy2, Paddy1}. 

\end{itemize}
Of course the field has now developed much further, see the various discussions below.

\subsection{Dynamical stability of horizons}
\label{S:Horizon-stability}

Although the two issues are very closely related, as we will soon see,
we have to carefully distinguish between the stability analysis of the
modes of a linear field theory (with or without modified dispersion
relations -- MDR) over a fixed background, and the stability analysis
of the background itself.

\subsubsection{Classical stability of the background (no MDR)}

Let us consider a three-dimensional irrotational and inviscid fluid
system with a stationary sink-type of flow -- a ``draining bathtub''
flow -- (see Figs.~\ref{F:cascade} and \ref{F:ondas}). The details
of the configuration are not important for the following discussion,
only the fact that there is a spherically-symmetric fluid flow
accelerating towards a central sink, that sink being surrounded by a
sphere acting as a sonic horizon. Then, as we have discussed in
Sect.~\ref{S:simple}, linearizing the Euler and continuity equations
leads to a massless scalar field theory over a black-hole--like
spacetime. (We are assuming that the hydrodynamic regime remains valid
up to arbitrarily-short length scales; for instance, we are neglecting
the existence of MDR.) To be specific, let us choose the geometry of
the canonical acoustic black-hole spacetime described
in \citet{Visser:1997ux}:
\begin{eqnarray} 
\d s^2=-c^2\left(1-{r_0^4 \over r^4}\right)\d\tau^2
+\left(1-{r_0^4 \over r^4}\right)^{-1}\d r^2
+r^2\left(\d\theta^2+\sin^2\theta \;\d\varphi^2 \right). 
\end{eqnarray} %
In this expression we have used the Schwarzschild time coordinate $\tau$
instead of the lab time $t$; $c$ is constant. If we expand the field in
spherical harmonics,
\begin{eqnarray} 
\phi_{lm}(\tau,r,\theta,\varphi) \equiv e^{-i\omega\tau} \;
{\chi_{lm}(r) \over r} \; Y_{lm}(\theta, \varphi), 
\end{eqnarray}
we obtain the following equation for the radial part of the field: 
\begin{eqnarray} 
{\omega^2 \over c^2} \; \chi = \left(-{\d^2 \over
\d{r^*}^2} + V_l(r) \right)\chi;
\end{eqnarray}
where
\begin{eqnarray} 
V_l(r)=\left(1- {r^4_0 \over r^4}\right)
\left[{l(l+1) \over r^2}+ {4r_0^4 \over r^6}\right]. 
\end{eqnarray}
Here 
\begin{equation} 
r^* \equiv r-{r_0 \over 4}\;\left\{\ln \left[ {(r+r_0) \over (r-r_0)}\right]+ 2 \arctan \left({r \over r_0}\right) \right\} 
\end{equation} 
is a ``tortoise'' coordinate. 

In a normal mode analysis one requires boundary conditions such that the
field is regular everywhere, even at infinity. However, if one is
analysing the solutions of the linear field theory as a way of probing
the stability of the background configuration, one can consider less
restrictive boundary conditions. For instance, one can consider the
typical boundary conditions that lead to quasinormal modes: These modes
are defined to be purely out-going at infinity and purely in-going at the
horizon; but one does not require, for example, the modes to be
normalizable. The quasinormal modes associated with this sink
configuration have been analysed in \citet{Berti:2004ju}.
The results found are qualitatively similar to those in the classical linear
stability analysis of the Schwarzschild black hole in general
relativity \citep{Vishveshwara:1970, Vishveshwara:1970cc, Regge:1957,
  Zerilli:1970se, Moncrief:1974}. Of course, the gravitational field
in general relativity has two dynamical degrees of freedom -- those
associated with gravitational waves -- that have to be analysed
separately; these are the ``axial'' and ``polar'' perturbations. In
contrast, in the present situation we only have scalar perturbations.
Nevertheless, the potentials associated with ``axial'' and ``polar''
perturbations of Schwarzschild spacetime, and that associated with
scalar perturbations of the canonical acoustic black hole, produce
qualitatively the same behaviour: There is a series of \emph{damped}
quasinormal modes -- proving the linear stability of the system -- with
higher and higher damping rates.
Since then, this particular subject has produced a steady flow of research efforts. 
Here we provide just a (incomplete) sample:
\citet{Cardoso:2004fi,Lepe:2004kv,Daghigh:2005ph,Okuzumi:2007hf, Xi:2007yb,Dolan:2010zza,Daghigh:2014mwa,Benone:2015jda,Oliveira:2018zen, Torres:2019sbr,Torres:2020tzs,Patrick:2021oqk,Campos:2023zmg,Liu:2024vde}.

An important point we have to highlight here is that, although in the
linear regime the dynamical behaviour of the acoustic system is similar
to general relativity, this is no longer true once one enters the
nonlinear regime. The underlying nonlinear equations in the two cases
are very different. The differences are so profound, that in the
general case of acoustic geometries constructed from compressible
fluids, there exist sets of perturbations that, independent of how
small they are initially, can lead to the development of shocks, a
situation completely absent in vacuum general relativity.

\subsubsection{Semiclassical stability of the background (no MDR)}

Now, given an approximately stationary, and at the very least
metastable, classical black-hole-like configuration, a standard quantum
mode analysis leads to the existence of Hawking radiation in the form of
phonon emission. This shows, among other things, that quantum
corrections to the classical behaviour of the system must make the
configuration with a sonic horizon dynamically unstable against Hawking
emission. As a consequence, in any system (analogue or general
relativistic) with quantum fluctuations that maintain strict adherence
to the equivalence principle (no MDR), it must then be impossible to
create an isolated \emph{truly stationary} horizon by merely setting
up external initial conditions and letting the system evolve by
itself. However, in an analogue system a truly stationary horizon can
be set up by providing an external power source to stabilise it
against Hawking emission. Once one compensates, by manipulating
external forces, for the backreaction effects that in a physical
general relativity scenario cause the horizon to shrink or evaporate,
one would be able to produce, in principle, an analogue system
exhibiting precisely a stationary horizon and a stationary Hawking
flux.

Let us describe what happens when one takes into account the existence
of MDR. Once again, a wonderful physical system that has MDR
explicitly incorporated in its description is the Bose--Einstein
condensate. The macroscopic wave function of the BEC behaves as a
classical irrotational fluid but with some deviations when short
length scales become involved. (For length scales on the order of, or
shorter than, the healing length.) What are the effects of the MDR on
the dynamical stability of a black-hole-like configuration in a BEC?
The stability of a sink configuration in a BEC has been analysed
in \citet{Garay:1999sk, Garay:2000jj} but taking the flow to be
effectively one-dimensional. What these authors found is that these
configurations are dynamically unstable: There are modes satisfying
the appropriate boundary conditions such that the imaginary parts of
their associated frequencies are positive. These instabilities are
associated basically with the bound states inside the black hole. The
dynamical tendency of the system to evolve is suggestively similar to
that in the standard evaporation process of a black hole in
semiclassical general relativity.

\subsubsection{Classical stability of the background (MDR in BECs)}

Before continuing with the discussion of the instability of configurations
with horizons, and in order not to cause confusion between the different
wording used when talking about the physics of BECs (and similar systems) and the emergent
gravitational notions on them, let us write down a quite loose but
useful translation dictionary:

\begin{itemize}
\item The ``classical'' or macroscopic wave function of the BEC
  represents the classical spacetime of GR, but only when probed at
  long-enough wavelengths such that it behaves as pure hydrodynamics.
\item The ``classical'' long-wavelength perturbations to a background
 solution of the Gross--Pitaev\-skii equation correspond to classical
 gravitational waves in GR. Of course, this analogy does not imply that
 these are spin 2 waves; it only points out that the perturbations
 are made from the same ``substance'' as the background configuration
 itself.
\item The macroscopic wave function of the BEC, without the restriction
 of being probed only at long wavelengths, corresponds to some sort of
 semiclassical vacuum gravity. Its ``classical'' behaviour (in the sense
 that does not involve any probability notion) is already taking into
 account, in the form of MDR, its underlying quantum origin.
\item The Bogoliubov quantum quasiparticles over the ``classical'' wave
 function correspond to a further step away from semiclassical gravity in
 that they are analogous to the existence of quantum gravitons over a
 (semiclassical) background spacetime.
\end{itemize} 
At this point we would like to remark, once again, that the analysis
based on the evolution of a BEC has to be used with care. For example,
they cannot directly serve to shed light on what happens in the final
stages of the evaporation of a black hole, as the BEC does not fulfil,
at any regime, the Einstein equations. Summarizing: 
\begin{itemize}
\item If the perturbations to the BEC background configuration have
 ``classical seeds'' (that is, are describable by the linearized
 Gross--Pitaevskii equation alone), then, one will have ``classical''
 instabilities.
\item If the perturbations have ``quantum seeds'' (that is, are
 described by the Bogoliubov equations), then, one will have ``quantum''
 instabilities.
\end{itemize}
%

\subsubsection{Black holes, white holes, and rings}

\begin{figure}[htpb]
   \centerline{\includegraphics[width=0.75\textwidth]{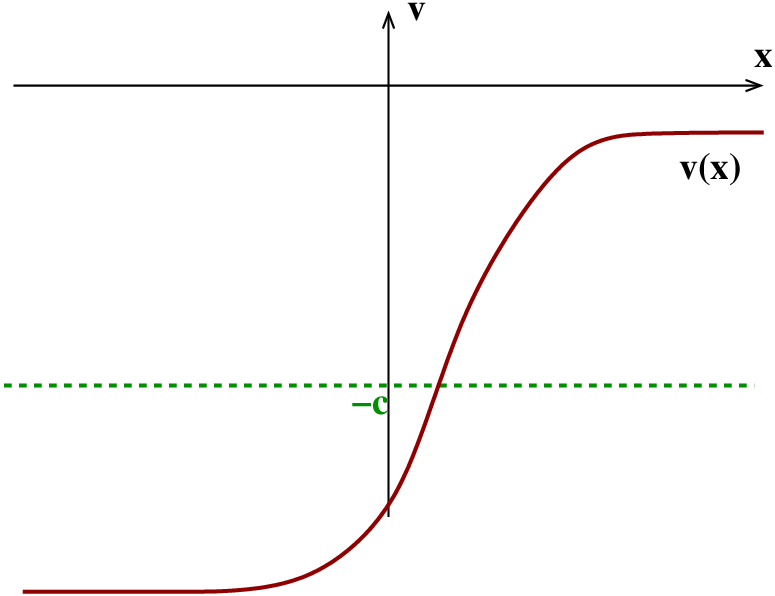}}
\caption[Black hole velocity profile]{One-dimensional
velocity profile with a black-hole horizon.}
\label{F:bh-flow}
\end{figure}

\begin{figure}[htpb]
   \centerline{\includegraphics[width=0.75\textwidth]{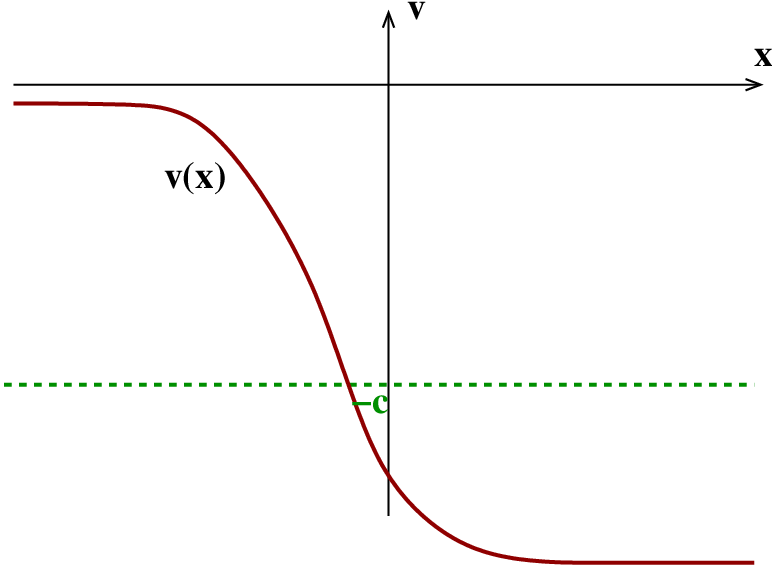}}
\caption[White hole velocity profile]{One-dimensional
velocity profile with a white-hole horizon.}
\label{F:wh-flow}
\end{figure}

\begin{figure}[htpb]
   \centerline{\includegraphics[width=0.75\textwidth]{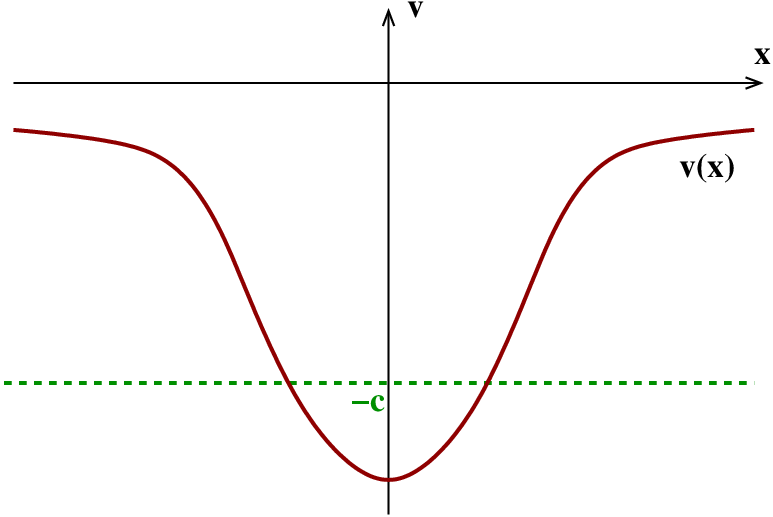}}
\caption[Black hole and white hole velocity profile]{One-dimensional
velocity profile with a black-hole horizon and a white-hole horizon.}
\label{F:bh-wh-flow}
\end{figure}

\begin{figure}[htpb]
   \centerline{\includegraphics[width=0.75\textwidth]{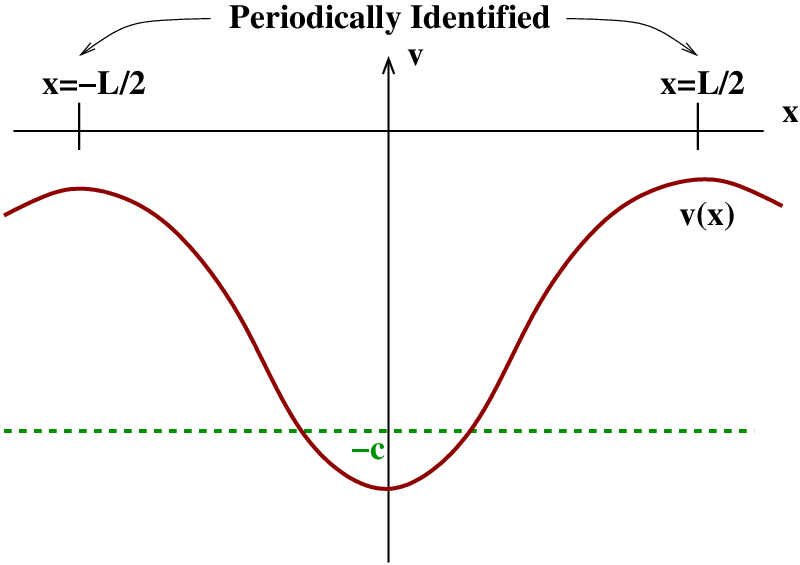}}
\caption[Black hole and white hole horizons in a ring]{One-dimensional
velocity profile in a ring; the fluid flow exhibits two sonic horizons,
one of black hole type and the other of white-hole type.}
\label{F:bh-wh-ring-flow}
\end{figure}

In the light of the acoustic analogies it is natural to ask whether
there are other geometric configurations with horizons of interest,
besides the sink type of configurations (these are the most similar to
the standard description of black holes in general relativity, but
probably not the simplest in terms of realizability in a real
laboratory; for an entire catalogue of them see \citealt{Barcelo:2004wz}).
Here, let us specifically 
mention four effectively one-dimensional configurations: a black hole
with two asymptotic regions, a white hole with two asymptotic regions,
a black-hole--white-hole in a straight line and the same in a ring (see
Figs.~\ref{F:bh-flow}, \ref{F:wh-flow}, \ref{F:bh-wh-flow} and
\ref{F:bh-wh-ring-flow}, respectively).

There are several classical instability analyses of these types of
configurations in the literature \citep{Garay:1999sk, Garay:2000jj,
  Leonhardt:2002yu, Barcelo:2006yi, Coutant:2009cu,
  Finazzi:2010nc}. In these analyses one looks for the presence or
absence of modes with a positive-imaginary-part eigenfrequency, under
certain appropriate boundary conditions. The boundary condition in
each asymptotic region can be described as outgoing, as in
quasi-normal modes, or as convergent, meaning that at a particular
instant of time the mode is exponentially damped towards the
asymptotic region. Let us mention that in Lorentz invariant theory
these two types of conditions are not independent: any unstable mode
is at the same time both convergent and outgoing. However, in general,
in dispersive theories, once the frequency is extended to the complex
plane, these two types of conditions become, at least in principle,
independent.

Under outgoing and convergent boundary conditions in both asymptotic
regions, in \citet{Barcelo:2006yi} it was concluded that there are no
instabilities in any of the straight line (non-ring)
configurations. If one relaxed the convergence condition in the
downstream asymptotic region, (the region that substitutes the unknown
internal region, and so the region that might require a different
treatment for more realistic black hole configurations), then the
black hole is still stable, while the white hole acquires a continuous
region of instability, and the black-hole--white-hole configuration
shows up as a discrete set of unstable modes. The white-hole instability
was previously identified in \citet{Leonhardt:2002yu}. Let us mention
here that the stable black-hole configuration has been also analyzed
in terms of stable or quasinormal modes in \citet{Barcelo:2007ru}. It
was found that, although the particular configurations analyzed
(containing idealised step-like discontinuities in the flow) did not
posses quasinormal modes in the acoustic approximation, the
introduction of dispersion produced a continuous set of quasinormal
modes at trans-Planckian frequencies. 

Continuing with the analysis of instabilities, in contrast
to \citet{Barcelo:2006yi}, the more recent analysis
in \citet{Coutant:2009cu, Finazzi:2010nc} consider only convergent
boundary conditions in both asymptotic regions. They argue that the
ingoing contributions that these modes sometimes have always
correspond to waves that do not carry energy, so that they have to be
kept in the analysis, as their ingoing character should not be
interpreted as an externally-provoked
instability\footnote{Personal communication by R.~Parentani}. 
{It appears that the appropriate boundary condition for
instability analysis under dispersion are just the convergence
conditions, as in non-dispersive theories.}

Under these convergence conditions, \citet{Coutant:2009cu} and 
\citet{Finazzi:2010nc} show that the previously-considered black-hole and
white-hole configurations in BECs are stable. (Let us remark that this
does not mean that configurations with a more complicated internal
region need be stable.) However, black-hole--white-hole configurations
do show a discrete spectrum of instabilities. In these papers, one can
find a detailed analysis of the strength of these instabilities,
depending on the form and size of the intermediate supersonic
region. For instance, it is necessary that the supersonic region
acquire a minimum size so that the first unstable mode appears. (This
feature was also observed in \citealt{Barcelo:2006yi}.) When the previous
mode analysis is used in the context of a quantum field theory, as we
mention in Sect.~\ref{S:Hawking-radiation}, one is led to the
conclusion that black-hole--white-hole configurations emit particles
in a self-amplified (or runaway)
manner \citep{Corley:1998rk, Coutant:2009cu, Finazzi:2010nc}. Although
related to Hawking's process, this phenomenon has a quite different
nature. For example, there is no temperature associated with it. 

When the black-hole--white-hole configuration is compactified in a ring,
it is found that there are regions of stability and instability,
depending on the parameters characterizing the
configuration \citep{Garay:1999sk, Garay:2000jj}
{(see also \citep{Jain:2007,Yatsuta:2020})}. We suspect that the
stability regions appear because of specific periodic arrangements of
the modes around the ring. Among other reasons, these arrangements are
interesting because they could be easier to create in the laboratory
with current technology, and their instabilities easier to detect than
Hawking radiation itself.

To conclude this subsection, we would like to highlight that there is
still much to be learned by studying the different levels of
description of an analogue system, and how they influence the
stability or instability of configurations with horizons.

 \subsection{Super-radiance} 
\label{S:Super-radiance}

Another phenomenon that has been (and is being) analysed from the
analogue gravity perspective is super-radiance. The rotational kinetic
energy accumulated in a rotating black hole can be extracted from it by
scattering into it waves of sufficiently low frequency and high angular
momentum. In general, in order that the wave can extract rotational
energy from the system, it has to satisfy the condition
\begin{eqnarray} 
\omega < m\;\Omega, 
\end{eqnarray} 
where $\Omega$ is the angular speed of the black hole at the event
horizon and $m$ is the harmonic azimuthal number of the wave. This is a
purely classical process first considered by
\citet{Penrose:1969pc}. When dealing with quantum fields, as
opposed to classical fields, this process can proceed spontaneously. Quantum
mechanically, a rotating black hole will tend to radiate away all of its
angular momentum, eventually approaching a non-rotating Schwarzschild
black hole \citep{Zeldovich:1971,Zeldovich:1971t}. This process is known
as super-radiance. (The term super-radiance was already used in
condensed matter to describe processes in which there was some coherent
emission of radiation from an otherwise disordered system.)

Again, these processes have a purely kinematical origin, so they are
perfectly suitable for being reproduced in an analogue
model. Regarding these processes, the simplest geometry that one can
reproduce, thinking of analogue models based on fluid flows, is that
of the draining bathtub of Sect.~\ref{S:simple}. Of course, this
metric does not exactly correspond to Kerr geometry, nor even to a
section of it \citep{Visser:2004zs, Visser:2007fj}. However, it is
\emph{qualitatively} similar. It can be used to simulate both Penrose's
classical process and quantum super-radiance, as these effects do not
depend on the specific multipole decomposition of Kerr's geometry, but
only on its rotating character. 

A specific experimental setup {was put forward} by Sch\"utzhold and Unruh 
using gravity waves in a shallow
basin mimicking an ideal draining bathtub \citep{Schutzhold:2002rf}.
Equivalent to what happens with Kerr black holes, this configuration
is classically stable in vacuum (in the linear regime; \citealt{Berti:2004ju}). 
A word of caution is in order here: Interactions of the gravity surface waves
with bulk waves (neglected in the analysis) could cause the system to
become unstable \citep{Volovik:2002ci}. This instability has no
counterpart in standard general relativity (though it might have one in
braneworld theories). Super-resonant scattering of waves in this
rotating sink configuration, or in a simple purely rotating vortex,
could in principle be observed in this and other analogue models. There
are already several articles dealing with this problem \citep{Basak:2005fv,
Basak:2002aw, Basak:2003uj, Cardoso:2004fi, Federici:2005ty,
Lepe:2004kv,Richartz:2009mi}. 

More recently, there have been advanced explicit proposals for the observation of the effect in different set ups: a rotating fluid~\citet{Cardoso:2016zvz} and an optical fiber~\citet{Prain:2019jqk}. 
Then, first observational evidence has been reported in a rotating fluid experiment~\citet{Torres:2016iee}; latter on in an optical experiment~\citet{Braidotti:2021nhw} (more on next section).

A related phenomenon one can consider is the black-hole bomb
mechanism \citep{Press:1972}. One would only have to surround the
rotating configuration by a mirror for it to become grossly
unstable. What causes the instability is that those in-going waves
that are amplified when reflected in the ergosphere would then in turn
be reflected back toward the ergoregion, due to the exterior mirror,
thus being amplified again, and so on.

An interesting phenomenon that appears in many condensed matter systems
is the existence of quantized  vortices. The angular momentum of these
vortices comes in multiples of some fundamental unit (typically $\hbar$
or something proportional to $\hbar$). The extraction of rotational
energy by a Penrose process in these cases could only proceed via
finite-energy transitions. This would supply an additional specific
signature to the process. In such a highly quantum configuration, it is
also important to look for the effect of having high-energy dispersion
relations. 

For example, in BECs, the radius of the ergoregion of a
single quantized  vortex is on the order of the healing length, so one
cannot directly associate an effective Lorentzian geometry with this
portion of the configuration. Any analysis that neglects the high-energy terms 
is not going to give any sensible result in these cases.
In this regard, it is worth highlighting a recent experiment reporting 
the construction of a giant vortex in an intermediate regime --- between 
the quantum and classical regimes~\citep{Svancara:2023yrf}.
Finally, it is also worth mentioning that quantum vortices in BECs have also been used as an indirect probe of the aforementioned black hole bomb mechanism~\citep{Patrick:2021oqk}.

\subsection{Cosmological particle production} 
\label{sec:cosmo}

Analogue model techniques have also been applied to cosmology.
Models considered to date focus on variants of the BEC-inspired
analogues:
\begin{itemize}

\item \citet{Fedichev:2003bv, Fedichev:2003id} have
  investigated WKB estimates of the cosmological particle production rate 
  and (1+1) dimensional cosmologies, both in expanding BECs.
  
\item \citet{Lidsey:2003ze}, and \citet{Fedichev:2003dj} have focussed on
  the behaviour of cigar-like condensates in grossly-asymmetric traps.
  
\item \citet{Barcelo:2003et, Barcelo:2003wu}
  have focussed on BECs and tried to mimic FLRW behaviour as closely
  as possible, both via free expansion, and via external control of
  the scattering length using a Feshbach resonance. 
  
\item \citet{Fischer:2004bf} propose the use
  of two-component BECs to simulate cosmic inflation.
  
\item \citet{Weinfurtner:2004in, Weinfurtner:2004mu} has
  concentrated on the approximate simulation of de~Sitter spacetimes.
  
\item Weinfurtner, Jain, et al.\ have undertaken both
  numerical \citep{Jain:2007gg} and general
  theoretical \citep{Weinfurtner:2008if, Weinfurtner:2008ns} analyses
  of cosmological particle production in a BEC-based FLRW universe.

\item \citet{Fagnocchi:2010sn} takes advantage of the disformal/Gordon 
structure of the acoustic geometry in relativistic BEC to extend the 
simulation of FLRW universes to the open case.

\item \citet{Prain:2010zq} investigated the structure of quantum 
correlations in an expanding BEC, both in 3+1 and 1+1 dimensions.

\item \citet{Steinhauer:2021fhb} reported the spontaneous creation 
of analogue cosmological particles in the laboratory, using a quenched 
3-dimensional quantum fluid of light.

\end{itemize}

\noindent
In all of these models the general expectations of the relativity
community have been borne out -- the theory definitely predicts particle
production, and the very interesting question is the extent to which the
formal predictions are going to be modified when working with real
systems experimentally \citep{Barcelo:2003wu}. We expect that these
analogue models provide us with new insights as to how their inherent
modified-dispersion relations affect cosmological processes such as the
generation of a primordial spectrum of perturbations (see, for example,
\citealt{Brandenberger:2000as, Brandenberger:2000zm,
  Brandenberger:2002sr, Brandenberger:2003vk, Brandenberger:2002ty,
  Brandenberger:2000wr, Brandenberger:2002hs, Casadio:2002dj,
  Easther:2001fi, Hassan:2002qk, Hu:1987rr, Hu:1994dx, Kempf:2001fa,
  Lemoine:2001ar, Lemoine:2002rf, Lubo:2003rs, Martin:2003kp,
  Martin:2000bv, Martin:2000xs, Martin:2002kt, Niemeyer:2002ze,
  Niemeyer:2001qe, Niemeyer:2002kh, Parentani:1996sz,
  Shankaranarayanan:2002ax, Paddy4, Starobinsky:2001kn,
  Starobinsky:2002rp, Tanaka:2000jw, Ford:2021syk} where analogue-like ideas are
applied to cosmological inflation).

An interesting side-effect of the original investigation, is that
birefringence can now be used to model ``variable speed of light'' (VSL)
geometries \citep{Bassett:2000wj, Ellis:2003pw}. Since analogue models
quite often lead to two or more ``excitation cones'', rather than one,
it is quite easy to obtain a bimetric or multi-metric model. If one of
these metrics is interpreted as the ``gravitational'' metric and the
other as the ``photon'' metric, then VSL cosmologies can be given a
mathematically well-defined and precise
meaning \citep{Bassett:2000wj, Ellis:2003pw}.

\subsection{Bose novae: an example of the reverse flow of information?}

As we have seen in the previous Sects.~\ref{S:Hawking-radiation},
  \ref{S:Horizon-stability}, \ref{S:Super-radiance}, and \ref{sec:cosmo}, analogue
models have in the past been very useful in providing new,
condensed-matter-physics--inspired ideas about how to solve
longstanding problems of semiclassical gravity. In closing this
section, it is interesting to briefly discuss 
a paradigmatic attempt
to use analogue models in the reverse
direction; that is to import well-known concepts of semiclassical
gravity into condensed matter frameworks.

The phenomenon we are referring to is the \emph{``Bose
nova''} \citep{Donley}. This is an experiment dealing with a gas of a few
million $^{85}$Rb atoms at a temperature of about 3~nK. The condensate
is rendered unstable by exploiting the possibility of tuning the
interaction (more precisely the scattering length) between the atoms via
a magnetic field. Reversing the sign of the interaction, making it
attractive, destabilises the condensate. After a brief waiting time
(generally called $t_\mathrm{collapse}$), the condensate implodes and
loses a sizable fraction of its atoms in the form of a ``nova burst''.
If left to evolve undisturbed, the number of atoms in the burst
stabilises and a remnant condensate is left. However, if the condensate
interaction is again made repulsive after some time $t_\mathrm{evolve}$,
before the condensate has sufficient time to stabilise, then the
formation of ``jets'' of atoms is observed, these jets being 
characterised by lower kinetic energy and a distinct shape with respect
to the burst emission.

Interestingly, an elegant explanation of such a phenomenology was
proposed in \citet{Calzetta:2002jz, Calzetta:2002zz}, based on the
well-known semiclassical gravity analysis of particle creation in an
expanding universe. In fact, the dynamics of quantum excitations over
the collapsing BEC were shown to closely mimic that for quantum
excitations in a time-reversed (collapsing instead of expanding)
scenario for cosmological particle creation. This is not so surprising
as the quantum excitations above the BEC ground state feel a
time-varying background during the collapse, and, as a consequence,
one then expects squeezing of the vacuum state and mode mixing, which
are characteristic of quantum field theory in variable external fields.

However, the analogy is even deeper than this. In fact,
in \citet{Calzetta:2002jz, Calzetta:2002zz} a key role in explaining the
observed burst and jets is played by the concepts of
``frozen'' versus ``oscillating'' modes -- borrowed from cosmology -- 
(although with a reverse
dynamics with respect to the standard (expanding) cosmological case). In
the case of Bose novae, the modes, which are amplified, are those for
which the physical frequency is smaller than the collapse rate, while
modes with higher frequencies remain basically unaffected and their
amplitudes obey a harmonic oscillator equation. As the collapse rate
decreases, more and more modes stop growing and start oscillating, which
is equivalent to a creation of particles from the quantum vacuum. In the
case of a sudden stop of the collapse by a new reversal of the sign of
the interaction, all of the previously growing modes are suddenly
converted into particles, explaining in this way the generation of jets
and their lower energy (they correspond to modes with lower frequencies
with respect to those generating the bursts).

Although this simple model cannot explain all the details of the Bose
novae phenomenology, we think it is remarkable how far it can go in
explaining several observed features by exploiting the language and
techniques so familiar to quantum cosmology. In this sense, the analysis
presented in \citet{Calzetta:2002jz, Calzetta:2002zz} primarily shows a
possible new application of analogue models, where they could be used to
lend ideas and techniques developed in the context of gravitational
physics to the explanation of condensed matter phenomena.

\subsection{Romulan cloaking devices} 

A wonderful application of analogue gravity techniques is the design
of \emph{cloaking devices} \citep{leonhardt-science, pendry-science,
  leonhardt-philbin-review}. How to achieve invisibility, or more
properly, low observability has been a matter of extensive study for
decades. With the appearance of a technology capable of producing and
controlling meta-materials and plasmonic
structures \citep{PhysRevLett.102.233901}, cloaking is becoming a real
possibility. 

To achieve cloaking, one needs to ensure that light rays [beyond
geometric optics it is impossible to produce perfect
invisibility \citep{Nachman1988, Wolf1993}] effectively behave as if
they were propagating in Minkowski spacetime, although in reality they
are bending around the invisible compact region. One way of producing
this is, precisely, to make rays propagate in Minkowski spacetime but
using non-Cartesian coordinates. Take the Minkowski metric in some
Cartesian coordinates $x'$, $\eta_{\mu\nu}$, and apply a
diffeomorphism, which is different from the identity only inside the
compact region. One then obtains a different representation
$g_{\mu\nu}(x)$ of the flat geometry. Now take the $x$ to be the
Cartesian coordinates of the real laboratory spacetime, and build this
metric with the meta-material. By construction, the scattering process
with the compact region will not change the directions of the asymptotic rays,
making everything within the compact region invisible. Recent
implementations of these ideas have investigated the concept of a
``spacetime cloak'' or ``history editor'' that cloaks a particular
event, not a particular region \citep{history-editor}.

To end this brief account we would like to highlight the broad scope
of application of these ideas: Essentially the cloaking techniques can
be applied to any sort of wave, from acoustic
cloaking \citep{chen:183518,Garcia-Meca:2013soa,Garcia-Meca:2014tpa} to earthquake damage
prevention \citep{PhysRevLett.103.024301}. More radically, and with
enough civil engineering, one might adapt the ideas of
\citet{Berry1, Berry2} to anti-tsunami cloaking.

\subsection{Going further} 

For more details on the trans-Planckian problem, 
super-radiance
and cosmological issues (especially particle production) 
one should carefully check
INSPIRE for the most recent articles as a great deal of activity is still going on.

\clearpage
\section{Experimental efforts}
\label{S:experiments}
After {an initial} period in which the field was driven almost entirely by theoretical ideas and speculation, some experimental groups started to take on the challenge of analogue gravity, and design experiments with gravitational motivation. The first experiments can be traced back to the water tank experiment by {the group led by} Germain Rousseaux~\citep{Rousseaux:2007is} (in Nice, France, at that time) and experiments with slow light realized at St. Andrews University~\citep{Philbin:2007jj, Philbin:2007ji}. After that and to this day, the experimental progress has been steadily increasing. We would like to highlight the landmark paper by Steinhauer~\citep{Steinhauer:2015saa} in which the much sought-after spontaneous Hawking radiation has finally been observed. But analogue gravity experiments do not end here. Quite the opposite, experiments continue and as mentioned in the introduction appear to enter into a new phase with renovated motivations. 

\subsection{Wave tank experiments} 
\label{S:water-tank}

As we have seen, waves in shallow water can be considered to be a
particularly simple analogue gravity system. (See Sect.~\ref{S:WW}.)
Experimentally, water basins are relatively cheap and easy to
construct and handle. In particular, shallow water basins (more
precisely, wave tanks or wave flumes) have acquired a prominent role
in recent years. In particular, such technology underlay the work
of \citet{badulin:782}, and such wave tanks have been used by groups in Nice, France \citep{Rousseaux:2007is}
and Vancouver, Canada \citep{Weinfurtner:2010nu}. A third experiment has 
more recently been running in Poitiers, France \citep{Euve:2018uyo,Rousseaux:2020cjo}.
Specifically, the Nice experiments
have been carried out using a large wave-tank 30~m long, 1.8~m wide
and 1.8~m deep. 

The simplest set-up with such a device is to send water
waves (e.g., produced by a piston) against a fluid flow produced by a
pump. To generate a water-wave horizon, a ramp is placed in the water,
with positive and negative slopes separated by a flat section. When a
train of waves is sent against the reverse fluid flow there will be a
place where the flow speed equals the group velocity of the waves --
there a group velocity horizon will be created. (Remember that in the
shallow-water regime the low momentum co-moving dispersion relation
for surface water waves is $\omega^2=gk\tanh (kh)$ where $g$ denotes the
gravitational acceleration of the Earth at the water surface and h is
the height of the channel.) For incident waves moving against the flow
it would be impossible to cross such a horizon, and in the sense
that the system is the analogue of a white-hole horizon, the time
reversal of a black-hole horizon. (In the engineering and fluid
mechanics literature this effect is typically referred to as ``wave
blocking''.)

Remarkably, \citet{Rousseaux:2007is}
reported the first direct observation of negative-frequency waves,
converted from positive-frequency waves in a moving medium, albeit the
degree of mode conversion appears to be significantly higher than that
expected from theory. The same group \cbs{also} set up a more compact
experiment based on the hydraulic jump, wherein measurements of the
``Froude cones'' convincingly demonstrate the presence of a
surface-wave white hole \citep{Jannes:2010sa}, (as described in
Sects.~\ref{S:WW} and \ref{S:WW-gen}, and possibly implicit in the
results of \citet{badulin:782}).

A related experiment, with a smaller water basin, performed under the
auspices of the gravitation theory group (Physics) and the fluid
mechanics group (Civil Engineering) at the University of British
Columbia, has reported (August 2010) the detection of
stimulated Hawking emission \citep{Weinfurtner:2010nu}. This stimulated
Hawking emission is a classical effect that (via the usual discussion
in terms of Einstein $A$ and $B$ coefficients) is rather closely
related to spontaneous quantum Hawking emission. A central result of
the Weinfurtner et~al.\ group is that the relevant Bogoliubov
coefficients have been experimentally measured and are observed to
satisfy the expected Boltzmann relation
\begin{equation}
{|\beta|^2\over|\alpha|^2} = \exp\left\{ - {2\pi \omega\over g_H/c_H} \right\}.
\end{equation}
More details of the experimental setup can be found
in \citet{Weinfurtner:2010nu}. 


This experiment was able to clearly show the presence of mode conversion, but even more importantly it gave an important boost to the experimental efforts in analogue gravity. Nonetheless, with hindsight its results are not as close to the observation of a pure (stimulated) Hawking effect as initially thought~\citep{Unruh:2014hua}. On the one hand, it appears that the actual setup only provided real blocking for a partial sector of the analyzed frequencies~\citep{Michel:2014zsa,Euve:2014aga,Coutant:2016bgk}. 
To construct a fluid flow with a more strict horizon is a complex matter that has reported to be attained only in the most recent experiments in Poitiers~\citep{Euve:2018uyo}, see also \citet{Bossard:2023zyq,Bossard:2024eys}. On the other hand, analyses in the Poitiers facility reproducing the Vancouver setup, are showing that also non-linear effects are playing an important role in the mode conversion~\citep{Euve:2021mnj}, something that has not been much analyzed in the gravitational realm.

Given these complexities it is still not clear how much of the constancy of the fitted temperature in the Vancouver experiment should be associated with being close to a horizon situation~\citep{Coutant:2016bgk}; in fact, the Planckian fit has not been reproduced in the Poitiers experiment~\citep{Euve:2018uyo,Euve:2021mnj}. In any case, it appears to us that all these experiments exemplify the richness of analogue gravity as a cross-fertilizing venture.

\subsection{Other classical fluid experiments}
\label{S:fluid-experiments}
 
In 2017 Weinfurtner and collaborators in Nottingham reported the observation of superresonant scattering \citep{Torres:2016iee}. 
In a steady draining-bathtub configuration, water swirls around until reaching a drainage hole. In the central core an ergoregion for surface waves can be generated. Once a steady flow is established, small surface waves are excited at one side of a rectangular tank and absorbed on the other side. The waves traverse the vortex configuration and produce a scattering pattern. By monitoring the behaviour of the waves, and applying a frequency filter and an azimuthal decomposition, the authors showed that the 
$m = 1,2$ co-rotating low-frequency modes experience amplification. This is interpreted as superradiant scattering and indicates that the superradiance condition $\omega-m\Omega(r)<0$, with $\Omega$ the angular velocity of the fluid, is satisfied in the central regions of the configuration. Other works related to this experiment are \citep{Torres:2016iee,Torres:2017vaz} and Torres' PhD thesis~\citep{TorresVicente:2019ahs}.

Later on, with an equivalent setup, the Nottingham group has also been able to observe quasinormal mode ringing in vortex flows driven away from equilibrium~\citep{Torres:2020tzs}. This is another instance of how these notions {can be transported across} to different realms. More recently the group has designed a rotating fluid experiment but now using superfluid helium 4~\citep{Svancara:2023yrf}. {We will briefly discuss  this setup in a subsequent subsection.}

As additional analogue gravity experiments with non-quantum fluids let us mention the Rousseaux' group hydraulic-jump experiment, wherein measurements of the ``Froude cones'' convincingly demonstrate the presence of a
surface-wave white hole \citep{Jannes:2010sa}, (as described in
Sects.~\ref{S:WW} and \ref{S:WW-gen} and possibly already implicit in the
results of \citep{badulin:782}).

We would also like to mention a new promising experimental setup based on flowing soap films~\citep{Rousseaux:2020cjo}. Some experiments have already been performed by Hamid Kellay and collaborators at the University of Bourdeaux, France~\citep{Kellay:2017}.

\bigskip
\subsection{Bose--Einstein condensate experiments} 
\label{S:Lahav}

We have already discussed quite extensively how one can find gravitational analogies using Bose--Einstein Condensates.
The potentialities of BECs as analogue systems were first identified in \citet{Garay:1999sk} and \citet{Garay:2000jj}, and in a series of follow up papers \citep{Barcelo:2001ca, Sakagami:2001ph, Furuhashi:2006dh, Okuzumi:2007hf}. To simulate black hole configurations, in these papers it was proposed to use linear traps with sinks, circular traps, as well as Laval-nozzle--shaped traps. The theoretical challenge was taken up 
by Jeff Steinhauer at Technion, Israel, and collaborators whom put together an entire roadmap to realize an analogue of Hawking emission.

The first milestone in the roadmap was the observation of horizon formation in a BEC~\citep{Lahav:2009wx}. The experiment used a BEC 
formed by {about} $10^5$ atoms of $^{87}$Rb. They generated a cigar-shaped effectively 1-dimensional configuration, using a confining anisotropic harmonic  potential with one broad direction. This condensate was then subject to an additional (narrow) step-like potential that swept at a constant velocity along the length of the condensate.  This potential {produces} a cascading effect: atoms accelerate in falling {through} the step, {so generating} a supersonic flow. The {achievement} of a supersonic region is also helped by the fact that the region with fast moving particles down the cascade, becomes, by continuity of the flow, a region of low density. {This low density region corresponds to a slower-than-normal speed of sound (roughly millimetres per second), and hence it facilitates the possibility for the flow speed to become supersonic}. The crossing point where the speed of the flow and that of sound coincide marks the location of a black horizon, which was kept stable for about 8~ms. Beyond the step the BEC slows down again, generating a second horizon, this time a white horizon.

After this first observation, they continued to further refine the experiment and their observational techniques. They analysed phonon propagation in~\citep{Shammass:2012}, and the thermal distribution of phonons in~\citep{Schley:2013}. In~\citep{Steinhauer:2014dra} it was reported the observation of stimulated and self-amplified Hawking radiation in a black horizon--white horizon configuration. Finally, in 2016 Steinhauer reported the first observation of the much sought-after pure Hawking effect: spontaneous particle production by a single black-hole horizon~\citep{Steinhauer:2015saa}. The observation was made through the determination of the correlation function~\citep{Balbinot:2007de}, which exhibits correlations between outgoing and ingoing Hawking pairs~\citep{Carusotto:2008ep}. In addition, the measured correlations were argued to correspond to genuine quantum correlations, namely from a non-separable quantum state (see~\citep{Busch:2014bza} for more theoretical details). This was {interpreted  as signaling} the presence of spontaneous Hawking emission. In our opinion Steinhauer's results are the quintessence of the analogue gravity programme as originally envisioned. This does not mean that they are completely free {from} controversies. In fact, we think that the presence of non-trivial controversies makes the results even more interesting.

The 2014 results {demonstrated} the generation of a series of oscillations of the BEC density in the trapped region and its growth in time. They were initially interpreted as the observation of stimulated Hawking radiation produced by a black-hole-laser mechanism~\citep{Corley:1998rk}. A laser effect should occur in a theory with superluminal dispersion (like BECs) whenever one generates a combination of a black and a white horizons. The negative energy Hawking partners produced at the black horizon are (partially) reflected back at the white horizon becoming high-frequency modes that are able to traverse the trapped region between the horizons, and to stimulate further pair generation at the black horizon. 

The interpretation of the observations as a laser effect was soon challenged in~\citep{Tettamanti:2016ntx} and~\citep{Wang:2016jaj,Wang:2017wnj}, with counter arguments in~\citep{Steinhauer:2016hfa}. 
They argued that the seed of the oscillation is not a laser mechanism but a Bogoliubov--\v{C}erenkov--Landau (BCL) effect~\citep{Carusotto:2006}.  This is a deterministic effect caused by the supersonic flow being reflected at a potential wall, in this
case the region in which the white horizon is formed. 

In fact, further numerical analysis in~\citep{Tettamanti:2020rvt} show that similar results appear even if one switches off the atomic repulsions in the supersonic region. In this case a (larger) interference pattern appears, just because  a Schr\"odinger wave function is being reflected at the internal potential wall
(this case represents an extrapolation of the BCL phase to a regime in which the phononic excitations are frozen). 

In 2021, Steinhauer group has reported a more refined experiment along the same lines~\citep{Kolobov:2019qfs}. With further insight, they are now confirming that in both experiments one is observing a stimulated Hawking effect, but that its main cause is not a laser but a BCL mechanism~\citep{Steinhauer:2021xxj}. Additional explanations {regarding} a plausible laser-BCL crossover are given in~\citep{deNova:2023yyu}.

Regarding the claimed observation of a pure Hawking effect in~\citep{Steinhauer:2015saa}, some debate needs to be reported. On the one hand, Michel \emph{et al.} analysed Steinhauer’s experiments from a theoretical perspective, and the agreement is substantial~\citep{Michel:2016tog}. On the other hand, there has been criticism against the claimed detection~\citep{Finke:2016wcn,Leonhardt:2016qdi}, criticism which has been responded to by the author~\citep{Steinhauer:2016ftg,Steinhauer:2018qzg}.
One of the sources of controversy is whether the observed correlations are purely quantum, or whether they have a classical contribution/origin.

Using BECs {another experimental} group has been able to simulate the generation of fluctuations by a rapid expansion of the universe~\citep{Hung:2013}. Precisely, they rapidly vary the interaction strength between Cesium atoms in a {lenticular-shaped} BEC (effectively 2-dimensional). After performing the quench they monitor how the fluctuation pattern just formed evolves in time. They observe oscillations in the pattern structure,  that they interpret in terms of cosmological Sakharov oscillations. The interactions can also be made attractive during some limited periods of time by invoking additional phenomenology~\citep{Chen:2021xhd}. In GR language this is analogous to a change of signature in the metric~\citep{Weinfurtner:2007dq,White:2008xr}. These are instances of the type of the cosmological problems that can be tackled with cold atoms~\citep{Schmiedmayer:2013xsa}. For example, these cosmological effects are tightly related \blue{to} the dynamical Casimir effect which has also been realized in BECs~\citep{Jaskula:2012}.

\subsection{Fibre-optic models} 
\label{SS:fibre}

A recent implementation of electrodynamic analogue models is
that based on fibre-optic engineering \citep{Philbin:2007jj,
  Philbin:2007ji}. The fundamental idea in this case is to use long
dispersive light pulses (solitons), generated with a suitable laser,
to create a propagating front at which the refractive index of the
fibre changes suddenly (albeit by a small amount). Basically, the
refractive index of the fibre, $n_0$, acquires a time- and
position-dependent correction $\delta n$, which is proportional to the
instantaneous pulse intensity $I$ at a give space-time position,
$\delta n\propto I(t,x)$. The wavefront at which this change in the
refractive index occurs will move naturally at a speed close to the
speed of light (and fibre optic engineering allows one to control this
feature). If one now sends a continuous wave of light, what we might
call a probe, along the fibre in such a manner that the probe group
velocity in the fibre is arranged to be slightly larger than the pulse
group velocity, then it will be possible to obtain horizon-like
effects. In fact, as the probe wave reaches the back of the pulse, the
increase in the refractive index will slow it down, until the probe
group velocity will match the pulse one. Effectively, the rear end of
the pulse will act as a white hole for the probe wave. Similarly, there
will be a point on the front side of the pulse where the two group
velocities will match. This will be the equivalent of a black-hole
horizon for the probe wave. 

In \citet{Philbin:2007jj,Philbin:2007ji} the behaviour of the probe
waves at the pulse was investigated, and it was shown for the white
hole case that the expected classical behaviour is theoretically
reproduced. Since this behaviour lies at the core of the mechanism
responsible for the mode conversion underlying the Hawking effect, it
is then expected that the quantum counterpart should also be
reproducible in this manner.
Indeed, in 2010 Belgiorno et~al.\ have reported
experimental detection of photons from a black-hole--white-hole
configuration possessing a ``phase velocity
horizon'' \citep{Belgiorno:2010xxx}. The underlying theory behind their
specific experiment is considered in \citet{Faccio:2009yw, Belgiorno:2010iz, Belgiorno:2010zz}.

Also in this case, the experiment capitalized on the Kerr effect, as in~\cite{Philbin:2007ji}, to induce a traveling perturbation in the refractive index within a transparent dielectric medium (fused silica glass) using an ultra-short laser pulse. Experimental evidence reported in \citet{Belgiorno:2010xxx, Rubino:2011zq} showcased unpolarized photon emission, interpreted as a Hawking-like radiation phenomenon, attributed to the presence of a blocking region mimicking a black hole-white hole system.

The details of the experimental setup were as follows: a laser pulse with a duration of 1 picosecond illuminated the fused silica glass, with the input energy varied within the range of $100-1200$ $\mu$J. Radiation emitted from the filament was then collected perpendicular to the laser pulse propagation axis, so minimizing known spurious effects (e.g., Cherenkov radiation, fluorescence). The emission spectra were fitted reasonably well with Gaussian profiles. These fits revealed an increase in peak wavelength and overall flux with rising pulse intensity. Key observations included emission predominantly centered around a peak wavelength of approximately $(850 \pm 25)$ nm, with a bandwidth seemingly contingent on pulse intensity. Though the precise bandwidth determination relied on fitting, it was evident that the majority of photon counts fell within the $700-1000$ nm range.

The interpretation of the observed spectrum as an optical analogue of spontaneous quantum Hawking radiation primarily hinged on the aforementioned window condition. While the observed radiation exhibited non-thermal characteristics, it was nonetheless argued to be linked to optical horizons (blocking regions).

While this observation can be listed among the most relevant ones in experimental realizations of analogue gravity, it is worth mentioning that also in this case the interpretation of the observed radiation has been controversial due to some oddities in the observed radiation. For instance, the observed photons were emitted parallel to the optical horizon, and the relevant optical horizon was itself defined in an unusual manner by combining group and phase velocities. This propelled some activity in finding alternative interpretations for the observed phenomenon, in particular see~\citep{PhysRevLett.107.149401} and the reply \citep{Belgiorno:2010hk} or \citep{Liberati:2011ep} where a dynamical Casimir effect interpretation is advanced. It is also worth mentioning that this experiment propelled intense theoretical activity in the simulation of the emission generated by varying refractive indexes, see e.g.~\citep{Lang:2019avs,PhysRevA.87.023803}.

In closing this section, let us mention that optic fibers models had several other implementations in recent years.
For example, again by using optical fibers, the researchers in~\citep{Vocke:2017tif} have been able to design an ingenious analogue gravity experiment in which a spiraling light signal, moving along the fiber, simulates a rotating superfluid vortex in 2+1 dimensions (the propagation direction plays the role of time). The experiment implements the theoretical ideas developed earlier in~\citep{Marino:2008kk,Marino:2009}. There, it was observed that the behaviour of light in the transversal dimensions as it travels through a self-defocussing fiber satisfies a Gross-Pitaevkii equation, and so behaves as a quantum fluid made of light. This fluid is then made to rotate. A first experiment has been realized~\citep{Vocke:2017tif} reporting the observation of an ergoregion separated from a more internal horizon. The experiment relies also on previous studies~\citep{Vocke:2015,Vocke:2016}.

More recently, and following previous theoretical studies~\citep{Prain:2019jqk,Braidotti:2020ize,Solnyshkov:2018dgq}, in~\citep{Braidotti:2021nhw} it has been reported the observation of superradiant scattering using an equivalent setup.
This is the second observation of superradiant scattering in an analogue gravity experiment, so pointing out the expected generality of the process.

\subsection{Experiments with quantum fluids of light} 
\label{S:6.5}

Quantum fluids of light have become an extensive and significant area of research with multiple applications~\citep{Carusotto:2013}. Essentially, the generic name ``quantum fluids of light'' can be attached to any situation involving a collective behaviour of photons with fluid-like characteristics (typically through interactions with some matter degrees of freedom). In many circumstances 
a relevant quantity satisfies a Gross--Pitaevskii-like equation and so exhibits an acoustic behaviour like that on a BEC.  Their potential advantages as analogue gravity systems with respect to atomic systems were appreciated already by Marino~\citep{Marino:2008kk}; see also the experimental designs envisioned in~\citep{Fouxon:2009be,Marino:2009,Solnyshkov:2011sf}. For instance, these systems allow stationary configurations to be set up with a single black horizon and larger Hawking temperatures (a few kelvin).

Besides the experiment in a self-defocusing wave fiber described above, there have also been designed and run two experiments with a exciton-polariton gas in a microcavity~\citep{Nguyen:2015,Jacquet:2020znq}. A microcavity is created in the space between two mirrors and with an internal potential well so that the energy of the fundamental mode of the photon between the plates is resonant with an excited state of the potential well. In this way an exciton-polariton degree of freedom is generated which satisfies a generalized Gross–Pitaevskii equation (with an external force and a dissipative term). In the Nguyen at al experiment they generate an accelerating fluid by introducing a transversal widening in the cavity~\citep{Nguyen:2015}. They managed to accelerate the fluid up to make it to transition to a supersonic regime. There have been several studies of an experiment along these lines, both before its actual realization~\citep{Solnyshkov:2011sf,Gerase:2012,Larre:2013tba}, and also after the experiment was performed~\citep{Grisins:2016gru}.

A slightly different experiment was the one of Jacquet et al.~where again a supersonic regime was generated but this time by applying an all-optical control over the flow~\citep{Jacquet:2020znq}. By doing so, it was possible to observe the expected negative energy modes in the supersonic region~\citep{Falque:2023ctx}. It is worth remarking, that the characteristics of the collective excitations of {such} a system are very rich and can be investigated using a high-resolution coherent probe spectroscopy method~\citep{Claude:2021rkt,Claude:2022wfm}. At present there is not yet {any} observation of Hawking-like radiation {in any} of these systems, but new promising avenues are currently being explored~\citep{Jacquet:2022vak,Jacquet:2021scv}.

In ending this section let us also mention {that} a similar quantum fluid of light has been used to simulate the generation of fluctuations by an expanding universe in 2+1 dimensions~\citep{Steinhauer:2021fhb}. 
In this case the set up was based on a laser traversing a vapor cell which provides the required repulsive interactions between the photons. The sudden switching on and off of the interactions, respectively, when the laser enters and exists the vapor cell, acts effectively as the analogue of a rapid contracting and expanding phase in the universe. The data fits the theoretical expectations on particle production based on quantum field theory. The initial contracting phase generates fluctuations spontaneously, while in the second (expanding) phase they found a combination of spontaneous and stimulated emission due to the particles already present. 

\bigskip

\subsection{Vortex flows in superfluid helium} 

To approach an explorational regime in which one could observe spontaneous superradiance in a vortex fluid flow, one should construct vortex flows in a quantum fluid (i.e. a superfluid). But then, the rotational characteristics typically appear as isolated quantized  vortices or vortex tangles, quite different from the GR situation one would like to simulate. A clever strategy to ameliorate this is to try to force oneself  into an intermediate regime, between a classical and a quantum flow --- a regime with a single vortex with a large, almost classical, circulation. This is precisely what has been attained by the Nottingham group~\citep{Svancara:2023yrf}.  

In that work the authors report the construction of a giant quantum vortex in superfluid He$^4$ (a record of $\sim 10^4$ circulation quanta). They have already been able to observe the presence of bound states and black hole ring-down signatures, so opening a new avenue to simulate rotating spacetimes.

\subsection{Differentially-rotating flows in superfluid helium} 

At the end of Sect.~\ref{S:helium} we briefly described an analogue
model based on the ripplons in the surface separating two
differentially-moving superfluids, in particular an AB-interphase in
$^{3}$He. These interphases are being produced in Helsinki's Low
Temperature Lab \citep{Blaauwgeers:2001ev, Blaauwgeers200357,
  Finne:2006a, Finne:2006bi}. The AB-interphase is prepared in a small
quartz cylinder (3~mm radius times 11~cm long) inside a rotating
cryostat. The $^{3}$He-A is rotating with the cryostat while the
$^{3}$He-B remains at rest with respect to the lab. Among other things,
in this setting the critical values at which instabilities appear as
functions of the temperature, and the nature of these instabilities,
are being investigated. These instabilities are related to the
appearance of an ergoregion in the analogue metric for ripplons and to
the Kelvin--Helmholtz instability \citep{Volovik:2002ci}. In
particular, this has been the first time that the Kelvin--Helmholtz
instability has been observed in
superfluids \citep{Blaauwgeers200357}. The nature of the instability in
these experiments is controlled by the difference in velocities
between the normal and superfluid components of the flow. It still
remains to further lower the ambient temperature so as to probe the
nature of the instabilities in the absence of any normal fluid
component.

\subsection{More experiments} 

In this section on experimental efforts we have highlighted those experiments that make a strong usage of spacetime gravitational analogies, and we have separated them by the type of analogue system used. To end this section we would like to also mention some other experiments that can also be considered within the scope of analogue gravity.

There is a cosmological particle production analogue experiment based on ions in a tramp~\citep{Wittemer:2019agm}. They change the size of the confining potential and observe how pairs of ions begin to oscillate. There are experiments demonstrating the dynamical Casimir effect~\citep{Macri:2017ent,Michael:2018nvi,Dodonov:2020eto} using superconducting circuits~\citep{Wilson:2011rsw,Johansson:2010vqd,Johansson:2009zz}, dispersive optical fibers~\citep{Vezzoli:2019}, Josephson metamaterials~\citep{Lahteenmaki:2011cwo}, and the already mentioned BECs~\citep{Jaskula:2012}. 
 
There is a plethora of work regarding the search for an experimental demonstration of the Unruh effect (see \emph{e.g.}~\citep{Biermann:2020bjh} and references therein). Although a complete demonstration is still lacking, there are already experiments reporting indirect observations of the effect~\citep{Leonhardt:2017lwm,Hu:2019}.

Overall, analogue gravity motivations seem to have no end: There are experimental observations of a curved causality in quasi-condensates~\citep{Tajik:2022lyt}, a verification of the entropy-area law in a quantum simulator~\citep{Tajik:2022ycs}, and a verification of the effective quantum field theory description of a quantum many-body system~\citep{Zache:2019xkx}, or simulations of false vacuum decay in two-BEC systems~\citep{Zenesini:2023afv, Cominotti:2022jrj}.

\subsection{Going further} 

There seems to be considerable ongoing interest in experimental probes
of analogue spacetimes, and quantum effects in analogue spacetimes,
and one should carefully check INSPIRE for the most recent articles.

\clearpage
\section{Hints towards a Theory of Quantum Gravity?} 
\label{S:towards}

The key question one should ask at this stage is this: ``Where can we
go from here?'' Apart from continuing with the analysis of the issues
described in the previous two sections, it is natural to explore
whether the analogue gravity programme could be extended to the point
of yielding a theory of ``Quantum Gravity''. In particular, the
following topics come to mind as steps on a path towards a possible
theory of Quantum Gravity:
\begin{itemize}
\item Backreaction.
\item Equivalence principle.
\item Diffeomorphism invariance.
\item Effective spin-two excitations.
\item Weinberg--Witten theorem.
\item Emergent gravity.
\item The cosmological constant problem.
\item Quantum gravity --- phenomenology.
\item Quantum gravity --- fundamental models.
\end{itemize}
Some work has already been done dealing with these topics in the context
of analogue gravity. Let us now expand on these issues a little.

\subsection{Backreaction}

There are important phenomena in gravitational physics whose
understanding needs analysis well beyond classical general relativity
and field theory on (fixed) curved background spacetimes. The black-hole
evaporation process can be considered as paradigmatic among these
phenomena. Here, we confine our discussion to this case. Since we
are currently unable to analyse the entire process of black-hole
evaporation within a complete quantum theory of gravity, a way of
proceeding is to analyse the simpler (but still extremely difficult)
problem of semiclassical backreaction (see, for
example, \citealt{Davies:1976ei, Christensen:1977jc, Birrell, Fulling,
Parentani:1994ij, Brout:1995rd, Massar:1999wg}). One takes a background black-hole
spacetime, calculates the expectation value of the quantum
energy-momentum tensor of matter fields in the appropriate quantum state
(the Unruh vacuum state for a radiating black hole), and then takes this
expectation value as a source for the perturbed Einstein equations. This
calculation gives us information about the tendency of spacetime to
evolve under vacuum polarization effects\footnote{For completeness, let us also mention a minority position in the literature arguing that backreaction effects should be incompatible with Hawking evaporation~\citep{Helfer:2000wh, Helfer:2003va,Helfer:2004jx} or with the very formation of a horizon~\citep{Terno:2019kwm,Baccetti:2018qrp}.}.
Semiclassical effect might even alter the way we understand the process of gravitational collapse~\citep{Barcelo:2022gii} introducing novel states of equilibrium~\citep{Arrechea:2021xkp} in the form of ultracompact stars (see~\citep{Arrechea:2023hmo} and references therein for a compelling description of these novel situations).

A nice feature of analogue models of general relativity is that, although
the underlying classical equations of motion have nothing to do with
Einstein equations, the tendency of the analogue geometry to evolve due
to quantum effects is formally equivalent (approximately, of course) to
that in semiclassical general relativity. Therefore, the onset of the
backreaction effects (if not their precise details) can be simulated
within the class of analogue models. An example of the type of
backreaction calculations one can perform are those
in \citet{Balbinot:2004dc, Balbinot:2004da}. These authors started from
an effectively one-dimensional acoustic analogue model, configured to have
an acoustic horizon by using a Laval nozzle to control the flow's speed.
They then considered the effect of quantizing the acoustic waves over
the background flow. To calculate the appropriate backreaction terms
they took advantage of the classical conformal invariance of the
(1+1)-dimensional reduction of the system. In this case, we know
explicitly the form of the expectation value of the energy-momentum
tensor trace (via the trace anomaly). The other two independent
components of the energy-momentum tensor were approximated by the
Polyakov stress tensor. In this way, what they found is that the
tendency of a left-moving flow with one horizon is for it to evolve in
such a manner as to push the horizon down-stream at the same time that
its surface gravity is decreased. This is a behaviour similar to what is
found for near-extremal Reissner--Nordstr\"om black holes. (However, we
should not conclude that acoustic black holes are, in general, closely
related to near-extremal Reissner--Nordstr\"om black holes, rather than
to Schwarzschild black holes. This result is quite specific to the
particular one-dimensional configuration analysed.)
{Other works on backreaction within analogue system are e.g.~\citet{Schutzhold:2005ex, Schutzhold:2008zzb,Rinaldi:2007hw,Girelli:2008gc, Liberati:2020mdr}.}

Can we expect to learn something new about gravitational physics by
analysing the problem of backreaction in different analogue models? As
we have repeatedly commented, the analyses based on analogue models
force us to consider the effects of modified high-energy dispersion
relations. For example, in BECs, they affect the ``classical''
behaviour of the background geometry as much as the behaviour of the
quantum fields living on the background. In seeking a semiclassical
description for the evolution of the geometry, one would have to compare
the effects caused by the modified dispersion relations to those caused
by pure semiclassical backreaction (which incorporates deviations from
standard general relativity as well). In other words, one would have to
understand the differences between the standard backreaction scheme in
general relativity, and that based on Eqs.~(\ref{bec-self-consistent1})
and (\ref{quantum-field}). 

To end this subsection, we would like to comment that one can go beyond
the semiclassical backreaction scheme by using the \emph{stochastic
semiclassical gravity programme} \citep{Hu:1999mm, Hu:2002jm, Hu:2003qn,Hu:2020luk}.
This programme aims to pave the way from semiclassical gravity toward a
complete quantum-gravitational description of gravitational phenomena.
This stochastic gravity approach not only considers the expectation
value of the energy-momentum tensor but also its fluctuations, encoded
in the semiclassical Einstein--Langevin equation. In a very interesting
paper, \citet{Parentani:2000ts} showed that the effects of the
fluctuations of the metric (due to the in-going flux of energy at the
horizon) on the out-going radiation led to a description of Hawking
radiation similar to that obtained with analogue models. It would be
interesting to develop the equivalent formalism for quantum analogue
models, and to investigate the different emerging approximate regimes.

\subsection{Equivalence principle} 

Analogue models are of particular interest in that the analogue
spacetimes that emerge often violate, to some extent, the Einstein
equivalence principle \citep{Barcelo:2001cp, Visser:2002sf}. This is
the heart and soul of any metric theory of gravity and is basically
the requirement of the universality of free fall, plus local Lorentz
invariance and local position invariance of non-gravitational
experiments.

As such, the Einstein equivalence principle is a
``principle of universality'' for the geometrical structure of
spacetime. Whatever the spacetime geometrical structure is, if all excitations
``see'' the same geometry, one is well on the way to satisfying the
observational and experimental constraints. In a metric theory, this
amounts to the demand of mono-metricity: A \emph{single} universal metric
must govern the propagation of all excitations.

Now it is this feature that is relatively difficult to arrange in
analogue models. If one is dealing with a single degree of freedom, then
mono-metricity is no great constraint. But with multiple degrees of
freedom, analogue spacetimes generally lead to refringence -- that is
the occurrence of Fresnel equations that often imply multiple
propagation speeds for distinct normal modes. To even obtain a bi-metric
model (or, more generally, a multi-metric model), requires an algebraic
constraint on the Fresnel equation that it completely factorises
into a product of quadratics in frequency and wavenumber. Only if this
algebraic constraint is satisfied can one assign a ``metric'' to each of
the quadratic factors. If one further wishes to impose mono-metricity,
then the Fresnel equation must be some integer power of some single
quadratic expression, an even stronger algebraic
statement \citep{Barcelo:2001cp, Visser:2002sf}.
Faced with this situation, there are two ways in which the analogue
gravity community might proceed:
\begin{enumerate} 
\item Try to find a broad class of analogue models (either physically
  based or mathematically idealised) that naturally lead to
  mono-metricity. Little work along these lines has yet been done; at
  least partially because it is not clear what features such a model should
  have in order to be ``clean'' and ``compelling''.
\item Accept refringence as a common feature of the analogue models and
  attempt to use refringence to ones benefit in one or more ways: 
  \begin{itemize}
  \item There are real physical phenomena in non-gravitational settings
    that definitely do exhibit refringence and sometimes multi-metricity.
    Though situations of this type are not directly relevant to the gravity
    community, there is significant hope that the mathematical and
    geometrical tools used by the general relativity community might in
    these situations, shed light on other branches of physics.
  \item Use the refringence that occurs in many analogue models as a way of
    ``breaking'' the Einstein equivalence principle, and indeed as a way of
    ``breaking'' even more fundamental symmetries and features of standard
    general relativity, with a view to exploring possible extensions of general
    relativity. 
    While the analogue models are not themselves primary physics, they can
    nevertheless be used as a way of providing \emph{hints} as to how more
    fundamental physics might work. 
  \end{itemize}
\end{enumerate}

\subsection{Nontrivial dispersion as Einstein-aether theory} 

There is a certain precise sense in which nontrivial dispersion
relations can effectively be viewed as implicitly introducing an
``aether field'', in the sense of providing a kinematic (but not
necessarily dynamic) implementation of Einstein-aether
theory \citep{Jacobson:2004ts, Eling:2004dk, Foster:2005dk,
  Jacobson:2008aj}. The point is that to define nontrivial dispersion
one needs to pick a rest frame $V^a$, and then assert $\omega^2 =
f(k^2)$ in this rest frame. But one can then re-write this dispersion
relation (in the eikonal approximation) as
\begin{equation}
  \left[ - (V^a \partial_a)^2 + f( [g^{ab}+V^a V^b] \partial_a \partial_b ) \right] \Psi(x) = 0.
\end{equation}
That is, using $f(w) = j(w) + w$,
\begin{equation}
 \left[ g^{ab} \partial_a \partial_b + j( [g^{ab}+V^a V^b] \partial_a \partial_b ) \right] \Psi(x) = 0.
\end{equation}
As long as the background is slowly varying, this can be re-written
as:
\begin{equation}
 \left[ \Delta_{d+1} + j( \Delta_d ) \right] \Psi(x) = 0,
\end{equation}
with $ \Delta_{d+1} = g^{ab} \nabla_a \nabla_b$ and with the aether
field $V^a$ hiding in the definition of the spatial Laplacian
$\Delta_d = [g^{ab}+V^a V^b] \nabla_a \nabla_b $. This procedure
allows us to take a quantity that is manifestly not Lorentz invariant,
the dispersion relation $\omega^2 = f(k^2)$, and nevertheless
``covariantise'' it via the introduction of new structure --- a
locally specified preferred frame defined by the (possibly position-
and time- dependent) aether 4-velocity $V^a$.

Of course, in standard analogue models such an aether field does not
come with its own dynamics: It is a background structure which breaks
the physically-relevant content of what is usually called
diffeomorphism invariance (see next
Sect.~\ref{S:internal}). However, in a gravitation theory context
one might still want to require background independence taking it as a
fundamental property of any gravity theory, even a Lorentz breaking
one. In this case one has to provide the aether field with a suitable
dynamics; we can then rephrase much of the analogue gravity discussion
in the presence of nontrivial dispersion relations in terms of a
variant of the Einstein-aether models \citep{Jacobson:2004ts,
  Eling:2004dk, Foster:2005dk, Jacobson:2008aj}.

\subsection{Diffeomorphism invariance} 
\label{S:internal}

When looking at the analogue metrics one problem immediately comes to
mind. The laboratory in which the condensed-matter system is set up
provides a privileged coordinate system. Thus, one is not really
reproducing a geometrical configuration but only a specific metrical
representation of it. This naturally raises the question of whether or
not some appropriate notion of diffeomorphism invariance is lost in the analogue spacetime
construction. Indeed, if all the degrees of freedom contained in the analogue
metric were directly observable,
as opposed to what happens in a general
relativistic context in which only the geometrical degrees of freedom
(metric modulo diffeomorphism gauge) are 
observable,
then diffeomorphism
invariance would be violated.
Here we are thinking of ``active'' diffeomorphisms, not ``passive''
diffeomorphisms (coordinate changes). As is well known, \emph{any}
theory can be made invariant under passive diffeomorphisms (coordinate
changes) by adding a sufficient number of
external/background/non-dynamical fields (prior structure). See, for
instance, \citet{doi:10.1142/S0217732302006801} or \citet{Giulini:2006yg}. 
{In this context,} invariance under active
diffeomorphisms is equivalent to the assertion that there is no
``prior geometry'' (or that the prior geometry is undetectable). Many
readers may prefer to re-phrase the current discussion in terms of the
undetectability of prior structure.
  
The answer to this question is that active {diffeomorphism} invariance
is {partially} maintained but only for (low-energy) internal observers, i.e.,
those observers who can only perform (low-energy) experiments
involving the propagation of the relativistic collective fields. See e.g.~\citet{Girelli:2008qp} for an example of this. 
(This situation is sometimes referred to as ``active kinemetical diffeomorphism invariance''.)
By
revisiting classic Lorentz--FitzGerald ideas on length contraction,
and analyzing the Michelson--Morley experiment in this context, it has
been explicitly shown in \citet{Barcelo:2007iu} that (low-energy)
Lorentz invariance is not broken, i.e., that an internal observer
cannot detect his absolute state of motion. (For earlier suggestions
along these lines, see, for example, \citealt{Liberati:2001sd}
and \citealt{Volovik:2003fe}.)

The argument is the same for curved spacetimes; the internal observer
would have no way to detect the ``absolute'' or fixed background. So
the apparent background dependence provided by the (non-relativistic)
condensed-matter system will not violate active {kinematic-diffeomorphism}
invariance, at least not for these internal inhabitants. These
internal observers will then have no way to collect any metric
information beyond what is coded into the intrinsic geometry (i.e.,
they only get metric information up to a gauge or diffeomorphism
equivalence factor). Internal observers would be able to write down
{kinematic-diffeomorphism invariant} Lagrangians for relativistic matter fields in a curved geometry. However, the dynamics of this geometry is a different issue {(i.e. whether it implements a dynamic-diffeomorphism invariance, {``full'' active diffeomorphisim invariance})}. It is a well-known issue that the expected
relativistic dynamics, i.e., the Einstein equations, have to date not
been reproduced in any known condensed-matter system.

\subsection{Effective spin-two particles} 

Related to the previous point is the possibility of having quantum
systems with no pre-geometric notions whatsoever (i.e., a
condensed-matter--like system) that still exhibit in their low-energy
spectrum effective massless spin-two excitations. This precise
question has been investigated in \citet{Gu:2006vw, Gu:2009jh,
  Xu:2006faa, Xu:2010eg} for abstract quantum systems based on the
underlying notion of qubits. Although not fully conclusive, these
works indicate the possible existence of systems exhibiting purely
helicity $\pm 2$ excitations. One crucial ingredient in these
constructions is the existence of a specific vacuum state with the
characteristics of a string-net condensate.

These authors also show that it is not easy to have just helicity $\pm
2$ excitations --- typically one would also generate helicity $\pm 1$
and massless scalar excitations. This is what would happen, for
instance, in the emergent gravity scenario inspired in the
phenomenology of $^{3}$He, which will be described below.

\subsection{Weinberg--Witten theorem} 

The Weinberg--Witten theorem \citep{Weinberg:1980kq} has often been
interpreted as an insurmountable obstacle for obtaining massless
spin-two excitations as effective degrees of freedom emerging from any
reasonable underlying quantum field theory.
However, the status of the Weinberg--Witten
theorem insofar as it applies to analogue
models is rather subtle. First, note that whenever one's main concern
is in developing an analogue spacetime at the purely kinematic level
of an effective metric, then the Weinberg--Witten theorem has nothing
to say. (This includes, for instance, all analogue experiments probing
the Hawking effect or cosmological particle production; these are
purely kinematic experiments that do not probe the dynamics of the
effective spacetime.) When one turns to the dynamics of the effective
spacetime, desiring, for instance, to investigate quantum fluctuations
of the effective geometry (gravitons), then one should bear in mind
that the Weinberg--Witten theorem is derived under specific technical
assumptions (strict Lorentz invariance in flat spacetime) that are not
applicable in the current context. Furthermore, even if the specific
technical assumptions are satisfied, then those authors
state \citep{Weinberg:1980kq}:
\begin{quote}
Of course, there are acceptable theories that have massless charged
particles with spin $j > 1/2$ (such as the massless version of the
original Yang--Mills theory), and also theories that have massless
particles with spin $j > 1$ (such as supersymmetry theories or general
relativity). Our theorem does not apply to these theories because they
do not have Lorentz-covariant conserved currents or energy-momentum
tensors, respectively.
\end{quote}
Furthermore, when it comes to Sakharov-style induced gravity those
authors explicitly state \citep{Weinberg:1980kq}:
\begin{quote}
However, the theorem clearly does not apply to theories in which the
gravitational field is a basic degree of freedom but the Einstein
action is induced by quantum effects.
\end{quote}
That is: The Weinberg--Witten theorem has no direct application to
analogue spacetimes -- at the kinematic level it has nothing to say,
at the dynamic level its applicability is rather limited by the
{extremely} stringent technical assumptions invoked -- specifically exact Lorentz
invariance at all scales -- and the fact that these technical
assumptions are not applicable in the current context. For careful
discussions of the technical assumptions see \citet{Sudarshan:1981cj,
  Kugo:1981fe, Flato:1982xb, Loebbert:2008zz}. Note particularly the
comment by \citet{Kugo:1981fe}
\begin{quote}
\ldots\,the powerful second part of the theorem becomes empty in the
presence of gravity\,\dots
\end{quote}
Finally, we mention that, though motivated by quite different concerns,
the review article by \citet{Bekaert:2010hw} gives a good overview of the
Weinberg--Witten theorem, and the ways in which it may be evaded. 

\subsection{Marolf theorem} 
\label{S:Marolf-theorem}

In what can be seen as an effort to further refine Weinberg--Witten theorem, 
\cite{Marolf:2014yga} has argued that analogue-like {systems, apt} for the emergence of an effective geometrodynamical regime similar to general relativity, should {either} be ``kinematically non-local", or {otherwise} would need to pass through an abrupt change in their notion of locality which is difficult to understand.
More precisely, Marolf presents a theorem that states:

``Consider any limit where the effective description of a local theory is a gravitational theory with universal coupling to energy, the same notion of time evolution, and a compatible definition of locality. In this limit, all local observables away from the boundary become independent of time."

So, first of all the theorem concerns gravitational theories with universal coupling to energy. These are theories wherein the Hamiltonian can, {at each instant of time},  be expressed as the integral over the boundary of space  of {some} local function, termed {the} gravitational flux. This gravitational flux must be a gauge-invariant function of the gravitational field and its derivatives, encompassing all diffeomorphism-invariant theories. Notably, non-Lorentz-invariant theories such as Hořava-Lifshitz gravity (see below) do not strictly fall into this category, although Marolf hints at the potential for a straightforward extension of the theorem to encompass such theories.

With regard to locality, the theorem adopts a notion akin to microcausality, labelled as kinematic locality. According to this delineation, a kinematically local theory is one where every pair of local bosonic gauge-invariant Heisenberg operators, assessed at spacelike points, commute.

Basically, the theorem works by {demonstrating} that if a fundamental (non-gravitat\-ional) kinematically local theory were to transition to an effective description in terms of a gravitational theory with universal coupling to energy, {then} all of its local bulk observables would cease to evolve unless there were alterations to the concept of time evolution or locality between the gravitational and non-gravitational depictions.
This is essentially an extension/modification of the ideas underpinning the ``problem of time'' more typically encountered in canonical general relativity (either classical or quantum).

Given these conclusions \cite{Marolf:2014yga} extrapolates that deriving an emergent gravitational description from microscopic kinematically local theories without abrupt shifts in locality appears arduous. Such emergence would entail peculiar behavior in the non-gravitational depiction. Consequently, it is advocated for the exploration of emergent theories stemming from microscopic frameworks already inherently kinematically non-local, as exemplified, for instance, by gauge-gravity duality.

However, this conclusion rests on the {(rather restrictive)} additional premise that all observables in the UV completion of the theory must stagnate at low energies or, equivalently, that all such observables must be gauge invariant—a presumption that may not hold in an emergent framework. Indeed, if at least one observable within the UV completion lacks gauge invariance, the notion of locality could be preserved, a scenario typically encountered if gauge symmetries {are also emergent,} {as} for example, {in the scenario explored} in~\cite{Barcelo:2021ryh}.

In conclusion, although {the} Marolf and Weinberg--Witten theorems pinpoint some of the particularities of gravity with respect to other gauge theories, they do not constitute a {clearly} insurmountable impediment for the emergent gravity programme if{, for instance,} gauge symmetries (diffeomorphisms) are as well emergent.

\bigskip
\subsection{Emergent gravity} 

One of the more fascinating approaches to ``quantum gravity'' is the
suggestion, typically attributed to
Sakharov \citep{Sakharov:1967pk, Visser:2002ew}, that gravity itself may
not be ``fundamental physics''. 
Sakharov had in mind a specific model in which gravity could be viewed
as an ``elasticity'' of the spacetime medium, and where gravity was
``induced'' via one-loop physics in the matter
sector \citep{Sakharov:1967pk, Visser:2002ew}. In this way, Sakharov had
hoped to relate the observed value of Newton's constant (and the
cosmological constant) to the spectrum of particle masses.
More recent contributions related to the notion of emergence include \citet{Volovik:2020bpn, Volovik:2021myq, Volovik:2022wgq, Volovik:2023faj}

Indeed it is now a relatively common
opinion, maybe not mainstream but definitely a strong minority opinion,
that gravity (and in particular the whole notion of spacetime and
spacetime geometry) might be no more ``fundamental'' than is fluid
dynamics. The word ``fundamental'' is here used in a rather technical
sense -- fluid mechanics is not fundamental because there is a known
underlying microphysics, that of molecular dynamics, of which fluid
mechanics is only the low-energy low-momentum limit. Indeed, the very
concepts of density and velocity field, which are so central to the
Euler and continuity equations, make no sense at the microphysical level
and emerge only as one averages over time-scales and distance-scales
larger than the mean free time and mean free path.

More generally, the phrase ``emergent gravity'' is now used to describe
the whole class of theories in which the spacetime metric arises as a
low-energy approximation, and in which the microphysical degrees of
freedom might be radically different. Analogue models, and in particular
analogue models based on fluid mechanics or the fluid dynamic
approximation to BECs, are specific examples of ``emergent physics'' in
which the microphysics is well understood. As such, they are useful for
providing \emph{hints} as to how such a procedure might work in a more
fundamental theory of quantum gravity.

In the same way, it is plausible (even though no specific and
compelling model of the relevant microphysics has yet emerged) that
the spacetime manifold and spacetime metric might arise only once one
averages over suitable microphysical degrees of freedom.

\subsection{One specific route to the Einstein equations?}

In fact, there is a specific route to reproduce Einstein equations
within a Fermi-liquid-like system advocated by
\citet{Volovik:2003fe}. In a Fermi liquid like $^{3}$He-A
phase, there exist two important energy scales. One is the energy
scale $E_B$ at which bosonization in the system start to develop. This
scale marks the onset of the superfluid behaviour of helium three. At
energies below $E_B$, the different bosons appearing in the systems
condense so that they can exhibit collective behaviour. The other
energy scale, $E_L$, is the Lorentz scale at which the quasiparticles
of the system start to behave relativistically (as Weyl spinors). As
we mention in Sect.~\ref{S:helium} discussing the ``heliocentric
universe'', this occurs in the $^{3}$He-A phase because the vacuum
has Fermi points. It is in the immediate surroundings of these Fermi
points where the relativistic behaviour shows up. There are additional
relevant scales in these systems, but in this section we are going to
talk exclusively of these two.

When one is below both energy scales, one can describe the system as a
set of Weyl spinors coupled to background electromagnetic and
gravitational fields. For a particular Fermi point, the
electromagnetic and gravitational fields encode, respectively, its
position and its ``light-cone'' structure through space and time. Both
electromagnetic and gravitational fields are built from bosonic
degrees of freedom, which have condensed. Apart from any predetermined
dynamics, these bosonic fields will acquire additional dynamical
properties through the Sakharov-induced gravity mechanism. Integrating
out the effect of quantum fluctuations in the Fermionic fields
\textit{\`a la} Sakharov, one obtains a one-loop effective action for
the geometric field, to be added to the tree-level contribution (if
any). This integration cannot be extended beyond $E_B$, as at that
energy scale the geometrical picture based on the bosonic condensate
disappears. Thus, $E_B$ will be the cut-off of the integration.

Now, in order that the geometrical degrees of freedom follows an
Einstein dynamics, we need three conditions (which we shall see
immediately are really just two):
\begin{enumerate}

\item $E_L > E_B$: For the induction mechanics to give rise to an
  Einstein--Hilbert term, $\sqrt{-g}R$, in the effective Lagrangian we
  have to be sure that the fluctuating Fermionic field ``feels'' the
  geometry (fulfilling a locally-Lorentz-invariant equation) at all
  scales up to the cut-off. The term $\sqrt{-g}R$ will appear
  multiplied by a constant proportional to $E_B^2$. That is why from now on
  we can called $E_B$ alternatively the Planck energy scale $E_P
  \equiv E_B$.

\item \emph{Special relativity dominance or $E_L \gg E_P$}: The
  $E_P^2$ dependence of the gravitational coupling constant tells us
  that the fluctuations that are more relevant in producing the
  Einstein--Hilbert term are those with energies close to the cut-off,
  that is, around the Planck scale. Therefore, to assure the induction
  of an Einstein--Hilbert term one needs the Fermionic fluctuation
  with energies close to the Planck scale to be perfectly Lorentzian
  to a high degree. This can only be assured if $E_L \gg E_B$.

\item \emph{Sakharov one-loop dominance}: Finally, one also needs
  the induced dynamical term to dominate over the pre-existing tree-level contribution (if any). 

\end{enumerate}
Unfortunately, what we have called special-relativity dominance is not
implemented in helium three, nor in any known condensed-matter
system. In helium three the opposite happens: $E_B \gg
E_L$. Therefore, the dynamics of the gravitational degrees of freedom
is non-relativistic but of fluid-mechanical type. That is, the
dynamics of the gravitational degrees of freedom is not Einstein, but
of fluid-mechanical type. 

An alternative route for the emergence of a general relativistic dynamics (also inspired by Volovik's ideas) is being explored in a series of papers~\citep{Barcelo:2016xhp,Barcelo:2021idt,Barcelo:2021ryh}. The crucial point here is not the inversion of the Lorentz breaking and Planck scales, and their relation to one-loop dominance, but the natural emergence of gauge symmetries once Lorentz invariance is established.

The possible emergence of gravitational
dynamics in the context of a condensed-matter system has also been
investigated for BECs \citep{Girelli:2008gc, Sindoni:2009fc}. It has
been shown that, for the simple gravitational dynamics in these
systems, one obtains a modified Poisson equation, and so it is completely
non-relativistic, giving place to a short range interaction (on the
order of the healing length). However, starting from abstract systems
of PDFs with \textit{a priori} no geometrical information, the emergence
of Nordstr\"om spin-0 gravity has been shown to be
possible \citep{PhysRevD.79.044019}; this is relativistic though not
Einstein. 

More recently, the very same reasoning used in~\citep{Girelli:2008gc} has been extended to relativistic BEC~\citep{Fagnocchi:2010sn}, showing that  the background dynamics can in this case be rewritten, in a suitable regime, as Nordstr\"om spin-0 gravity with a cosmological constant~\citep{Belenchia:2014hga}.

In counterpoint, in \Horava gravity the graviton appears to be
fundamental, and need not be emergent \citep{Horava:2009uw,
  Horava:2009if, Horava:2010zj}. Additionally, the Lorentz breaking
scale and the Planck scale are in this class of models distinct and
unconnected, with the possibility of driving the Lorentz breaking
scale arbitrarily high \citep{Sotiriou:2009bx, Sotiriou:2009gy,
  Visser:2009ys, Weinfurtner:2010hz}. In this sense, the \Horava models
are a useful antidote to the usual feeling that Lorentz violation is
typically Planck-scale.

\subsection{The cosmological constant problem}

The condensed matter analogies offer us an important lesson
concerning the cosmological constant
problem \citep{Volovik:2003fe}. Sakharov's induced gravity not only can
give rise to an Einstein--Hilbert term under certain conditions, it
also gives rise to a cosmological term. This contribution would
depend on the cut-off, (naively as $E_P^4$), so if there were not additional
contributions counterbalancing this term, emergent gravity in
condensed-matter-like systems will always give place to an enormous
cosmological constant inducing a strongly-repulsive force between
quasiparticles.

However, we know that at low temperatures, depending on the
microscopic characteristics of the system we can have quite different
situations. Remarkably, liquid systems (as opposed to gases) can
remain stable on their own, without requiring any external
pressure. Their total internal pressure at equilibrium is (modulo
finite-size effects) always zero. This implies that, at zero
temperature, if gravity emerges from a liquid-like system, the total
vacuum energy $\Lambda \propto \rho_V=-p_V$ will be automatically
forced to be (relatively) small, and not a large number. The $E_P^4$
contribution coming from quasiparticle fluctuations will be exactly
balanced by contributions from the microphysics or ``trans-Planckian''
contributions.

If the temperature is not zero there will be a pressure $p_M$
associated with the thermal distribution of quasiparticles, which
constitute the matter field of the system. Then, at equilibrium one
will have $p_M+p_V=0$, so that there will be a small vacuum energy
$\Lambda \propto p_M$. This value is not expected to match exactly the
preferred value of $\Lambda$ obtained in the standard cosmological
model ($\Lambda$CDM) as we are certain to be out of thermodynamic
equilibrium. However, it is remarkable that it matches its order of
magnitude, (at least for the current epoch), albeit any dynamical
model implementing this idea will probably have to do so only at late
times to avoid possible tension with the observational data. Guided by
these lessons, there are already a number of heuristic investigations
about how a cosmological term could dynamically adapt to the evolution
of the matter content, and which implications it could have for the
evolution of the universe \citep{Barcelo:2006cs, Klinkhamer:2007pe,
  Klinkhamer:2008ns, Klinkhamer:2009gm}.

As final cautionary remark let us add that consideration of an
explicit toy model for emergent gravity \citep{Girelli:2008gc} shows
that the quantity that actually gravitates cannot be so easily
predicted without an explicit derivation of the analogue gravitational
equations. In particular in \citet{new-article} it was shown that the
relevant quantity entering the analogue of the cosmological constant
is a contribution coming only from the excitations above the
condensate, while in \citet{Belenchia:2014hga} the cosmological constant is determined (classically) by the strength of the atomic coupling constant.

\subsection{The information-loss problem}

The information loss problem \citep{PhysRevD.14.2460} has been, even more than the \blue{trans-Planckian} one, a major line of investigation in theoretical physics, for almost fifty years by now.

In black hole physics, Hawking pair production can be described by the action of a squeezing operator on the initial vacuum state, and as such it is associated to a unitary preserving evolution. This implies that the initial (pure) state can be always reconstructed as long as one can (at least in principle) measure both members of each pair.
Of course, this is operationally impossible when they are on opposite sides 
of a horizon, {but} it is nonetheless conceptually possible to see that the unitary evolution is preserved. However, {standard general relativistic} black holes are endowed with singularities and eventually evaporate. Indeed, in the standard scenario, after the complete evaporation of the black hole, one is left with flat spacetime and a thermal bath (as the complete evaporation forces {one} to trace out all the Hawking partners which fell into the hole). This implies that, after a sufficient time, not even ideally {could} one recover the correlations carried by the infalling Hawking partners, so that unitary of the quantum evolution appears  to be fundamentally lost. This is in loose words the ``information loss problem".

While analogue black hole geometrodynamics is clearly not ruled by the Einstein equations, it is nonetheless possible to devise configurations for which Hawking radiation would drive a black hole evaporation~\citep{Liberati:2020mdr} or in which a flow can simulate an interior singularity~\citep{Barcelo:2004wz}. In this case one could be lead to the erroneous conclusion that an information loss issue is present also in this system.

This is clearly impossible, as e.g.~an acoustic black hole in a BECs exists only in {the} hydrodynamic limit of {a} perfectly well defined and unitary evolving quantum gas of atoms.
In this sense it is then instructive to understand how the apparent information loss, perceived say by a ``phononic observer", is resolved by a full knowledge of the substratum from which spacetime and matter emerge in this system.

This was investigated in~\citep{Liberati:2019fse} by adopting a number preserving description of the BEC, beyond the Bogolyubov {description}, in which one does not need to introduce a mean field approximation for treating the condensate background. This is the so called natural orbital formalism~\citep{Penrose:1956zz,BECBOOK}.

Unfortunately, a full calculation has been so far carried out only in a cosmological setting, nonetheless its lesson appears sufficient to draw lessons applicable to the black hole case. The resulting answer is that in {a BEC there} are always subtle correlations between the atoms in the condensate (the emergent spacetime constituents) and Hawking quanta, which become increasingly non-negligible as the evaporation process goes on~\citep{Liberati:2019fse}.

In the end, the evaporation process is fully unitary on the whole Hilbert space (spacetime atoms plus Hawking quanta) but becomes non-unitary once one decide to replace the spacetime quantum degrees of freedom with their mean field (i.e.~to move to a description {analogous} to that of semiclassical gravity) as in this case the above mentioned correlatorions are completely lost.

In conclusion, analogue gravity seems to suggest that in an emergent gravity scenario, the information loss {could be} a mere artifact of the classical treatment of gravity which ends up preventing addressing the unitary evolution given that in the case of black holes the latter is realised on the full Hilbert space (gravity plus matter) rather than solely on the matter one.

In any case, let us finish this note on the information loss problem recalling that it could also be just a hypothetical (not real) problem, associated with the standard way in which gravitational collapse is understood within general relativity. Quantum effects regularizing the singularity combined with dispersive and semiclassical gravity effects might result in a different scenario with no information loss problem to start with (see e.g. the discussions on~\cite{Barcelo:2015noa,Barcelo:2018ynq,Arrechea:2023hmo}).

\subsection{Quantum gravity -- phenomenology}

Over the last few years a widespread consensus has emerged that
observational tests of quantum gravity are for the foreseeable future
likely to be limited to precision tests of dispersion relations and
their possible deviations from Lorentz
invariance \citep{Mattingly:2005re, Jacobson:2005bg}. The key point is
that at low energies (well below the Planck energy) one expects the
locally-Minkowskian structure of the spacetime manifold to guarantee
that one sees only special relativistic effects; general relativistic
effects are negligible at short distances. However, as ultra-high
energies are approached (although still below Planck-scale energies)
several quantum-gravity models seem to predict that the locally
Euclidean geometry of the spacetime manifold will break down. There
are several scenarios for the origin of this breakdown ranging from
string theory \citep{Kostelecky:1988zi, Ellis:1999sf} to brane
worlds \citep{Burgess:2002tb} and loop quantum
gravity \citep{Gambini:1998it}. Common to all such scenarios is that
the microscopic structure of spacetime is likely to show up in the
form of a violation of Lorentz invariance leading to modified
dispersion relations for elementary particles. Such dispersion
relations are characterised by extra terms (with respect to the
standard relativistic form), which are generally expected to be
suppressed by powers of the Planck energy. Remarkably, the last years
have seen a large wealth of work in testing the effects of such
dispersion relations and in particular strong constraints have been
cast by making use of high energy astrophysics observations (see, for
example, \citealt{Amelino-Camelia:1997gz, Coleman:1998ti,
  Jacobson:2002hd, Jacobson:2002ye, Jacobson:2004rj, Jacobson:2004qt,
  Jacobson:2005bg, Mattingly:2005re, Smolin:1995ai, Liberati:2009pf}
and references therein).

Several of the analogue models are known to exhibit similar behaviour,
with a low-momentum effective Lorentz invariance eventually breaking
down at high momentum once the microphysics is explored.%
\footnote{However, it is
important to keep in mind that not all the above-cited quantum
  gravity models violate the Lorentz symmetry in the same manner. The
  discreteness of spacetime at short scales is not the only way of
  breaking Lorentz invariance.} 
Thus, some of the analogue models provide
controlled theoretical laboratories in which at least some forms of
subtle high-momentum breakdown of Lorentz invariance can be explored. As
such, the analogue models provide us with \emph{hints} as to what sort of
modified dispersion relation might be natural to expect given some
general characteristics of the microscopic physics. Hopefully, an
investigation of appropriate analogue models might be able to illuminate
possible mechanisms leading to this kind of quantum gravity
phenomenology, and so might be able to provide us with new ideas about other
effects of physical quantum gravity that might be observable at
sub-Planckian energies.

\subsection{Quantum gravity -- fundamental models}

When it comes to dealing with ``fundamental'' theories of quantum
gravity, the analogue models play an interesting role which is
complementary to the more standard approaches. The search for a quantum
theory of gravity is fundamentally a search for an appropriate
mathematical structure in which to simultaneously phrase both quantum
questions and gravitational questions. More precisely, one is searching
for a mathematical framework in which to develop an abstract quantum
theory which then itself encompasses classical Einstein gravity (the
general relativity), and reduces to it in an appropriate
limit \citep{Carlip:2001wq, Smolin:2003rk, Friedan:2002aa}.

The three main approaches to quantum gravity still dominating the theoretical landscape,
``string models'' (also known as ``M-models''), ``loop space'' (and the
related ``spin foams''), and ``lattice models'' (Euclidean or
Lorentzian) all share one feature: They attempt to develop a
``pre-geometry'' as a replacement for classical differential geometry
(which is the natural and very successful mathematical language used to
describe Einstein gravity) \citep{Carlip:2001wq, Smolin:2003rk,
Friedan:2002aa, Rovelli, Rovelli:2003cu, Bousso:2004fc,
Piazza:2005wm}. The basic idea is that the mooted replacement for
differential geometry would be relevant at extremely small distances
(where the quantum aspects of the theory would be expected to
dominate), while at larger distances (where the classical aspects
dominate) one would hope to recover both ordinary differential
geometry and specifically Einstein gravity or possibly some
generalization of it. The ``string'', ``loop'', and ``lattice''
approaches to quantum gravity differ in detail in that they emphasise
different features of the long-distance model, and so obtain rather
different short-distance replacements for classical differential
geometry. Because the relevant mathematics is extremely difficult, and
by and large not particularly well understood, it is far from clear
which, if any, of these three standard approaches will be preferable
in the long run \citep{Smolin:2003rk}.

A more recent (Jan.~2009) development is the appearance of \Horava
gravity \citep{Horava:2009uw, Horava:2009if, Horava:2010zj}. This
model is partially motivated by condensed matter notions such as
(deeply non-perturbative) anomalous scaling and the existence of a
``Lifshitz point'', and additionally shares with most of the analogue
spacetimes the presence of modified dispersion relations and
high-energy deviations from Lorentz invariance \citep{Horava:2009uw,
  Horava:2009if, Horava:2010zj, Visser:2009fg, Visser:2009ys,
  Sotiriou:2009bx, Sotiriou:2009gy, Weinfurtner:2010hz,Blas:2018flu,Herrero-Valea:2023zex}.
  Projectable \Horava gravity was shown to be renormalizable~\citep{Barvinsky:2015kil}, see also the recent proof of renormalizebility for the non-projectable case in~\citep{Bellorin:2024qyy}. This shows the power of dispersion in solving deep conceptual problems in gravitational theories.

Though \Horava gravity is not directly an analogue model \emph{per se}, there
are deep connections -- with some steps toward an explicit connection
being presented in \citet{Xu:2010eg}. Furthermore, the analysis of 
Hawking radiation in \Horava gravity share many elements with the one of acoustic horizons~\citet{DelPorro:2023lbv} albeit black holes in this theory present characteristic structures, the so-called universal horizons, which are not present in standard acoustic geometries~\cite{Cropp:2013sea} albeit they can in principle be as well simulated in sufficiently sophisticated analogue gravity systems~\cite{Cropp:2016teb}.

We feel it likely that analogue models can {continue shedding} new light on this very confusing field by providing a concrete specific situation in which the transition from the short-distance ``discrete'' or ``quantum'' theory to
the long-distance ``continuum'' theory is both well understood and
non-controversial {(see for an example~\citet{Barcelo:2021nhs})}. 
Here we are specifically referring to fluid
mechanics, where, at short distances, the system must be treated using
discrete atoms or molecules as the basic building blocks, while, at large
distances, there is a well-defined continuum limit that leads to the
Euler and continuity equations. Furthermore, once one is in the continuum
limit, there is a well-defined manner in which a notion of ``Lorentzian
differential geometry'', and in particular a ``Lorentzian effective
spacetime metric'' can be assigned to any particular fluid
flow \citep{Unruh:1981cg, Visser:1997ux, Novello:2002qg}. Indeed, the
``analogue gravity programme'' is extremely successful in this regard,
providing a specific and explicit example of a ``discrete'' $\to$
``continuum'' $\to$ ``differential geometry'' chain of development. What
the ``analogue gravity programme'' does not seem to do as easily is to
provide a natural direct route to the Einstein equations of general
relativity, but that merely indicates that current analogies have their limits
and therefore, one should not take them too
literally \citep{Visser:1997ux,Novello:2002qg}. 
Fluid mechanics is a guide to the mathematical possibilities, not an end in
itself. The 
parts of the analogy that do work well are precisely the steps where the
standard approaches to quantum gravity have the most difficulty, and so
it would seem useful to develop an abstract mathematical theory of the
``discrete'' $\to$ ``continuum'' $\to$ ``differential geometry'' chain
using this fluid mechanical analogy (and related analogies) as
inspiration.
For examples of abstract constructions the reader can check~\citet{Konopka:2006hu,Konopka:2008hp, Wilkinson:2015fja,Oriti:2016acw,Oriti:2018qty,Oriti:2023yjj}.

\subsection{Going further}
\label{SS:going-further-7}

We think that the potential of the analogue gravity framework to bring new ideas into the gravitational realm is far from being exhausted. We hope to see in the future more specific models of gravitational phenomena incorporating some of the learned lessons.

\clearpage
\section{Conclusions}
\label{S:conclusions}

Almost 25 years have now passed since the first version of this Living Review appeared, and these authors feel it important to remark how prolific and energetic this field of research has become over this lapse of time. Developments have amounted to a virtuoso performance by the research community.

Indeed, by looking back at the previous version of this work, it is clear not only that the field has been growing in both scope and reach, but has also vastly changed in the composition of its community, attracting a wide range of practitioners with competencies well beyond those of the original researchers. What used to be mainly a community of researchers with a gravitation theory background, has now become a multi-faceted community in which experimentalists and theoreticians from both the GR and condensed matter communities meet.

We feel this is now the defining trait of Analogue Gravity, as it represents one of the few fields of modern physics where the general trend towards specialization and narrowing of communities has been successfully reverted. Analogue gravity has indeed grown from a largely theoretical endeavour into a field where table-top experiments are nowadays just as relevant, if not dominant, in settling the agenda and in driving very fruitful interdisciplinary collaborations. 

This synergy has proven extremely successful: We nowadays have a much better understanding of the nature of Hawking radiation, and how and why it can be robust against the UV structure of spacetime. Furthermore, we have been able to reproduce it experimentally, together with several other phenomena (from cosmological particle creation to superradiance) which are pillars of modern gravitation theory. At the same time analogue gravity has been able to stimulate new work and investigations in condensed matter theory by providing a new framework within which methods and intuition provided from gravitational phenomenal can be used. It is a pity that so many insights have sometimes been neglected by researchers in theoretical physics working on the same issues, merely on the basis that analogue gravity is not ``real" gravity.

Indeed, the fact that analogue models are not perfect mimics of the standard phenomena (e.g. because of the breakdown of Lorentz invariance in the UV) has proven to be a strength, rather than a weakness: These new features allowed one to break degeneracies, and exposed subtleties of these phenomena, so greatly enhancing our understanding of them.

Also, analogue gravity has provided many toy models for the possible interplay between large scale, classical and continuum structure of spacetime and the microphysics from which it might emerge. All useful lessons that are there to be used by the quantum gravity community (as they have been in rare cases).

In conclusion, following the research trajectory we hope we have properly conveyed by this work, these authors expect that interest in analogue models will continue unabated in the years to come, and
suspect that several key, unexpected, issues will come up in the future. Analogue gravity can by now be viewed as a mature field, but being supported by a large, interdisciplinary and vibrant community,
we expect that advances will continue to be very significant in the years to come.

\subsection{Going further}
\label{SS:going-further-8}

Though every practicing scientist already knows this, for the sake
of any student reading this we mention the following resources:

\begin{itemize}
 \item \url{https://inspirehep.net/} -- the INSPIRE bibliographic
   database for keeping track of (almost all of) the general relativity
   and particle physics aspects of the relevant literature.
 
 \item \url{https://ui.adsabs.harvard.edu/} -- the SAO/NASA Astrophysics
   Data System (ADS) for keeping track of (almost all of) the astrophysical
   aspects of the relevant literature.
   
 \item \url{https://arxiv.org/} -- the electronic-preprint (e-print) 
  database for accessing the text of (almost all, post 1992) relevant
  articles.

 \item \url{https://philsci-archive.pitt.edu} --- the Philosophy of Science e-print archive.

 \item \url{https://www.springer.com/journal/41114} -- the \textit{Living
  Reviews in Relativity} journal.

\end{itemize}
Those five access points should allow you to keep abreast of what is
going on in the field.


\clearpage
\begin{acknowledgements}

The work of Matt Visser was supported by the Marsden fund administered
by the Royal Society of New Zealand. MV also wishes to thank both
SISSA (Trieste) and the IAA--CSIC (Granada) for hospitality during
various stages of this work. 
Carlos Barcel\'o wishes to thank Victoria University of Wellington for hospitality during the realization of (part of) this update. 
CB has received founding from ``Ministerio de Universidades para estancias de movilidad de profesores e investigadores en centros extranjeros de ense\~anza superior e investigaci\'on".
CB has been also supported by the Spanish Government through projects PID2020-118159GB-C43 (with FEDER contribution), and by the Junta de Andaluc\'{\i}a through project FQM219. 
CB also acknowledges financial support from the grant CEX2021-001131-S funded by MCIN/AEI/ 10.13039/501100011033.

The authors also wish to specifically thank Enrique Arilla for
providing Figs.~\ref{F:cascade} and \ref{F:ondas}, and Silke
Weinfurtner for providing Fig.~\ref{F:horizon} of the artwork.
Additionally, the authors wish to thank Renaud Parentani for helpful
comments, specifically with respect to the question of which notion of
surface gravity is the most important for Hawking radiation.
Finally, the authors also wish to thank Germain Rousseaux for bringing
several historically-important references to our attention.

\end{acknowledgements}


\clearpage
\phantomsection
\addcontentsline{toc}{section}{References}
\bibliographystyle{spbasic-FS}
\bibliography{references-LRR-master}
\end{document}